%% file: main.tex
\documentclass[twocolumn,onecolappendix]{aastex63}
\usepackage{amsmath}

\shortauthors{Sheehan et al.}
\shorttitle{VANDAM: Orion - Radiative Transfer Modeling}

\begin{document}

\title{\bf The VLA/ALMA Nascent Disk and Multiplicity (VANDAM) Survey of Orion Protostars VI. Insights from Radiative Transfer Modeling}

\author[0000-0002-9209-8708]{Patrick D. Sheehan}
\altaffiliation{NSF Astronomy \& Astrophysics Postdoctoral Fellow}
\affiliation{Center for Interdisciplinary Exploration and Research in Astronomy, Northwestern University, 1800 Sherman Rd., Evanston, IL 60202, USA}
\affiliation{National Radio Astronomy Observatory, 520 Edgemont Rd., Charlottesville, VA 22901, USA}

\author[0000-0002-6195-0152]{John J. Tobin}
\affiliation{National Radio Astronomy Observatory, 520 Edgemont Rd., Charlottesville, VA 22901, USA}

\author[0000-0002-4540-6587]{Leslie W. Looney}
\affiliation{Department of Astronomy, University of Illinois, Urbana, IL 61801, USA}

\author[0000-0001-7629-3573]{S. Thomas Megeath}
\affiliation{Ritter Astrophsical Research Center, Department of Physics and Astronomy, University of Toledo, Toledo, OH 43606, USA}

\newcommand{\JT}[1]{{\color{blue}  JT; \bf #1}}

\begin{abstract}
We present Markov Chain Monte Carlo radiative transfer modeling of a joint ALMA 345 GHz and spectral energy distribution dataset for a sample of 97 protostellar disks from the VLA and ALMA Nascent Disk and Multiplicity Survey of Orion Protostars. From this modeling, we derive disk and envelope properties for each protostar, allowing us to examine the bulk properties of a population of young protostars. We find that disks are small, with a median dust radius of $ 29.4^{+  4.1}_{-  2.7}$ au and a median dust mass of $  5.8^{+  4.6}_{-  2.7}$ M$_{\oplus}$. We find no statistically significant difference between most properties of Class 0, I, and Flat Spectrum sources with the exception of envelope dust mass and inclination. The distinction between inclination is an indication that the Class 0/I/Flat Spectrum system may be difficult to tie uniquely to the evolutionary state of protostars. When comparing with Class II disk dust masses in Taurus from similar radiative transfer modeling, we further find that the trend of disk dust mass decreasing from Class 0 to Class II disks is no longer present, though it remains unclear whether such a comparison is fair due to differences in star forming region and modeling techniques. Moreover, the disks we model are broadly gravitationally stable. Finally, we compare disk masses and radii with simulations of disk formation and find that magnetohydrodynamical effects may be important for reproducing the observed properties of disks.
\end{abstract}

\keywords{star formation, planet formation, protoplanetary disks}

\section{Introduction}

Stars form from clouds of gas and dust that collapse under the force of their own gravity. Because of the initial angular momentum of the collapsing material, much of the material in the cloud forms into a disk rather than collapsing straight onto the star forming at the center. Material is then accreted through the disk onto the star, which in turn regulates much of the final build up of stellar mass. Moreover, it is in these protostellar disks that planets are expected to form, and so understanding their properties throughout their evolution is crucial for understanding how planets form.

A classification scheme has been developed for young stellar objects, initially based on observational properties such as the near-infrared spectral index \citep[$\alpha$; e.g.][]{Myers1987}, the bolometric temperature \citep[$T_{bol}$; e.g.][]{Chen1995}, or the submillimeter luminosity of the source \citep[e.g.][]{Andre1993}; however this scheme has also been mapped to proposed evolutionary picture of star formation. In this picture, forming stars progress from young sources with ``protostellar" disks, still embedded in substantial envelopes of infalling material  that obscure much of the shorter wavelength light (Class 0), to systems with more mature disks but still embedded in less massive envelopes (Class I or Flat Spectrum), to envelope-free pre-main sequence stars with ``protoplanetary" or ``planet-forming" disks (Class II), to stars with only a small amount of, if any, remnant disk material (Class III) \citep[e.g.][]{Robitaille2006,Crapsi2008}. The ages of sources in the Class 0 and I stages have been estimated at roughly 0.2 Myr and 0.5 Myr, respectively, using counting statistics \citep[e.g][]{Evans2009,Dunham2015}, though estimates of the half-lives suggest potentially shorter typical timescales \citep[e.g.][]{Kristensen2018ProtostellarEstimates}. Class II sources have typically been estimated to have ages of a few $\times1$ Myr by comparing sources with evolutionary track models \citep[e.g.][]{Hillenbrand2000}.

Measurements of the properties of the youngest disks, the protostellar or Class 0/I disks, are keys to informing much of our understanding of the processes that drive star and planet formation. For many years it was unclear whether a disk could even form during the initial collapse of cloud material, as ``magnetic braking" found in ideal magneto-hydrodynamics (MHD) simulations arrested the formation of a disk \citep[e.g.][]{Allen2003,Mellon2008,Li2014a}. In recent years, a number of Keplerian-rotating Class 0/I disks have been identified \citep[e.g.][]{Tobin2012,Murillo2013,Codella2014,Harsono2014,Yen2017}, even more compact, often elongated, continuum disk candidates have been detected around Class 0/I sources \citep[e.g.][]{Segura-Cox2016THECLOUD,Segura-Cox2018,Sheehan2017,Maury2018,Tobin2020}, and simulations have also found that the inclusion of non-ideal MHD effects, turbulence, and/or misalignment between the axis of rotation and the magnetic field can overcome the magnetic braking catastrophe \citep[e.g.][]{Dapp2012,Joos2012,Li2013,Masson2016,Hennebelle2016MAGNETICALLYDISKS,Wurster2019}. So while the question of whether disks can form early in the star formation process may no longer remain, the relative importance of these different pieces of physics remains unclear. The properties of actual protostellar disks are therefore critical for placing constraints on these simulations.

Moreover, isolating disk properties, particularly of young, embedded protostars, is also crucial for understanding the early stages of planet formation. There is a significant amount of evidence that the amount of material available in protoplanetary disks is insufficient to form the masses of planetary systems that we observe \citep[e.g.][]{Najita2014,Ansdell2016,Manara2018,Tychoniec2020}, suggesting that by the protoplanetary disk phase, much of the solid material in the disk may be already locked up in larger bodies that are not detectable with ALMA. Lending support to this idea are the wide array of substructures that have been found in nearly every protoplanetary disk observed with high enough spatial resolution \citep[e.g.][]{Brogan2015,Andrews2016,Isella2016,Andrews2018,Huang2018,Long2018,vanderMarel2019}. Moreover, purported protoplanets, or direct signatures of the presence of protoplanets, have been found in high contrast near-infrared imaging as well as submillimeter continuum and spectral line imaging of a few of these substructured disks \citep[e.g.][]{Sallum2015,Keppler2018,Pinte2018KinematicDisk,Teague2018A163296,Isella2019DetectionProtoplanets,Pinte2020NineGaps}.

Despite their importance, the properties of protostellar disks have long been difficult to study due to the dense envelopes of material in which they are embedded. Though millimeter observations are traditionally thought to be optically thin to dust, moderate resolution millimeter images can still have a significant amount of envelope emission entangled with that of the disks \citep[e.g.][]{Dunham2014} making early millimeter studies of protostars difficult to interpret \citep[e.g.][]{Jrgensen2009PROSAC:Protostars,Andersen2019}. The VLA and ALMA Nacent Disk and Multiplicity (VANDAM) Survey in Perseus \citep{Tobin2016a,Tychoniec2018} and subsequent VANDAM: Orion Survey \citep{Tobin2020} provided the first high resolution imaging surveys (spatial resolution of $\sim30-40$ au) of entire populations of protostellar disks, and put together a comprehensive picture of protostellar disk dust masses and radii, but even at these spatial resolutions envelopes can contribute to the images in difficult to disentangle ways.

One way to combat these issues is to use radiative transfer modeling to account for how the density, temperature, optical depth, viewing angle, and other effects ultimately come together to produce the emergent intensity distribution of young systems. \citet{Sheehan2017} applied rigorous radiative transfer forward modeling to CARMA + SED observations for a sample of 10 Class 0/I protostars in Taurus, and from this modeling found that embedded disks are, on average, more massive than protoplanetary disks by a factor of a few, though with only 10 sources the significance could be improved. Moreover, conclusions about other disk properties (e.g. radius) could not be drawn. Still, that work provided a blueprint for understanding the properties of these young, embedded disks through careful modeling.

In this work, we build upon that blueprint by applying the same radiative transfer modeling framework to protostellar disks in the Orion Molecular Cloud Complex, observed as a part of the VANDAM: Orion \citep{Tobin2020} and HOPS \citep{Manoj2013Herschel/pacsEmission,Stutz2013AProtostars,Furlan2016,Fischer2020TheB} surveys. These surveys collectively observed more than 300 protostellar disks with ALMA, $Spitzer$, and $Herschel$ to provide a rich dataset including high spatial resolution continuum imaging and broadband SEDs. With 97 protostellar disks modeled, this constitutes the largest sample of sources with such careful modeling to date, by over an order of magnitude. As such, it presents an opportunity to begin to put together a picture of the structures of protostellar disks at early times \citep[$\lesssim0.5-1$ Myr; e.g.][]{Evans2009,Dunham2015}.

In Section \ref{section:obs} we discuss our observations, sample selection, and data reduction processes. Then, in Section \ref{section:modeling} we give an overview of our modeling procedure, and present the results of this modeling in Section \ref{section:results}. Finally, we compare the disk properties derived from our modeling to observationally derived disk properties for Class II disks, theoretical simulations of disk formation, along with system properties derived from other observational methods (e.g. bolometric temperature, millimeter flux), and discuss the implications of these in Section \ref{section:discussion}, and wrap up with our conclusions in Section \ref{section:conclusions}.

\section{Observations \& Sample}
\label{section:obs}

The data we use for our modeling is drawn from the VLA and ALMA Nascent Disk and Multiplicity (VANDAM) Survey of Orion Protostars \citep{Tobin2020}, which surveyed 328 protostars in the Orion Molecular Cloud Complex. These observations include ALMA 345 GHz continuum and spectral line observations at 0.1" spatial resolution, along with VLA 33 GHz continuum observations at 0.06", for all protostars surveyed. The data reduction for these observations, including self-calibration when the signal-to-noise ratio was high enough, is described in detail in \citet{Tobin2020}.

Due to the computational cost of running our modeling, we were unable to model all 328 protostars with the available computing time, so we necessarily had to make cuts to the full sample to make the modeling more tractable. From this sample, we excluded $\sim100$ protostellar multiple systems, as modeling of multiple systems is significantly more challenging \citep[e.g.][]{Sheehan2014}. We also excluded $\sim30$ non-detections from the pool of protostars. Of the remaining $\sim200$ single protostars, we randomly selected 90 protostars for our modeling analysis, regardless of their signal-to-noise ratios. By selecting protostars completely at random, we hope to remove any potential biases that may be induced by not using the full sample. We do, however, also try to correct for our exclusion of the non-detections whenever appropriate, as we will describe below.

For our final sample of 97 protostars, including 25 Class 0's, 44 Class I's and 28 Flat Spectrum sources, we collect the ALMA 345 GHz continuum observations for our modeling analysis. We exclude the VLA 33 GHz continuum observations as there was not uniform coverage of all VANDAM: Orion targets \citep{Tobin2020} but also because modeling multiple millimeter observations would stretch our already thin computational resources and because our modeling code is not fully tested in that mode. In addition to our ALMA dataset, we collect archival photometry and spectroscopy to include in our modeling analysis, primarily from the Herschel Orion Protostar Survey \citep[HOPS;][]{Furlan2016}. This includes Spitzer IRAC and MIPS photometry and IRS spectroscopy, 2MASS photometry, Herschel PACS and SPIRE photometry, when available. To account for systematics across different instruments, we assume a 10\% flux calibration uncertainty for all flux measurements in our modeling.

We apply our modeling directly to the ALMA 345 GHz two dimensional visibilities; however we do show images of sources and their best-fit models for easier by-eye comparison. The images we show typically were made using Briggs weighting with a robust parameter of 0.5, to balance resolution and sensitivity, though for some faint sources we used a robust parameter of 2 to improve the sensitivity of the images. To ensure that in our modeling we are deriving reasonable parameter estimates and uncertainties, we checked the uncertainties on the observed visibility data by comparing $\sigma_{tot} = 1 / \sqrt{\sum_i \, (1 / \sigma_i^2)}$ with the root mean square of a naturally weighted image and found that we should reduce the weights by a factor of 0.25 for the two to match. Though this is not a perfect comparison, as we are reducing the weights this should provide more conservative estimates of the uncertainties on the data, and therefore on the best-fit parameters.

\section{Radiative Transfer Modeling}
\label{section:modeling}

To model our sample we use the \texttt{pdspy} code \citep{Sheehan2018}, which follows the framework outlined in \citep{Sheehan2017} to fit full two-dimensional, axisymmetric, radiative transfer models simultaneously to a multi-wavelength dataset using Markov Chain Monte Carlo (MCMC) fitting. Our radiative transfer model takes a parameterized density structure for a protostellar system as an input to radiative transfer codes, which then generate synthetic observations of the system to compare with the observations. We describe the details of the model used below, but also note that extensive documentation of the \texttt{pdspy} code is available online.\footnote{Documentation available at pdspy.readthdocs.io}

Our model includes a protostar, disk, and envelope with an outflow cavity. For simplicity, and because protostellar properties are unknown for the majority, if not entirety, of our sample, we assume that the central protostar has a temperature of 4000 K, reasonable for a generic young, low-mass protostar. We do, however, leave the luminosity of that protostar, $L_*$ as a free parameter in our fit. As we assume a blackbody spectrum for simplicity, $L_*$ is primarily controlled by varying the protostellar radius, and so this may not exactly emulate the true spectrum of a protostar, which may have significant accretion luminosity as well. As most of this emission is reprocessed by the envelope, this should not substantially affect the results of the modeling.

The model also includes a protostellar disk following the perscription for a viscously accreting disk, with the surface density described by
\begin{equation}
    \Sigma = \Sigma_0 \left(\frac{R}{R_c}\right)^{-\gamma} \, \exp\left[-\left(\frac{R}{R_c}\right)^{2-\gamma}\right],
    \label{eq:disk_surface_density}
\end{equation}
\citep{Lynden-Bell1974} where R is the radius within the disk as typically defined in a cylindrical coordinate system. $R_c$ is the radius in the disk at which the surface density is exponentially tapered, and $\gamma$ controls the surface density power-law shape, along with how the disk is tapered. $\Sigma_0$ is the surface density at $R_c$, but we instead parameterize the model in terms of the total disk mass, $M_{disk}$, from which the surface density normalization can be calculated as,
\begin{equation}
    \Sigma_0 = \frac{(2-\gamma) M_{disk}}{2 \pi R_c^2}.
\end{equation}
The vertical structure of the disk is given by the typical structure assumed for a flared accretion disk, with the density defined as,
\begin{equation}
    \rho = \frac{\Sigma(R)}{\sqrt{2 \pi} h(R)} \, \exp\left[-\frac{1}{2}\left(\frac{z}{h(R)}\right)^2\right].
\end{equation}
The function $h(r)$ defines the scale height of the disk as a function of radius, as
\begin{equation}
    h(r) = h_0 \left(\frac{R}{1 \, \mathrm{au}}\right)^{\psi},
\end{equation}
with $h_0$ specifying the scale height of the disk at 1 au, and $\psi$ defining the flaring of the disk. Finally, we also truncate the disk at an inner radius of $R_{in}$, inside of which the density drops to zero.

\begin{figure*}[t]
\centering
\includegraphics[width=7in]{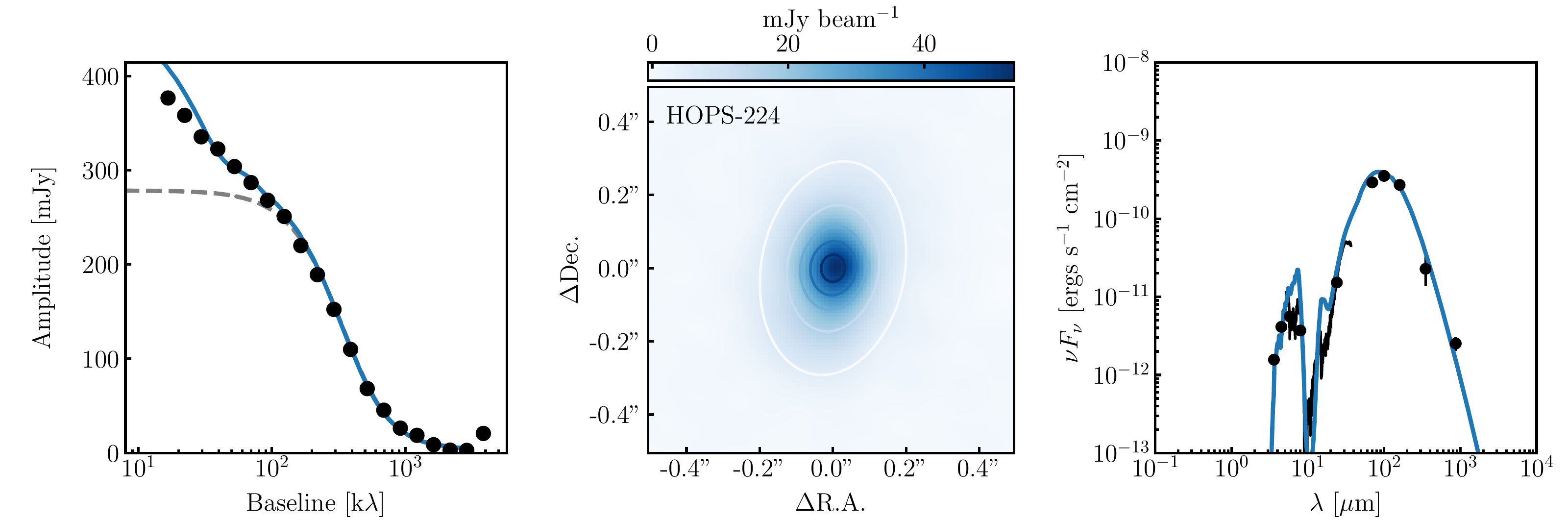}
\includegraphics[width=7in]{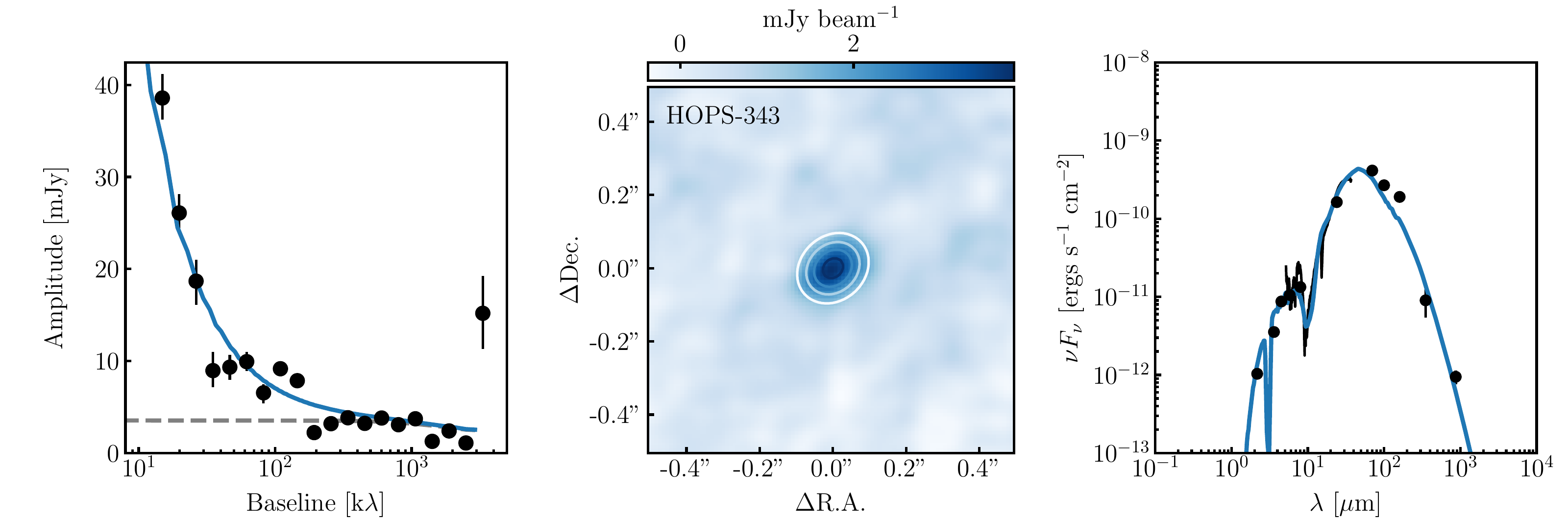}
\caption{Examples of radiative transfer models compared with our observational dataset. In the left column, we show the one-dimensional, azimuthally averaged 345 GHz visibility profiles with the best-fit model shown in blue and the disk contribution to that model as gray dashed lines. For ease of viewing, we show the azimuthally averaged visibilities, but the fit is done to the full, two dimensional dataset. In the center column we show the 345 GHz image with the model shown in contours. The contours are at levels of 10\%, 30\%, 50\%, 70\%, and 90\% of the peak flux value, with any contours that fall below $3\sigma$ excluded, and are meant primarily to demonstrate that the model profile matches the data in two dimensions. Finally, on the right we show the SED with the best fit model in blue. The complete figure set (97 images) is available in the online journal.}
\label{fig:rt_fits}
\end{figure*}

\input{model_plots.tex}

We also include an envelope in our model based off of the rotating collapsing model from \citet{Ulrich1976}. The density of the envelope is given by,
\begin{equation}
\small
\rho = \frac{\dot{M}}{4\pi}\left(G M_* r^3\right)^{-\frac{1}{2}} \left(1+\frac{\mu}{\mu_0} \right)^{-\frac{1}{2}} \left(\frac{\mu}{\mu_0}+2\mu_0^2\frac{R_c}{r}\right)^{-1},
\end{equation}
where $r$ is defined as is typical in a spherical coordinate system, and $\mu = \cos{\theta}$. $R_c$ is the critical radius, inside of which the density profile begins to flatten into a disk-like structure. In our model, we truncate the envelope at an outer radius $R_{env}$ beyond which the density drops to zero and define $R_c = R_{disk}$. We also truncate the density at the same inner radius as the disk, $R_{in}$. Finally, rather than parameterize the envelope in terms of the accretion rate, $\dot{M}$, we integrate over the entire density structure to calculate the total mass, $M_{env}$, and normalize the density properly from this value.

Finally, we include in our envelope model an outflow cavity in which the density of the envelope is reduced by some factor, $f_{cav}$. The structure of this cavity is defined by,
\begin{equation}
    z > 1 \, \mathrm{au} + R^{\xi},
\end{equation}
where $\xi$ defines the shape of the cavity and opening angle. The opening angle, defined as the full angle across the cavity at $R_{env}$ from the position of the protostar, is therefore,
\begin{equation}
    \theta_{open} = 2 \, \tan^{-1}\left(\frac{r}{1\,\mathrm{au} + {R_{env}}^{\xi}}\right).
\end{equation}
We show a simple schematic of this geometry in Appendix \ref{section:outflow_shape}.

SEDs can also be sensitive to extinction from foreground cloud emission that is not a part of the envelope. In previous works we have included foreground extinction as a parameter in our fit when well motivated \citep[e.g.][]{Sheehan2017}. Here, for simplicity, because of computational constraints and because foreground extinction can be very degenerate with envelope properties, we leave this out of our model. The spatial information provided by the millimeter visibilities should help to mitigate this effect, but it should be noted that for some sources this could impact the envelope properties.

\begin{figure*}[t]
\centering
\includegraphics[width=7in]{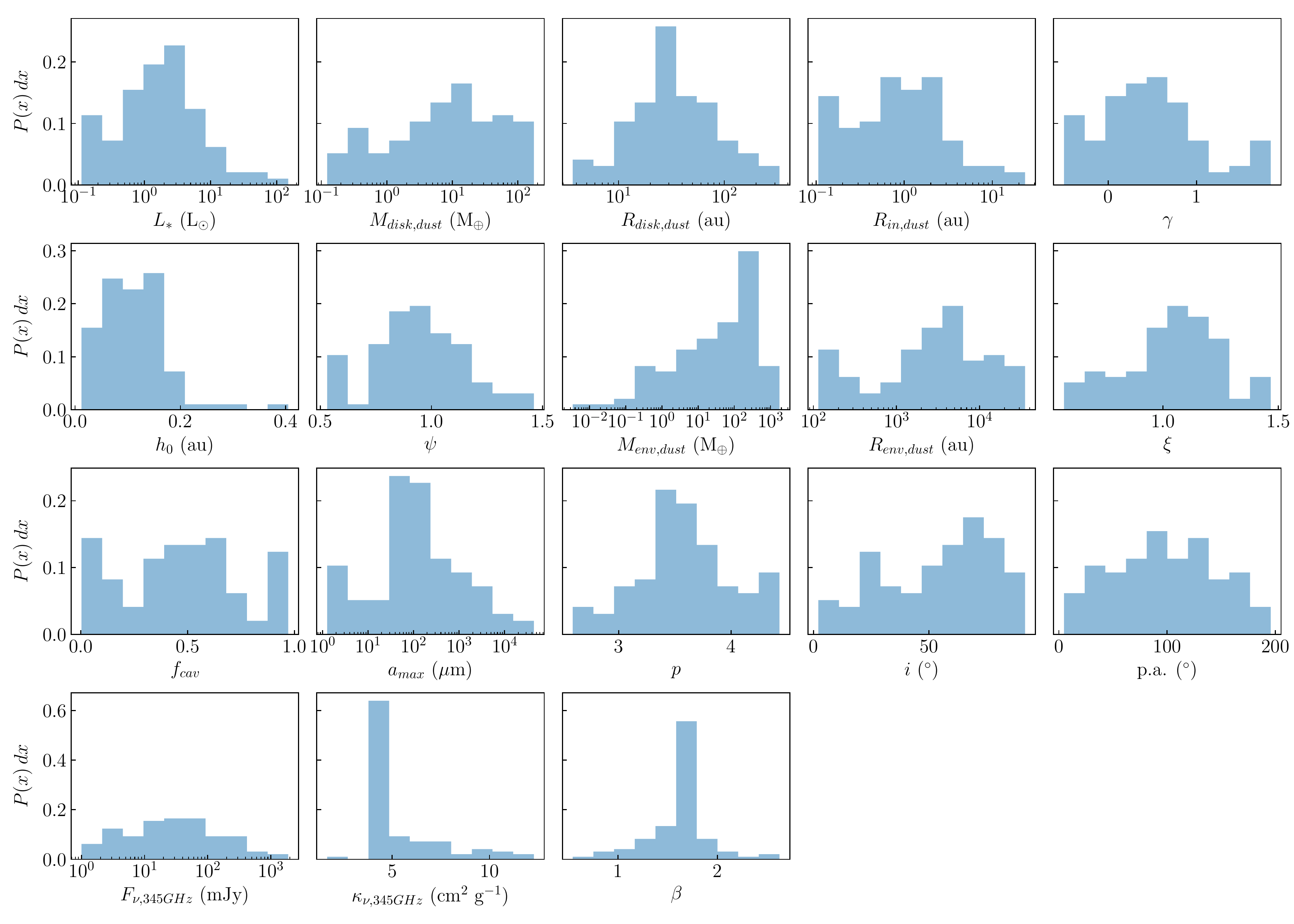}
\caption{Histograms of the distribution of best-fit parameter values found from our modeling for our sample of 90 protostellar disks. The last row shows quantities that are not formally parameters of our model, but are derived from the best fit models.}
\label{fig:parameter_distributions}
\end{figure*}

We provide our model with dust opacities following a similar recipe to \citep{Woitke2016}, with dust grains initially composed of 70\% astronomical silicates \citep{Draine2003} and 30\% carbonaceous material \citep{Zubko1996OpticalEllipsoids}, though we then add water ice \citep{Hudgins1993} at 50\% the level of silicates+carbon, for final abundances of 47\% silicates, 20\% carbonaceous material, and 33\% water ice. We follow the distribution of hollow spheres prescription \citep{Min2005ModelingSpheres}. Finally, we assume that the grains have a power-law size distribution, $n \propto a^{-p}$ with a minimum size of 0.05 $\mu$m. In the envelope, where dust grain growth is likely less advanced, we assume the maximum dust grain size is $a_{max} = 1$ $\mu$m and $p = 3.5$. In the disk, however, we allow both $a_{max}$ and $p$ to vary.

\begin{figure*}[t]
\centering
\includegraphics[width=7in]{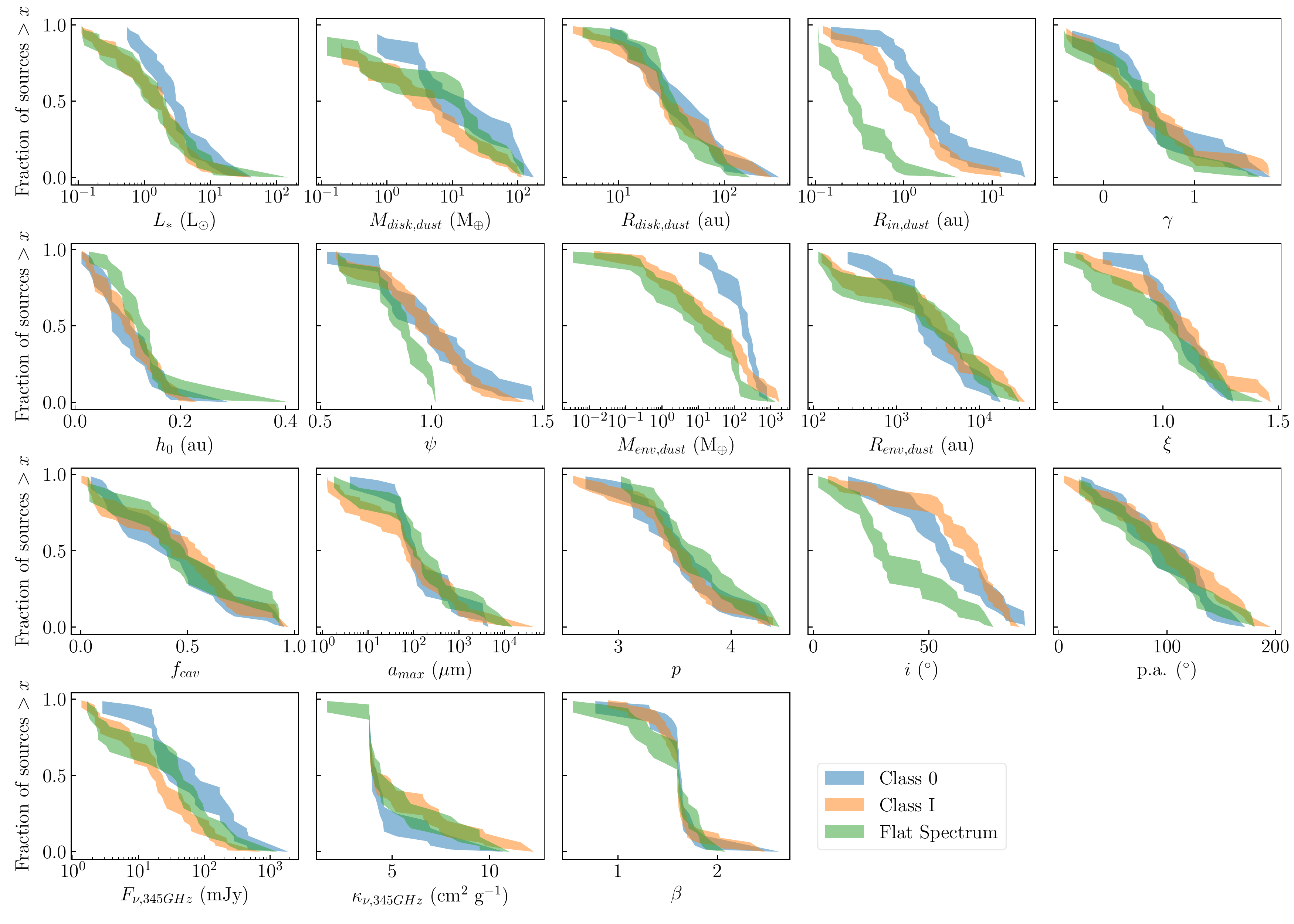}
\caption{Cumulative distributions showing the fraction of protostellar disks with a measured parameter value above a given value. We separately show the cumulative distributions for sources identified as Class 0 ($blue$), Class I ($orange$) and Flat Spectrum ($green$) to highlight any potential differences between populations at different evolutionary stages.}
\label{fig:class_comp}
\end{figure*}

This density structure is provided to the RADMC-3D Monte Carlo radiative transfer code \citep{Dullemond2012}, which is first used to calculate the thermal structure of the disk + envelope system. Then using the density and thermal structure, we use RADMC-3D to generate synthetic spectral energy distributions (SEDs) and 345 GHz millimeter images. We then Fourier transform the synthetic millimeter images to compare directly with our two dimensional ALMA visibility data. The inclusion of synthetic observations adds two additional parameters to our model, the position angle $p.a.$ and inclination $i$, for a total of 15 free parameters for each fit: $\hat{\theta} = \{L_*,M_{disk},R_{disk},R_{in}, h_0,\gamma,\psi,M_{env},R_{env},f_{cav},\xi,a_{max},\\ p,i,p.a.\}$.

To compare these synthetic models with our observational dataset for each source, we use the Markov Chain Monte Carlo fitting code $emcee$ \citep{Foreman-Mackey2013}, which employs an affine-invariant Monte Carlo sampler to sample parameter space and converge on regions that best fit the data. In our runs we use 200 walkers spread out over a wide range of parameter space. To simultaneously fit our two independent datasets (SED and ALMA 2D visibilities) we use a goodness-of-fit metric that is a linear combination of the $\chi^2$ values for each dataset separately,
\begin{equation}
    X^2 = w_{SED} \, \chi^2_{SED} + w_{vis} \, \chi^2_{vis}.
\end{equation}
In general we use $w_{SED} = w_{vis} = 1$, though in some cases we increase $w_{SED}$ to help improve the fit to the SED in the model, as the SED typically has many fewer data points to fit than the visibilities. Moreover, because the SED is more sensitive to the envelope, particularly at short wavelengths, while the visibilities are particularly sensitive to the disk and the two dimensional structure of the system, it is possible to achieve good fits to both datasets, despite this difference. Because of our adjustments to the weights, however, the uncertainties from our modeling should not be treated as true statistical uncertainties, but they do provide a reasonable estimate of the range of parameter values that can provide a good fit to the data.

As radiative transfer models are computationally expensive to generate, we use a range of supercomputing resources to do our model fitting. The majority of the modeling was done on Bridges at the Pittsburgh Supercomputing Center and Comet at the San Diego Supercomputer Center with a 5 million core-hour allocation through the National Science Foundation (NSF) XSEDE program. Additional models were run on Schooner at the Oklahoma Supercomputing Center for Education \& Research (OSCER), and testing was also done on Stampede2 at the Texas Advanced Computing Center (TACC). Fits with $emcee$ can be parallelized because for each step, the 200 walkers can have their likelihood functions evaluated independently. As such the radiative transfer models for multiple walkers can be computed in parallel. This parallelization can be further tuned by running the RADMC-3D radiative transfer models using multiple cores, creating a hybrid MPI-OpenMPI parallelization scheme. After some testing, we found that the most efficient way to run models was using $\sim5$ supercomputer nodes, each with $20-28$ cores per node. We would then run the models for 5 walkers simultaneously on each node (for a total of $\sim25$ walkers running models simultaneously), with each walker using $4-6$ cores to run RADMC-3D (with $\sim100 - 150$ cores in total).

\section{Results}
\label{section:results}

We show example model fits, using the maximum a posteriori model for each source, in Figure \ref{fig:rt_fits}. We note that in Figure \ref{fig:rt_fits} we show the one-dimensional, azimuthally averaged visibility profile for ease of interpretation, but the models are fit to the full, two dimensional visibility dataset. We also list a subset of the most relevant best-fit parameter values in Table \ref{table:rt_best_fits}, with the full table available online and in an accompanying machine-readable table with all best fit values and derived quantities discussed throughout the remainder of this work. The best-fit parameters are calculated using the maximum likelihoods of marginalized one-dimensional posteriors for each parameter, after the burn-in steps are discarded. To do so, we use a kernel density estimation to estimate the probability density function (PDF) from the marginalized samples on a fine one-dimensional grid, and then report the grid value where the PDF is maximized. The uncertainties on those values are determined by the range around those values containing 95\% of the post-burn-in walkers.

We have also provided the full data resulting from our model fits in an online repository\footnote{https://doi.org/10.5281/zenodo.5842333} so that the community might make use of our results in their own work. This includes the full posterior distributions for each parameter of each source in our model along with those additional values derived from our fit parameters as discussed throughout the remainder of the text. We also include, for each source, the \texttt{pdspy} configuration file and data files modeled, along with a script that demonstrates how to use \texttt{pdspy} to work with the models. With these tools, the resulting models should be easily accessible for anyone to generate and use for further studies.

\input{best_fits_table}

We note that throughout the remainder of this discussion, all masses, both disk and envelope, account for only the dust and ignore the gas in the system. We follow this convention throughout our analysis, unless otherwise noted, as our radiative transfer modeling is most directly sensitive to the dust content of the system. We therefore refer to disk and envelope properties derived from our modeling as $M_{disk,dust}$, $R_{disk,dust}$, $M_{env,dust}$, and $R_{env,dust}$ in the relevant tables, figures and discussion and list masses in units of $M_{\oplus}$ to remind readers of this. For readers interested in the gas-mass, we do however include estimates of the total mass assuming a gas-to-dust ratio of 100:1 in the online version of Table \ref{table:rt_best_fits} and the machine-readable table accompanying this work, and we refer to these parameters, for example, as $M_{disk}$ to indicate the total mass.

In Figure \ref{fig:parameter_distributions}, we show the distributions of parameter values from our modeling over our full sample, and therefore a picture of the demographics of protostellar disk properties. We also show the distribution of three additional properties derived from the modeling that are relevant for comparing with other studies of disk demographics: the 345 GHz flux of the disk in our models ($F_{\nu,345GHz}$), the 345 GHz opacity from our models ($\kappa_{\nu,345GHz}$), and the spectral index between 230 GHz and 345 GHz ($\beta$). We also list the median value for each property measured by our modeling over our sample of protostars in Table \ref{table:class_comps}. Median values were calculated using the Kaplan-Meier fitter in the \texttt{lifelines} package \citep{Davidson-Pilon2019}. To estimate the uncertainty on these median values, we generated 1000 realizations of the distribution of protostellar system (star + disk + envelope) properties by randomly sampling a set of parameters from the posterior for each protostar for each realization. We calculated the median parameter values for each of these 1000 realizations, and the uncertainty reported in Table \ref{table:class_comps} is given by adding the 68\% inclusion range of medians from the realizations around the median value calculated from the Kaplan-Meier fitter, and adding this in quadrature with the counting uncertainty reported by the Kaplan-Meier fitter.

Of particular note, we find the protostellar disks are small, with a median dust radius of $ 29.4^{+  4.1}_{-  3.2}$ au. Large disks with $R_{disk} > 100$ au do exist, though only account for ${11.3\%}^{+ 4.6\%}_{- 3.4\%}$ of our sample. This is in reasonable agreement with \citet{Maury2019} and \citet{Tobin2020}, both of which found that large protostellar disks are less common. 
The median embedded disk dust mass across all different classifications is $  5.8^{+  4.6}_{-  2.7}$ $M_{\oplus}$, and the median envelope dust mass is $ 75.9^{+ 13.7}_{- 51.3}$ $M_{\oplus}$.

The total envelope masses, assuming a gas-to-dust ratio of 100, range from $\sim10^{-4} - 1$ M$_{\odot}$ with a handful of sources with even lower masses. This range range is in reasonably good agreement with envelope masses estimated by \citet{Furlan2016}, who matched a grid of radiative transfer models with the SEDs for the same set of sources considered here. They are somewhat lower than what has been found from single dish observations of other star forming regions, which tend to fall in the 0.1 -- 10 M$_{\odot}$ range \citep[e.g.][]{Enoch2008,Sadavoy2014ClassDistribution,Pezzuto2021PhysicalObservations}. Those surveys, however, tend to find preferentially young, likely Class 0 sources, which tend to have more massive envelopes. Even considering just the Class 0 sources in our sample, however, we are still lacking the envelopes with $>1$ M$_{\odot}$ of material. Though this could be due to true differences in the envelope mass distributions between regions, it is also very likely that the difference could be due to the scales considered. Though we do fit single dish fluxes from the HOPS Survey when available, our ALMA observations are primarily tracing emission on scales up to a few thousand au, and may be missing out on larger scale cloud emission.

\input{class_comps_table}

To explore how protostellar disk and envelope properties change with evolutionary stage, we show the cumulative distribution of parameter values split into the three classes of sources examined here, Class 0, I, and Flat Spectrum, in Figure \ref{fig:class_comp}. We note that for most parameters, we have no a priori reason to believe that our observations are biased towards missing a particular range of values and so we assume that the sample is complete. The exception to this is the disk dust mass, as we selected only sources that were detected with ALMA, and the millimeter brightness is related to dust dust mass. Though in reality it is more complicated than this, for simplicity we add additional sources with upper limits on their disk dust mass equivalent to the lowest dust mass disk that we modeled until the fraction of detected sources in our sample matches the overall survey \citep[0.88 for all sources, 0.91 for Class 0's, 0.85 for Class Is, and 0.88 for Flat Spectrum sources;][]{Tobin2020}, when generating cumulative distributions or running two sample tests for disk dust masses.

To test whether there are any significant differences in the distribution of parameters between the separate classes, we run two-sample tests using the log-rank test in the \texttt{lifelines} package to compare the distribution of parameter values of each of the three classes against each other. To account for uncertainties in the measured parameters, instead of using the ``best-fit" parameters, we create 1000 realizations of the source parameters by randomly selecting parameters from the posterior for each source for each realization. We then calculate the p-value associated with the two sample test for each realization, and consider the fraction of times for which the test reported a statistically significant difference in the distributions (p $<$ 0.05). The results of this calculation are reported in the last three columns of Table \ref{table:class_comps}, with comparisons that we consider significant shown in bold.

For most parameters, we find that the distributions for each Class are consistent with being drawn from the same underlying distribution. 
One exception to this is that Flat Spectrum sources have lower envelope dust masses than Class 0 sources (100\% of realizations with p$<0.05$), which is what might be expected if the Class 0/I/Flat Spectrum scheme represents an evolutionary sequence. There are some realizations where Class I sources have lower envelope dust masses than Class 0 sources (15\% of realizations with p$<0.05$) and Flat Spectrum sources have lower envelope dust masses than Class I sources (6\% of realizations with p$<0.05$), but these are not significant. Furthermore, we find some evidence that Class I sources have disk dust masses that are lower than Class 0 sources (69\% of realizations with p$<0.05$). We cannot, however, confidently distinguish Class 0 disk dust masses from Flat Spectrum disk dust masses or Class I disk dust masses from Flat Spectrum disk dust masses ($0$\% of realizations with p$<0.05$).

Our inability to distinguish between Class 0 and Flat Spectrum disk dust masses is in contrast with \citet{Tobin2020}, who found that their masses were drawn from different underlying distributions. This appears to be due to a knee in the the cumulative distribution of Flat Spectrum disk dust masses, where the Flat Spectrum distribution actually crosses the Class 0 distribution. 

Interestingly, we find that Flat Spectrum sources preferentially have lower inclinations than Class 0 or I sources ($>99\%$ of realizations with p$<0.05$). If the classification of a source was purely determined by its evolutionary stage, we would not expect the inclination to be significantly different from class to class, as we find here. One potential reason for this difference could be that sources with lower inclination have lower extinction to the hotter central regions of the disk and the protostar itself, and so we see more near-infrared emission from these sources. As such, they are more likely to be classified as Flat Spectrum. Conversely, higher inclination sources will have more near-infrared extinction, and so they are more likely to be classified as Class 0/I.

Flat Spectrum sources also have $\psi$ and $R_{in}$ values that are smaller than those of Class 0 or I sources ($>93$\% of realizations with p$<0.05$). It is possible that this may point to underlying physical differences. For example, the difference in $\psi$ may indicate that Flat Spectrum disks are flatter, which may in turn hint that large dust grains have settled more in older disks. These differences may instead, however, be related to systematics in our relatively simple model. As Flat Spectrum sources tend to have more near-infrared emission, they may require smaller inner dust radii to ensure that there is hot material to produce this emission, and flatter disks to ensure that material is not obscured by optical depth effects. Higher resolution ALMA observations and/or more detailed models may be needed to better determine whether these are not simply systematics of our model.

\section{Discussion}
\label{section:discussion}

\subsection{Comparison with Simple Estimators of Disk Structure}
\label{section:simple-estimators}

\begin{figure*}
    \centering
    \includegraphics[width=2.3in]{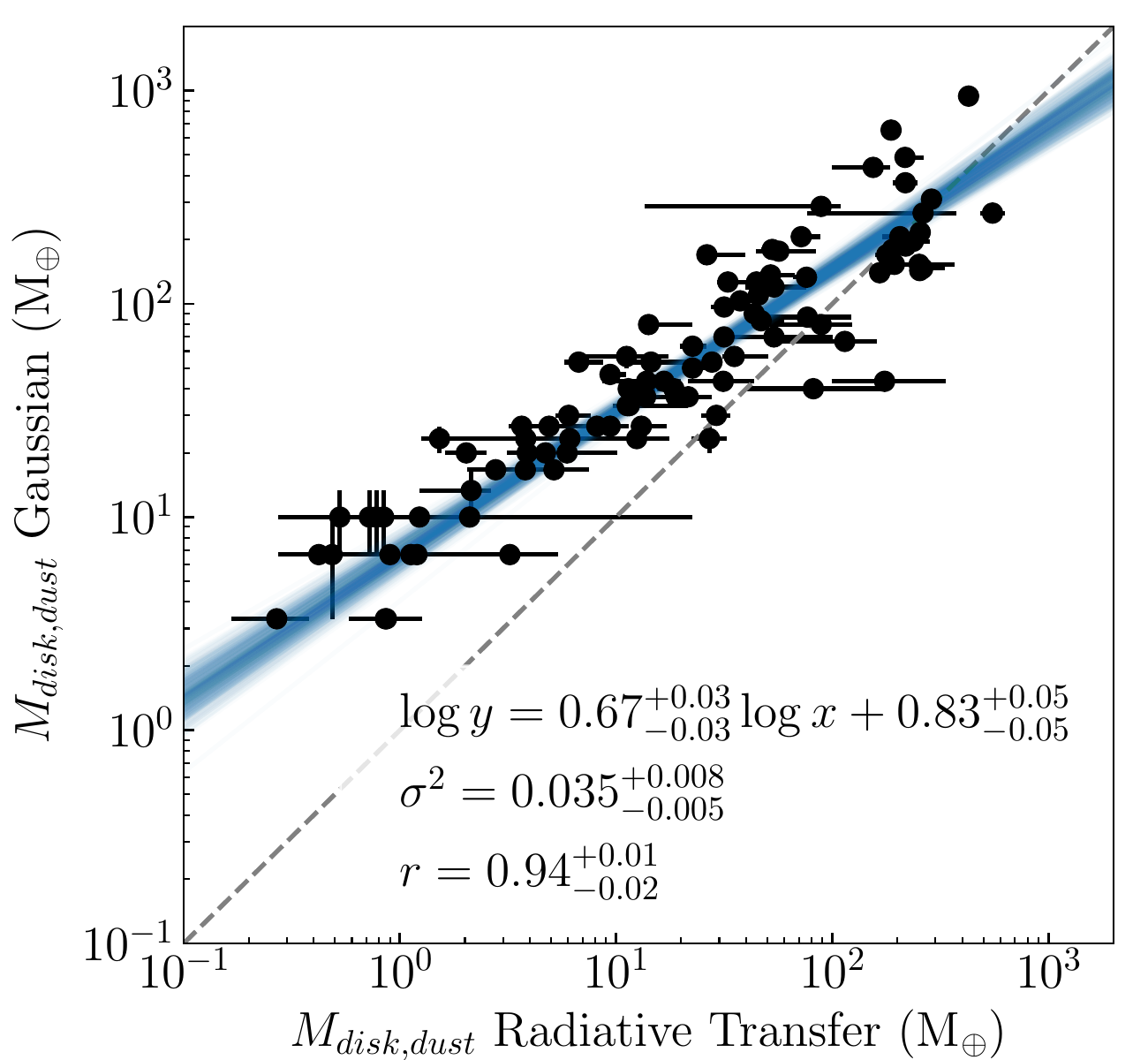}
    \includegraphics[width=2.3in]{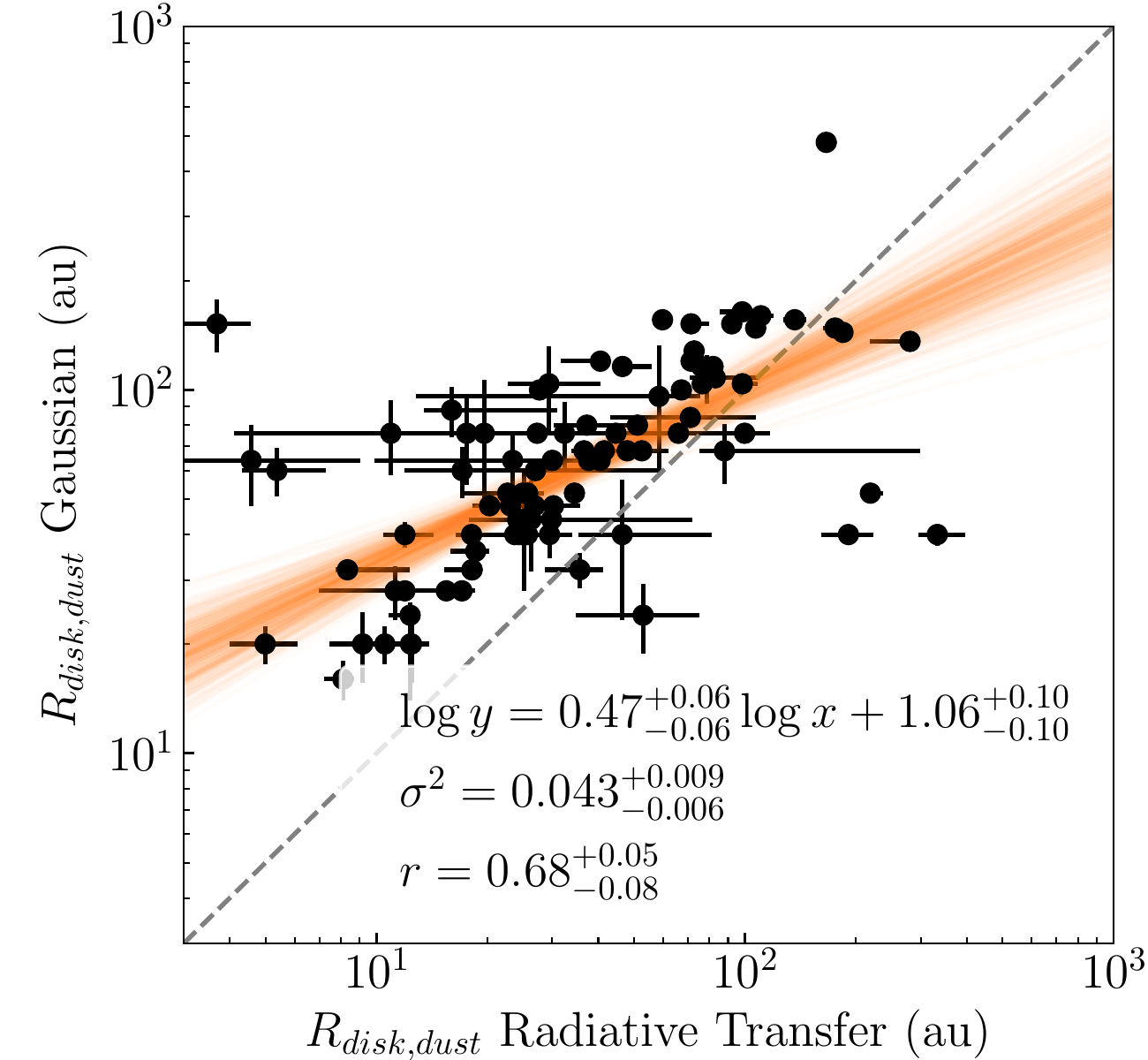}
    \includegraphics[width=2.3in]{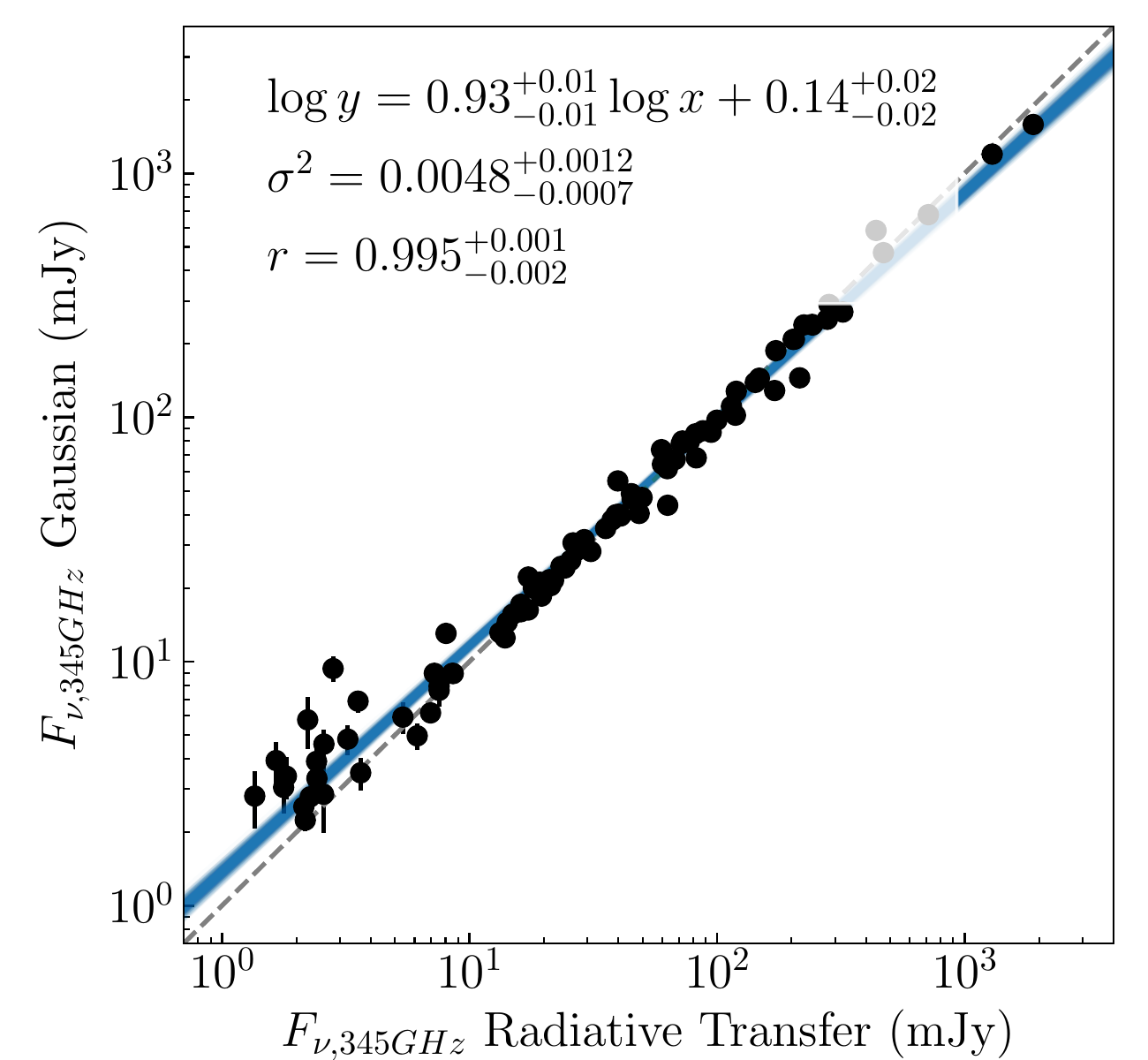}
    \caption{Comparisons between the disk dust masses ($left$), radii ($center$), and 345GHz fluxes ($right$) measured by our radiative transfer modeling and as measured by fitting a two dimensional Gaussian fit to the VANDAM: Orion images of our sources. We also show the results of a log-linear fit to each dataset and list the best fit parameter values, as well as show a dashed line where the values on each axis are equal for comparison.}
    \label{fig:mdisk_rdisk_comp}
\end{figure*}

Disk surveys have traditionally relied on simple estimates to measure disk properties owing, in large part, to the complexity and computational costs involved in doing radiative transfer modeling of protostellar and protoplanetary disks. For example, disk dust masses are often calculated from the submillimeter flux by assuming optically thin dust emission \citep{Hildebrand1983} with a uniform temperature of either 20 K \citep[e.g.][]{Pascucci2016,Ansdell2016} or that scales with protostellar luminosity \citep[e.g.][]{Andrews2013,Tobin2020}, and an estimate of the dust opacity at the relevant submillimeter wavelength \citep[e.g.][]{Beckwith1990}. Disk sizes are typically calculated by fitting images with two-dimensional Gaussian functions \citep[e.g.][]{Tobin2020}, Nuker profiles \citep[e.g.][]{Tripathi2017} or by using the curve-of-growth method to calculate the radius enclosing 95\% of the total flux \citep[e.g.][]{Ansdell2016}.

These simple estimates of disk dust masses and radii are useful for providing quick measurements of disk properties for large samples of disks without the need for the significant computational resources we use in this work. These estimates are, however, also limited by the fact that disk temperatures and densities are inherently three dimensional, coupled with optical depth effects, the viewing angle of the system, and the scattering of light by dust grains \citep[e.g][]{Zhu2019,Liu2019}. These difficulties are compounded when considering protostellar disks, which are embedded within an envelope of infalling dusty material that can also be bright in the submillimeter, making uniquely identifying the disk difficult. Some previous works have attempted to remove envelope emission by considering only the flux at a specific interferometer baseline that is thought to be large enough to resolve out typical size scales of envelopes \citep[e.g.][]{Jrgensen2009PROSAC:Protostars,Andersen2019}, though even those estimates can still be subject to significant envelope contamination \citep[e.g.,][]{Dunham2014}. Other works have extended this by fitting simple analytic disk intensity models directly to the visibility data, sometimes even including an envelope component to help separate disk from larger scale emission, \citep[e.g.][]{Maury2014,Harsono2014,Segura-Cox2016THECLOUD,Segura-Cox2018,Maury2019}, but are still limited by the relatively simple conversion from flux to dust mass.

\citet{Tobin2020} presented estimates of protostellar disk dust masses and radii for sources in the VANDAM: Orion survey by assuming that the emission fit in the image plane with a two-dimensional Gaussian function could be attributed uniquely to emission from a protostellar disk. They used the typical conversion from submillimeter flux to dust mass,
\begin{equation}
    M_d = \frac{d^2 \, F_{\nu}}{\kappa_{\nu} \, B_{\nu}(T)},
    \label{eq:disk_mass}
\end{equation}
\citep[e.g.][]{Hildebrand1983}, assuming a 345 GHz opacity of 1.84 cm$^2$ g$^{-1}$ and a temperature that was scaled based on the protostellar luminosity, $T = 43 \, \mathrm{K} \, (L / L_{\odot})^{0.25}$. This temperature scaling was derived from a grid of radiative transfer models, in an effort to account for the effects of protostellar luminosity on the disks average temperature. They also reported disk sizes using the 2$\sigma$, deconvolved, major axis size from two dimensional Gaussian fits to each source in the image plane.

As we have now modeled 97 of the sources from \citet{Tobin2020} using our disk radiative transfer modeling infrastructure, we can directly test how 
these simple calculations of protostellar disk properties   
compare with models that account for more physics in deriving properties.
In Figure \ref{fig:mdisk_rdisk_comp}, we show a comparison of the disk dust masses and radii, as measured by our radiative transfer modeling framework, compared with the dust masses and radii presented in \citet{Tobin2020}. For reference, we also show the best fit log-linear model for a fit to each set of data using the \texttt{linmix} package,\footnote{https://github.com/jmeyers314/linmix} which enables us to account for errors on multiple independent variables \citep{Kelly2007SomeData}, along with the $x = y$ line as a dashed line.

We find that the disk dust masses derived from our modeling differ substantially from the dust masses found with simple estimates from the flux. While there is a strong log-linear correlation between $M_{disk,Gaussian}$ and $M_{disk,RT}$, the disk dust masses measured in \citet{Tobin2020} over-estimate the mass compared to radiative transfer modeling, with the severity of the discrepancy increasing for lower-mass disks.

To test whether this discrepancy is the result of improperly separating disk and envelope, we also compare the 345 GHz flux inferred from our best fit disk model for each source with the 345 GHz flux measured in \citet{Tobin2020} in the rightmost panel of Figure \ref{fig:mdisk_rdisk_comp}. From a log-linear fit, we find a tight correlation with a slope very close to 1, indicating that there is very good agreement between the fluxes measured in \citet{Tobin2020} and the fluxes we recover from our modeling. Though the slope is nominally different from 1 with high significance, this appears to be due to a handful of faint sources for which the Gaussian fit over-estimated the flux. If we only fit sources with $F_{\nu,345GHz,RT} > 5$ mJy, we find that the slope is perfectly consistent with 1.

\begin{figure*}
    \centering
    \includegraphics[width=2.3in]{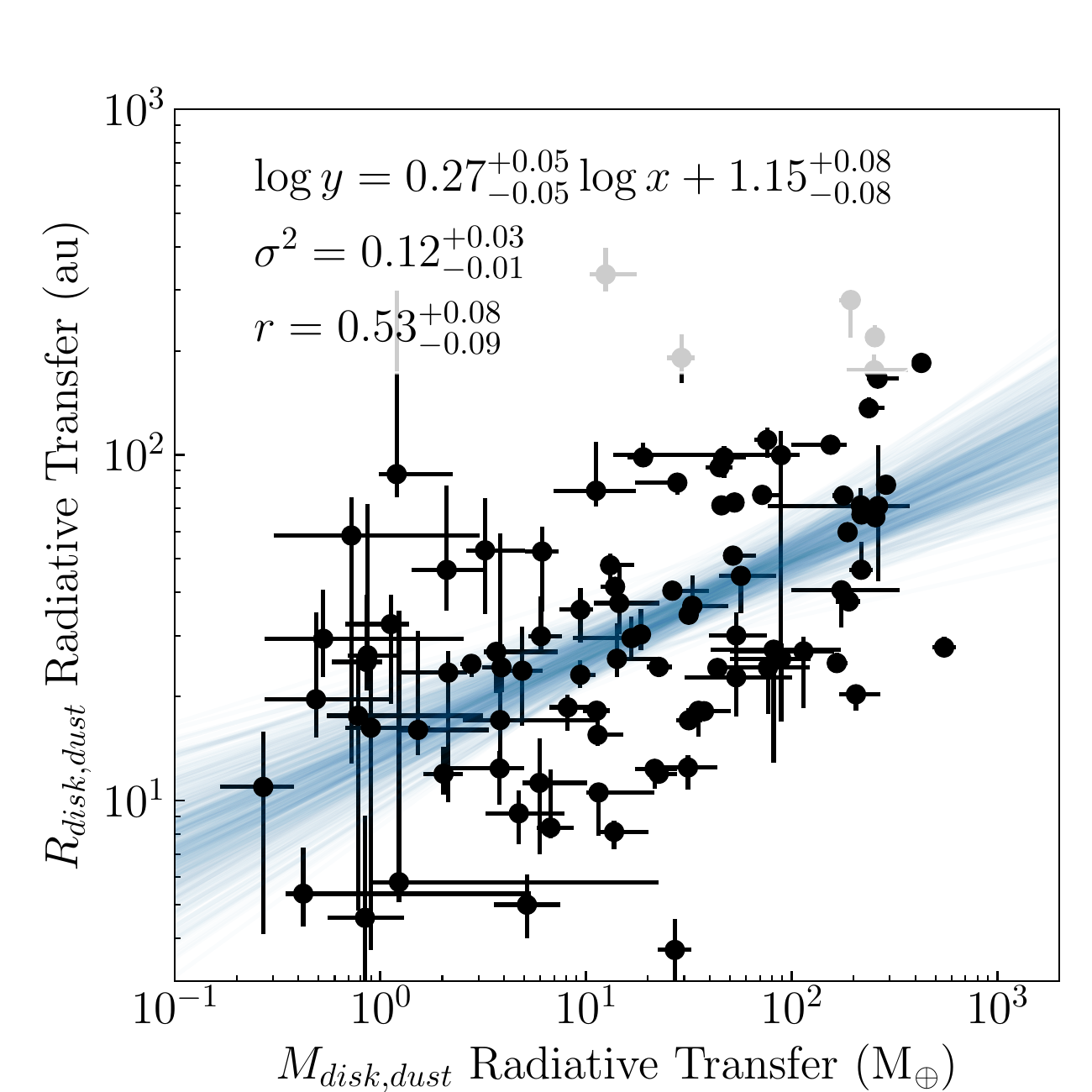}
    \includegraphics[width=2.3in]{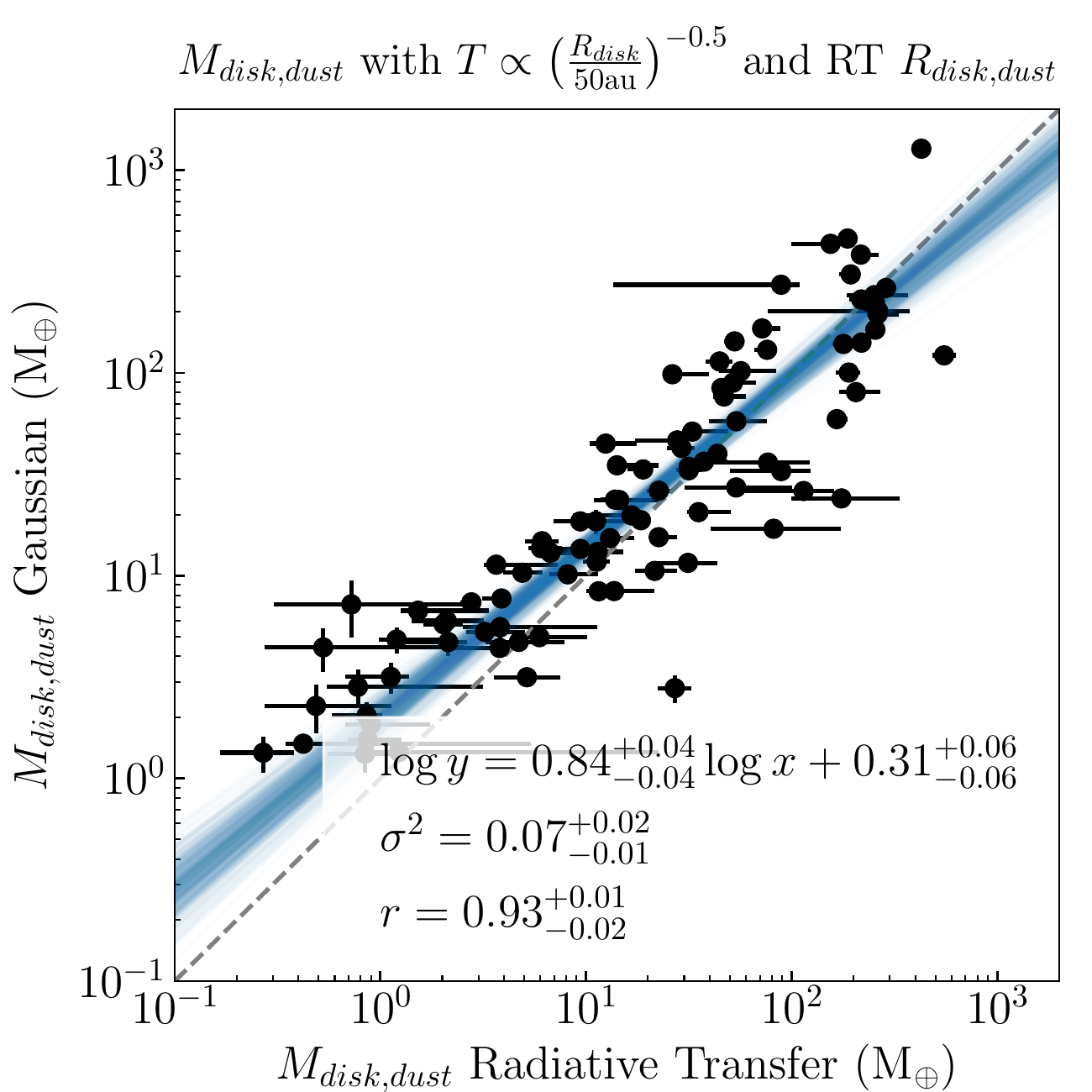}
    \includegraphics[width=2.3in]{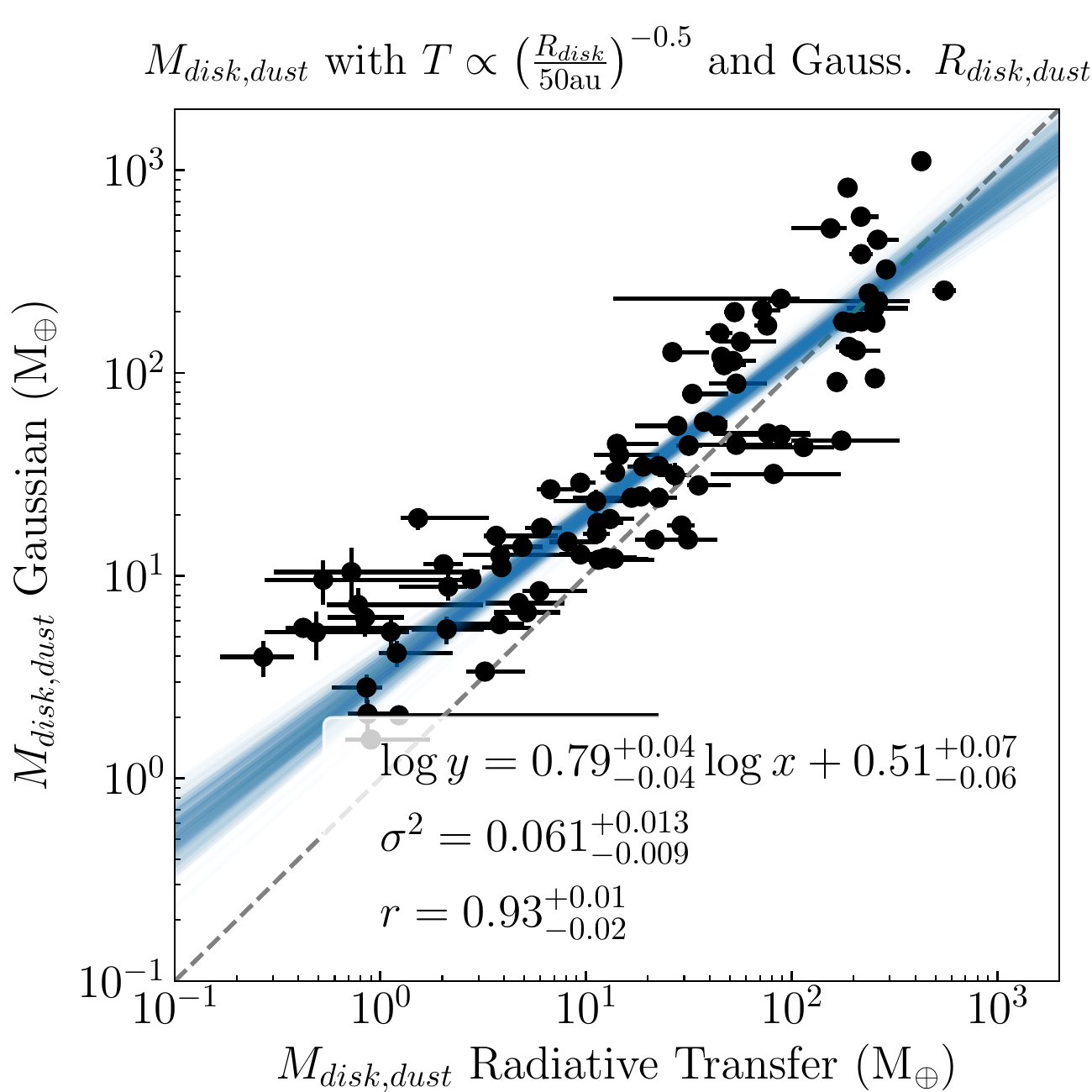}
    \caption{($Left$) Comparison of the disk dust masses and radii measured using our radiative transfer modeling framework, demonstrating that the two are correlated, though with significant scatter. ($Center$) A comparison of the disk dust masses found from our modeling with disk dust masses derived from fitting the disk with a two-dimensional Gaussian in the image plane \citep[e.g.][]{Tobin2020} and using Equation \ref{eq:disk_mass} with $T = 43\,\mathrm{K} \, (R_{disk} / \mathrm{50\,au})^{-0.5} \, (L / 1\,\mathrm{L_{\odot}})^{0.25}$ and $R_{disk}$ measured from our modeling. ($Right$) The same as in the center, but using $R_{disk}$ measured in \citet{Tobin2020}. We also show the results of a log-linear fit to each dataset and list the best fit parameter values, as well as show a dashed line where the values on each axis are equal for comparison.}
    \label{fig:mdisk_corrected}
\end{figure*}

This tight one-to-one correspondence between the flux measured from the different methods suggests that the discrepancy between disk dust masses measured from the different methods is not due to difficulty separating disk from envelope. Though this may still be a problem for lower resolution observations, with the resolution of $\sim30-40$ au that we have here, we can confidently identify the disk flux from millimeter images with a two dimensional Gaussian fit. The good agreement between the fluxes measured in different ways therefore instead indicates that 
the difference is in the conversion from millimeter flux to dust mass. In particular, this suggests that there may be physics missing from the simple estimates presented in Equation \ref{eq:disk_mass}, even with the modifications to the temperature from \citet{Tobin2020}.

That said, \citet{Tobin2020} also found from the radiative transfer modeling upon which their luminosity correction was based that the average disk temperature should scale like $R^{-0.5}$, though ultimately used a fiducial radius of 50 au in their calculations for simplicity and to avoid reliance on possibly inaccurate radii from Gaussian fits. We do, however, find that there is a correlation between $M_{disk,dust}$ and $R_{disk,dust}$ in our sample, as we show in Figure \ref{fig:mdisk_corrected}, so it is possible that this correction is indeed important. If we re-calculate dust masses from the ALMA fluxes while including the $R^{-0.5}$ scaling in calculating the temperature, using $R_{disk,dust}$ from our modeling, as shown in the center panel of Figure \ref{fig:mdisk_corrected}, we find better agreement with our radiative transfer derived disk dust masses. The best-fit line still has a slope shallower than 1, but this may again be due to the discrepancy between fluxes for the faintest sources, as these are presumably generally the lowest mass disks. Above $\sim5$ M$_{\oplus}$ the slope appears to be closer to 1. There is still a significant amount of scatter in the relationship, though, likely because we are marginalizing over many disk parameters that contribute to radiative transfer effects. As such, while this correction helps to reproduce  
the radiative-transfer modeling measured
distribution of disk dust masses on average, individual disks may still have significant differences from their radiative-transfer-measured mass. In the right panel of Figure \ref{fig:mdisk_corrected} we show the same comparison but using $R_{disk,dust}$ measured from Gaussian fitting, as this is more analogous to what would be done by an observer not using radiative transfer modeling. We find that this also reduces the discrepancy between the two mass measurements, though the slope is still shallower than one and is not corrected quite as well as in the case of using the radiative transfer modeling measured radii. This is likely because of the discrepancy between the disk radii we measure and those found from Gaussian fitting, as can be seen in Figure \ref{fig:mdisk_rdisk_comp} and discussed below.

Ultimately, we suggest that future works use the relation
\begin{equation}
    T = 43 \, \mathrm{K} \, \left(\frac{L_*}{1 \, L_{\odot}}\right)^{0.25} \, \left(\frac{R_{disk}}{50 \, \mathrm{au}}\right)^{-0.5}
\end{equation}
to calculate the average disk temperature, as it includes additional physical effects into simple estimates of disk dust masses. We do caution, however, that this relationship should only be applied to {\it embedded} disks, as the models it is based on included an envelope that can serve to keep the disk warmer than it would be otherwise. For protoplanetary disks, the temperature pre-factor may be different, though the scaling with luminosity and disk radius would likely be similar. We also note that these corrections are model-dependent, and could still produce incorrect disk dust masses if the assumptions underlying the model are also incorrect. That said, the particular corrections applied here, that brighter stars produce warmer disks and that smaller disks with more material close to the star are on average warmer, should be reasonably model agnostic.

We also find that the disk dust radii measured by the two-dimensional Gaussian fits also frequently disagree with the disk dust radii measured from our modeling.
The Gaussian fits tend to more severely overestimate disk radii for smaller disks, while they actually $underestimate$ the sizes of a number of the largest disks, some by quite a lot (see Figure \ref{fig:mdisk_rdisk_comp}). The discrepancy between radiative-transfer-measured disk radii and Gaussian-fitting-measured disk radii is likely due to the low spatial resolution, of only $\sim40$ au. As such, smaller disk radii are increasingly difficult to accurately deconvolve from the beam. Additionally, disks with massive envelopes may appear larger in the image plane, though the strong correlation between fluxes measured from our modeling and from two dimensional Gaussian fitting suggests that this may not be a major effect. On the other hand, for larger disks, two dimensional Gaussian functions become increasingly poor representations of the resolved disk brightness profile and often tend to fit the more compact central peak rather than the extended, low surface brightness disk structure. By fitting directly in the $uv$-plane we avoid the difficulties of measuring deconvolved sizes, and the more flexible geometry of our disk model allows us to better match the full brightness profiles of the disks in our sample.

\subsection{Protostellar Disk \& Envelope Evolutionary Trends}
\label{section:evolutionary_trends}

In Section \ref{section:results}, we discussed how the distributions of disk properties from our best-fit models change as a function of the protostellar classification, but for the most part found only marginal evidence that the distributions were distinct for different classes. In this section, we further examine trends as a function of protostellar ``age" by comparing a selection of disk and envelope properties with three separate quantities that have been proposed as tracers of protostellar evolution. In particular, we consider the bolometric temperature ($T_{bol}$), the ratio of envelope mass to total mass ($M_{env}$), and the age as inferred from simple models of protostellar evolution \citep[e.g.][]{Andre1994FromCloud,Saraceno1996AnObjects.,Molinari2008TheObjects,Andre2008FirstCamera,Fischer2017Evolution}. For each comparison, we run a log-linear fit to the data using the \texttt{linmix} package, which accounts for errors on both variables being fit \citep{Kelly2007SomeData}, in order to test whether there is a trend with protostellar age in each quantity.

\subsubsection{Bolometric Temperature ($T_{bol}$)}

Protostars have traditionally been classified into Class 0/I/II/III and Flat Spectrum sources according to their bolometric temperature \citep[e.g.][]{Chen1995} or near-infrared spectral index \citep[e.g.][]{Myers1987}. Though these classifications have been defined observationally, they are also thought to roughly follow the evolutionary state of the young star through the effect the envelope has on the SED of a protostar \citep[e.g.][]{Whitney2003TwodimensionalSequence,Crapsi2008}. Sources with low bolometric temperatures and (steeply) rising SEDs in the infrared are best explained by significant dust obscuration, presumably by a massive envelope of material while they are still young. Conversely, sources with higher bolometric temperatures or SEDs that are shallower in the infrared or even decreasing likely do not suffer from as much extinction, and therefore do not have as substantial envelope obscuration and are older. Properties such as the bolometric temperature or the classification as Class 0/I/Flat Spectrum have therefore sometimes been treated as estimators of the age of a protostellar system \citep[e.g.][]{Evans2009,Dunham2015,Kristensen2018ProtostellarEstimates, Andersen2019}.

\begin{figure*}
    \centering
    \includegraphics[width=2.3in]{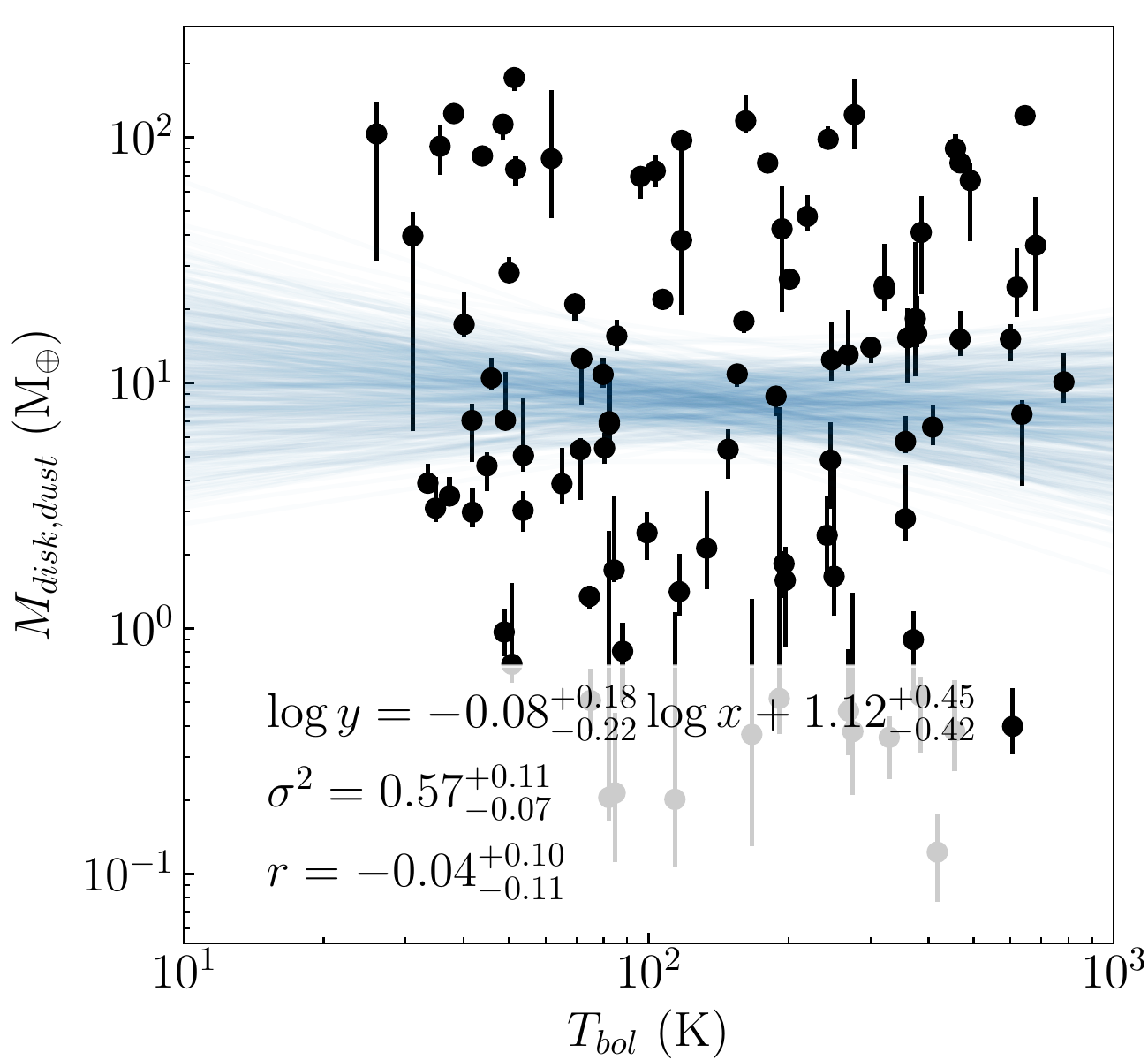}
    \includegraphics[width=2.3in]{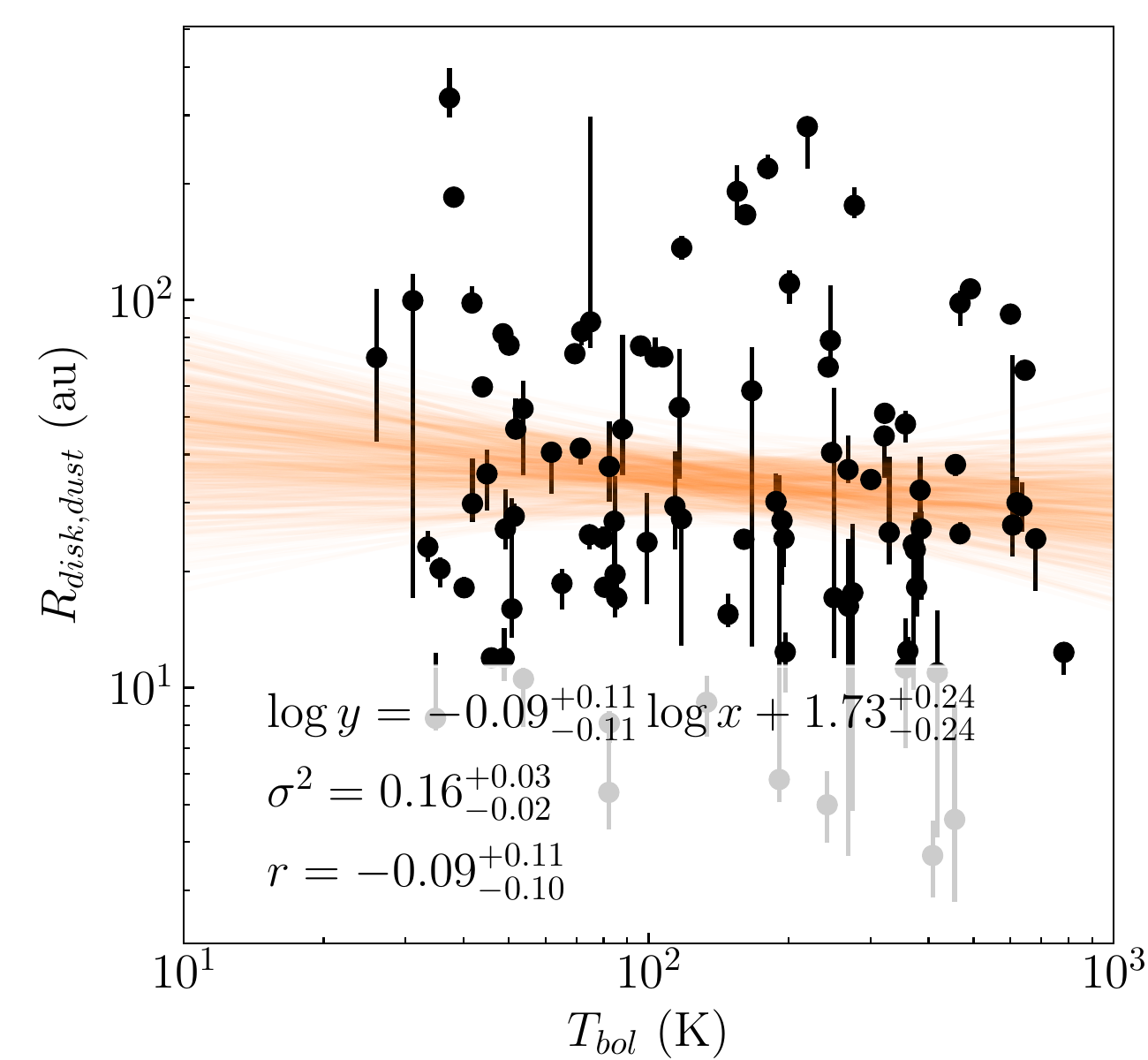}
    \includegraphics[width=2.3in]{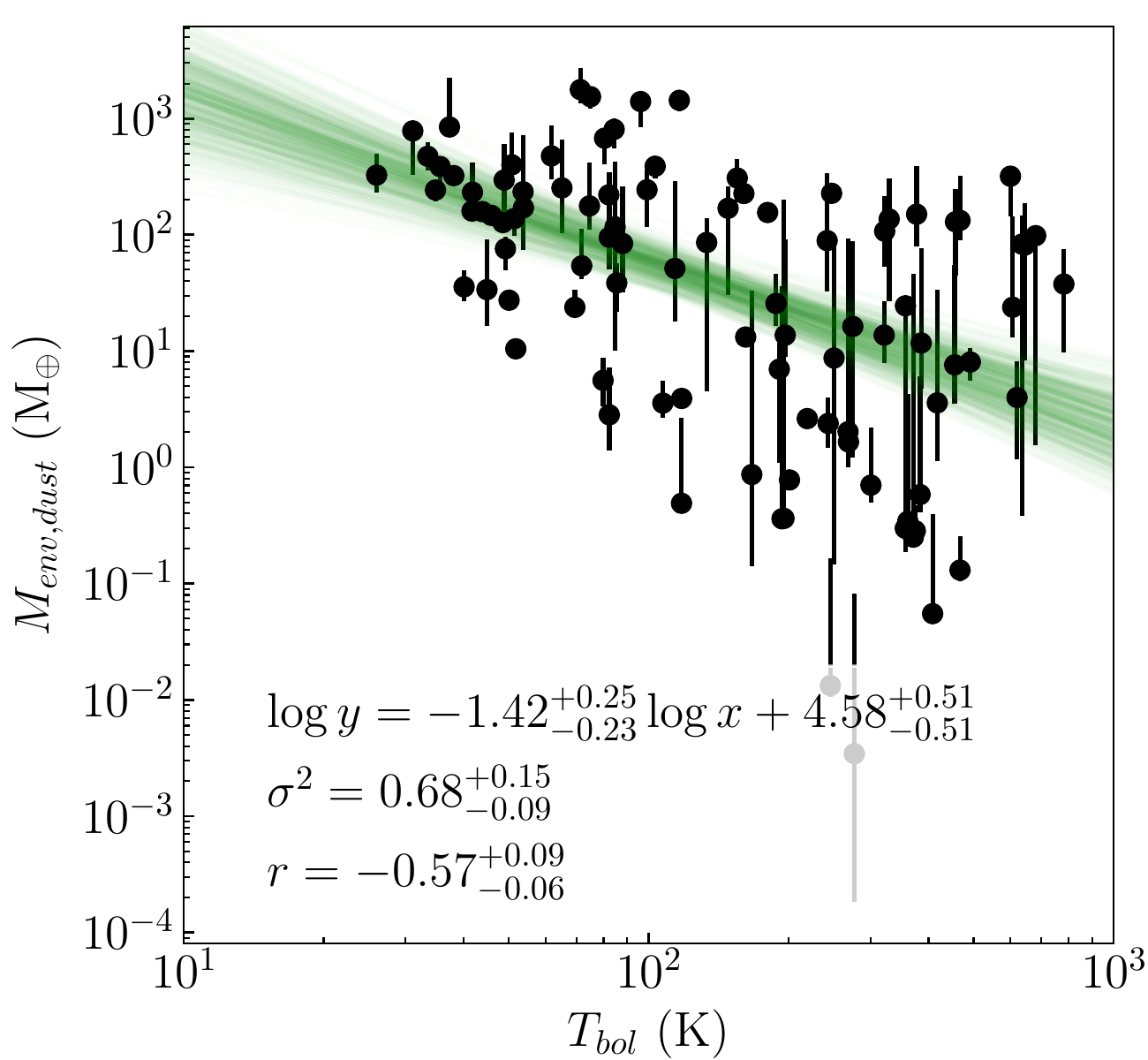}
    \includegraphics[width=2.3in]{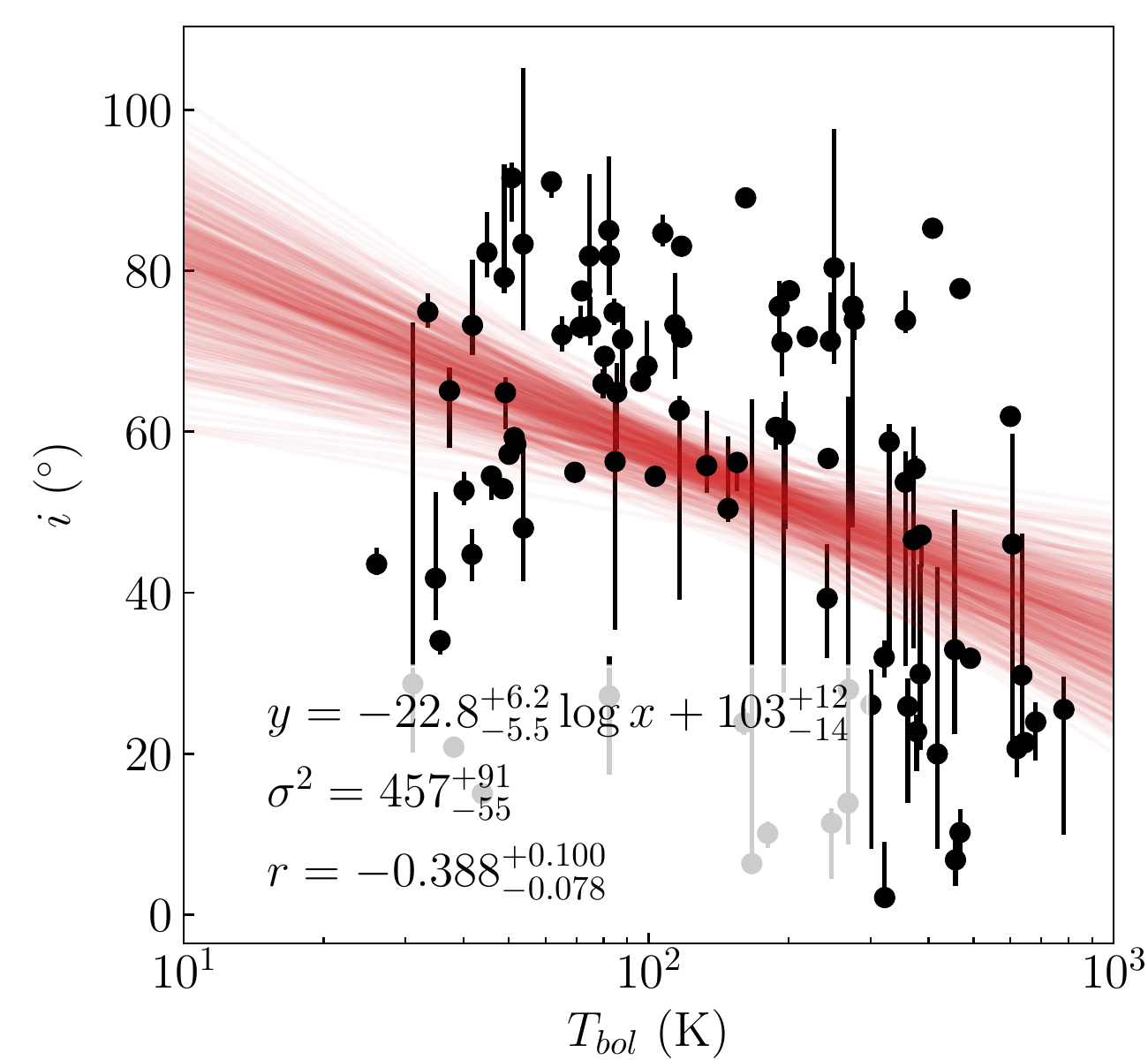}
    \includegraphics[width=2.3in]{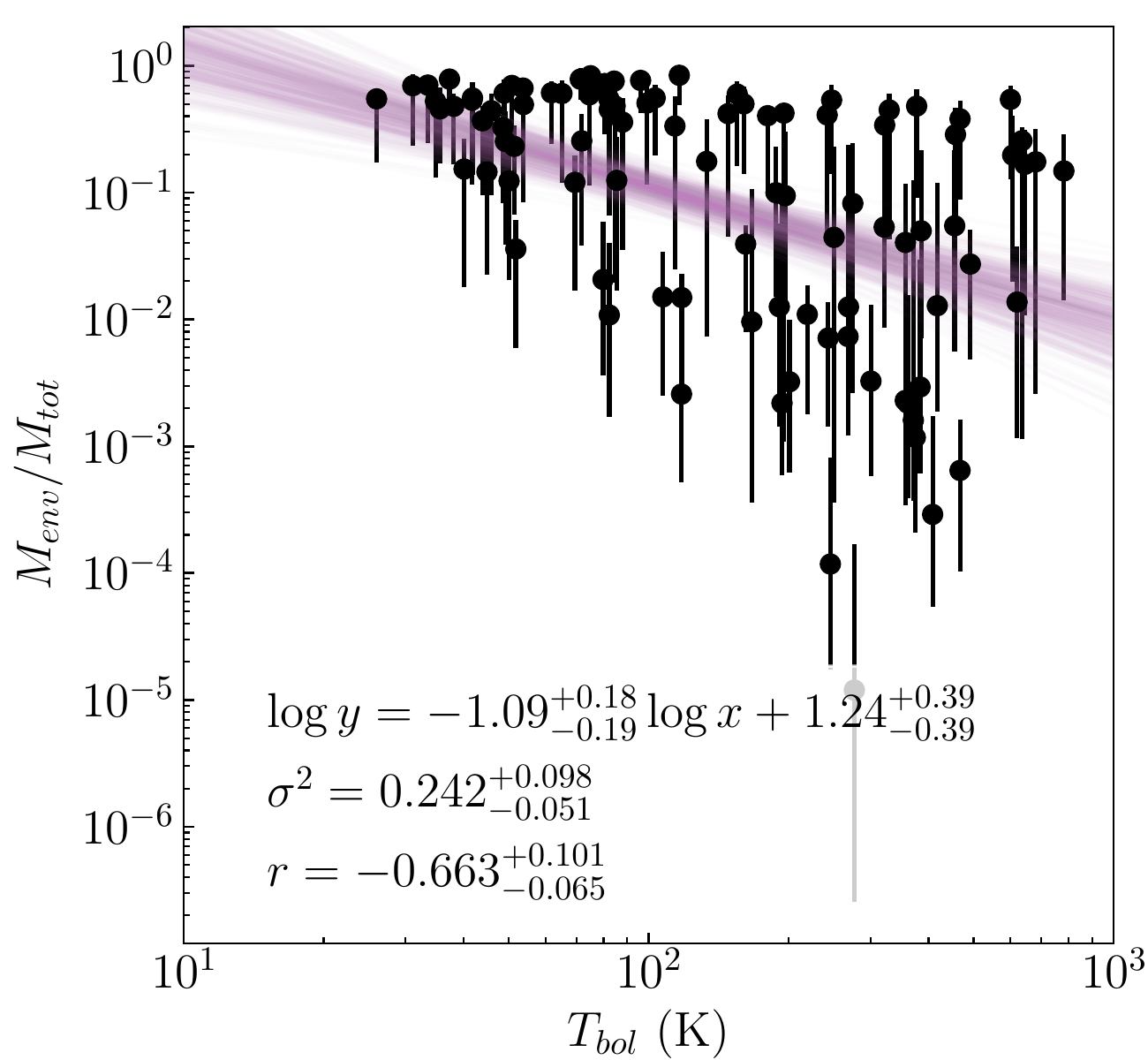}
    \caption{Distributions of a sample of parameters measured from our radiative transfer modeling framework compared with the bolometric temperature measured for each source. We also show the best fit log-linear, or semi-log-linear in the case of inclination, model for each comparison by plotting 100 samples drawn from the posterior of a Markov Chain Monte Carlo fit, and list the best fit parameters, slope, intercept, and variance, for the fit in each figure along with the correlation coefficient.}
    \label{fig:tbol_properties}
\end{figure*}

The comparisons of disk properties with $T_{bol}$ is shown in Figure \ref{fig:tbol_properties} and largely reinforces the trends, or lack thereof found in Section \ref{section:results}. Disk dust mass and radius show no statistically significant correlation with $T_{bol}$. While we do see that envelope dust mass decreases with $T_{bol}$, there remains a number of sources with large envelope dust masses for larger values of $T_{bol}$, particularly when compared with their disk dust mass. And though the relative importance of the envelope as compared with the disk, as measured by the ratio of their masses, $M_{env,dust}/M_{disk,dust}$ does decrease with increasing $T_{bol}$, there remain sources with high $T_{bol}$ that have very substantial envelopes, with $M_{env,dust}/M_{disk,dust} > 10$ or low $T_{bol}$ sources for which the importance of the envelope is perhaps less than expected ($M_{env,dust}/M_{disk,dust} < 1$).

Finally, as was discussed in Section \ref{section:results}, we find a statistically significant dependence of source inclination on bolometric temperature. As a whole, these results would seem to suggest that classifications of young stars that rely on SED diagnostics such as $T_{bol}$, may to some degree trace protostellar evolution, but are also contaminated by difficult to disentangle radiative transfer effects. Though this has, to some degree, always been known to be a difficulty of the classification scheme \citep[e.g.][]{Calvet1994FlatInfall,Chiang1999,Furlan2016}, this work provides the first clear demonstration of this for a large sample of protostars.

\subsubsection{Envelope Mass to Total Mass Ratio ($M_{env}/M_{tot}$)}
\label{section:menvmtot}

In response to the degeneracies involved in relating SED-derived properties back to underlying physical properties, \citet{Robitaille2006} instead suggested using the physical properties of a system as a more reliable method for estimating the evolutionary stage of protostars. One would naively expect that as the natal cloud collapses and material from the envelope accretes onto the disk and star, the amount of matter present in each component (envelope, star, and disk), should evolve with time. Initially, the bulk of the system material should be present in the envelope, but over time the mass of the central protostar and the disk should increase while the mass of the envelope should decrease.

This ratio may not be a perfect representation of the age of a protostellar system; $M_{env}$ (and therefore $M_{env}/M_{tot}$) may not decrease linearly, and possibly not even monotonically, with time. Simulations of global cloud collapse demonstrate that the star and disk formation process can be significantly impacted by environment, including the late accretion of envelope material onto the disk and star \citep[e.g.][]{Kuffmeier2018,Bate2018,Kuffmeier2020} and how the outflow interacts with envelope material \citep[e.g.][]{Offner2014InvestigationsEmission}. Nonetheless, $M_{env}/M_{tot}$ should roughly track the age of the system as the material in the envelope depletes and the central protostar becomes less and less obscured. More importantly, it should more exactly measure the physical properties of the system that quantities such as $T_{bol}$ are attempting to emulate, namely the ``embeddedness" of the system or how substantial of an envelope the system has.

$M_{env}/M_{tot}$ has not traditionally been used as a tracer of protostellar evolution, however, largely because of the difficulty of measuring disk and envelope masses compared with the ease of measuring bolometric temperatures and near- to mid-infrared spectral indices. However, with our detailed physical modeling we are able to measure disk and envelope masses for a sizeable population of protostars.

\begin{figure*}
    \centering
    \includegraphics[width=2.3in]{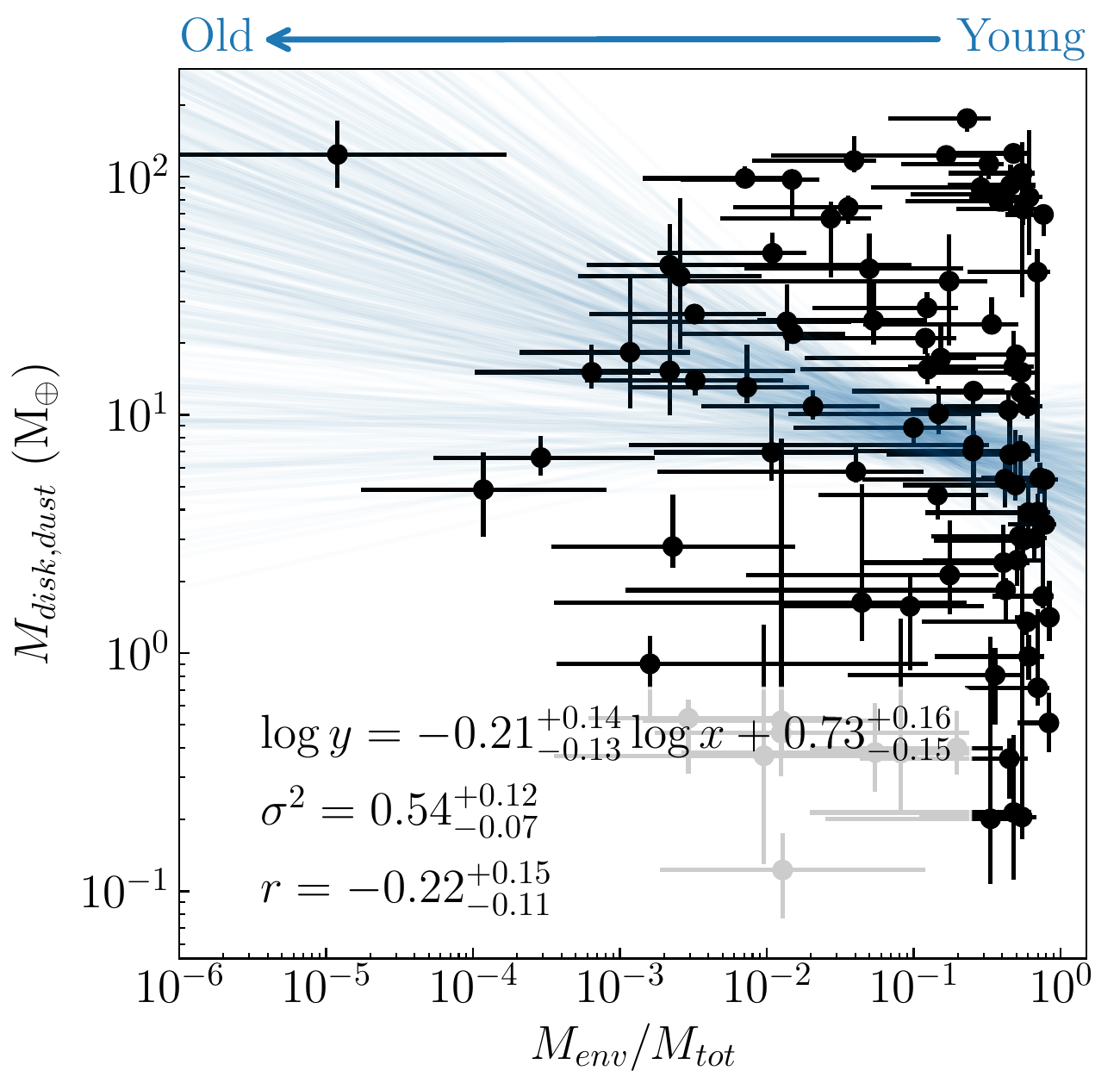}
    \includegraphics[width=2.3in]{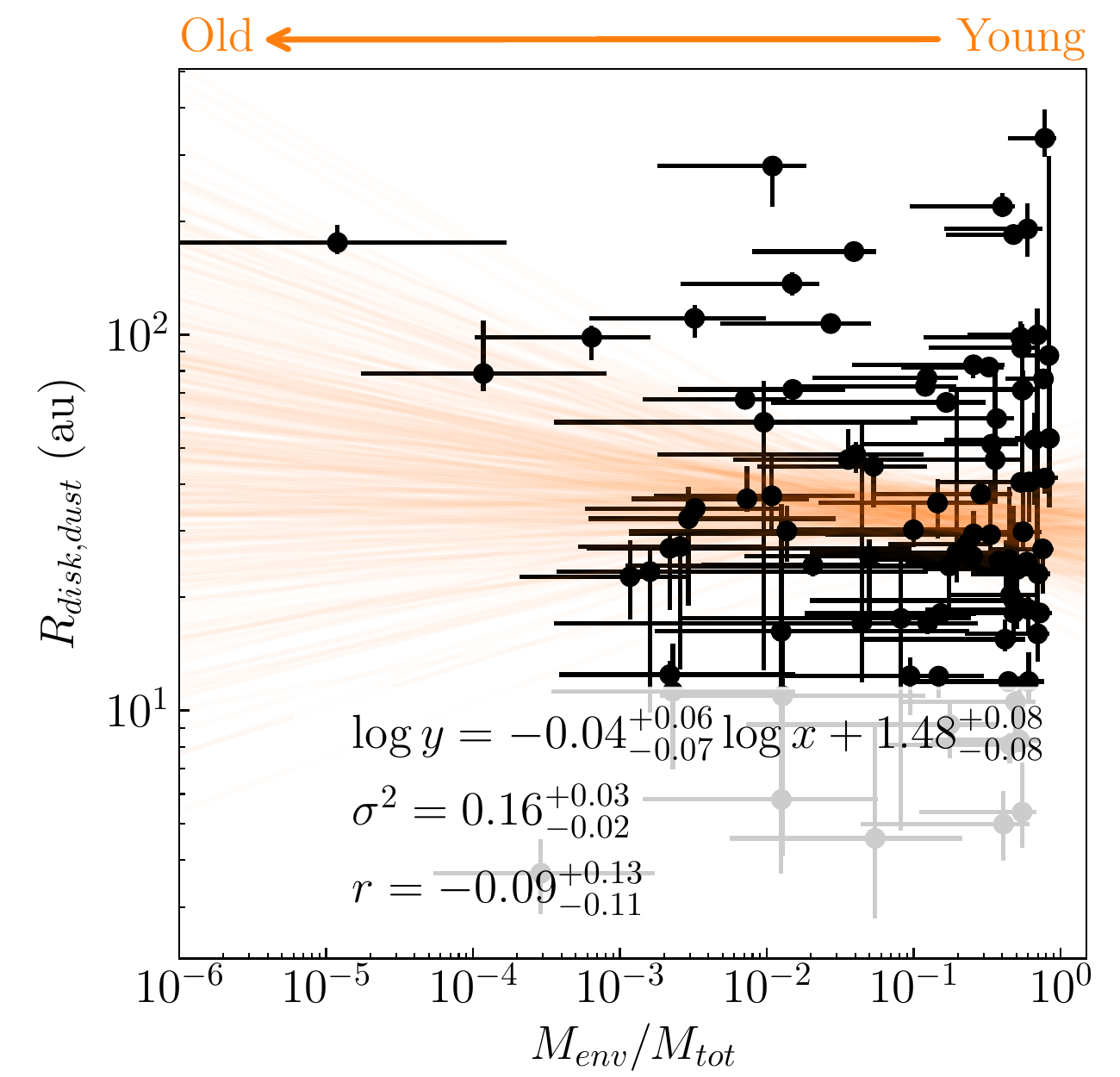}
    \includegraphics[width=2.3in]{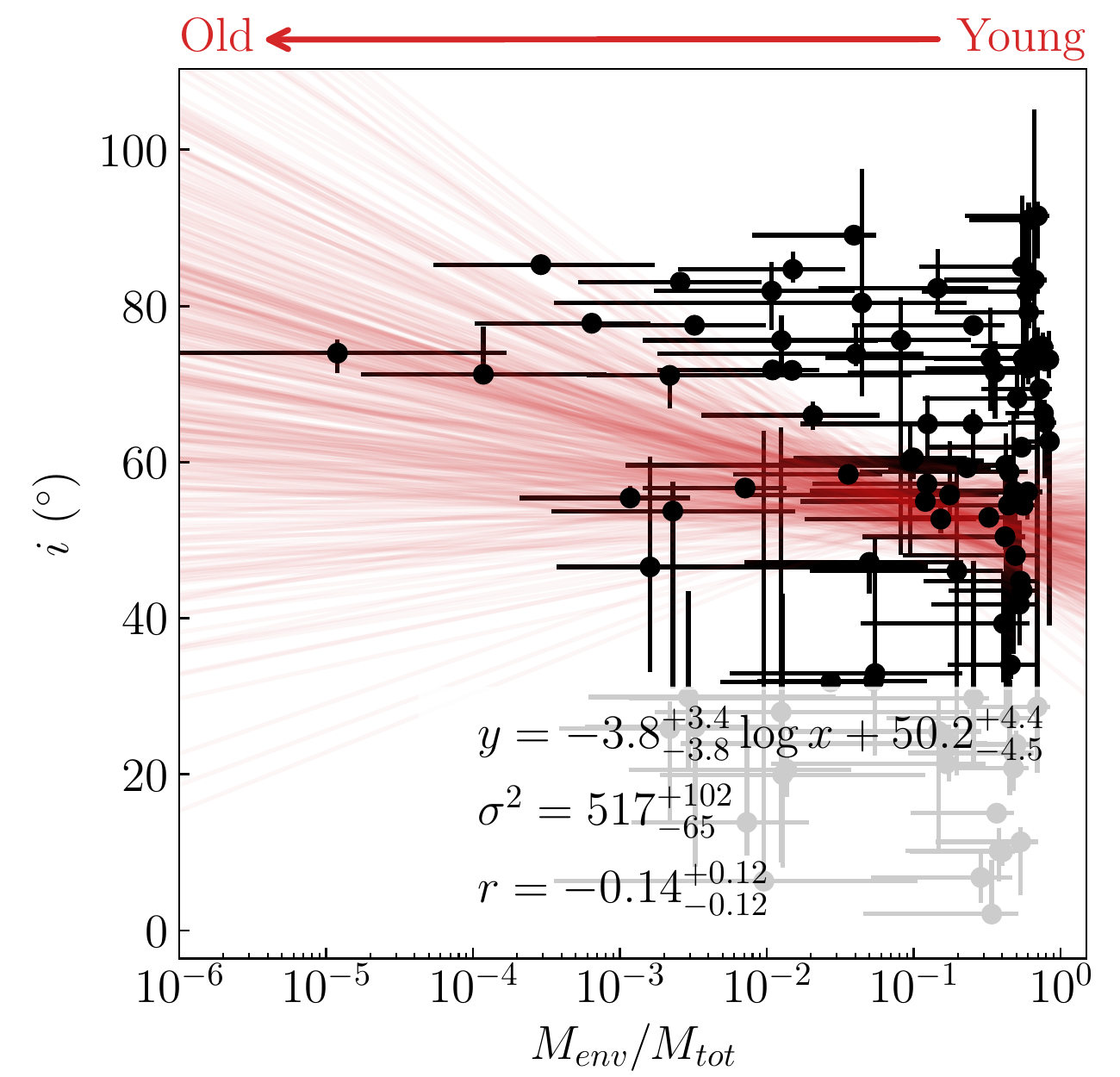}
    \caption{Distributions of a sample of parameters measured from our radiative transfer modeling framework compared with the ratio of envelope mass to total mass measured for each source. We also show the best fit log-linear, or semi-log-linear in the case of inclination, model for each comparison by plotting 100 samples drawn from the posterior of a Markov Chain Monte Carlo fit, and list the best fit parameters, slope, intercept, and variance, for the fit in each figure along with the correlation coefficient.}
    \label{fig:menvmdisk_evolution}
\end{figure*}

In Figure \ref{fig:menvmdisk_evolution}, we show how protostellar disk and envelope properties scale with $M_{env}/M_{tot}$. We remind readers that because stellar masses are inherently gas masses, we use the total (gas + dust) masses assuming a gas-to-dust ratio of 100 to calculate this ratio. Though protostellar masses have been measured kinematically for a few sources in our sample \citep[e.g.][]{Tobin2020TheOMC2-FIR3/HOPS-370}, they are unknown for the vast majority of our sample owing to the difficulty of detecting photospheric properties of embedded sources through the heavy extinction from their envelopes \citep[e.g.][]{Greene2018}. Instead, to properly and probabilistically account for the unknown masses of the majority of the sources in our sample, for each source we match the posterior distributions from our radiative transfer modeling with a sample of masses drawn from the full Chabrier initial mass function (IMF) \citep{Chabrier2003}. As such, the error bars shown in Figure \ref{fig:menvmdisk_evolution} should account for the large uncertainty in the stellar masses, using the IMF as a prior. We also assume a gas-to-dust ratio of 100 to convert the dust masses that we measure here to total masses.

We also show the results of log-linear fits to the data in Figure \ref{fig:menvmdisk_evolution} to investigate whether there are any statistically significant trends in disk properties with protostellar evolution. As the VANDAM: Orion sample was selected on the basis of $T_{bol}$ and not $M_{env}/M_{tot}$, it is likely that below some value of $M_{env}/M_{tot}$ our sample is not complete. As such, we exclude sources with $M_{env}/M_{tot} < 0.01$ from our fits, as these sources appear to be older sources with high inclinations that led to them being classified as protostars. As it is unclear where the appropriate boundary in $M_{env}/M_{tot}$ should lie we have explored more stringent cuts in $M_{env}/M_{tot}$ and find that our fits do not change substantially.

In contrast with the comparison with $T_{bol}$, we find that there is no statistically significant correlation between inclination and $M_{env}/M_{tot}$, which is to be expected as there is no reason to believe that sources at differing stages of evolution should have differing inclination distributions. Likewise, for most parameters of our model we find no statistically significant correlation with $M_{env}/M_{tot}$. 

We do find hints, at the not quite 2$\sigma$ significant level, of a trend of increasing $M_{disk,dust}$ as $M_{env}/M_{tot}$ decreases ($r = -0.22^{+0.15}_{-0.11}$), regardless of exact cutoff value for $M_{env}/M_{tot}$ we choose. Though the trend is somewhat weak, this may be an indication that disk dust mass $increases$ with time during the early stages of star formation as material from the envelope continues to accrete. We note that this is opposite the weak trends found by \citet{Tobin2020} or \citet{Andersen2019}, who used $T_{bol}$ as a proxy for evolution. 
Still, it is important to note that this trend is weak, if it exists at all, and exhibits a large amount of scatter. Modeling of a larger sample of protostellar disks may help to better determine whether the trend is real.

We again see no trend in disk dust radius with $M_{env}/M_{tot}$. This lack of a clear trend of increasing disk size with time might underscore the effects of dust radial drift relative to the gas \citep[e.g.][]{Weidenschilling1977AerodynamicsNebula,Birnstiel2010Gas-andDisks}, which may wash out the growth of the gas disk with time in observations tracing dust. Alternatively, or perhaps in addition, weak or no growth of protostellar disks with time may hint at the importance of magnetic fields in regulating the angular momentum of infalling material and preventing large disks from forming \citep[e.g.][]{Wurster2019b,Hennebelle2020WhatDiscs}. Observations of the sizes of gas disks, difficult as they may be, might be needed to help further illuminate which physics regulates the sizes and evolution of young disks.

It is important to note that while the IMF is an appropriate representation of the final masses of stars, it may not be an appropriate distribution to represent the masses of protostars in our sample that are young and still forming. It would be ideal, instead, to sample from a protostellar mass function (PMF) that accounts for the masses of these protostars as they form \citep[e.g.][]{McKee2010}, or better yet to use directly measured dynamical masses \citep[e.g.][]{Tobin2012,Tobin2020TheOMC2-FIR3/HOPS-370} for the sample. In the absence of either, however, we hope that the use of the IMF enables a reasonable estimate for the protostellar mass with broad errorbars to account for and propagate the large uncertainty in the estimate. Nonetheless, the lack of measured masses and our use of the IMF in their place could have an impact on the correlations, or lack thereof, found here. Future efforts to provide more realistic estimates of the sources in this sample are crucial for refining this analysis to search for evolutionary trends.

\subsubsection{A Simple Model of Protostellar Evolution}
\label{section:evolutionary_tracks}

Finally, we also consider a simple model of protostellar evolution \citep[e.g.][]{Andre1994,Saraceno1996AnObjects.,Andre2008FirstCamera,Fischer2017Evolution} to produce a set of evolutionary tracks in the $L_{bol} - M_{env}$ plane and compare the tracks with those values as inferred from our modeling. The model uses a simple prescription for how the envelope mass, stellar mass, and mass accretion rate evolve with time, and then ties to pre-main sequence evolutionary tracks to estimate properties of the protostar such as radius or luminosity.

\begin{figure}[b]
    \centering
    \includegraphics[width=3.3in]{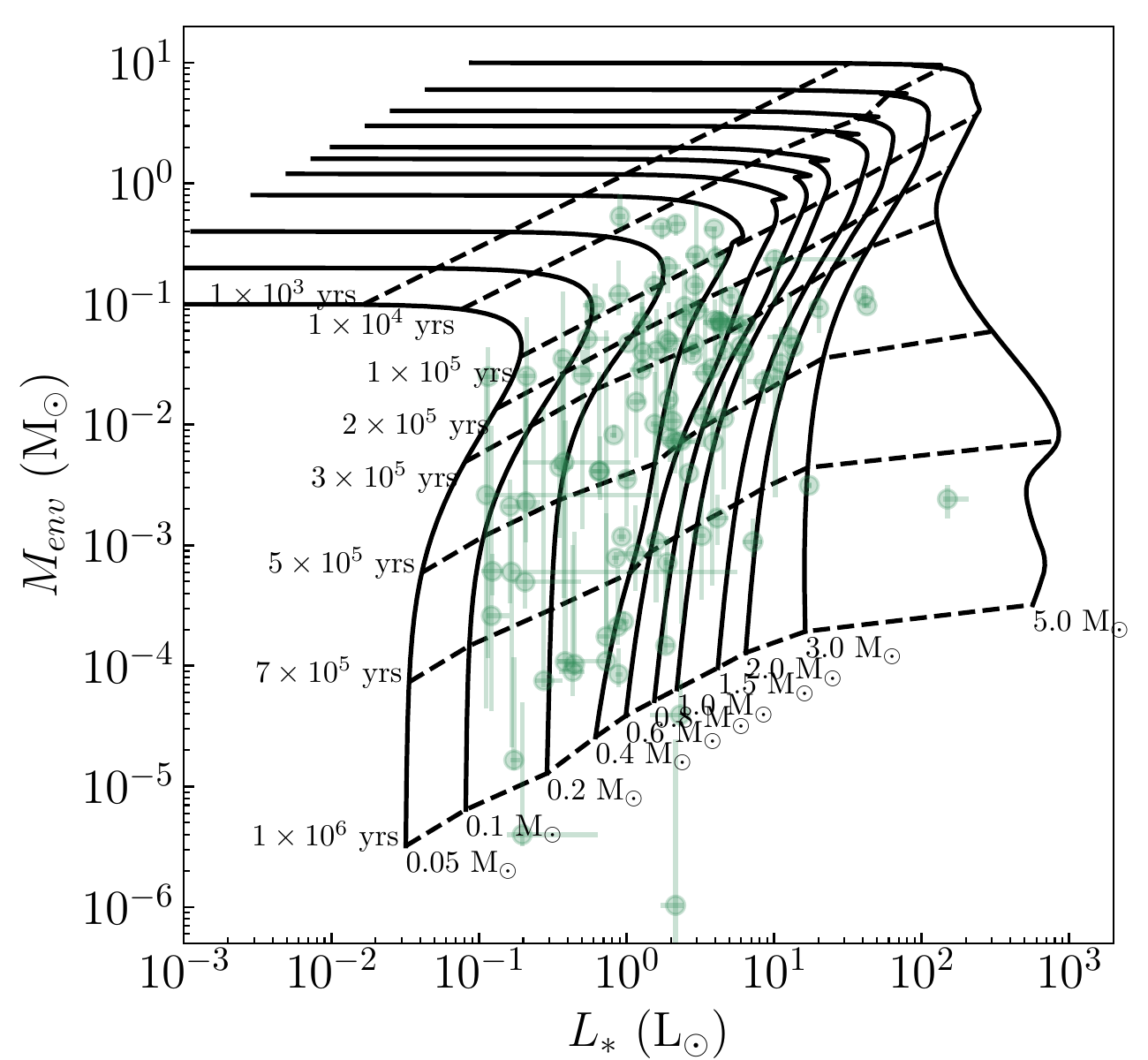}
    \caption{A demonstration of the $L_{*} - M_{env}$ evolutionary tracks described in Section \ref{section:evolutionary_tracks}. The protostellar evolutionary tracks are shown in black, with the curves of a constant final mass shown as solid lines and isochrones shown as dashed lines. The sources in our modeled sample are shown in as green points.}
    \label{fig:evolutionary_tracks}
\end{figure}

\begin{figure*}
    \centering
    \includegraphics[width=2.3in]{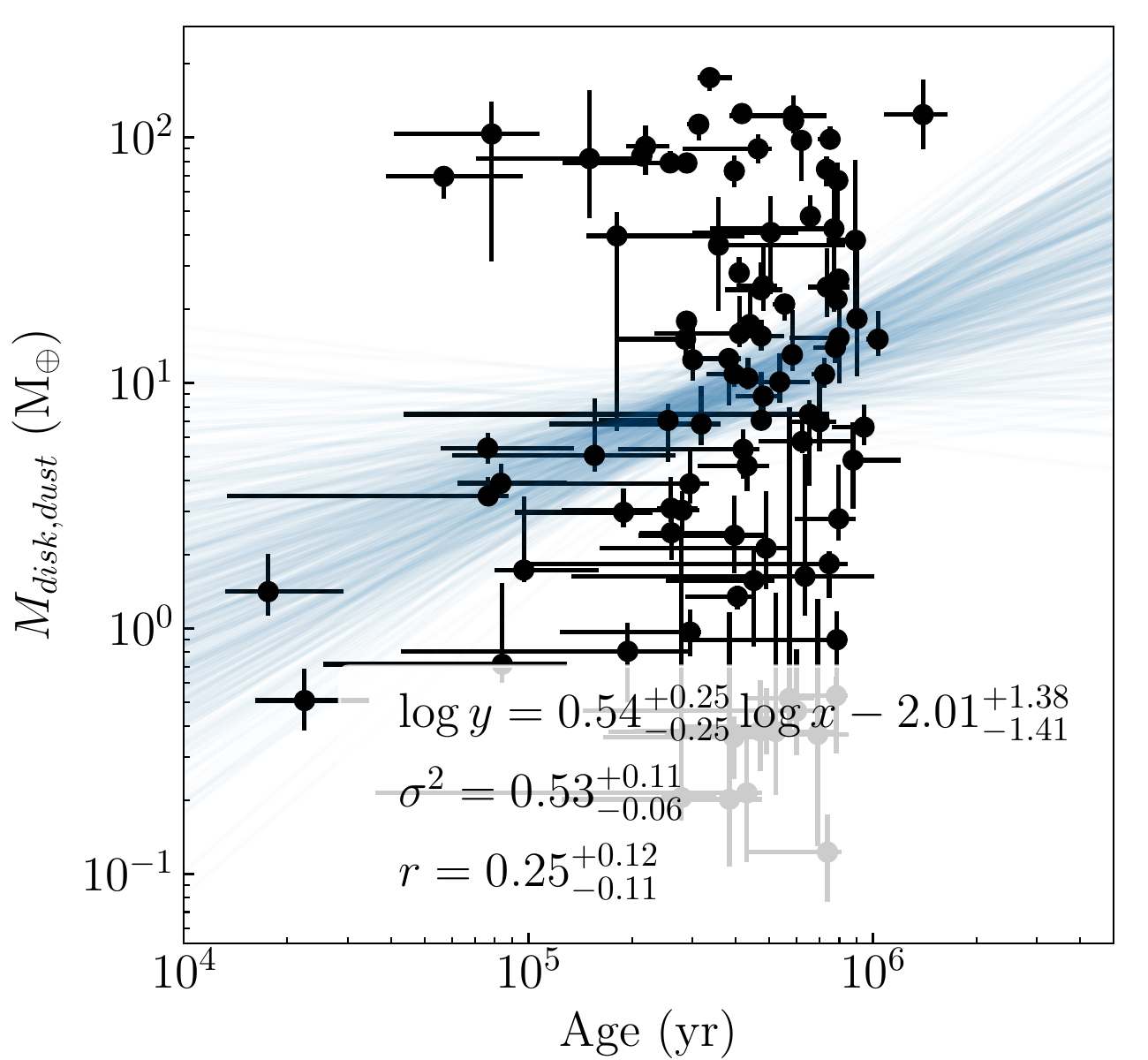}
    \includegraphics[width=2.3in]{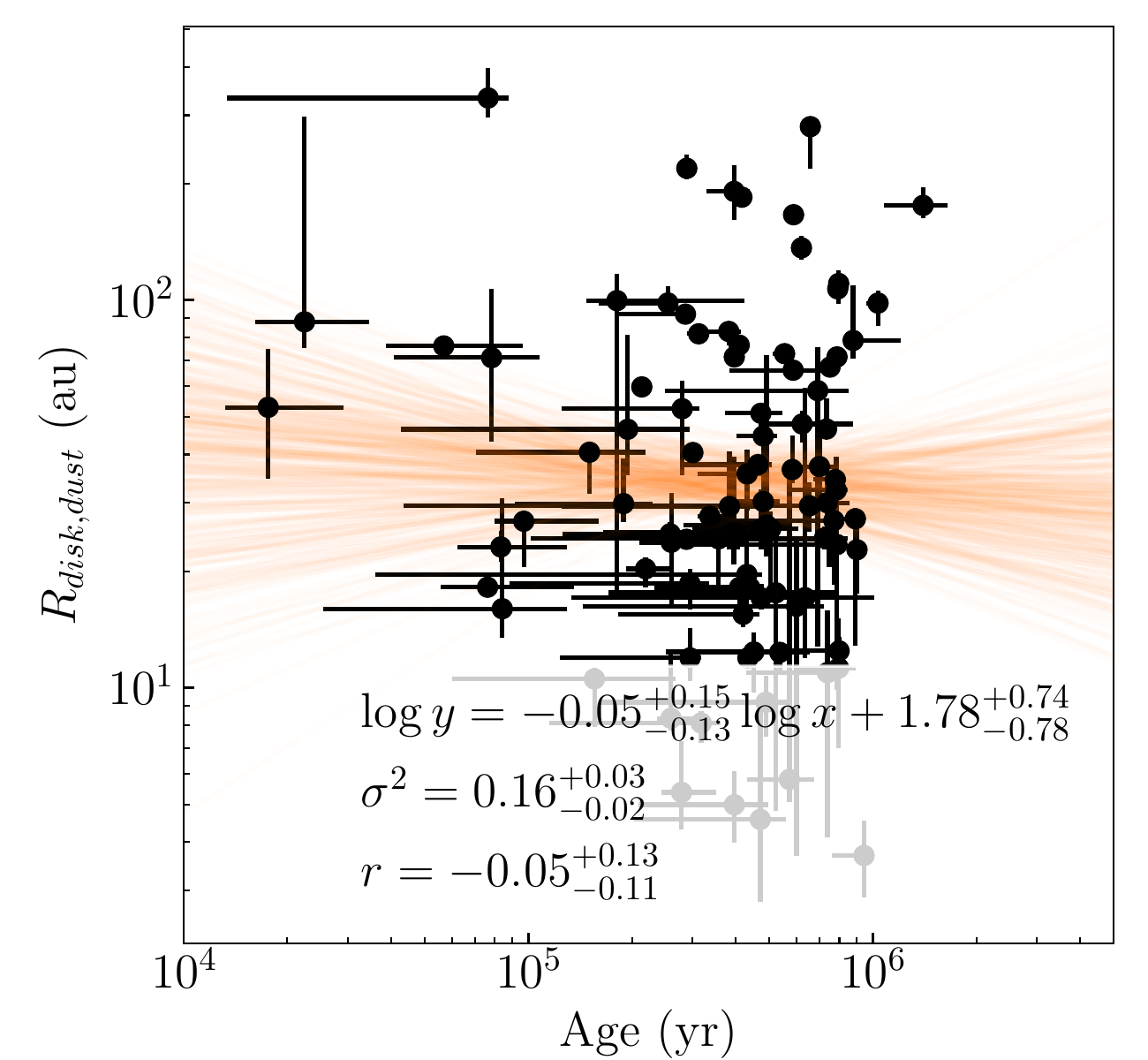}
    \includegraphics[width=2.3in]{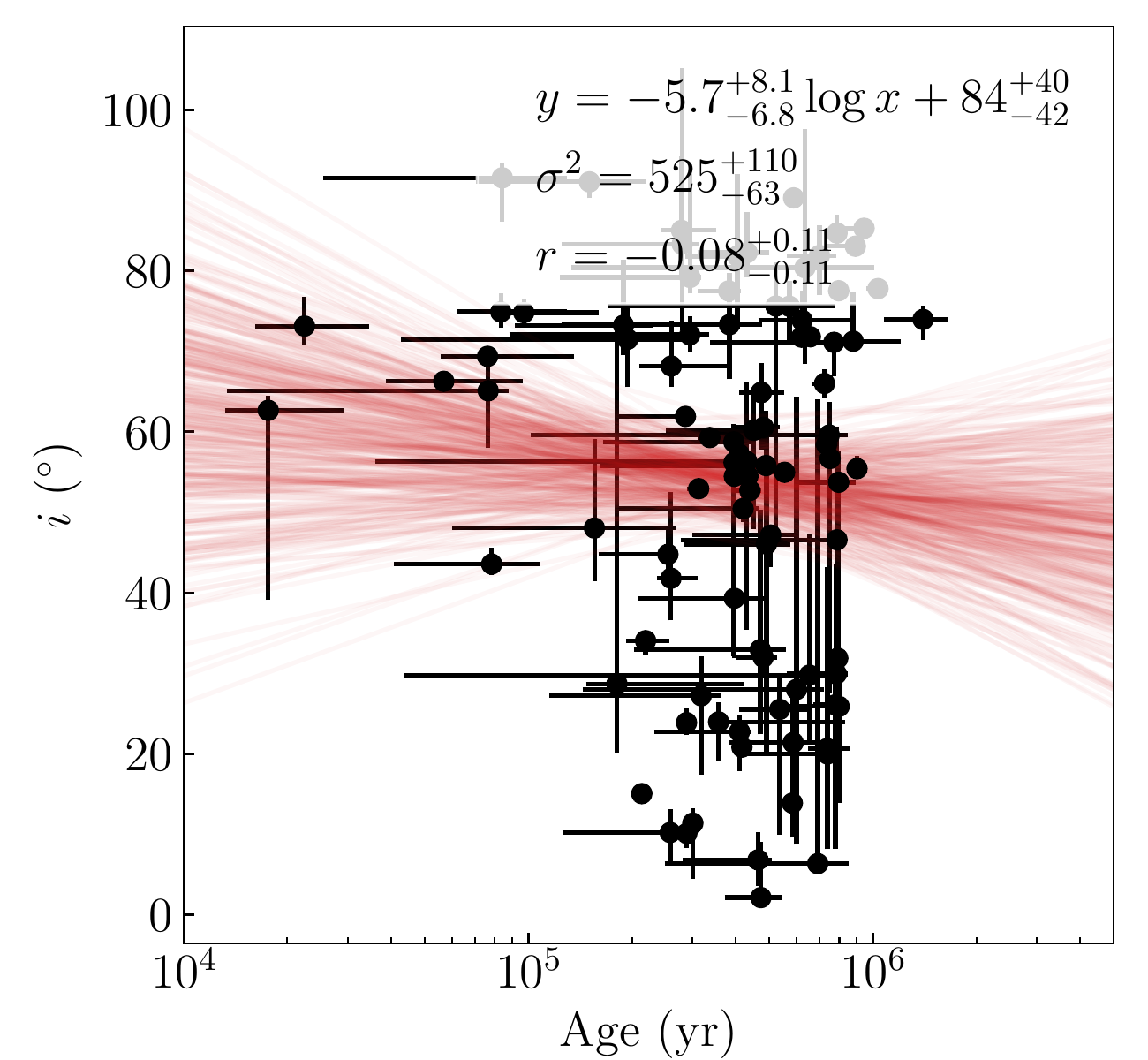}
    \caption{Distributions of a sample of parameters measured from our radiative transfer modeling framework compared with the age measured for each source using the simple protostar evolutionary tracks shown in Figure \ref{fig:evolutionary_tracks}. We also show the best fit log-linear, or semi-log-linear in the case of inclination, model for each comparison by plotting 100 samples drawn from the posterior of a Markov Chain Monte Carlo fit, and list the best fit parameters, slope, intercept, and variance, for the fit in each figure along with the correlation coefficient.}
    \label{fig:tracks_comps}
\end{figure*}

To generate the tracks, we follow the prescription of \citep{Andre2008FirstCamera}. For a given final stellar mass, $M_{star}$, we assume that star is formed from a core with mass $M_{env,0} = M_*/\epsilon$, where $\epsilon$ is the star formation efficiency of individual cores. At any given time, the instantaneous mass accretion rate is given by $\dot{M}_{acc} = \epsilon M_{env}/\tau$, where $\tau$ is the characteristic timescale for protostellar evolution. We create an array of times from $t = 0$ to $t = 10$ Myr, spaced logarithmically. At any given time, $t_i$, the mass of the star, the mass of the envelope, and the accretion rate can be calculated from the masses and accretion rate at the previous time step from 
\begin{equation}
    \label{eq:mstar}
    M_{star,i} = M_{star,i-1} + \dot{M}_{acc,i-1} (t_i - t_{i-1}),
\end{equation}
\begin{equation}
    \label{eq:menv}
    M_{env,i} = M_{env,i-1} - \dot{M}_{acc,i-1}/\epsilon (t_i - t_{i-1}),
\end{equation}
\begin{equation}
    \label{eq:macc}
    \dot{M}_{acc,i} = \epsilon M_{env,i} / \tau.
\end{equation}
Following \citet{Andre2008FirstCamera}, we assume that $\epsilon = 0.5$ \citep[e.g.][]{Matzner2000EfficienciesFormation,McKee2002TheCores} and $\tau = 1 \times 10^{5}$ years.

To determine the total luminosity at any given time, we calculate the contributions of the accretion luminosity and protostellar luminosity ($L_{*} = L_{acc} + L_{star}$). In this way, $L_{*}$ is analogous to the same parameter in our radiative transfer models, which accounts for the total luminosity that is reprocessed and re-emitted across the entire spectrum, as discussed at the beginning of Section \ref{section:modeling}. The accretion luminosity at any given time is given by
\begin{equation}
    L_{acc} = \eta \frac{G M_{star} \dot{M}_{acc}}{R_{star}}.
\end{equation}
In this equation, $M_{star}$ and $\dot{M}_{acc}$ are calculated in Equations \ref{eq:mstar} -- \ref{eq:macc}, $G$ is the gravitational constant, and we use $\eta = 1$ in line with \citet{Andre2008FirstCamera}. To calculate $R_{star}$ and $L_{star}$ at any given time, we use the evolutionary tracks from \citep{Feiden2016} from the current mass ($M_{star,i}$) and age ($t_i$). When stars have mass $<0.1$ M$_{\odot}$, below the lower limit of the evolutionary tracks, and also of hydrogen burning, we use the scaling relationships found by \citet[][$R_* \propto {M_{star}}^{0.34}$ and $L_{star} \propto {M_{star}}^{1.34}$]{Fischer2017Evolution} to scale the evolutionary tracks from 0.1 M$_{\odot}$ to the relevant stellar mass. Though approximate, this should give a rough idea of the properties of stars that fall outside of the boundaries of the evolutionary tracks.

With this setup, we calculate evolutionary tracks in the $L_{bol} - M_{env}$ plane for a range of protostar masses from 0.01 -- 5 M$_{\odot}$. We show these evolutionary tracks and how the sources in our sample compare in Figure \ref{fig:evolutionary_tracks}. As these tracks involve stellar mass, we again use the total envelope mass, assuming a gas-to-dust ratio of 100, from our modeling for comparison with the tracks. Our tracks are qualitatively similar to previous works \citep[e.g.][]{Andre1994,Saraceno1996AnObjects.,Andre2008FirstCamera,Fischer2017Evolution}, though we note that their exact values may differ from work to work due to the slightly different assumptions used by each study.

To estimate the age of the protostars in our sample with these evolutionary tracks, we use two dimensional linear interpolation to map any given $(L_{*},M_{env})$ pair onto the tracks and thereby infer the age, as demonstrated in Figure \ref{fig:evolutionary_tracks}. For each source, we repeat this process for every sample point in the posterior distribution to produce a distribution of ages for the source. The best-fit age of the source is then determined as the peak of the posterior distribution of ages, with the uncertainties from the 95\% confidence interval around those ages.

We show how a selection of protostellar disk and envelope properties, as inferred from our modeling, compare with the age inferred from these evolutionary tracks in Figure \ref{fig:tracks_comps}. We also show the results of a log-linear fit to the data in each comparison. To ensure that the comparison is not affected by completeness of the sample in age, as the sample was selected based on $T_{bol}$, we only consider sources with inferred ages $>10^{4.5}$ years to exclude sources that are outliers in age, though find that the choice of cutoff does not significantly affect our results.

The results presented in Figure \ref{fig:tracks_comps} are very similar to what we found when using $M_{env}/M_{tot}$ as an evolutionary tracer. We find no statistically significant correlation between inclination and age inferred from the evolutionary tracks, which is, again, sensible as there is no reason to believe that sources of different ages should have preferentially different inclinations. We also again find that the disk dust mass actually increases with time, albeit with a very large spread, suggesting that the dust mass of the disk tends to grow as material continues to fall onto the disk from the envelope. We do note that the correlation is only slightly more than $\sim2\sigma$ significant ($r = 0.25^{+0.12}_{-0.11}$), though, so it would be useful to add additional sources to this analysis to confirm the trend with higher significance. Finally, we again do not find a statistically significant trend in disk dust radius as a function of age, indicating little or no growth of the disk with time, though whether this lack of growth is due to radial drift, magnetic fields sapping angular momentum of infalling material, or something else remains open.

\begin{figure*}[t]
    \centering
    \includegraphics[width=3.5in]{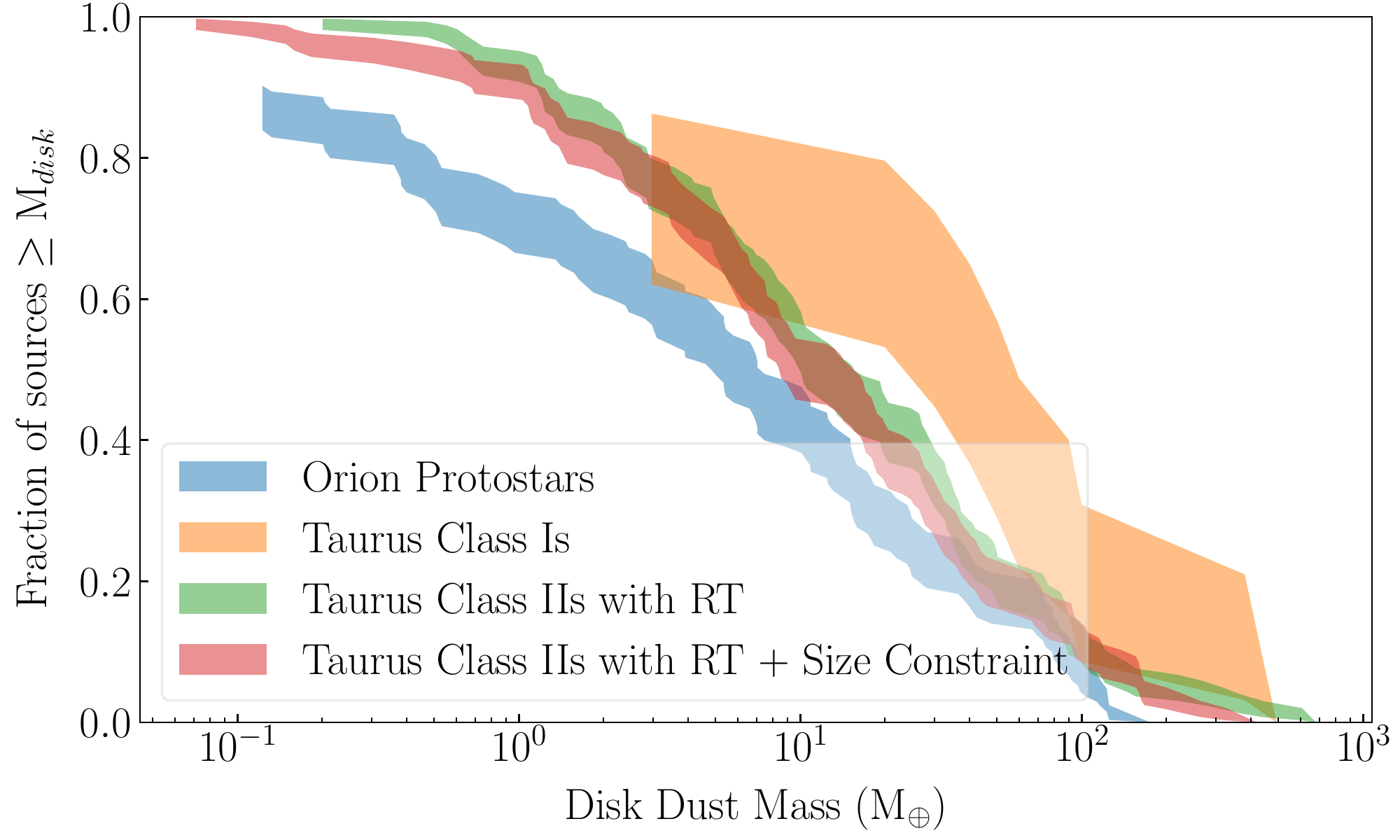}
    \includegraphics[width=3.5in]{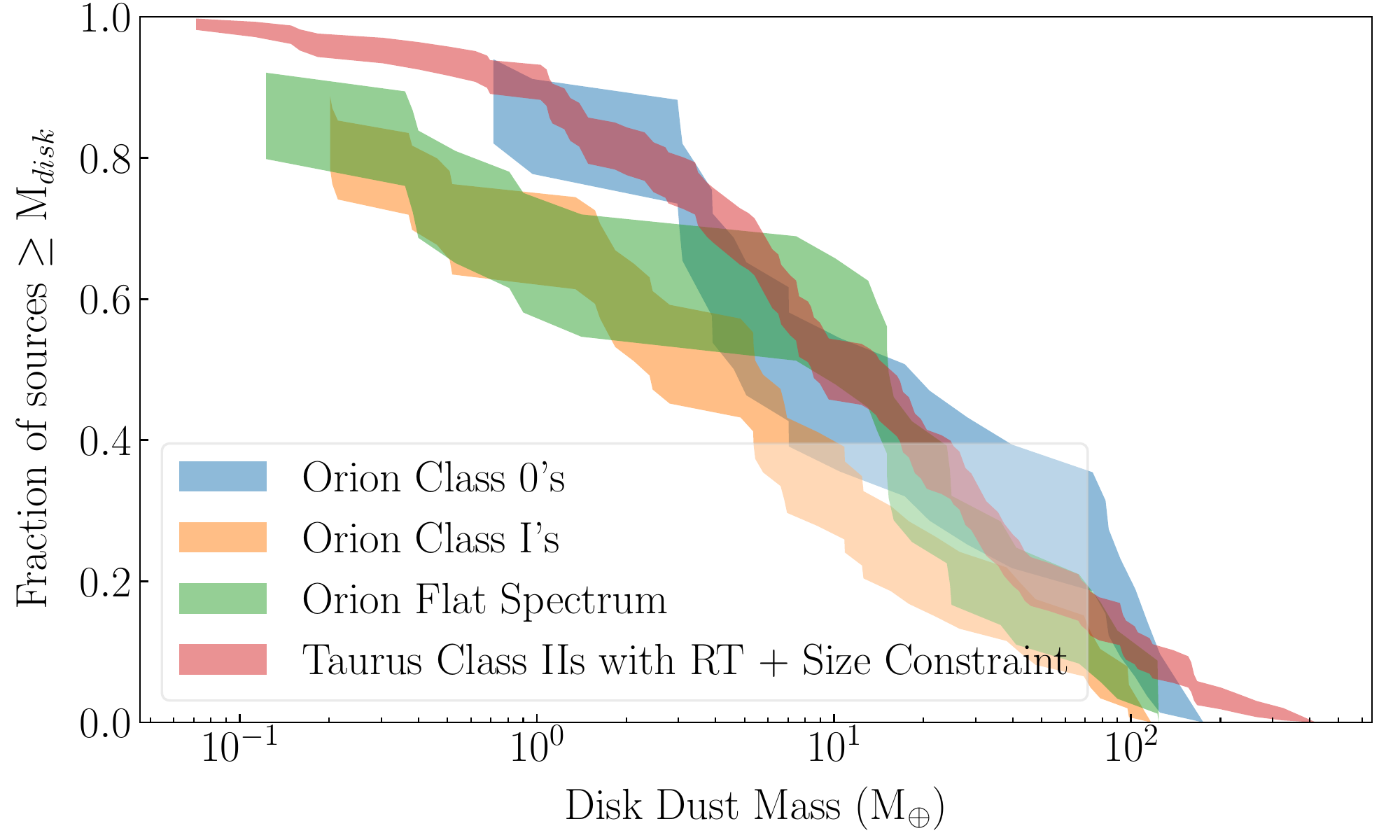}
    \caption{($left$) Cumulative distributions of protostellar disk dust masses as found from the samples of this work, \citet{Sheehan2017}, and \citet{Ballering2019ProtoplanetaryTaurus} using radiative transfer modeling. The distributions from \citet{Ballering2019ProtoplanetaryTaurus} correspond to the disk dust masses derived with and without using the disk mass-disk radius relation found by \citet{Tripathi2017} as a prior in the modeling. ($right$) Cumulative distribution of protostellar disk dust masses from this work, split by classification, compared with the Class II disk dust mass distribution for Taurus, using the disk mass-disk radius relation as a size constraint in the modeling.}
    \label{fig:rt_comps}
\end{figure*}

We do caution, however, that the ages inferred from these evolutionary tracks should be treated as rough estimates of the ages of these sources and not absolute values. The model used to generate the tracks is quite simple and ignores the potentially complicated accretion history of protostellar envelopes. Moreover it does not account for the possibility of episodic accretion \citep[e.g.][]{Kenyon1990,Dunham2010}. Outbursts in young sources are known to occur \citep[e.g.][]{Safron2015,Yoo2017,Lee2020}, but it is, for the most part, unknown whether any of the sources in our sample could be in an outburst state. If any were, they would appear more luminous than these tracks would predict and therefore we would infer the wrong age. Nonetheless, they should give some idea of the rough age of the systems in a physically motivated way.

\subsection{Comparison with Other Star Forming Regions}

Continuum surveys of protoplanetary disks have become routine with ALMA over the past decade, with a number of different star forming regions encompassing a range of ages and environments having been surveyed \citep{Pascucci2016,Barenfeld2016,Ansdell2016,Eisner2018,Ansdell2018,Cieza2019}. Such surveys have dramatically improved our understanding of the bulk properties of protoplanetary disks, altered our view of the planet forming potential of these disks \citep[e.g.][]{Najita2014,Manara2018}, demonstrated that protoplanetary disk dust masses decrease with age \citep[e.g.][]{Pascucci2016,Ansdell2018,Williams2019} with some interesting exceptions \citep[e.g.][]{Cazzoletti2019ALMAMasses}, and have also found evidence that environment may play some role in disk evolution \citep[e.g.][]{Eisner2018,VanTerwisga2020ProtoplanetaryPopulations,Ansdell2020AnFormation}.

Surveys have also begun to push towards younger, Class 0/I protostellar disks, including the VANDAM: Orion Survey studied here, but also as a part of the ODISEA Survey \citep[][]{Cieza2019}, the MASSES Survey \citep[][]{Stephens2018MassRelease}, and the VANDAM: Perseus Survey \citep[e.g.][]{Tobin2016} and subsequent ALMA follow-up \citep[][]{Tychoniec2020}. These surveys have painted a picture of protostellar disks as more massive than protoplanetary disks, or at least brighter \citep[e.g.][]{Sheehan2017,Williams2019,Andersen2019,Tobin2020}, suggesting that they may better represent the initial mass budget in disks for forming planets \citep[][]{Sheehan2017,Tychoniec2020}.

\citet{Tobin2020} compared the disk dust masses and radii measured for the VANDAM: Orion survey using two-dimenstional Gaussian fits with the results of existing millimeter surveys of protoplanetary disks employing similar methods in great detail. As we demonstrated in Section \ref{section:simple-estimators}, the fluxes measured by \citet{Tobin2020} are generally in good agreement with the disk fluxes derived by our modeling here, and as the surveys described above generally use similar methodologies to \citet{Tobin2020}, we opt not to rehash those comparisons here as we would not expect those results to change substantially other than from our smaller sample size. Instead, we compare the results of our modeling with the results of two other studies, \citet{Sheehan2017} and \citet{Ballering2019ProtoplanetaryTaurus}, that have done similar radiative transfer modeling 
to account for more physics in deriving disk properties. As such, comparing with these surveys should provide the most fair comparison of disk properties across different star forming regions and states of evolution.

\citet{Sheehan2017} employed the same radiative transfer framework we use here to model a sample of 10 Class 0/I protostars in Taurus observed with the Combined Array for Research in Millimeter-wave Astronomy (CARMA) and separate disk and envelope contributions. They found that the disk-only fluxes were on-average greater than the fluxes of Class II disks by a factor of a few, though with only 10 sources the result was only significant at the 2.5$\sigma$ significance level. \citet{Ballering2019ProtoplanetaryTaurus} used a similar radiative transfer modeling framework to model 132 Class II disks, also in the Taurus star forming region, finding that the dust masses that they derived from their modeling were $1-5\times$ higher than the dust masses found from the simple flux-to-mass conversion used more generally. Though their modeling did not include spatial information by fitting resolved millimeter observations directly, they did repeat their fits both including the disk mass-disk size relation measured by \citet{Tripathi2017} as a prior and also leaving the disk radius unconstrained. We also note that \citet{Ribas2020ModelingMasses} found a similar result an independent modeling analysis of a subset of 23 of these same disks.

In Figure \ref{fig:rt_comps}, we show a comparison between the distributions of disk dust masses derived here as well as in \citet{Sheehan2017} and in \citet{Ballering2019ProtoplanetaryTaurus}. For the results from \citet{Ballering2019ProtoplanetaryTaurus}, we show the distribution both with and without the size constraint. To rigorously compare the distributions, we use the \texttt{lifelines} package to run a log-rank two sample test. We find that the Orion protostellar disk dust masses can be statistically distinguished from both the Taurus Class Is (p-value = 0.009) as well as the Taurus Class IIs whether a size constraint is included (p-value = 0.0003) or not (p-value = $1.7\times10^{-6}$).

\begin{figure*}[t]
    \centering
    \includegraphics[width=3.5in]{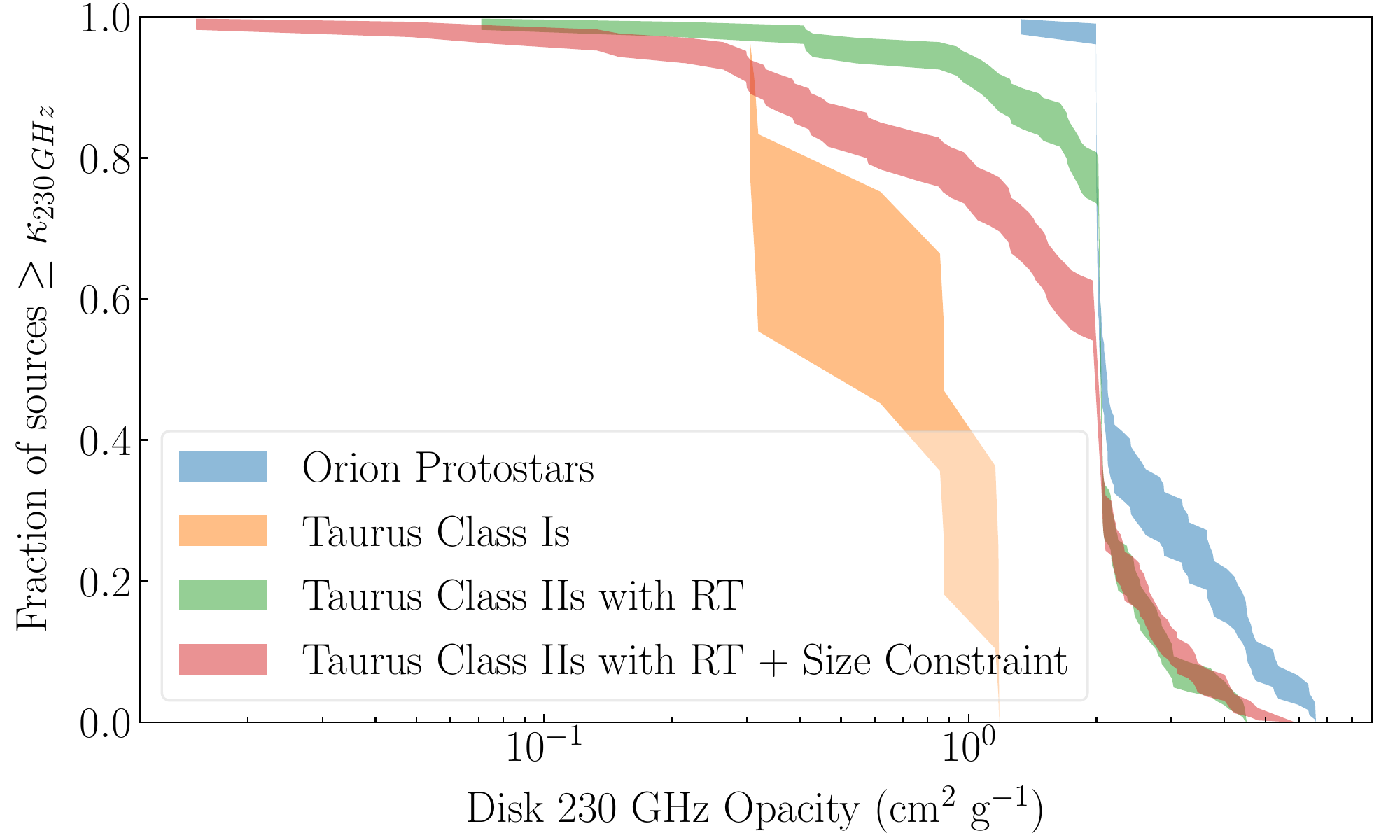}
    \includegraphics[width=3.5in]{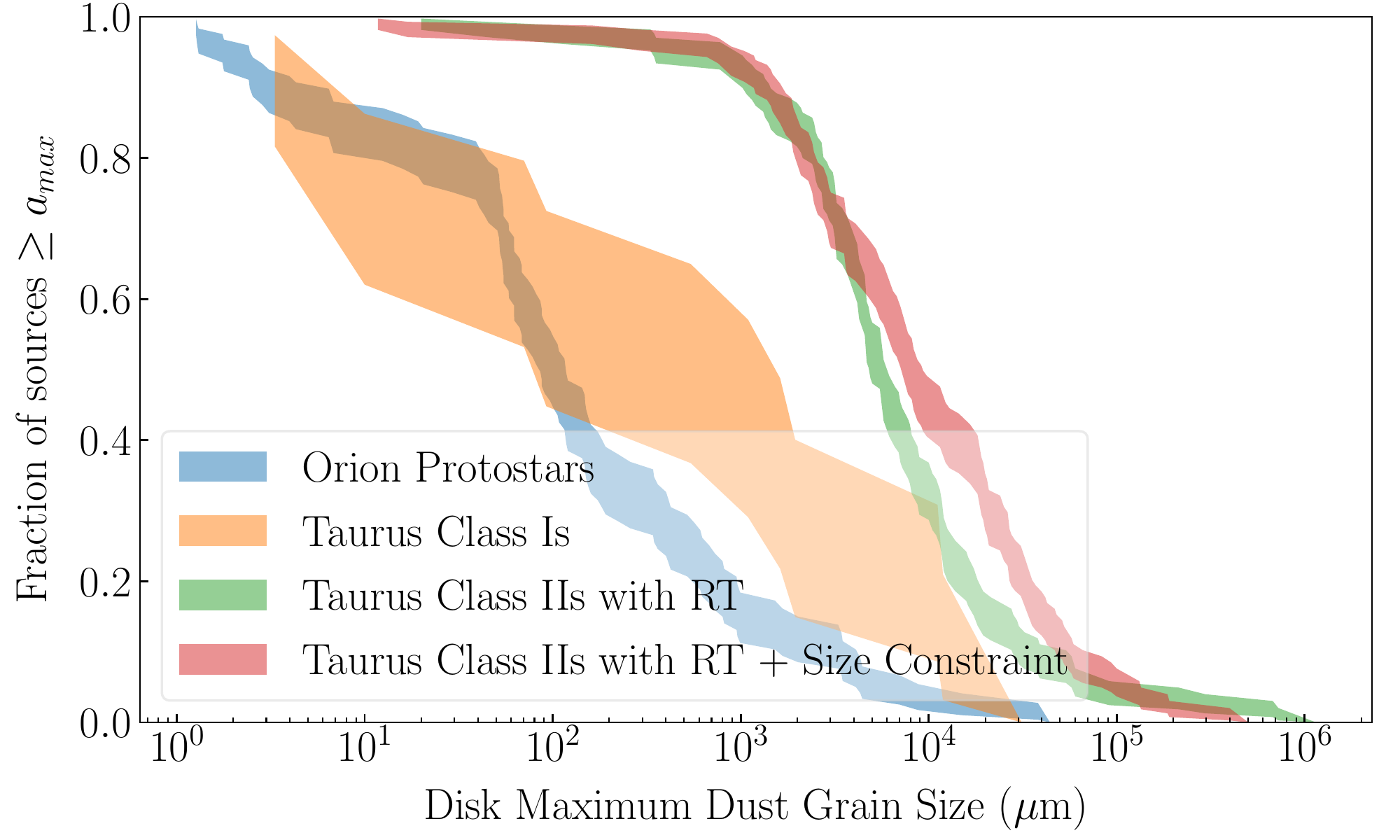}
    \caption{($left$) Cumulative distributions of protostellar disk 230 GHz dust opacity as found from the samples of this work, \citet{Sheehan2017}, and \citet{Ballering2019ProtoplanetaryTaurus} using radiative transfer modeling. The distributions from \citet{Ballering2019ProtoplanetaryTaurus} correspond to the fits that did and did not use the disk mass-disk radius relation found by \citet{Tripathi2017} as a prior in the modeling. ($right$) Same as on the left, but for maximum dust grain size.}
    \label{fig:rt_comps_dust}
\end{figure*}

If we split the protostellar disks modeled here into their Class 0/I/Flat Spectrum designations, we find that Class 0 disks have a distribution of disk dust masses that is consistent with the Class II disks (p-value = 0.58), while both the Class I and Flat Spectrum distributions are distinct (p-values of 0.0003 and 0.03, respectively). The Class I distribution, interestingly, mirrors the Class 0/Taurus Class II distributions at high dust masses but then has a knee that better follows the Class I disk dust masses.

Interestingly, and in contrast to previous works \citep[e.g.][]{Sheehan2017,Tychoniec2018,Andersen2019,Tobin2020},  
we do not find a monotonic trend of disk dust masses decreasing from Class 0 to Class II stages. Instead, protostellar disks as a whole are less massive than Class II disks, and disk dust mass decreases from the Class 0 to Class I stages before seemingly rising again while progressing to the Flat Spectrum and Class II stages. We do note that the latter effect may be mitigated by the issues with the Class 0/I/Flat Spectrum system, as discussed in Section \ref{section:evolutionary_trends}. This difference is the result of Class II disk dust masses from radiative transfer modeling being several times larger than dust masses from estimates from fluxes, in combination with the lower disk dust masses found in our radiative transfer modeling when compared with disk dust mass estimates from submillimeter fluxes.

It is important to note, however, that this comparison may not be entirely fair, as the embedded protostars in Taurus, a relatively low mass star forming region, may be substantially different from those in Orion, which is substantially more massive. Indeed, if we compare the Taurus protostars with the Taurus Class II sources we find that they cannot be statistically distinguished (p-values $\sim0.16$), though even this is in contrast with previous results due to the larger Taurus disk dust masses when done with radiative transfer modeling. A more apt comparison for the disks in our sample may be Class II disks also in Orion. Though a comprehensive survey has not been published \citep[van Terwisga et al., in prep, but see also][for a survey of $Herschel$-detected protoplanetary disks]{Grant2021} particularly not with radiative transfer modeling, when comparing submillimeter fluxes the disks in the Orion Nebular Cluster (ONC) do seem to be lower mass than typical Class II disks \citep[e.g.][]{Eisner2018}.

Another aspect of these comparisons that needs to be accounted for is the underlying protostellar/pre-main sequence mass distribution. As disk dust masses are correlated with stellar masses \citep[e.g.][]{Andrews2013,Pascucci2016,Barenfeld2016,Ansdell2016}, to properly compare disk dust masses it is imperative to ensure that the disk masses are drawn from identical populations of stars. As stellar masses are known only for a handful of protostars in the VANDAM: Orion survey \citep[e.g.][]{Tobin2020TheOMC2-FIR3/HOPS-370} we cannot meet this requirement and are instead forced to assume that it is met. But if the distributions are indeed drawn from differing underlying stellar mass distributions then the comparisons would be invalid.

Finally, though each of these studies employed radiative transfer modeling to measure disk dust masses, there are differences that could potentially lead to systematic offsets between the properties measured. Though our work and \citet{Sheehan2017} employ the same code, we use slightly different dust opacities in our modeling and model interferometric data at a different wavelength (230 GHz vs 345 GHz). On the other hand, we use similar dust opacities to \citet{Ballering2019ProtoplanetaryTaurus}, but their modeling did not directly model any imaging data to provide spatial constraints on the model. To test how much these differences in opacity may be playing a role, we show cumulative distributions for the 230 GHz dust opacity and maximum dust grain sizes in Figure \ref{fig:rt_comps_dust}. We do indeed find a fairly significant difference, of a factor of $\sim2\times$ between typical dust opacities from \citet{Sheehan2017} as compared with this work, which may help to explain the difference in dust masses. Though there are some small differences in the 230 GHz opacity between our work and \citet{Ballering2019ProtoplanetaryTaurus}, we overall have similar values with the majority falling near the canonical opacity of $\sim2$ cm$^{2}$ g$^{-1}$. That said, our sample has very different maximum dust grain sizes and therefore likely different broadband opacity curves that could influence the disk dust mass through the temperature structure and thereby hinder a fair comparison of the samples. On the other hand, these differences could reflect real differences in the properties of disks across differing evolutionary stages and environments. Ultimately it would be ideal to model all of these samples, and more, in a uniform manner to ensure that disk properties are being compared in as fair a manner as possible.

Assuming, however, that our comparisons are fair, these results raise interesting questions about the potential for planet formation in disks. The onset of disk surveys with ALMA has raised questions about whether Class II disks have enough dust mass in them to form the population of planets that are being found by various exoplanet surveys \citep[e.g.][]{Najita2014,Manara2018}. Early studies of Class 0/I/Flat Spectrum disks suggested that they were, on average, more massive than Class II disks, suggesting that these young disks might be a better representation of the initial reservoir of material for forming planets \citep[e.g.][]{Sheehan2017,Tychoniec2018,Tychoniec2020}. In comparing the disk dust masses measured using detailed radiative transfer modeling, however, it is less clear that young disks are indeed more massive and more readily capable of forming planets.

One potential resolution to this problem is that disk dust masses measured at 870 $\mu$m or 1.3 mm are systematically being underestimated. Disk dust masses measured from longer wavelength observations do indeed find higher disk masses for both embedded disks as well as protoplanetary disks \citep[e.g.][]{Tobin2020,Tychoniec2020,Tazzari2020}. Whether this discrepancy is the result of dust grain growth \citep[e.g.][]{Ricci2010,Tazzari2020}, optical depth, or some combination of the two, and how that plays in to the disk dust masses, however, likely requires detailed modeling to account for all of these effects simultaneously. Even with modeling this picture may be complicated by the effects of dust grain scattering \citep[e.g.][]{Liu2019,Zhu2019} across unseen substructures that can produce a low spectral index even in the face small dust grains \citep[e.g.][]{Lin2020}.

Overall, this work provides a new look at how disk properties evolve from protostellar to protoplanetary disks that contrasts results from earlier works. It is, however, likely that this picture will continue to evolve as improved models are confronted with better data.

\begin{figure*}
    \centering
    \includegraphics[width=3.5in]{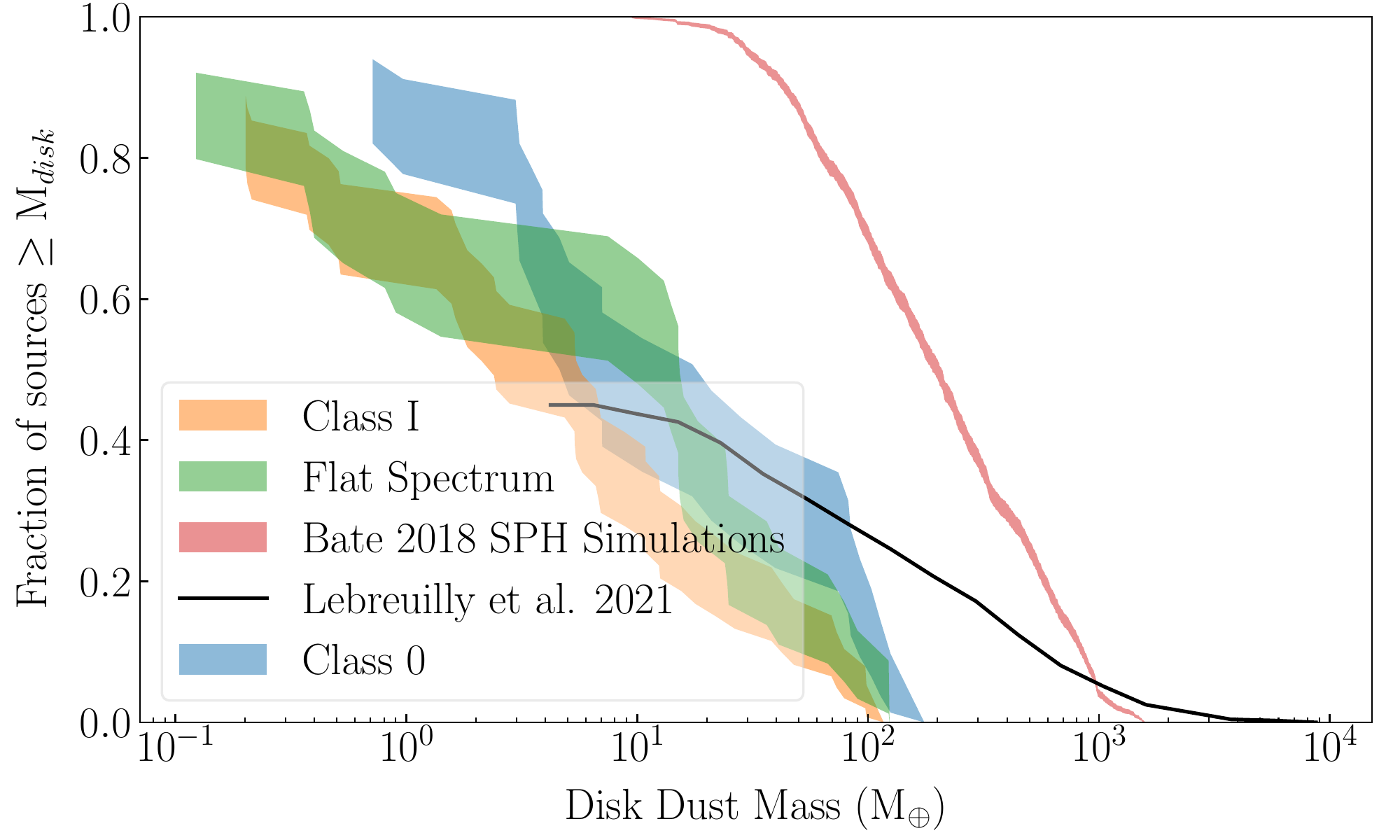}
    \includegraphics[width=3.5in]{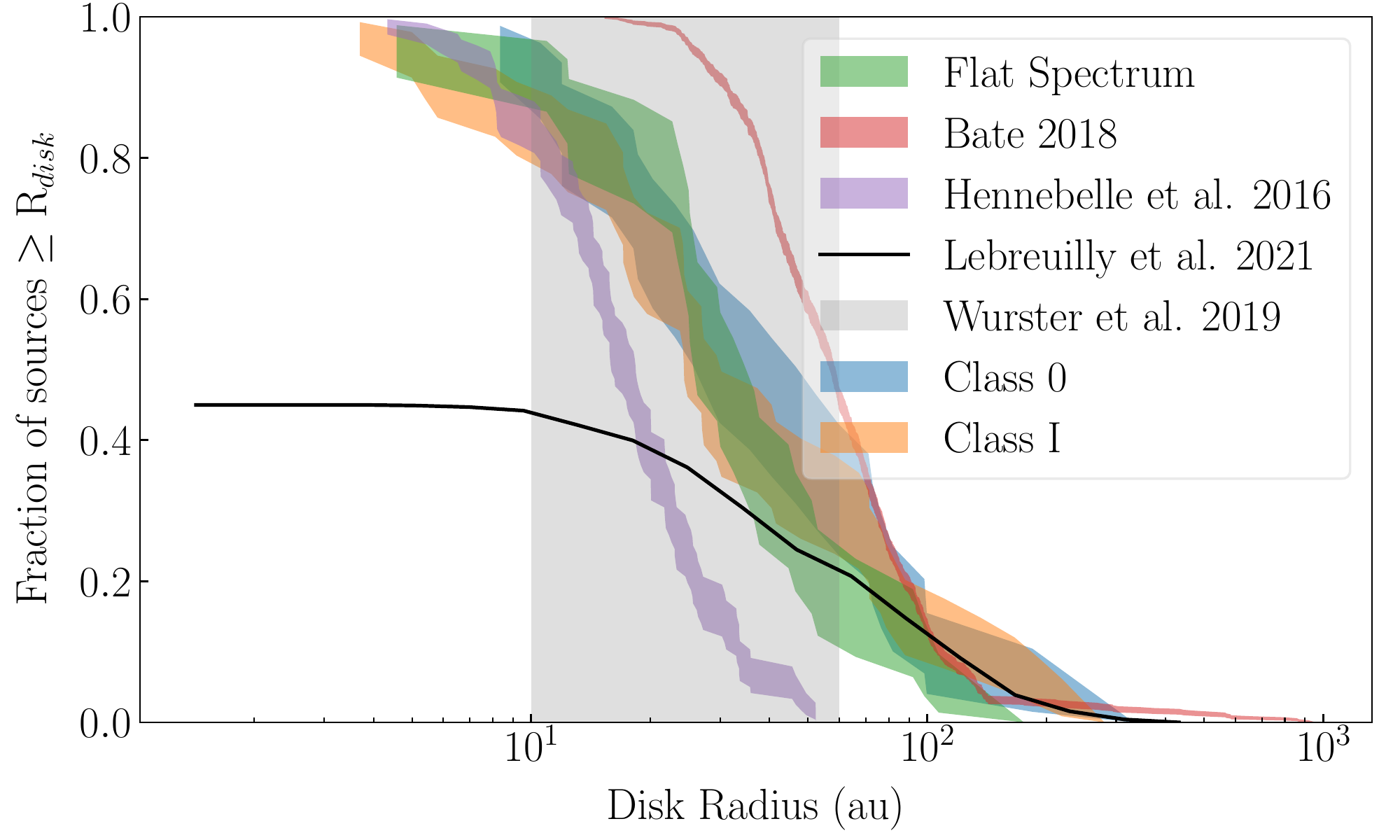}
    \caption{Cumulative distributions of disk dust mass ($left$) and radius ($right$) of disks in our sample, as compared with the masses and radii of disks formed in simulations from \citet{Bate2018}, \citet{Hennebelle2016MAGNETICALLYDISKS}, \citet{Wurster2019}, and \citet{Lebreuilly2021}. The masses and radii from \citet{Bate2018} come from SPH simulations of the global collapse of a molecular cloud to form a population of stars and disks, whereas the the distribution of radii from \citet{Hennebelle2016MAGNETICALLYDISKS} are the result of applying Equation \ref{eq:hennebelle_relation} to the disks in our sample. \citet{Wurster2019} and \citet{Lebreuilly2021} both run simulations similar to \citet{Bate2018} but include the effects of non-ideal MHD in their models. We note that our observations probe dust masses and radii, while these simulations probe gas-disk properties. To compare masses directly, we divide the simulation gas masses by a gas-to-dust ratio of 100, while we assume that gas and dust disks are of comparable size.}
    \label{fig:sims_comp}
\end{figure*}

\subsection{Constraints on Disk Formation Theory}

When and how protostellar disks form during the star formation process has been a longstanding problem in our understanding of star and planet formation. For many years, it was unclear whether protostellar disks even could form during the early stages of cloud collapse and star formation, as ideal magneto-hydrodynamic (MHD) simulations suggested that magnetic fields may prevent entirely the formation of disks \citep[e.g.][]{Allen2003,Galli2006,Mellon2008,Hennebelle2008MagneticCore,Li2014a}. In more recent years, the inclusion of non-ideal MHD effects such as ambipolar diffusion, Hall effect, and Ohmic resistivity \citep{Dapp2011AvertingDissipation,Li2011,Machida2011,Tsukamoto2015,Masson2016,Hennebelle2016MAGNETICALLYDISKS,Tsukamoto2017,Wurster2019a} as well as misalignments between magnetic field and angular momentum axis of the cloud \citep{Joos2012,Seifried2012DiscCatastrophe,Hennebelle2009DiskCores} or turbulance \citep[e.g.][]{Santos-Lima2012TheDisks} can lead to the formation of disks early in the star formation process, and the question has now switched to the importance of these effects and the characteristics of the disks that are formed \citep{Machida2011,Masson2016,Wurster2020Non-idealFormation,Wurster2020Non-idealFormationb,Hennebelle2020WhatDiscs,Xu2021FormationConvergence}. For example, if the Hall effect is dominant, it may lead to a bimodal distribution in disk radii as disk formation is enhanced for configurations where angular momentum is anti-aligned with the magnetic field and suppressed when they are aligned \citep[e.g.][]{Tsukamoto2015,Tsukamoto2017TheEvolution}, though whether such bimodality will persist through subsequent evolution is unclear \citep[e.g.][]{Zhao2020HallEvolution}.

In recent years, computational capacity has recently reached the point where global simulations of cloud collapse that follow gas from the molecular cloud scale down to the size-scales of disks as they form are now becoming feasible \citep[e.g.][]{Bate2018,Kuffmeier2017,Wurster2019b}, enabling the demographics of entire populations of protostellar disks to be simulated. Comparisons between the demographics derived from such simulations that include different suites of input physics with the observed demographics then provide a powerful constraint on the physics of disk formation.

In Figure \ref{fig:sims_comp}, we show comparisons of the cumulative distributions of protostellar disk dust masses and radii with the cumulative distribution of protostellar disk properties from simulations by \citep{Bate2018}. \citet{Bate2018} used smoothed-particle hydrodynamics to consider pure hydrodynamics during the collapse of a 500 M$_{\odot}$ cloud to form a population of over 100 protostars and disks. Though both single stars and multiple systems are formed in their simulations, in Figure \ref{fig:sims_comp} we consider only the single stars to best match with our own sample. We also note that the disk radii in \citet{Bate2018} are calculated using the radius inside of which 63.2\% of the total disk mass is enclosed based on their observation that for the density profile in Equation \ref{eq:disk_surface_density}, for $\gamma < 2$ then $R_c$ always encloses a fraction $(1 - 1/e)$ of the total disk mass, or 63.2\%. As their criterion is based off the surface density model that we use in our analysis, our radii should be directly comparable. Finally, as \citet{Bate2018} considered gas simulations while we have measured dust properties, we note that in comparing disk masses we have divided the simulated disk masses by a gas-to-dust ratio of 100.

As can be seen in Figure \ref{fig:sims_comp}, the disks produced in the \citet{Bate2018} simulations tend to predict masses and sizes that are too large when compared with the disk dust masses and radii in our sample. This discrepancy may reflect the lack of magnetic fields in their simulations, which are known to remove angular momentum from the cloud, which should in turn lead to smaller disks. Indeed, using simple relations related to the timescales of magnetized collapse, \citet{Hennebelle2016MAGNETICALLYDISKS} derived that young disks should have sizes of
\begin{equation}
    R \approx 18\,\mathrm{au}\left(\frac{\eta_{AD}}{0.1\,\mathrm{s}}\right)^\frac{2}{9}\left(\frac{B_z}{0.1\,\mathrm{G}}\right)^{-\frac{4}{9}}\left(\frac{M_* + M_{disk}}{0.1\,\mathrm{M}_{\odot}}\right)^\frac{1}{3},
    \label{eq:hennebelle_relation}
\end{equation}
where $\eta_{AD}$ is the ambipolar diffusion coefficient and $B_z$ is the magnetic field in the inner region of the core. \citet{Hennebelle2020WhatDiscs} later showed that this relation does a reasonably good job of predicting the disk sizes from a suite of MHD simulations within a factor of a few. Though exact values of $\eta_{AD}$, $B_z$, and $M_*$ are unknown, we can estimate the range of radii we would expect assuming their standard values for $\eta_{AD}$ and $B_z$, $M_*$ sampled randomly from a Chabrier IMF, and using the disk mass calculated for each disk individually via our modeling. We show the distribution of disk radii calculated in Figure \ref{fig:sims_comp}. 

Interestingly, the disk sizes predicted by \citet{Hennebelle2016MAGNETICALLYDISKS} using fiducial values for $B_z$ and $\eta_{AD}$ appear to be too $small$ as compared with the the disks in our sample, possibly suggesting that magnetic braking might be too efficient at sapping angular momentum from the cloud. We note that there may be correlations between $M_*$, $M_{disk}$, and $R_{disk}$ that we are not accounting for by randomly assigning stellar masses and that might impact the exact distribution. Alternatively, magnetic fields were systematically weaker in our real disks than their fiducial values, this could also help to bring the two into agreement. For example, a reduction in the typical field strength by a factor of $3-4\times$ would bring the distributions into better agreement. An increase in the strength of ambipolar diffusion by a factor of $10-20\times$ could also help remedy this disagreement. This might be achieved by altering the the cosmic ray ionization rate, to which ambipolar diffusion is sensitive \citep[e.g.][]{Marchand2016}, which has numerically been shown to regulate the sizes of protostellar disks that are formed \citep[e.g.][]{Kuffmeier2020Ionization:Regions}. In both cases, the change from the fiducial values of $B_z$ or $\eta_{AD}$ in concert with assuming a log-normal distribution of their values with a standard deviation of $\sim1$ dex also helps to improve the agreement with the spread of observed disk radii.

To further examine the role of magnetic fields on disk formation in a global environment, \citet{Wurster2019} followed on the work of \citet{Bate2018} by simulating the global collapse of a cloud to form stars and disks including MHD, while varying initial conditions such as the strength of the magnetic field and whether non-ideal MHD effects were included in the calculations. For expediency, the simulations followed a cloud of only 50 M$_{\odot}$ and so fewer disks were formed overall, making the comparison of demographics more challenging. The disks that did form around single protostars, though, had typical sizes of 10 -- 60 au, in reasonable agreement with the typical sizes of disks in our sample.

Perhaps most realistically, \citet{Lebreuilly2021} recently ran simulations of the collapse of a 1000 M$_{\odot}$ cloud to form a population of 191 protostars and 42 disks when including non-ideal MHD effects along with radiative transfer and stellar feedback. We show the cumulative distributions of the properties of those disks in Figure \ref{fig:sims_comp}. We note that we follow \citet{Lebreuilly2021} in scaling the distribution according to the fraction of stars with disks (particularly their Figure 4, bottom-middle panel), as they note that the smaller disks in their simulation were likely disrupted by numerical effects. We find the at the large radius end of the distribution, their simulated disks are in good agreement with the distribution of disk dust radii that we measure. At smaller radii our distributions do diverge, but this may again be due to numerical effects in the simulations. The disk masses from \citet{Lebreuilly2021}, on the other hand, tend to predict that the most massive disks are significantly more massive than what we find in our sample, but come to be in better agreement at the lower end of the mass distribution.

It is important to note, however, that while these comparisons may provide insights into disk formation physics, there are a number of limitations that must be kept in mind. Most notably these figures compare dust disk radii derived from observations with gas disk radii as measured by simulations. As dust particles tend to decouple from the gas and experience aerodynamic drag that leads to radial drift of the dust \citep{Weidenschilling1977AerodynamicsNebula,Birnstiel2010Gas-andDisks}, it is not necessarily the case that the sizes being compared should be in agreement \citep{Lebreuilly2020ProtostellarDisks}. Observations of Class II protoplanetary disks have indeed demonstrated that gas disk radii are, on average, $2.5\times$ larger when measured with gas observations than when measured with dust observations \citep[e.g.][]{Ansdell2018,Boyden2020,Sanchis2021}. There is some question as to whether this is a true difference in size, as opposed to an optical depth effect \citep[e.g.][]{Trapman2019}, and Class II disks are significantly older than the disks considered here, but nonetheless the difference between gas and dust radii may limit the utility of such comparisons. Moreover, while differing levels of physics are included in the simulations we consider, none include the effects of stellar feedback, which has been shown to have a significant impact on cloud collapse and presumably therefore disk formation and evolution \citep[e.g.][]{Guszejnov2021OUPManuscript}. 

Regardless of these caveats, such simulations provide a starting point for placing constraints on the physics that is important in setting the size scales of protostellar disks that will be improved upon as both simulations and our observational data improve.

\subsection{Gravitational Stability of Protostellar Disks}

Self-gravity has long been thought to be an important driver of star and planet formation and the evolution of protoplanetary disks. In young, massive disks it has been suggested that gravitational instabilities might be a source of angular momentum transport \citep{Vorobyov2007Self-regulatedDiscs,Vorobyov2008SecularDiscs,Machida2010FormationFormation,Zhu2010Long-termInfall}. Instabilities in disks might also be a source of companion stars in multiple systems \citep{Adams1989,Kratter2006FragmentationDisks,Stamatellos2009TheFragmentation}. Moreover, it has been suggested that gravitational instabilities might form planets in the outer regions of disks where core accretion is less effective \citep[e.g.][]{Boss1997,Durisen2006,Boss2011}, though the general consensus seems to be that planetary-mass companions formed via gravitational instabilities are difficult to keep from growing into brown dwarfs or stars \citep{Kratter2010,Zhu2012,Forgan2013TowardsDownsizing}. Nonetheless, gravitational instabilities may drive spiral arms that could be conducive to dust trapping that would otherwise enhance planet formation \citep[e.g.][]{Dipierro2015}.

\begin{figure}[t]
    \centering
    \includegraphics[width=3.3in]{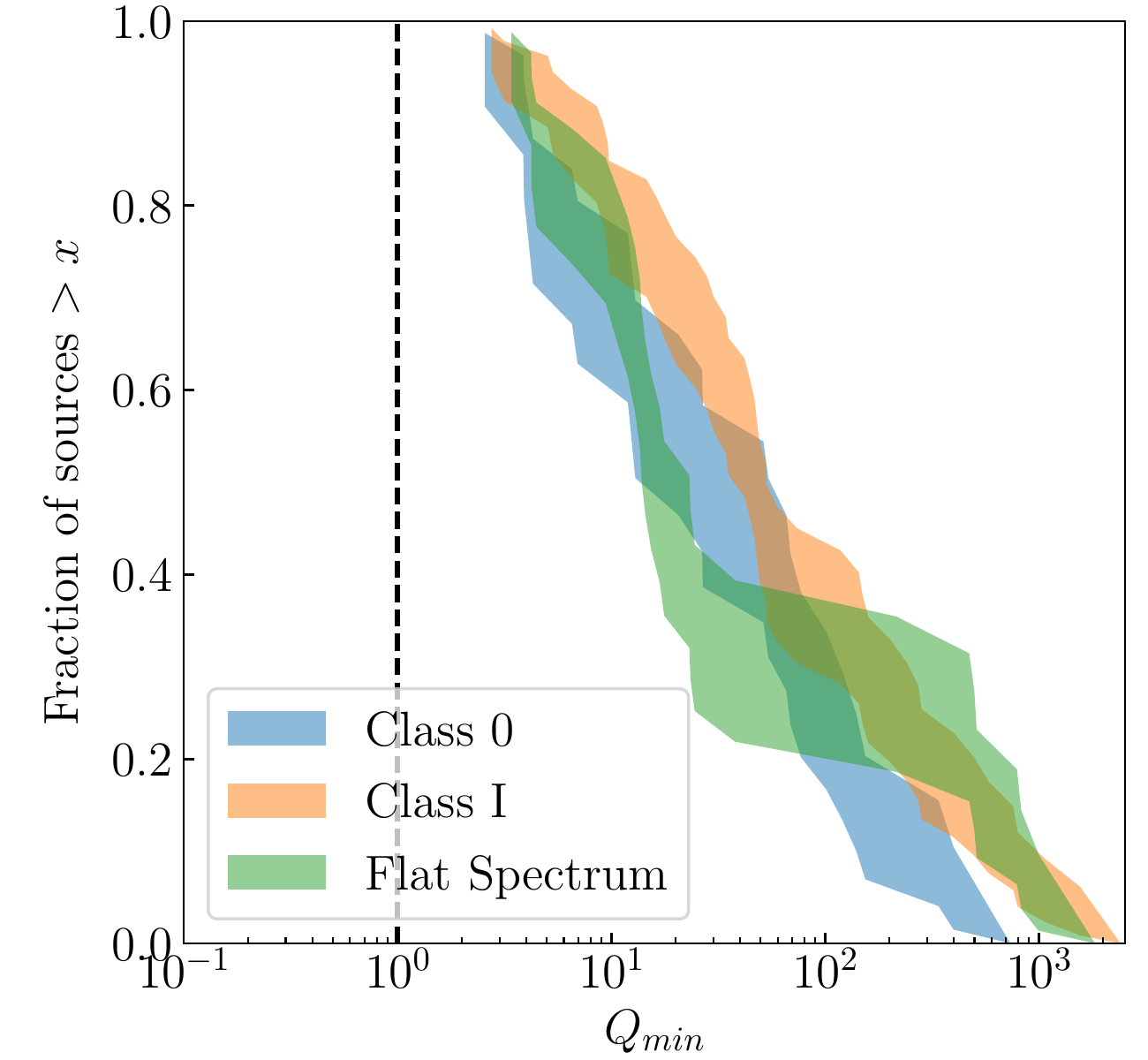}
    \caption{Comparison of the minimum value of Toomre's Q for the disks in our sample as a function of protostellar classification.}
    \label{fig:toomreq}
\end{figure}

\begin{figure*}
    \centering
    \includegraphics[width=2.3in]{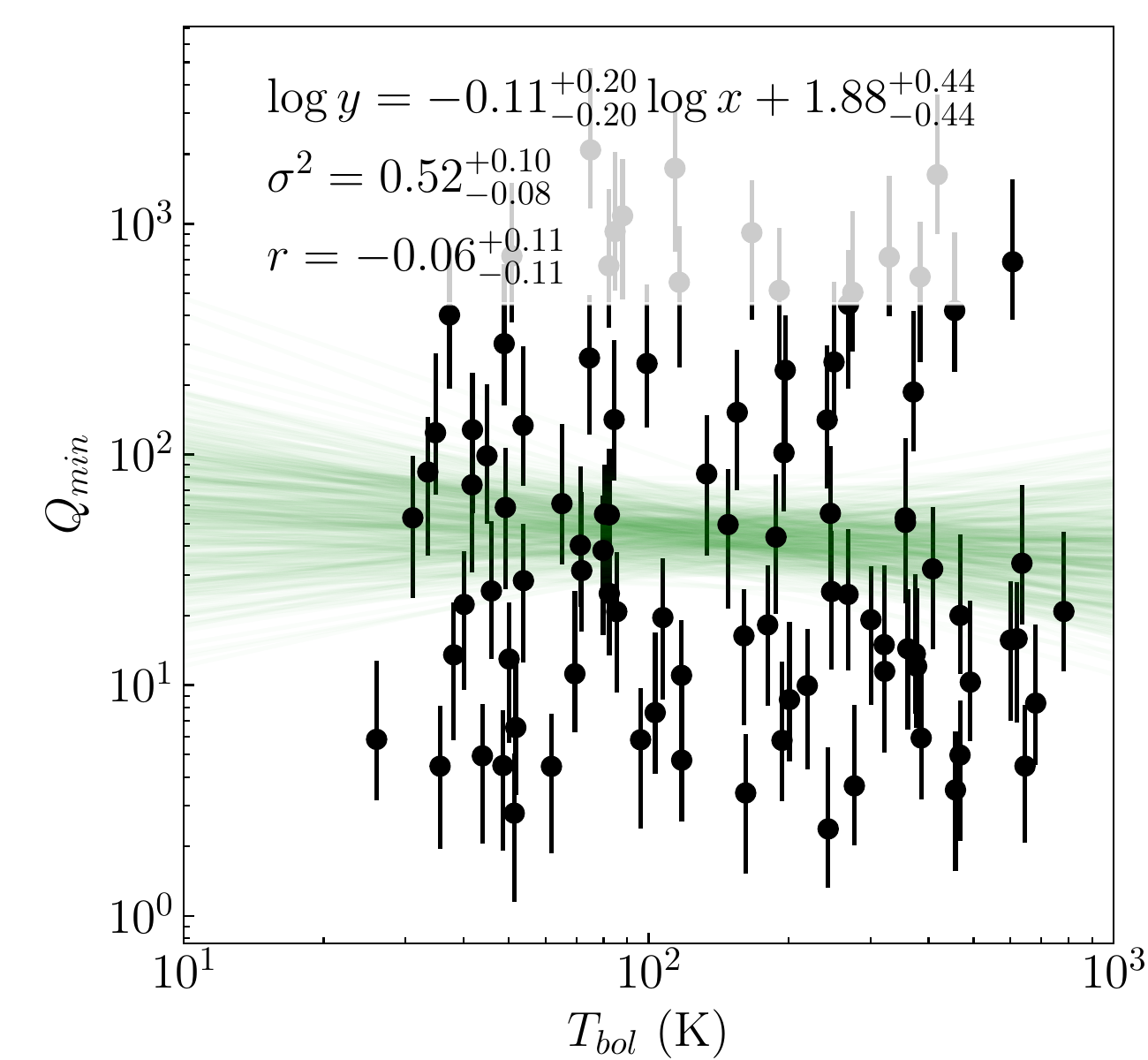}
    \includegraphics[width=2.3in]{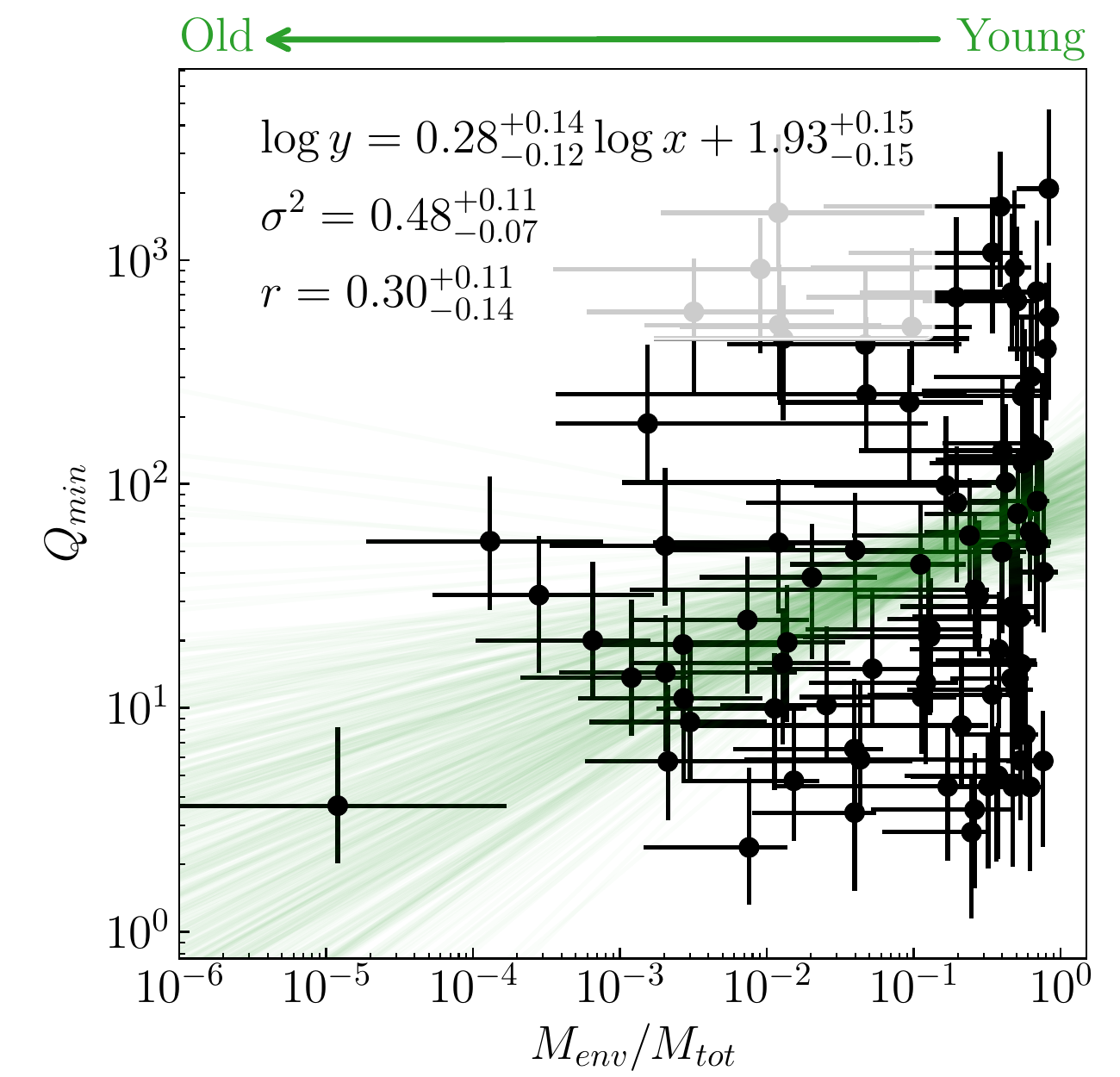}
    \includegraphics[width=2.3in]{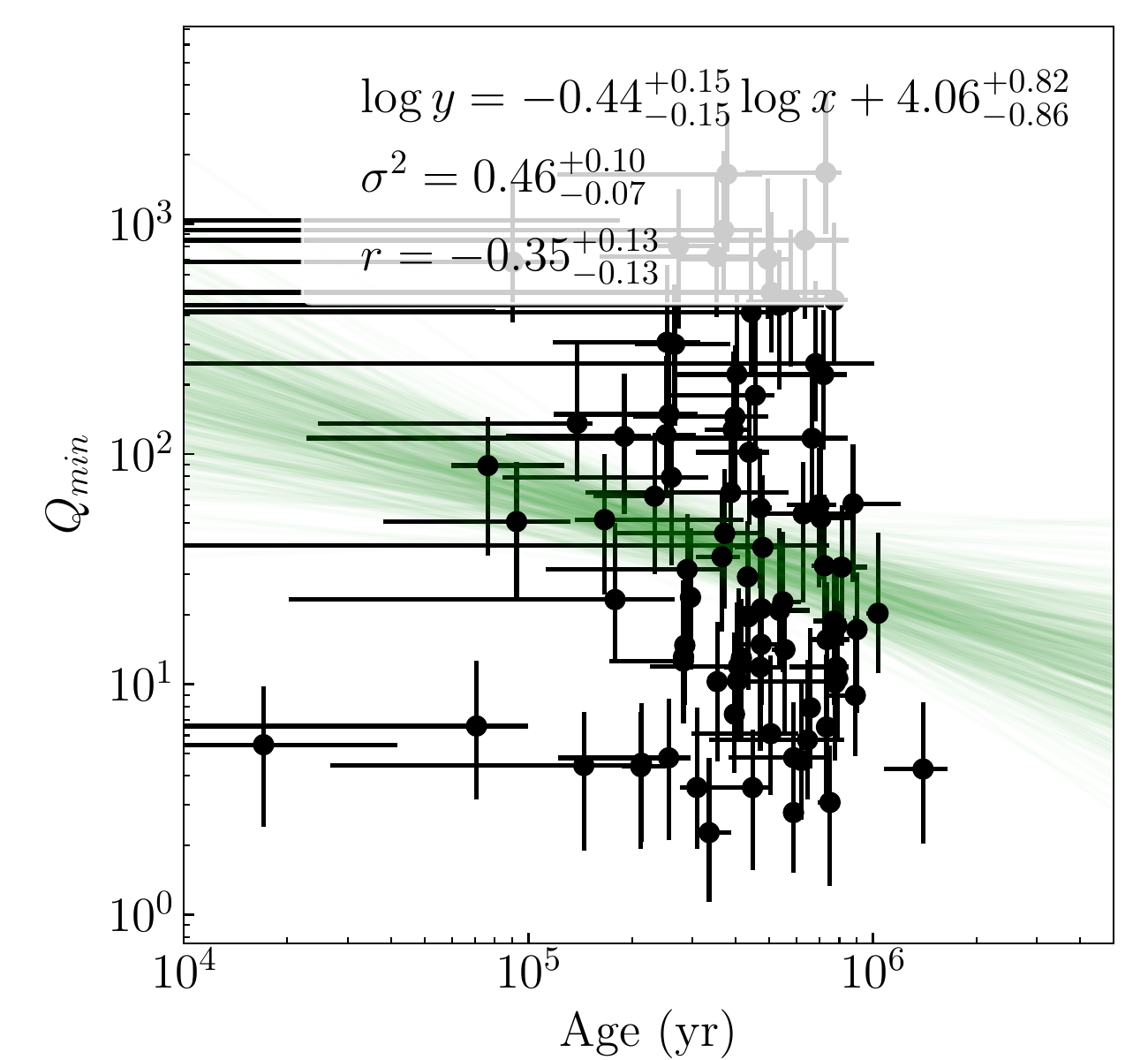}
    \caption{Comparison of the minimum value of Toomre's Q for the disks in our sample as a function of protostellar classification ($left$),  the bolometric temperature ($center$), and the ratio of envelope mass to total mass ($right$).}
    \label{fig:toomreq_evolution}
\end{figure*}

To test the gravitational stability of the disks in our sample and determine whether self-gravity is important, we use our radiative transfer modeling framework to calculate the Toomre Q profile for each disk in our sample. We do this via the equation,
\begin{equation}
    Q(R) = \frac{c_s(R) \, \Omega(R)}{\pi \, G \, \Sigma(R)},
\end{equation}
where $c_s$ is the sound speed, $\Omega$ is the rotational frequency, and $\Sigma$ is the surface density, all of which vary as a function of radius $R$ in the disk. Rather than calculate the Toomre Q profile for each sample of parameters in our posterior, which would be quite computationally expensive ($\sim10,000$ radiative transfer models per source), we calculate the profile only for the best-fit model. We use midplane temperature, averaged vertically over the five cells closest to the midplane, to calculate the sound speed for a gas with a mean molecular mass of 2.37, and assume a gas-to-dust ratio of 100. Rather than assume a mass for the central source, we randomly draw $\sim$10000 masses from the Charbrier IMF and use those to estimate the range of $Q$ values for each protostar in our sample.

In Figure \ref{fig:toomreq}, we show the minimum Toomre's Q value found in each disk as cumulative distributions split by source classification as well as compared with $T_{bol}$ and $M_{env}/M_{tot}$. We note that for simplicity we calculate the cumulative distributions using only $M_* = 0.1$ M$_{\odot}$, the peak of the Charbrier IMF. Though accounting for the stellar masses may alter the shape of these distributions, without direct knowledge of the stellar masses it may be difficult to account for their effect in a self-consistent way. The uncertainties shown in the comparisons with $T_{bol}$ and $M_{env}/M_{disk}$ are derived from the range of values calculated by drawing stellar masses randomly from the IMF.

We find that, collectively, protostellar disks do not seem to be broadly gravitationally unstable. About 20\% of the disks in our sample have $Q_{min} < 10$, with a handful approaching $Q_{min} \sim 1$ for low-mass protostars, but few seem to be definitively subject to self-gravity. Disks with more modestly low values, of $Q\sim2$, may still be capable of forming spirals and transporting angular momentum \citep[e.g.][]{Tomida2017}, but there still remain few disks that approach this level of instability.

In Figure \ref{fig:toomreq_evolution}, we explore how $Q_{min}$ compares with the simple estimators of protostellar evolution discussed in Section \ref{section:evolutionary_trends}. If we consider the Class 0/I/Flat Spectrum identification as an indicator of evolution, or equivalently the bolometric temperature, we do not see much evidence that the gravitational stability of the disks changes with time. On the other hand, if we again consider the ratio of envelope mass to total mass, $M_{env}/M_{tot}$, or use the age inferred from simple models of protostellar evolution, we do find a trend of decreasing $Q_{min}$ for older systems at the $\sim2\sigma$ level. This decrease towards gravitationally unstable disks with time may reflect the weak trend of growth of the disk dust mass that was seen in Section \ref{section:menvmtot}.

Ultimately, though the lack of disks that appear to be gravitationally unstable, or perhaps even close, may provide commentary on the importance of gravitational instability in these disks, it is also possible that we do not see a substantial number of gravitationally unstable disks because the timescales for instabilities are short, and mass is drained quickly from the disk when this state is reached \citep[e.g.][]{Stamatellos2009TheFragmentation}. We do find some disks that appear to have been caught in the act \citep[e.g.][]{Tobin2016,Tobin2018a,Reynolds2021}. So it may simply be the case that these disks remain in this state for short periods of time and are therefore difficult to directly observe, or have already formed multiple systems that would have been excluded from our modeling.

\section{Conclusions}
\label{section:conclusions}

In this work we have fit 97 protostars from the VANDAM: Orion sample with two-dimensional, axisymmetric radiative transfer models. We fit the VANDAM: Orion ALMA 345 GHz visibilities along with the HOPS Survey SEDs simultaneously for each source using MCMC fitting to provide a comprehensive picture of disk and envelope structure. Our main results are:

\begin{itemize}

\item We find a median protostellar disk dust mass of $  5.8^{+  4.6}_{-  2.7}$ M$_{\oplus}$ and a median disk dust radius of $ 29.4^{+  4.1}_{-  3.2}$ au. Only ${11.3\%}^{+ 4.6\%}_{- 3.4\%}$ of disks have dust radii larger than 100 au. If we group our sample by observational class, we find median disk dust masses of $  7.1^{+ 14.3}_{-  2.0}$ M$_{\oplus}$, $  4.9^{+  1.0}_{-  2.7}$ M$_{\oplus}$, and $ 14.0^{+  1.3}_{-  7.0}$ M$_{\oplus}$ and median disk dust radii of $ 35.6^{+ 17.1}_{-  10.0}$ au, $ 26.9^{+  4.5}_{-  3.3}$ au, and $ 29.5^{+  6.1}_{-  4.4}$ au for Class 0, I, and Flat Spectrum sources, respectively.

\item {Disk dust masses measured from the frequently used simple flux-based measurement disagree with the disk dust masses measured using our models, and the difference is primarily due to the additional physics included in our radiative transfer modeling that impacts the temperature of the disk. A treatment of the temperature that accounts for disk radius does substantially improve the discrepancy, though there remains substantial scatter. We do note that {\it disk fluxes} can be accurately recovered with simple methods like two dimensional Gaussian fitting.}

\item We find evidence that Class I disk dust masses can be distinguished from Class 0 disk dust masses, and that the envelope dust masses of each class are distinct, though only the distinction between Flat Spectrum and Class 0 is statistically significant. Otherwise, we find little evidence that most disk and envelope properties evolve with time, whether we consider the Class 0/I/Flat Spectrum identification or the bolometric temperature as the tracer of evolution.

\item The distribution of Flat Spectrum source inclinations is distinct from both Class 0 and Class I sources, and also that inclination is correlated with bolometric temperature, demonstrating that the bolometric temperature is contaminated by viewing angle, and does not directly trace the evolutionary state of these young systems.

\item We use {both $M_{env}/M_{tot} = M_{env}/(M_* + M_{disk} + M_{env})$ and simple evolutionary tracks in the $L_{bol}-M_{env}$ plane as alternate ways to trace protostellar evolution} and find weak evidence that disk dust masses may actually $increase$ with time.

\item A comparison of the disk dust masses derived from our radiative transfer modeling with dust masses derived from other, similar, radiative transfer modeling studies of disks makes it less clear that Class 0/I/Flat Spectrum disks are more massive than Class II disks, though acknowledge that this difference may be due to environment or systematic differences in modeling details.

\item We compare the bulk properties of the disks in our sample with simulations of protostellar disk formation. We find that simulations that follow the collapse of a molecular cloud to form a population of protostars and disks using pure hydrodynamics \citep{Bate2018} produce disks that are too large in both radius and mass. Simulations that include the effects of magnetic fields produce smaller disks, perhaps in better agreement with our results, though similar simulations of global cloud collapse are as of yet still much smaller in scale and number of protostars (and disks) formed \citep[e.g.][]{Wurster2019}.

\item We use our radiative transfer models to construct the Toomre Q profile for all of the disks in our sample, and find that very few disks are gravitationally unstable.

\end{itemize}

Though it may be expensive, these results demonstrate the importance of detailed radiative transfer modeling in interpreting surveys of protostellar and protoplanetary disks with ALMA. It is clear that further studies employing more uniform techniques across disks in a range of environments and at different ages will be critical for evaluating the evolution of disk properties, and their effects on planet formation.

\software{pdspy \citep{Sheehan2018}, CASA \citep{McMullin2007}, RADMC-3D \citep{Dullemond2012}, emcee \citep{Foreman-Mackey2013}, matplotlib \citep{Hunter2007}, corner \citep{Foreman-Mackey2016}, GALARIO \citep{Tazzari2017a}, lifelines \citep{Davidson-Pilon2019}}

\acknowledgements {The authors would like to thank the anonymous referee for feedback that helped to focus and improve the manuscript. P.D.S would also like to thank Leonardo Testi and the referee for pointing him to the protostellar evolutionary tracks used in Section \ref{section:evolutionary_tracks}, that added an interesting new piece of analysis to the work.} P.D.S is supported by a National Science Foundation Astronomy \& Astrophysics Postdoctoral Fellowship under Award No. 2001830. JJT acknowledges support from NSF AST-1814762 and past support from the Homer L. Dodge Endowed Chair at the University of Oklahoma.
LWL acknowledges support from NSF AST-1910364 and AST-2108794.
This work used the Extreme Science and Engineering Discovery Environment (XSEDE), which is supported by National Science Foundation grant number ACI-1548562, for the majority of the computing done. Additional computing was performed at the OU Supercomputing Center for Education \& Research (OSCER) at the University of Oklahoma (OU), when needed. This paper makes use of the following ALMA data: ADS/JAO.ALMA\#2015.1.00041.S. ALMA is a partnership of ESO (representing its member states), NSF (USA) and NINS (Japan), together with NRC (Canada), NSC and ASIAA (Taiwan), and KASI (Republic of Korea), in cooperation with the Republic of Chile. The Joint ALMA Observatory is operated by ESO, AUI/NRAO and NAOJ. The National Radio Astronomy Observatory is a facility of the National Science Foundation operated under cooperative agreement by Associated Universities, Inc.

\bibliography{references.bib}

\appendix

\onecolumngrid
\section{Diagram of Envelope Cavity Shape}
\label{section:outflow_shape}

\begin{figure*}[h]
    \centering
    \includegraphics[width=6in]{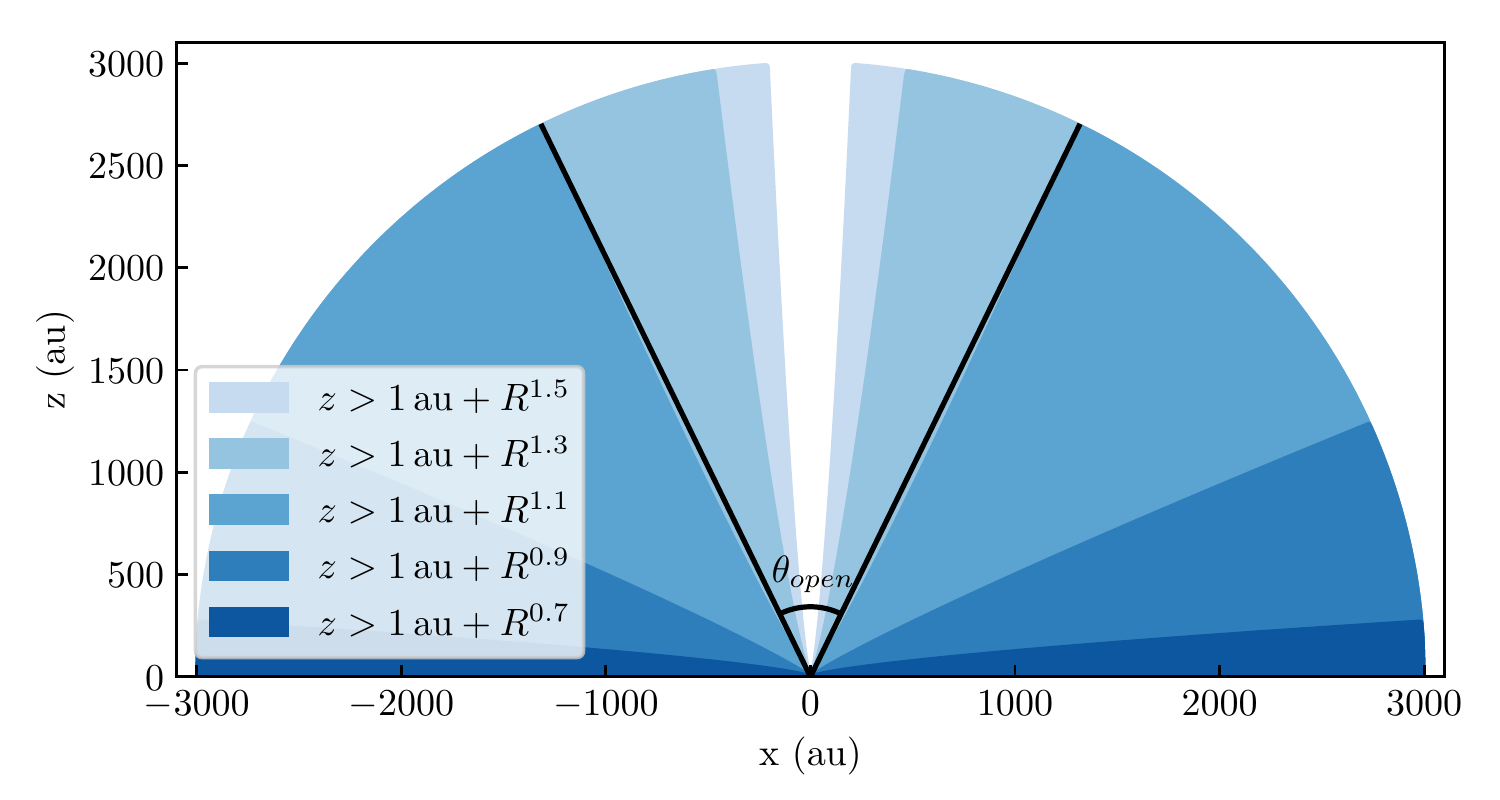}
    \caption{Diagram showing the shape of the outflow cavity, as defined by $z > 1 \mathrm{au} + R^{\xi}$, for a range of values of $\xi$. The black lines show the definition of $\theta_{open}$ for $\xi = 1.1$.}
    \label{fig:outflow_shape}
\end{figure*}

We show a diagram demonstrating how the $\xi$ parameter affects the shape of the outflow cavity in our model in Figure \ref{fig:outflow_shape}.

\end{document}

%% file: model_plots.tex
\figsetstart
\figsetnum{1}
\figsettitle{Best-fit Radiative Transfer Models}

\figsetgrpstart
\figsetgrpnum{1.1}
\figsetgrptitle{HOPS-2}
\figsetplot{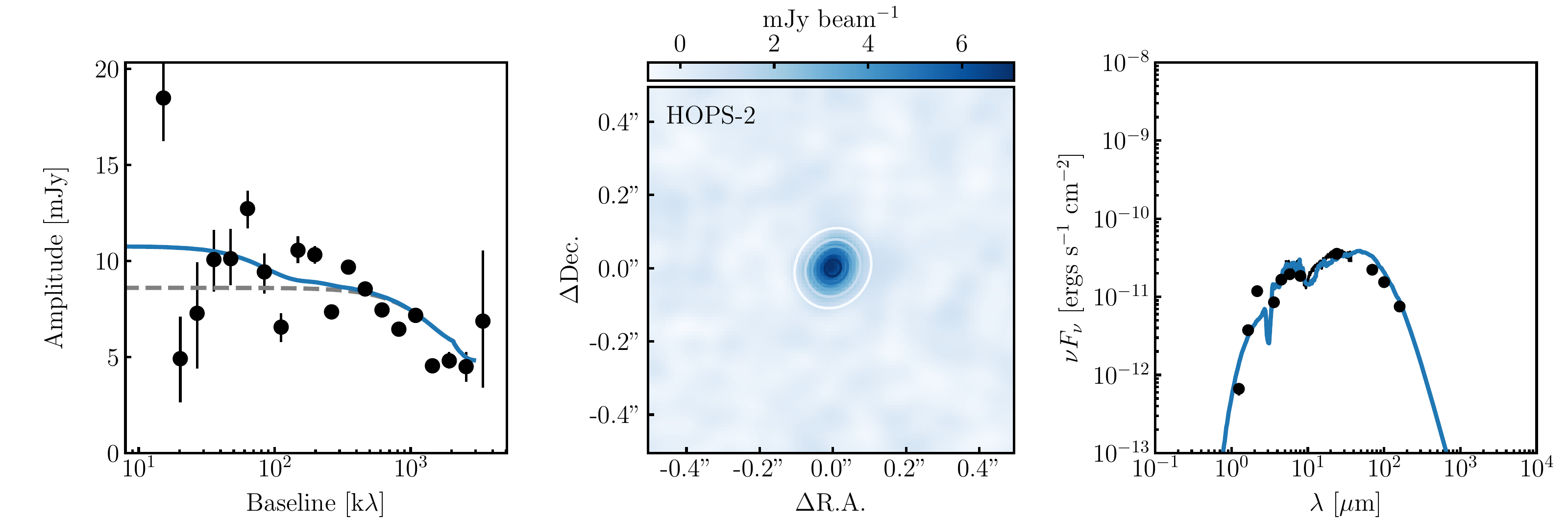}
\figsetgrpnote{A continuation of Figure \ref{fig:rt_fits} showing the best-fit model for HOPS-2. As in Figure \ref{fig:rt_fits}, black points ({\it left/right}) or color scale ({\it center}) show the data, while the blue lines ({\it left/right}) or contours ({\it center}) show the model, and the gray dashed line shows the disk contribution to the model.}
\figsetgrpend

\figsetgrpstart
\figsetgrpnum{1.2}
\figsetgrptitle{HOPS-3}
\figsetplot{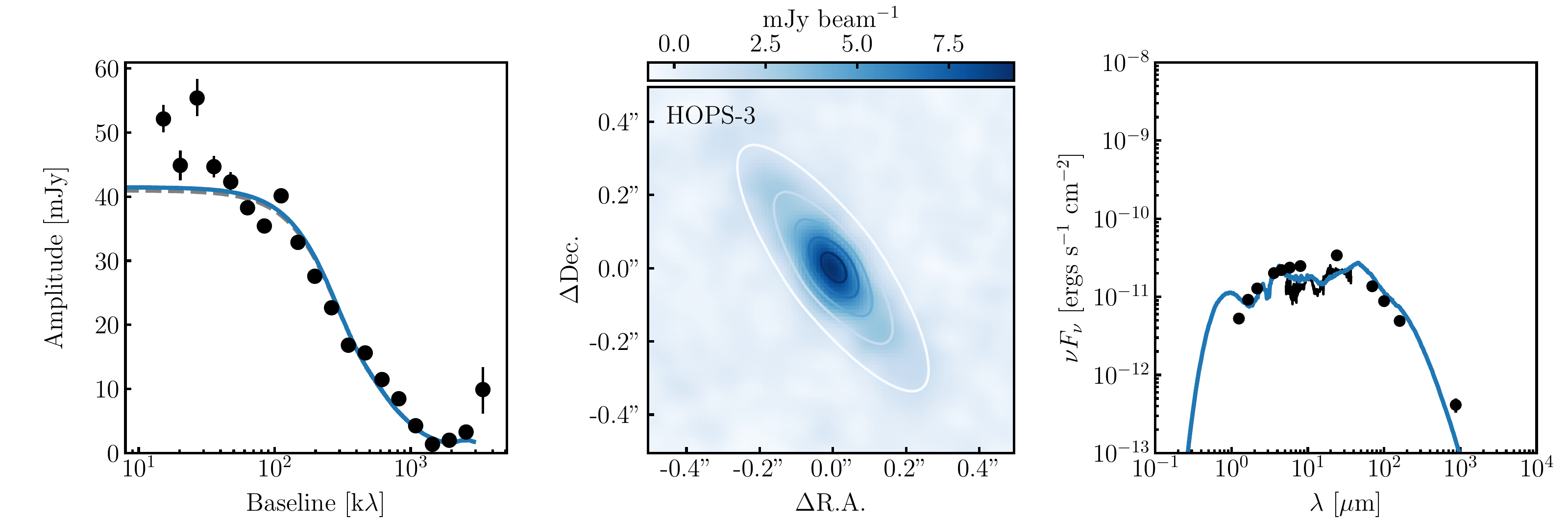}
\figsetgrpnote{A continuation of Figure \ref{fig:rt_fits} showing the best-fit model for HOPS-3. As in Figure \ref{fig:rt_fits}, black points ({\it left/right}) or color scale ({\it center}) show the data, while the blue lines ({\it left/right}) or contours ({\it center}) show the model, and the gray dashed line shows the disk contribution to the model.}
\figsetgrpend

\figsetgrpstart
\figsetgrpnum{1.3}
\figsetgrptitle{HOPS-13}
\figsetplot{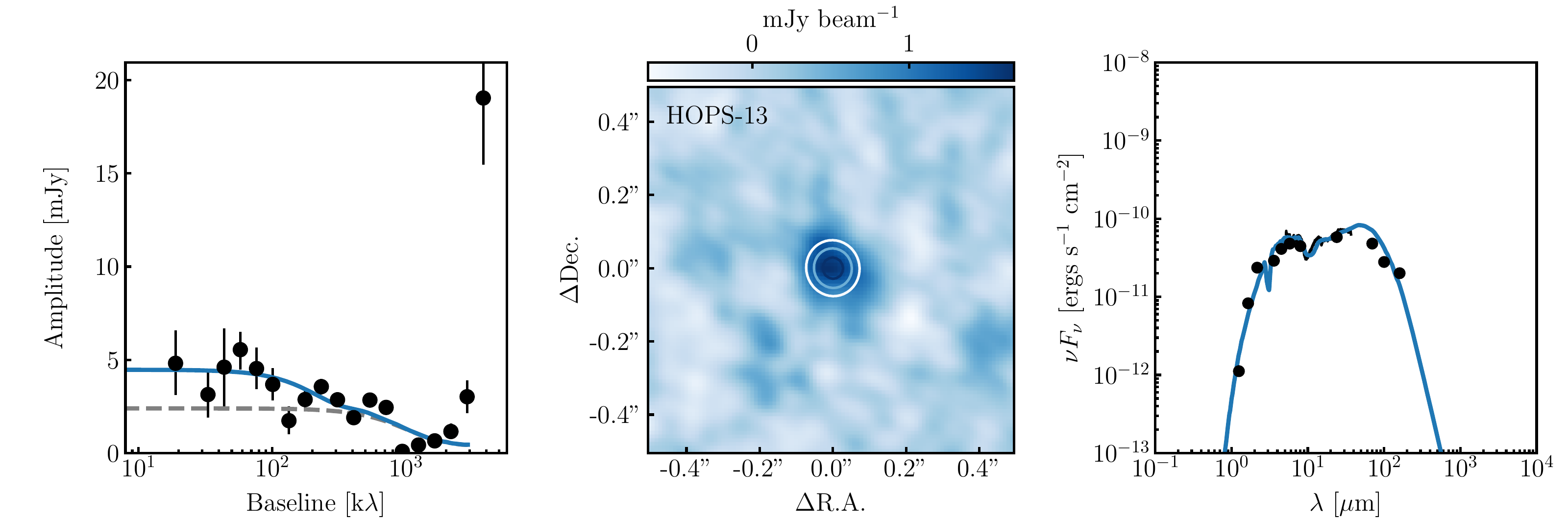}
\figsetgrpnote{A continuation of Figure \ref{fig:rt_fits} showing the best-fit model for HOPS-13. As in Figure \ref{fig:rt_fits}, black points ({\it left/right}) or color scale ({\it center}) show the data, while the blue lines ({\it left/right}) or contours ({\it center}) show the model, and the gray dashed line shows the disk contribution to the model.}
\figsetgrpend

\figsetgrpstart
\figsetgrpnum{1.4}
\figsetgrptitle{HOPS-16}
\figsetplot{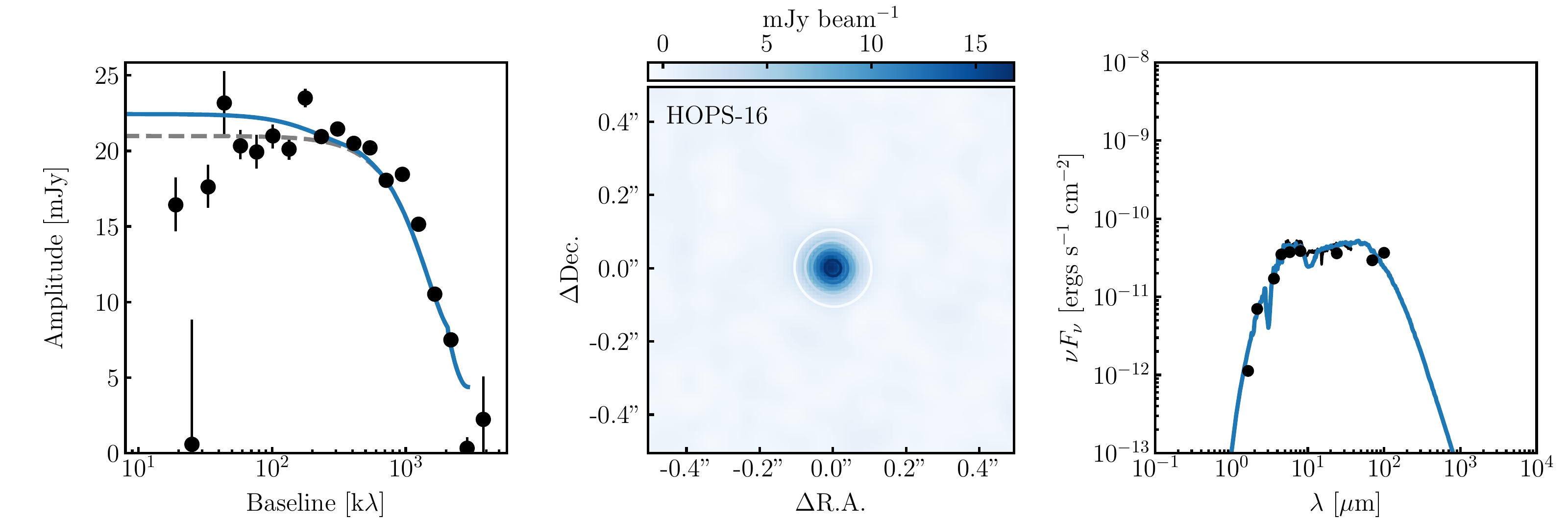}
\figsetgrpnote{A continuation of Figure \ref{fig:rt_fits} showing the best-fit model for HOPS-16. As in Figure \ref{fig:rt_fits}, black points ({\it left/right}) or color scale ({\it center}) show the data, while the blue lines ({\it left/right}) or contours ({\it center}) show the model, and the gray dashed line shows the disk contribution to the model.}
\figsetgrpend

\figsetgrpstart
\figsetgrpnum{1.5}
\figsetgrptitle{HOPS-18}
\figsetplot{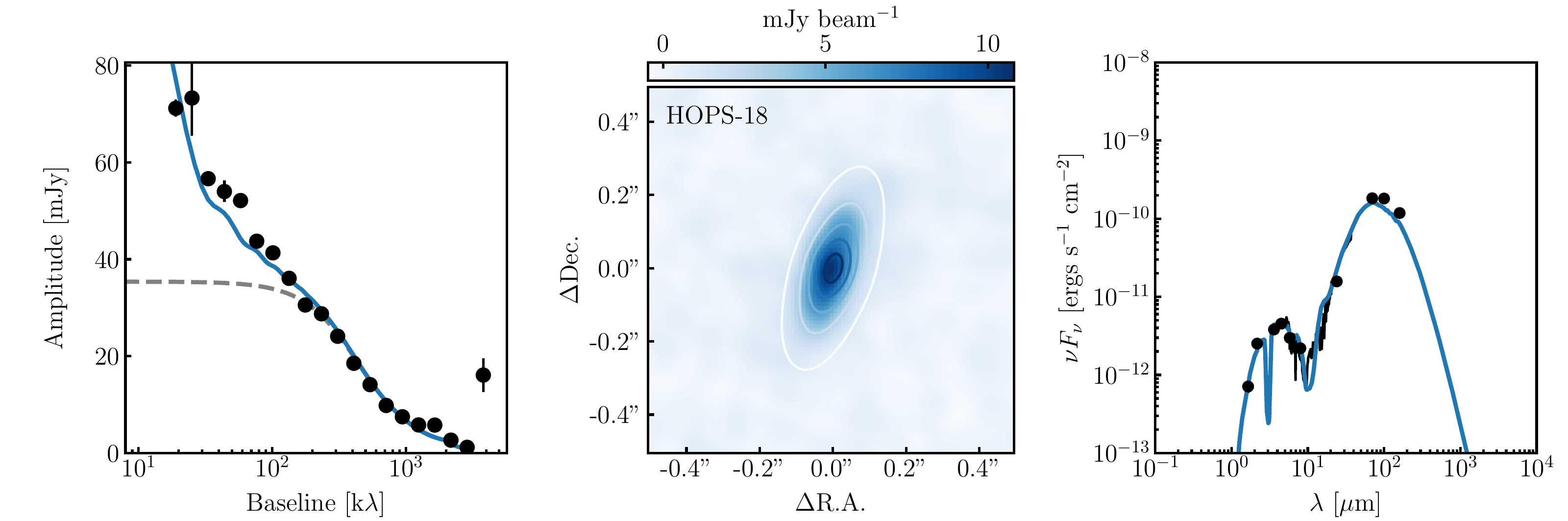}
\figsetgrpnote{A continuation of Figure \ref{fig:rt_fits} showing the best-fit model for HOPS-18. As in Figure \ref{fig:rt_fits}, black points ({\it left/right}) or color scale ({\it center}) show the data, while the blue lines ({\it left/right}) or contours ({\it center}) show the model, and the gray dashed line shows the disk contribution to the model.}
\figsetgrpend

\figsetgrpstart
\figsetgrpnum{1.6}
\figsetgrptitle{HOPS-29}
\figsetplot{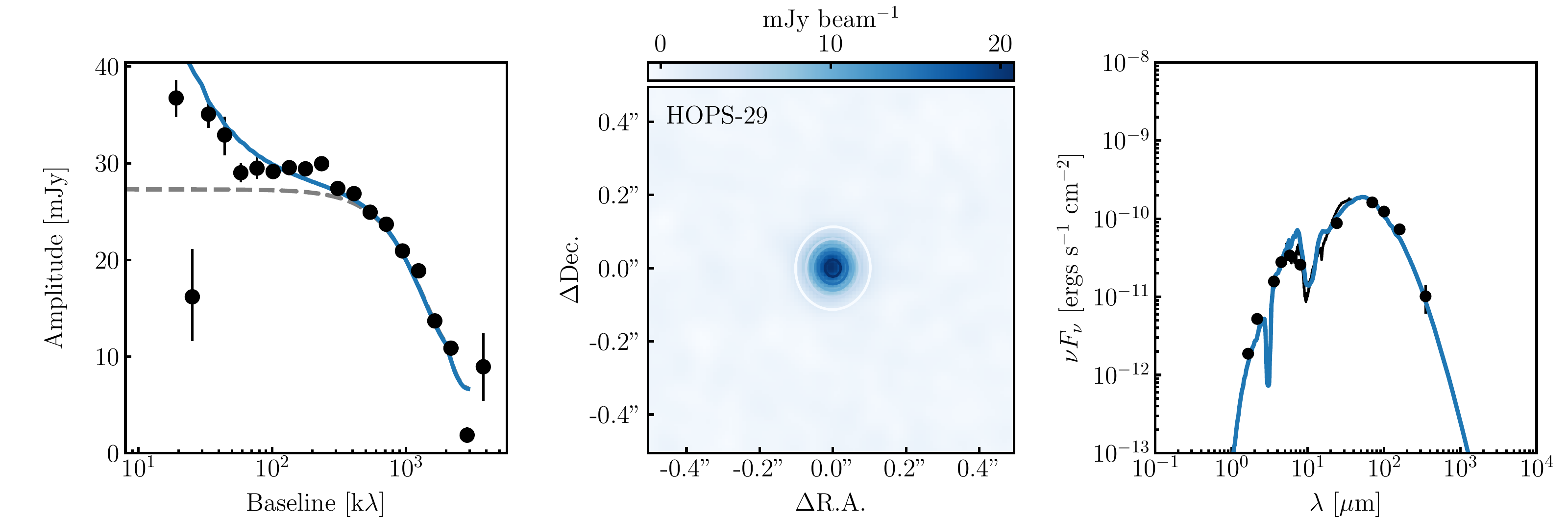}
\figsetgrpnote{A continuation of Figure \ref{fig:rt_fits} showing the best-fit model for HOPS-29. As in Figure \ref{fig:rt_fits}, black points ({\it left/right}) or color scale ({\it center}) show the data, while the blue lines ({\it left/right}) or contours ({\it center}) show the model, and the gray dashed line shows the disk contribution to the model.}
\figsetgrpend

\figsetgrpstart
\figsetgrpnum{1.7}
\figsetgrptitle{HOPS-36}
\figsetplot{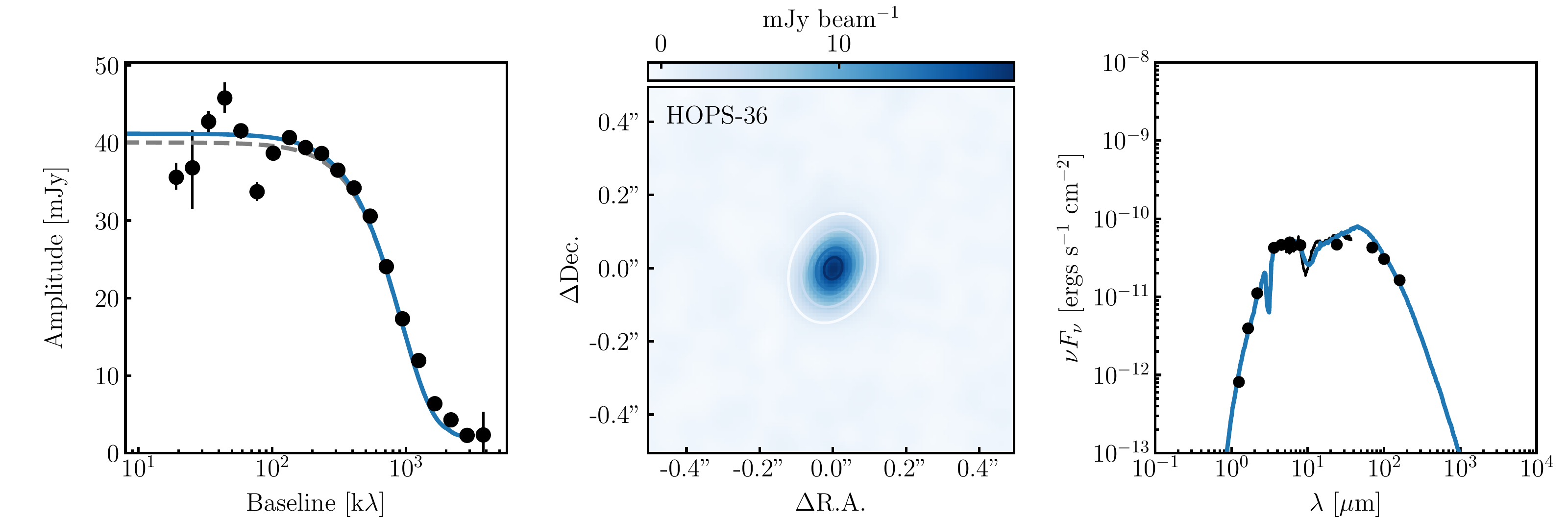}
\figsetgrpnote{A continuation of Figure \ref{fig:rt_fits} showing the best-fit model for HOPS-36. As in Figure \ref{fig:rt_fits}, black points ({\it left/right}) or color scale ({\it center}) show the data, while the blue lines ({\it left/right}) or contours ({\it center}) show the model, and the gray dashed line shows the disk contribution to the model.}
\figsetgrpend

\figsetgrpstart
\figsetgrpnum{1.8}
\figsetgrptitle{HOPS-41}
\figsetplot{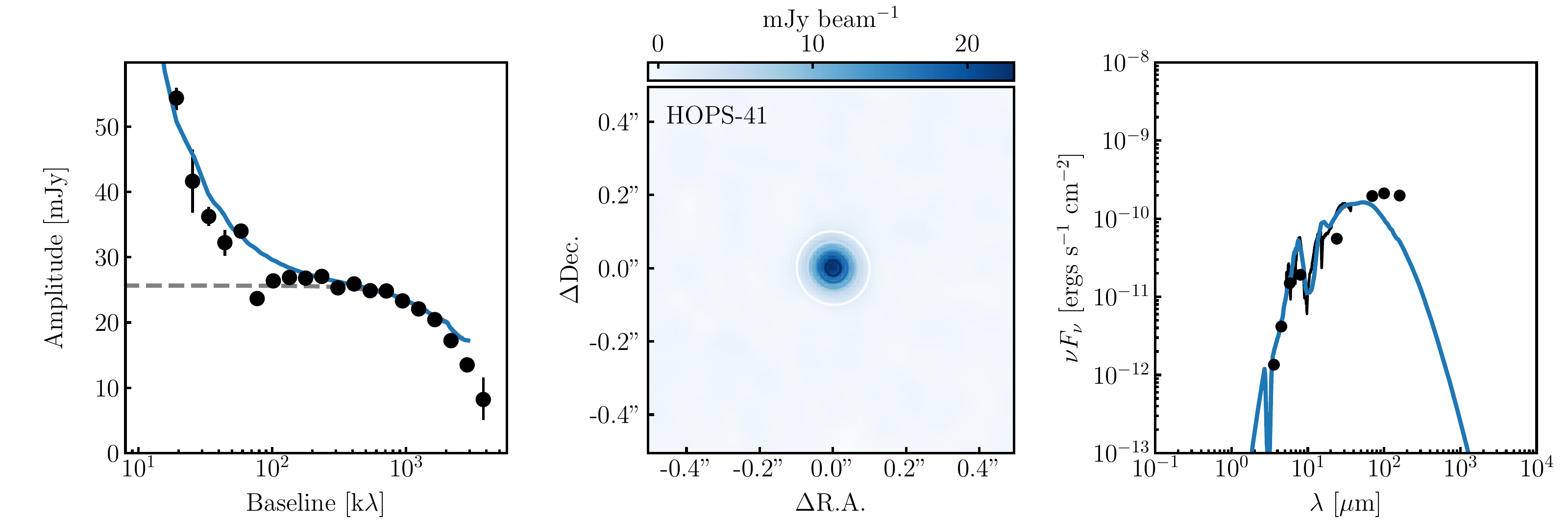}
\figsetgrpnote{A continuation of Figure \ref{fig:rt_fits} showing the best-fit model for HOPS-41. As in Figure \ref{fig:rt_fits}, black points ({\it left/right}) or color scale ({\it center}) show the data, while the blue lines ({\it left/right}) or contours ({\it center}) show the model, and the gray dashed line shows the disk contribution to the model.}
\figsetgrpend

\figsetgrpstart
\figsetgrpnum{1.9}
\figsetgrptitle{HOPS-42}
\figsetplot{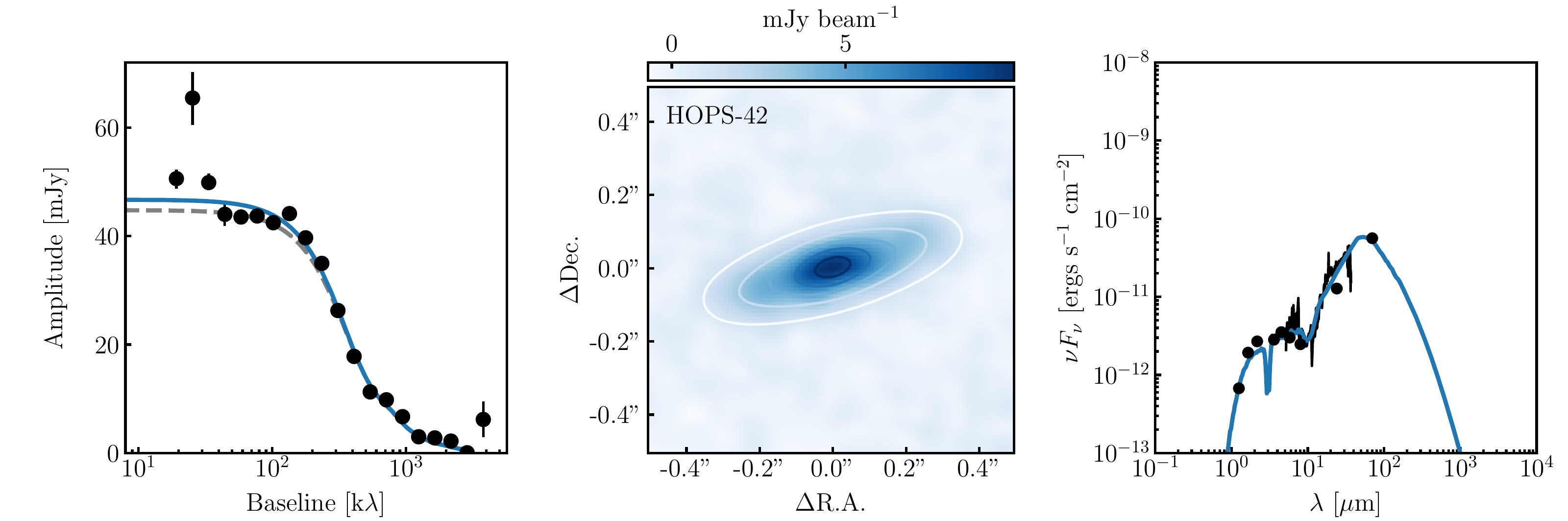}
\figsetgrpnote{A continuation of Figure \ref{fig:rt_fits} showing the best-fit model for HOPS-42. As in Figure \ref{fig:rt_fits}, black points ({\it left/right}) or color scale ({\it center}) show the data, while the blue lines ({\it left/right}) or contours ({\it center}) show the model, and the gray dashed line shows the disk contribution to the model.}
\figsetgrpend

\figsetgrpstart
\figsetgrpnum{1.10}
\figsetgrptitle{HOPS-43}
\figsetplot{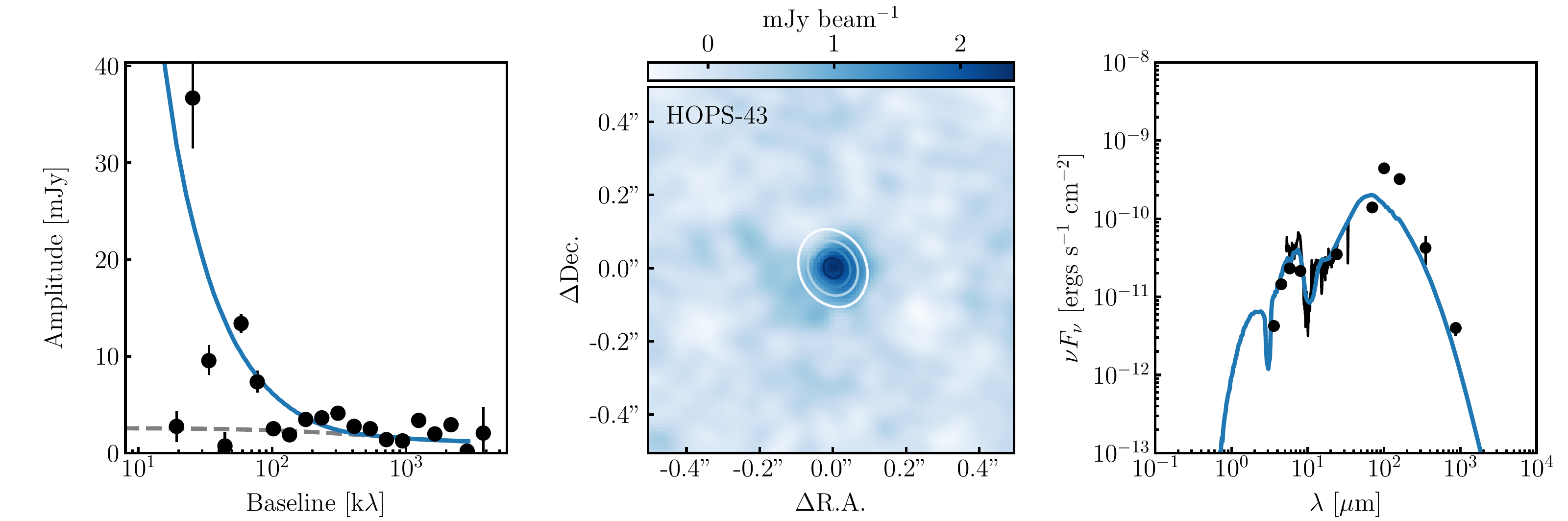}
\figsetgrpnote{A continuation of Figure \ref{fig:rt_fits} showing the best-fit model for HOPS-43. As in Figure \ref{fig:rt_fits}, black points ({\it left/right}) or color scale ({\it center}) show the data, while the blue lines ({\it left/right}) or contours ({\it center}) show the model, and the gray dashed line shows the disk contribution to the model.}
\figsetgrpend

\figsetgrpstart
\figsetgrpnum{1.11}
\figsetgrptitle{HOPS-49}
\figsetplot{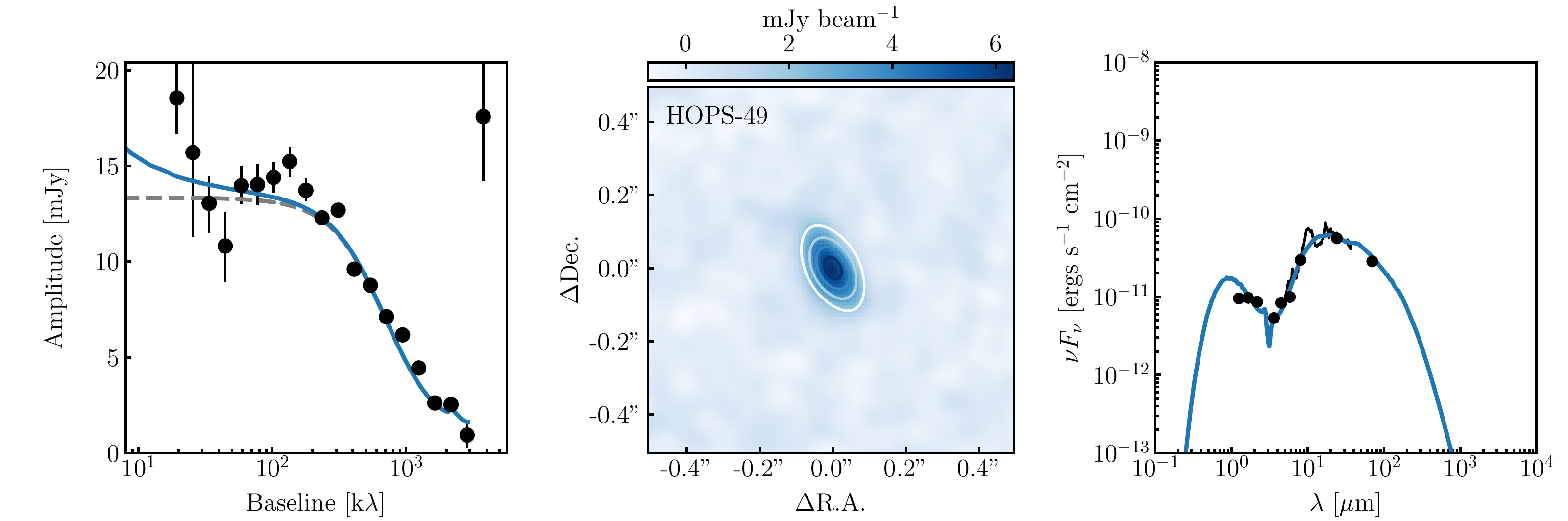}
\figsetgrpnote{A continuation of Figure \ref{fig:rt_fits} showing the best-fit model for HOPS-49. As in Figure \ref{fig:rt_fits}, black points ({\it left/right}) or color scale ({\it center}) show the data, while the blue lines ({\it left/right}) or contours ({\it center}) show the model, and the gray dashed line shows the disk contribution to the model.}
\figsetgrpend

\figsetgrpstart
\figsetgrpnum{1.12}
\figsetgrptitle{HOPS-50}
\figsetplot{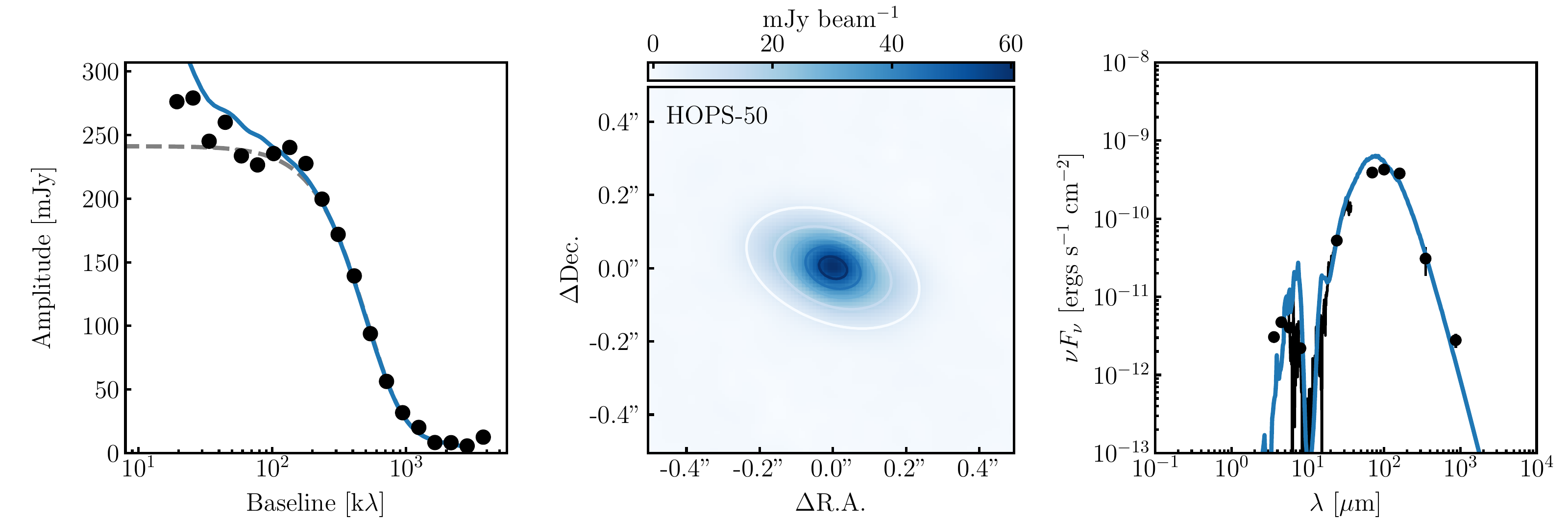}
\figsetgrpnote{A continuation of Figure \ref{fig:rt_fits} showing the best-fit model for HOPS-50. As in Figure \ref{fig:rt_fits}, black points ({\it left/right}) or color scale ({\it center}) show the data, while the blue lines ({\it left/right}) or contours ({\it center}) show the model, and the gray dashed line shows the disk contribution to the model.}
\figsetgrpend

\figsetgrpstart
\figsetgrpnum{1.13}
\figsetgrptitle{HOPS-53}
\figsetplot{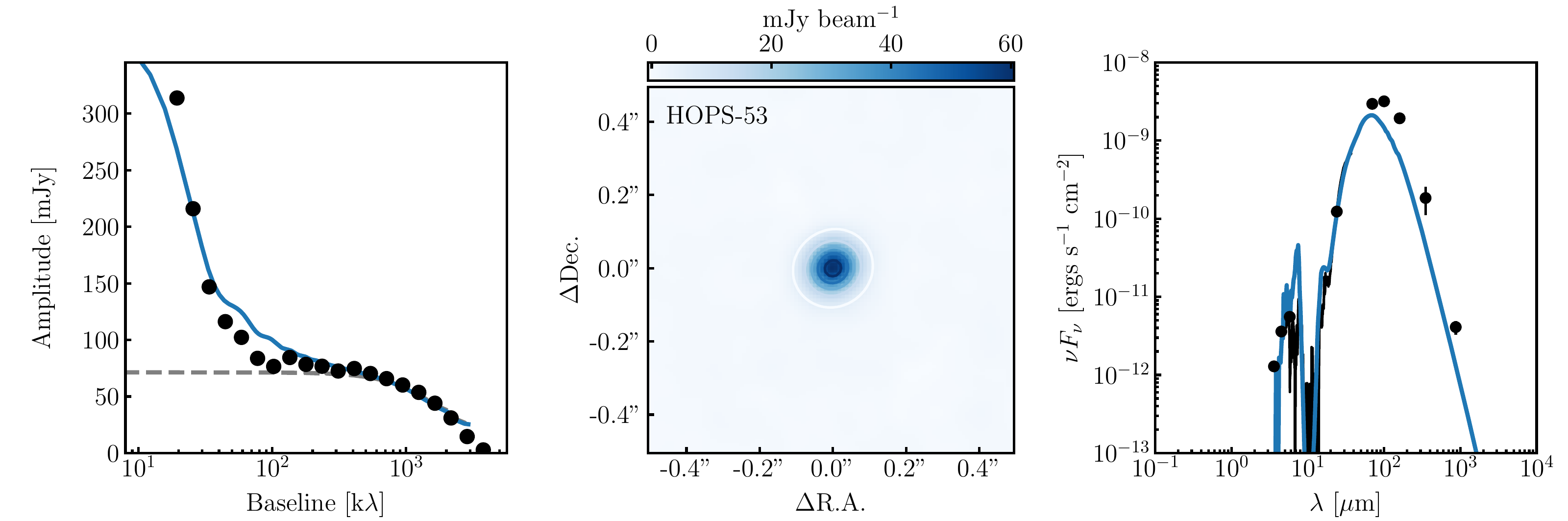}
\figsetgrpnote{A continuation of Figure \ref{fig:rt_fits} showing the best-fit model for HOPS-53. As in Figure \ref{fig:rt_fits}, black points ({\it left/right}) or color scale ({\it center}) show the data, while the blue lines ({\it left/right}) or contours ({\it center}) show the model, and the gray dashed line shows the disk contribution to the model.}
\figsetgrpend

\figsetgrpstart
\figsetgrpnum{1.14}
\figsetgrptitle{HOPS-58}
\figsetplot{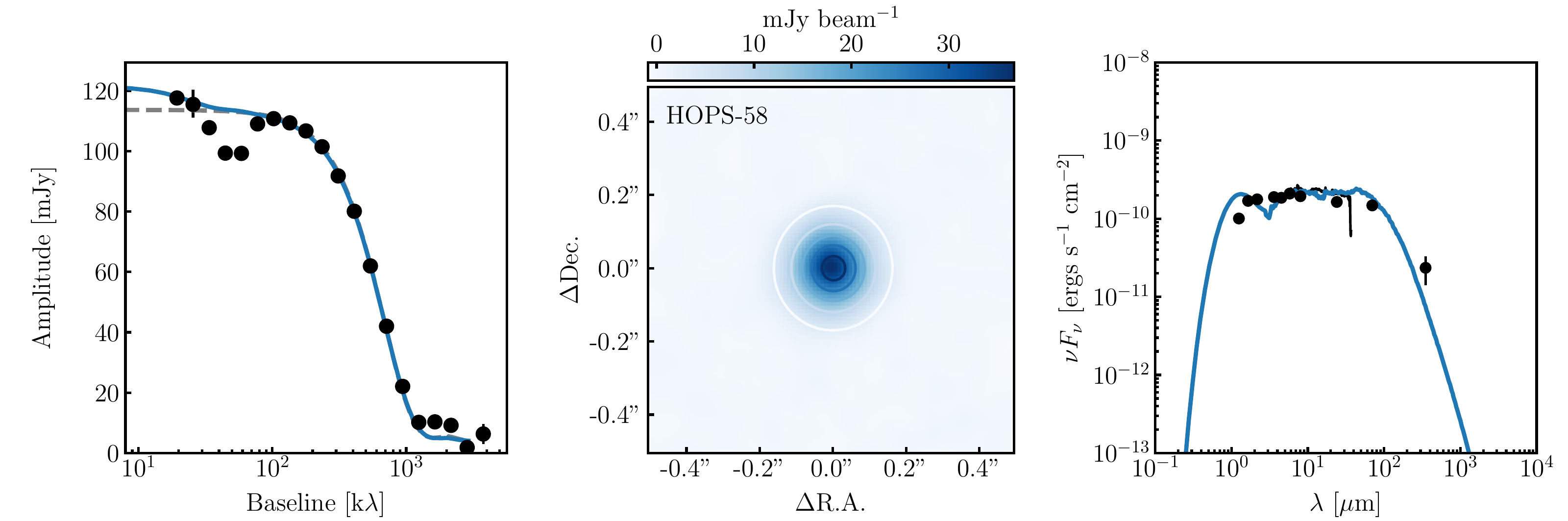}
\figsetgrpnote{A continuation of Figure \ref{fig:rt_fits} showing the best-fit model for HOPS-58. As in Figure \ref{fig:rt_fits}, black points ({\it left/right}) or color scale ({\it center}) show the data, while the blue lines ({\it left/right}) or contours ({\it center}) show the model, and the gray dashed line shows the disk contribution to the model.}
\figsetgrpend

\figsetgrpstart
\figsetgrpnum{1.15}
\figsetgrptitle{HOPS-81}
\figsetplot{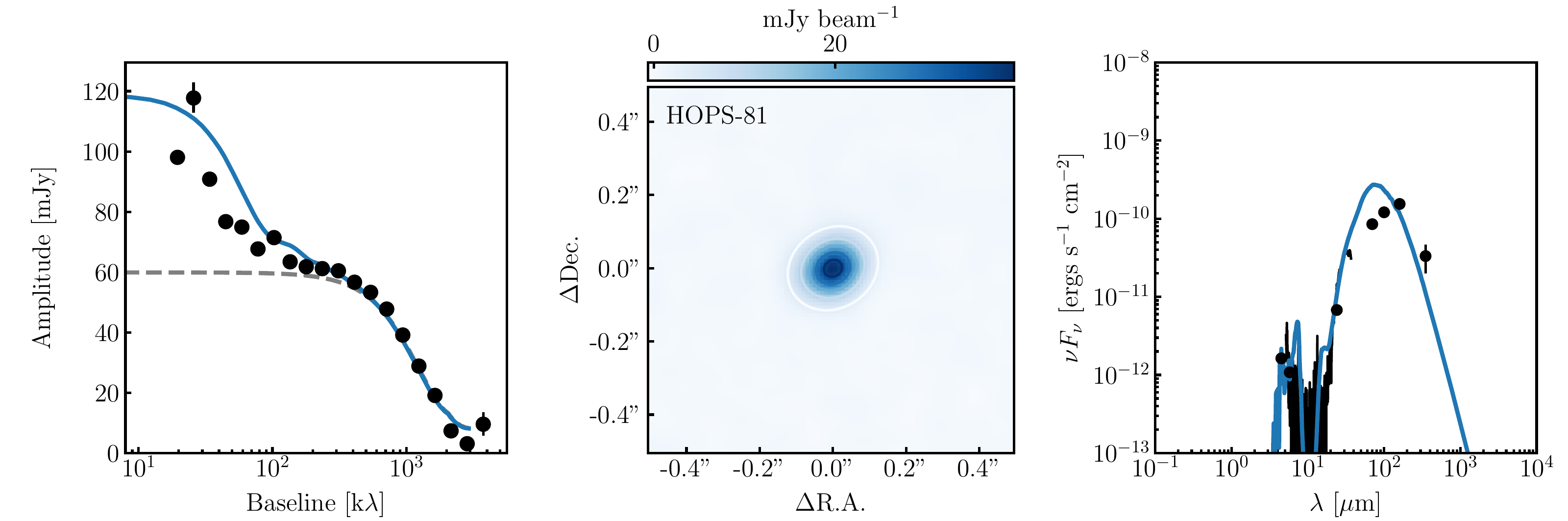}
\figsetgrpnote{A continuation of Figure \ref{fig:rt_fits} showing the best-fit model for HOPS-81. As in Figure \ref{fig:rt_fits}, black points ({\it left/right}) or color scale ({\it center}) show the data, while the blue lines ({\it left/right}) or contours ({\it center}) show the model, and the gray dashed line shows the disk contribution to the model.}
\figsetgrpend

\figsetgrpstart
\figsetgrpnum{1.16}
\figsetgrptitle{HOPS-82}
\figsetplot{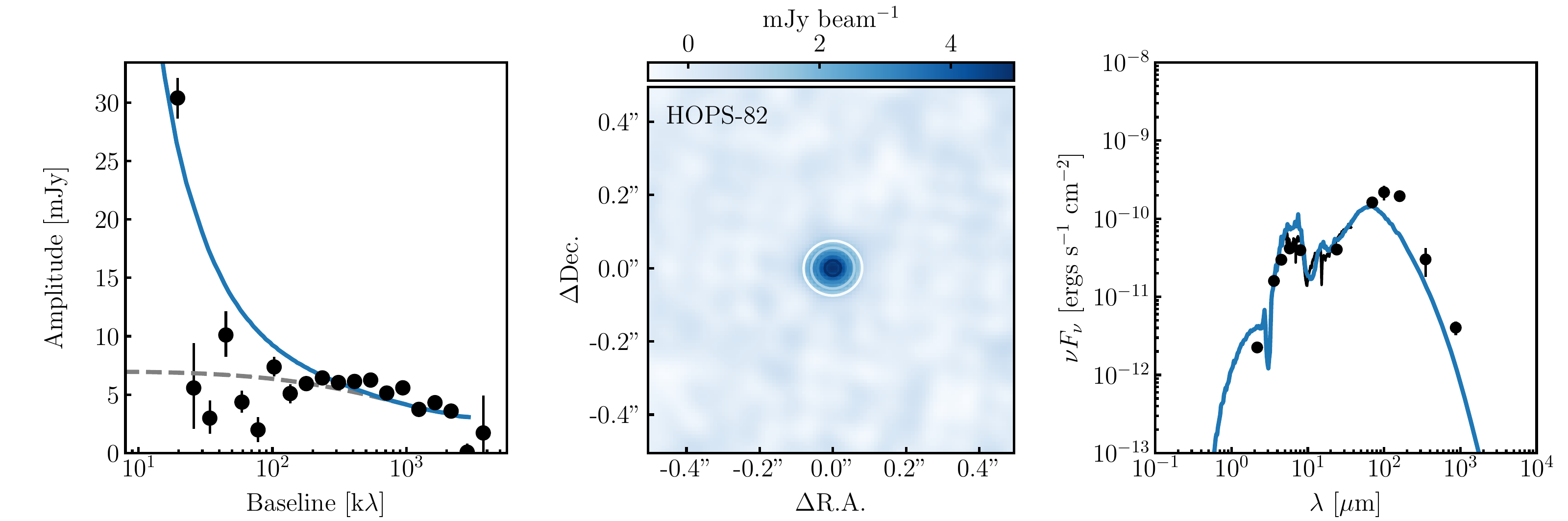}
\figsetgrpnote{A continuation of Figure \ref{fig:rt_fits} showing the best-fit model for HOPS-82. As in Figure \ref{fig:rt_fits}, black points ({\it left/right}) or color scale ({\it center}) show the data, while the blue lines ({\it left/right}) or contours ({\it center}) show the model, and the gray dashed line shows the disk contribution to the model.}
\figsetgrpend

\figsetgrpstart
\figsetgrpnum{1.17}
\figsetgrptitle{HOPS-87}
\figsetplot{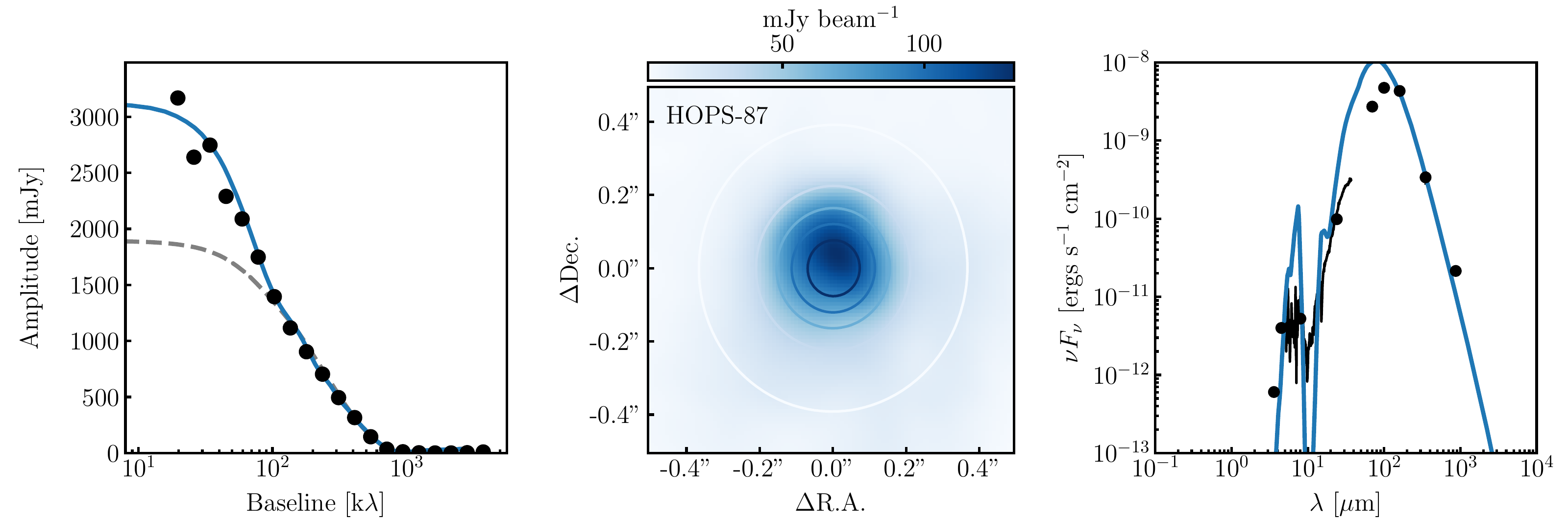}
\figsetgrpnote{A continuation of Figure \ref{fig:rt_fits} showing the best-fit model for HOPS-87. As in Figure \ref{fig:rt_fits}, black points ({\it left/right}) or color scale ({\it center}) show the data, while the blue lines ({\it left/right}) or contours ({\it center}) show the model, and the gray dashed line shows the disk contribution to the model.}
\figsetgrpend

\figsetgrpstart
\figsetgrpnum{1.18}
\figsetgrptitle{HOPS-90}
\figsetplot{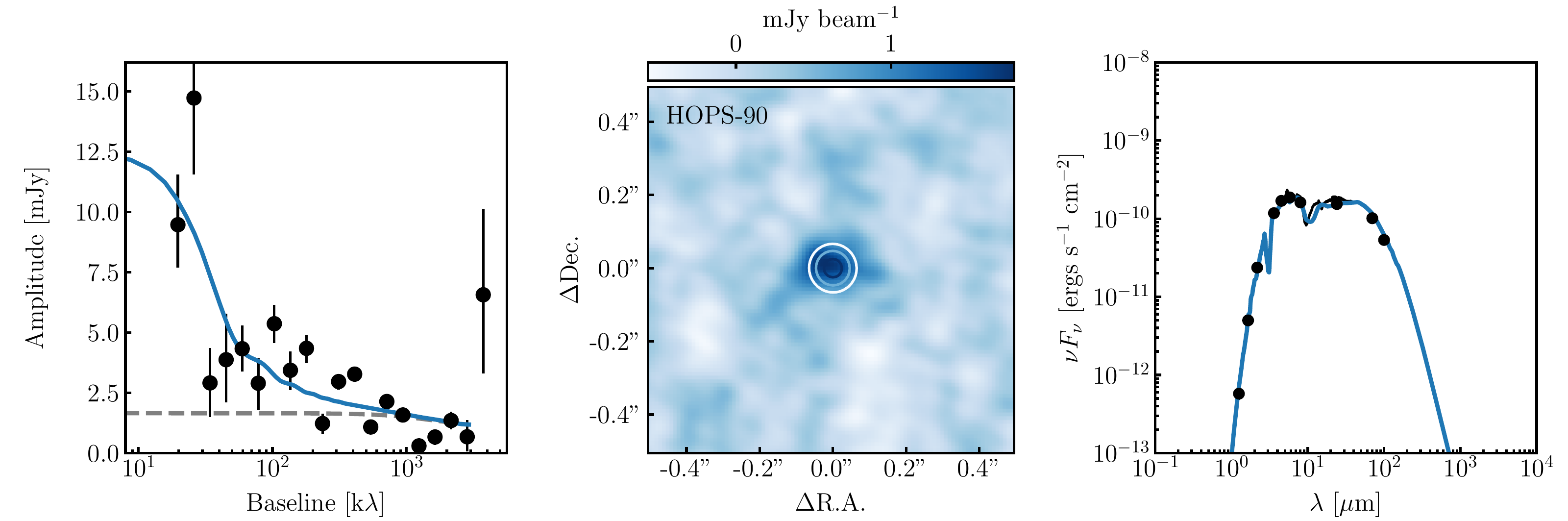}
\figsetgrpnote{A continuation of Figure \ref{fig:rt_fits} showing the best-fit model for HOPS-90. As in Figure \ref{fig:rt_fits}, black points ({\it left/right}) or color scale ({\it center}) show the data, while the blue lines ({\it left/right}) or contours ({\it center}) show the model, and the gray dashed line shows the disk contribution to the model.}
\figsetgrpend

\figsetgrpstart
\figsetgrpnum{1.19}
\figsetgrptitle{HOPS-91}
\figsetplot{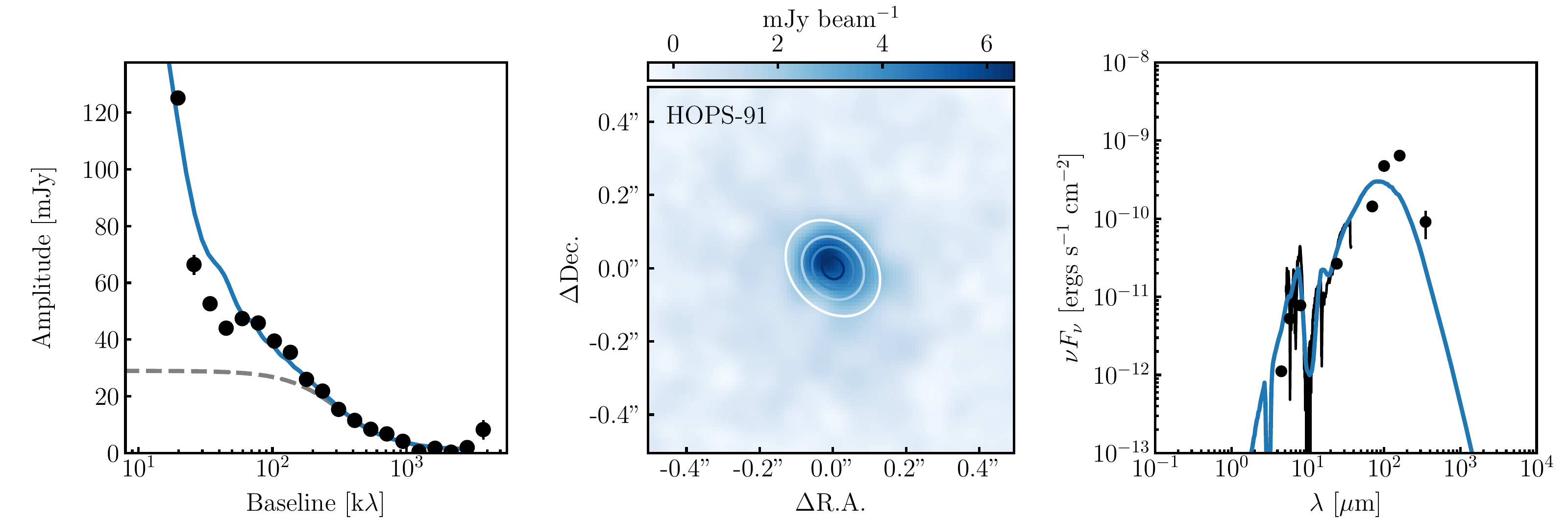}
\figsetgrpnote{A continuation of Figure \ref{fig:rt_fits} showing the best-fit model for HOPS-91. As in Figure \ref{fig:rt_fits}, black points ({\it left/right}) or color scale ({\it center}) show the data, while the blue lines ({\it left/right}) or contours ({\it center}) show the model, and the gray dashed line shows the disk contribution to the model.}
\figsetgrpend

\figsetgrpstart
\figsetgrpnum{1.20}
\figsetgrptitle{HOPS-93}
\figsetplot{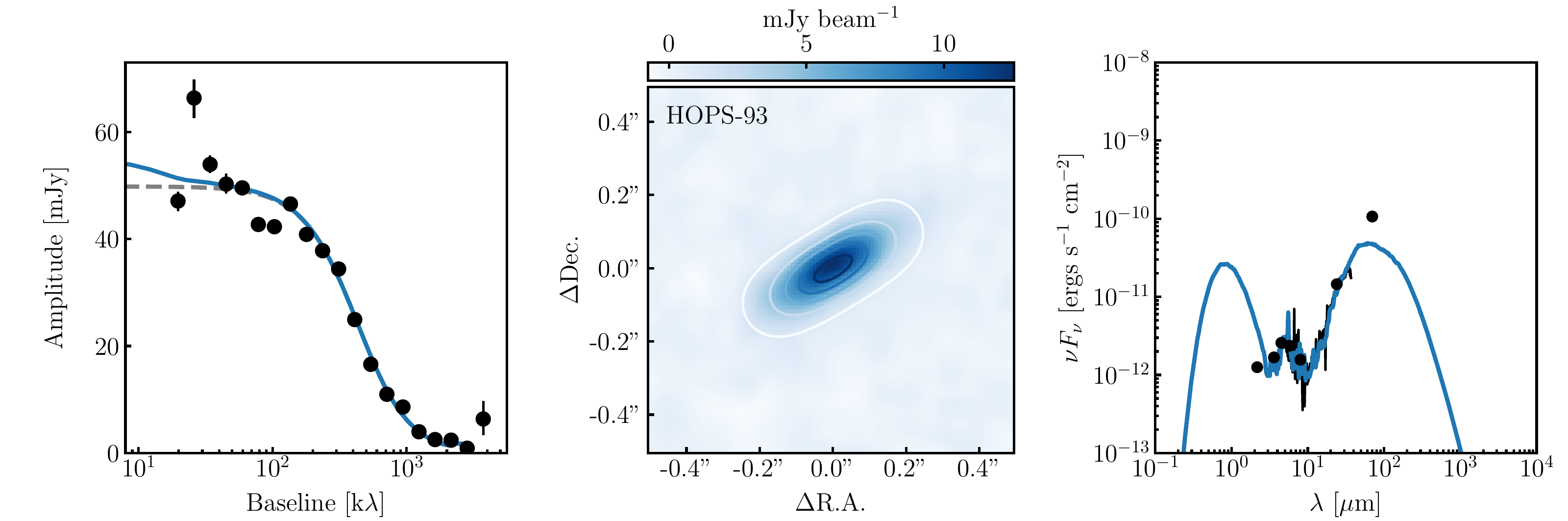}
\figsetgrpnote{A continuation of Figure \ref{fig:rt_fits} showing the best-fit model for HOPS-93. As in Figure \ref{fig:rt_fits}, black points ({\it left/right}) or color scale ({\it center}) show the data, while the blue lines ({\it left/right}) or contours ({\it center}) show the model, and the gray dashed line shows the disk contribution to the model.}
\figsetgrpend

\figsetgrpstart
\figsetgrpnum{1.21}
\figsetgrptitle{HOPS-95}
\figsetplot{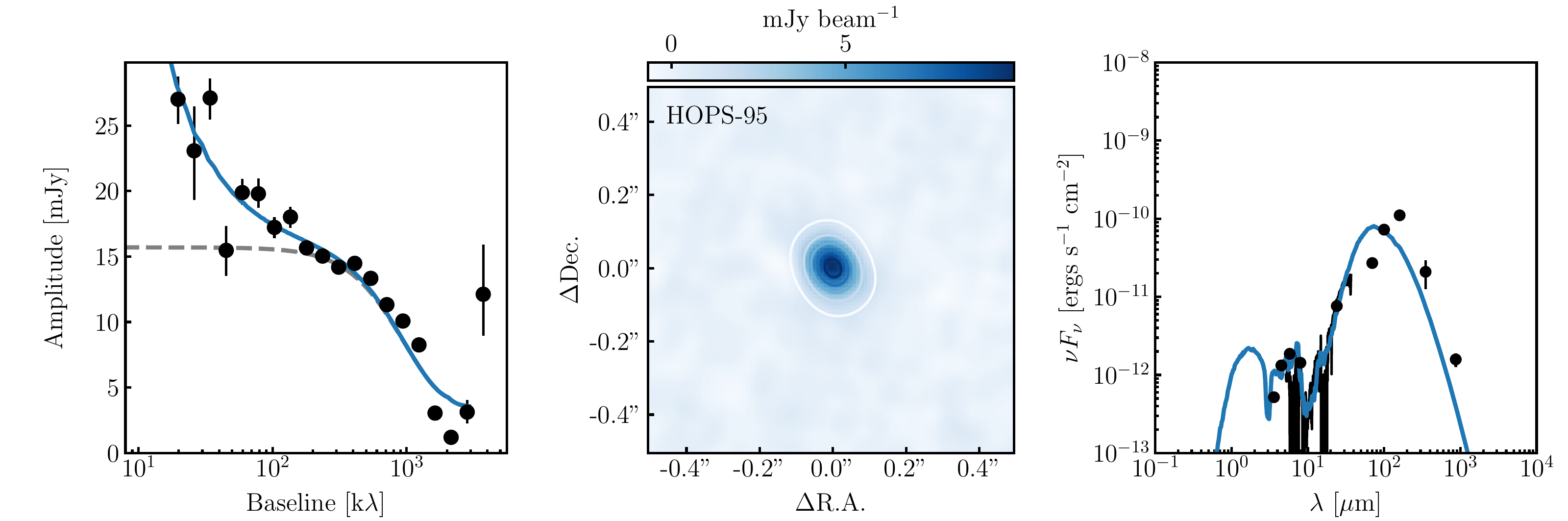}
\figsetgrpnote{A continuation of Figure \ref{fig:rt_fits} showing the best-fit model for HOPS-95. As in Figure \ref{fig:rt_fits}, black points ({\it left/right}) or color scale ({\it center}) show the data, while the blue lines ({\it left/right}) or contours ({\it center}) show the model, and the gray dashed line shows the disk contribution to the model.}
\figsetgrpend

\figsetgrpstart
\figsetgrpnum{1.22}
\figsetgrptitle{HOPS-96}
\figsetplot{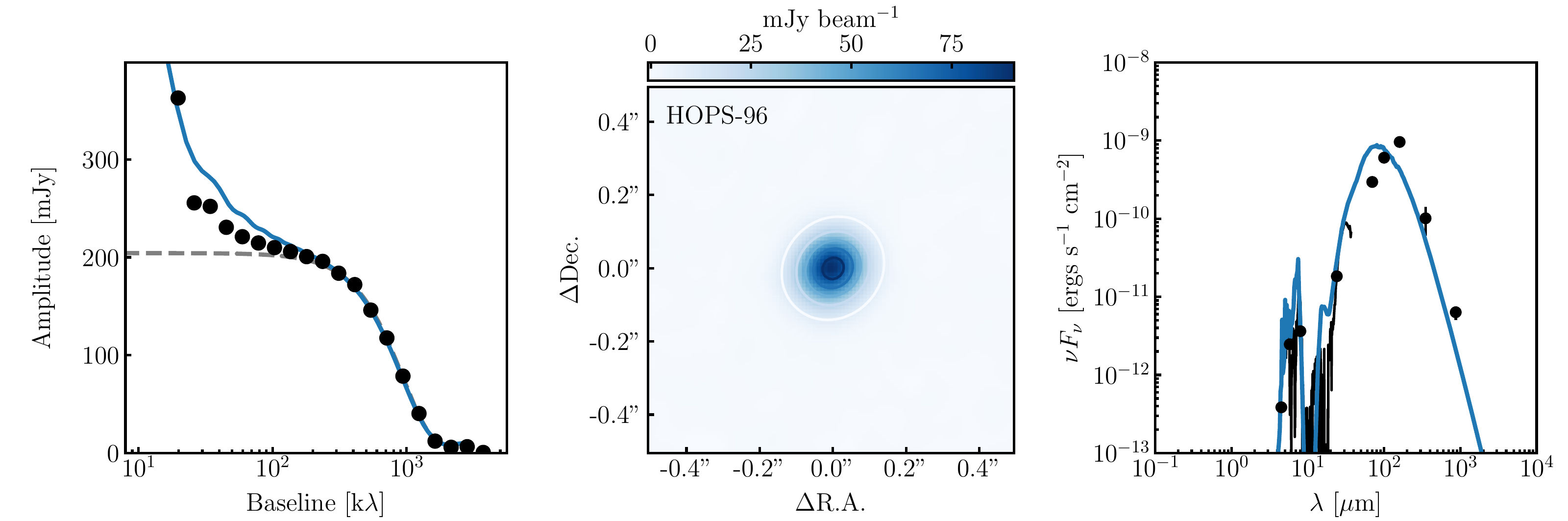}
\figsetgrpnote{A continuation of Figure \ref{fig:rt_fits} showing the best-fit model for HOPS-96. As in Figure \ref{fig:rt_fits}, black points ({\it left/right}) or color scale ({\it center}) show the data, while the blue lines ({\it left/right}) or contours ({\it center}) show the model, and the gray dashed line shows the disk contribution to the model.}
\figsetgrpend

\figsetgrpstart
\figsetgrpnum{1.23}
\figsetgrptitle{HOPS-99}
\figsetplot{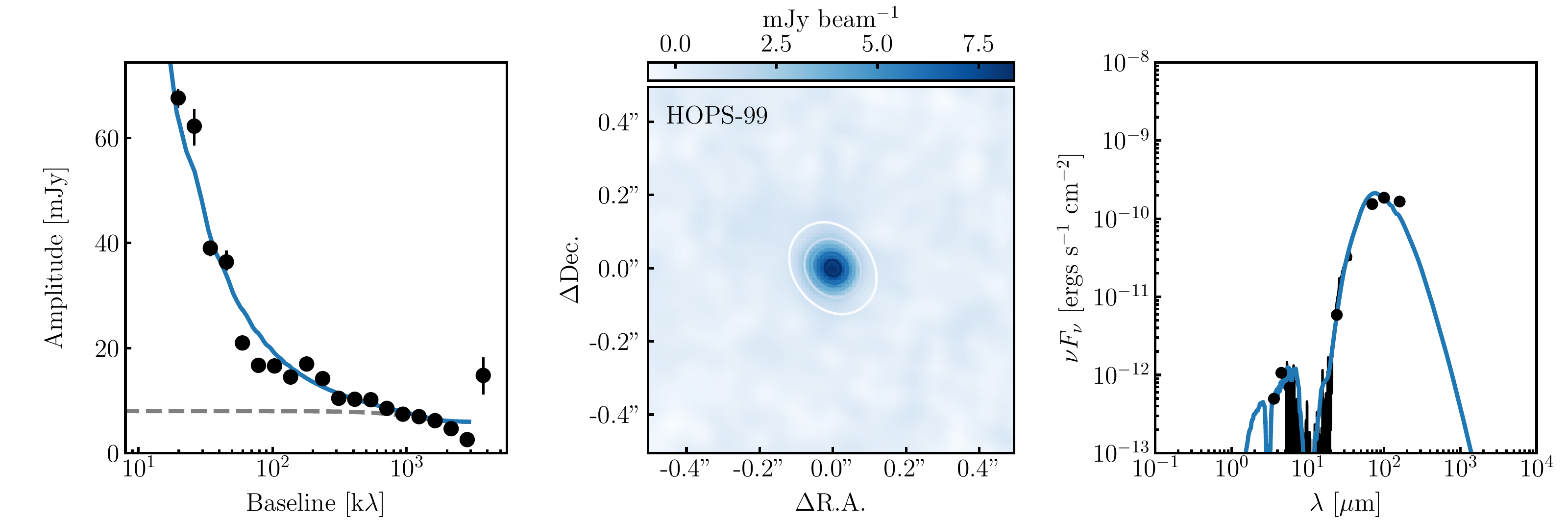}
\figsetgrpnote{A continuation of Figure \ref{fig:rt_fits} showing the best-fit model for HOPS-99. As in Figure \ref{fig:rt_fits}, black points ({\it left/right}) or color scale ({\it center}) show the data, while the blue lines ({\it left/right}) or contours ({\it center}) show the model, and the gray dashed line shows the disk contribution to the model.}
\figsetgrpend

\figsetgrpstart
\figsetgrpnum{1.24}
\figsetgrptitle{HOPS-117}
\figsetplot{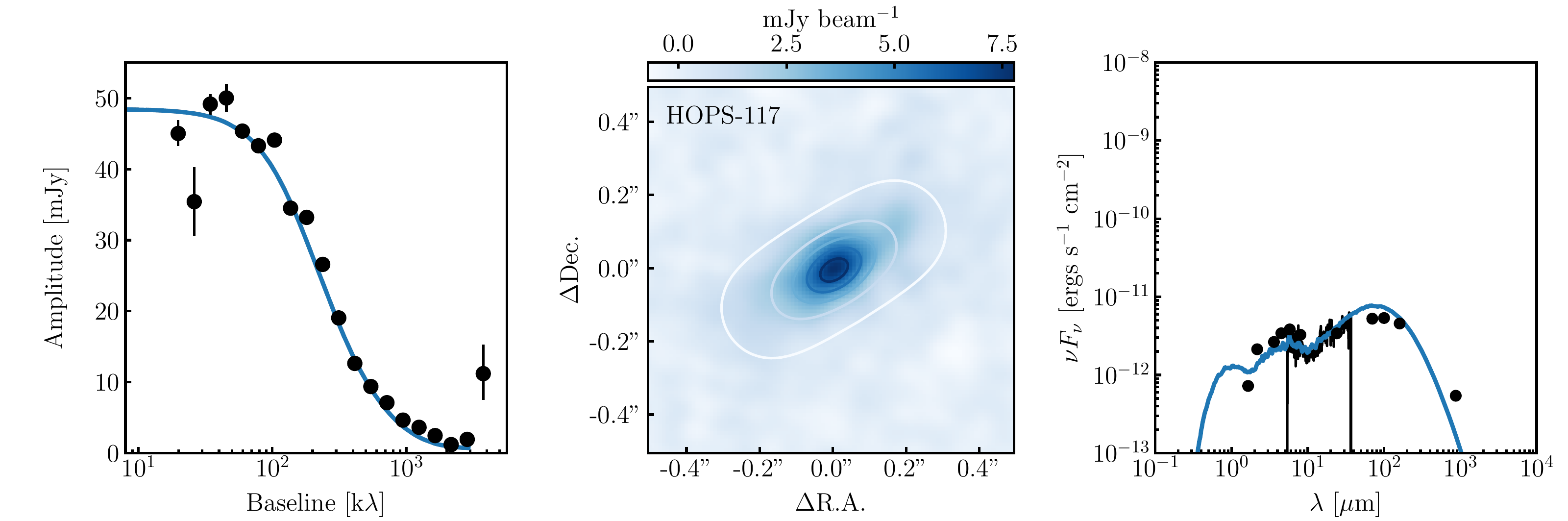}
\figsetgrpnote{A continuation of Figure \ref{fig:rt_fits} showing the best-fit model for HOPS-117. As in Figure \ref{fig:rt_fits}, black points ({\it left/right}) or color scale ({\it center}) show the data, while the blue lines ({\it left/right}) or contours ({\it center}) show the model, and the gray dashed line shows the disk contribution to the model.}
\figsetgrpend

\figsetgrpstart
\figsetgrpnum{1.25}
\figsetgrptitle{HOPS-120}
\figsetplot{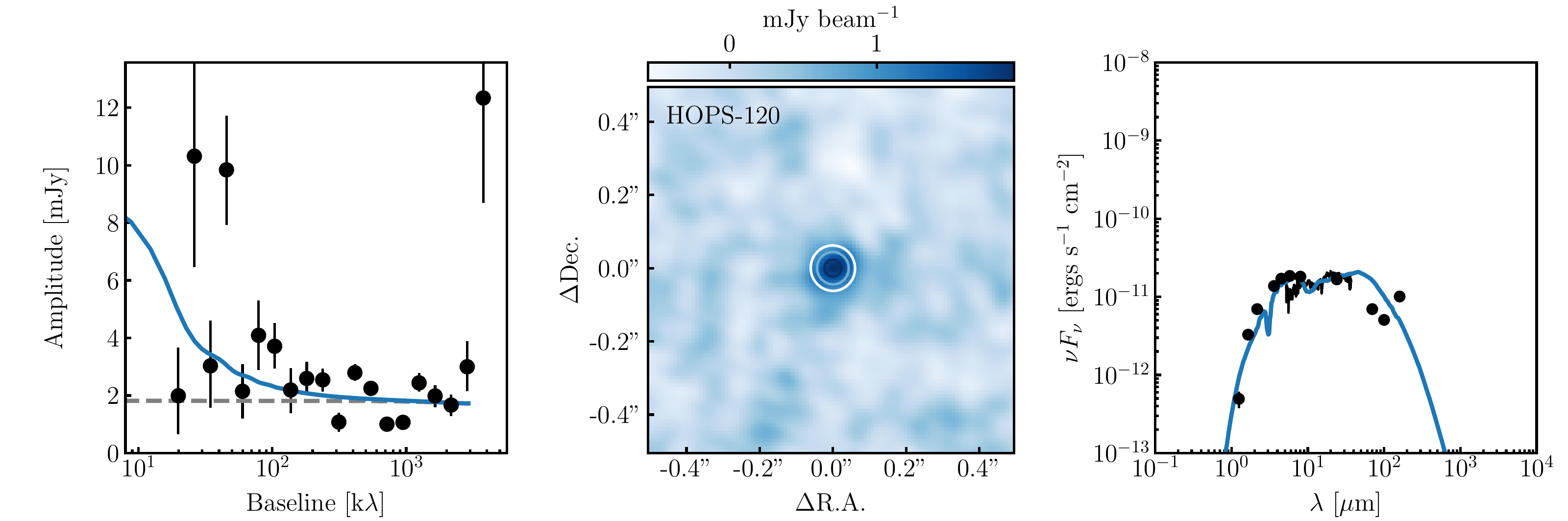}
\figsetgrpnote{A continuation of Figure \ref{fig:rt_fits} showing the best-fit model for HOPS-120. As in Figure \ref{fig:rt_fits}, black points ({\it left/right}) or color scale ({\it center}) show the data, while the blue lines ({\it left/right}) or contours ({\it center}) show the model, and the gray dashed line shows the disk contribution to the model.}
\figsetgrpend

\figsetgrpstart
\figsetgrpnum{1.26}
\figsetgrptitle{HOPS-122}
\figsetplot{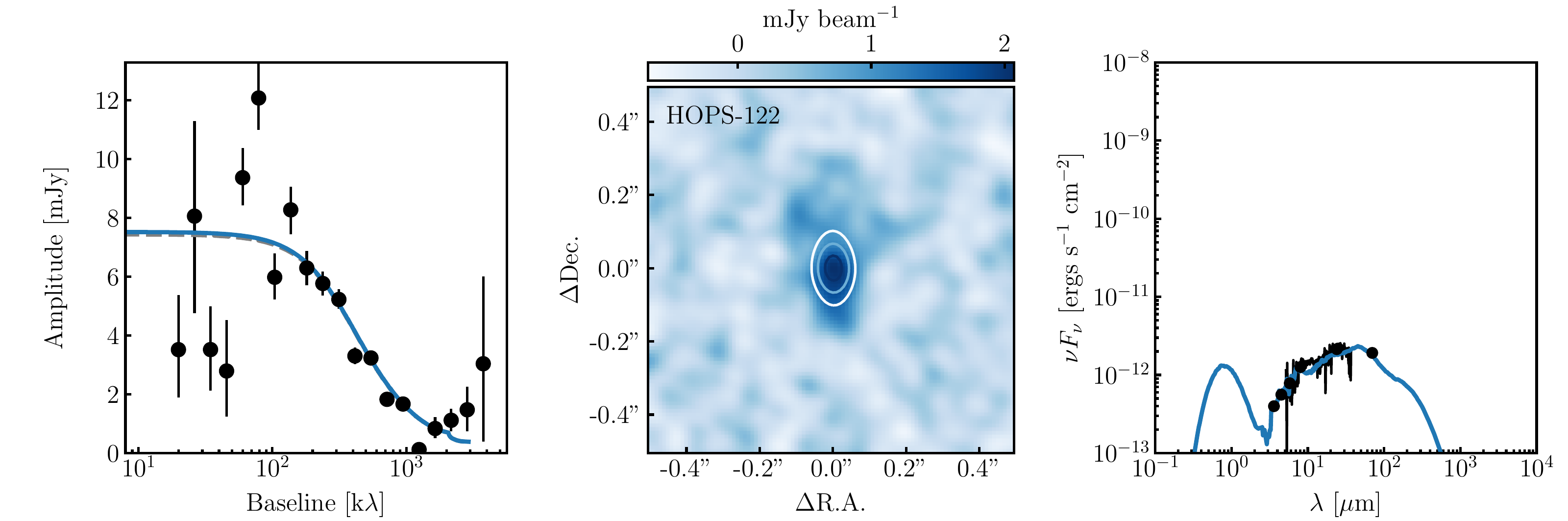}
\figsetgrpnote{A continuation of Figure \ref{fig:rt_fits} showing the best-fit model for HOPS-122. As in Figure \ref{fig:rt_fits}, black points ({\it left/right}) or color scale ({\it center}) show the data, while the blue lines ({\it left/right}) or contours ({\it center}) show the model, and the gray dashed line shows the disk contribution to the model.}
\figsetgrpend

\figsetgrpstart
\figsetgrpnum{1.27}
\figsetgrptitle{HOPS-123}
\figsetplot{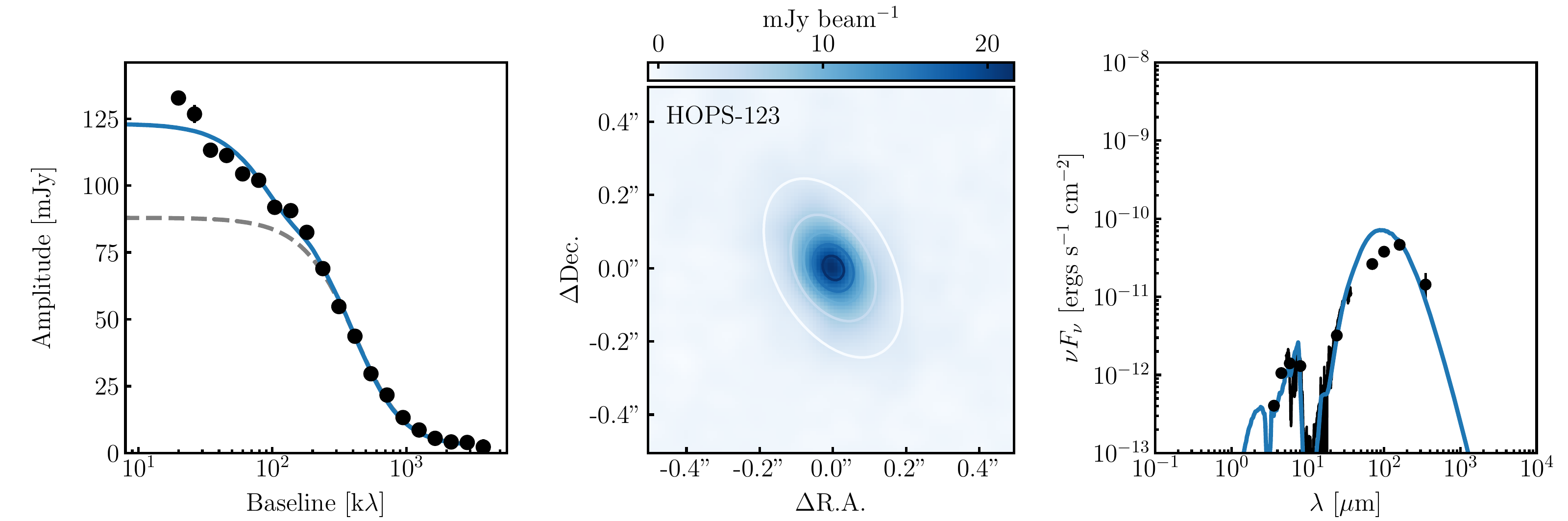}
\figsetgrpnote{A continuation of Figure \ref{fig:rt_fits} showing the best-fit model for HOPS-123. As in Figure \ref{fig:rt_fits}, black points ({\it left/right}) or color scale ({\it center}) show the data, while the blue lines ({\it left/right}) or contours ({\it center}) show the model, and the gray dashed line shows the disk contribution to the model.}
\figsetgrpend

\figsetgrpstart
\figsetgrpnum{1.28}
\figsetgrptitle{HOPS-127}
\figsetplot{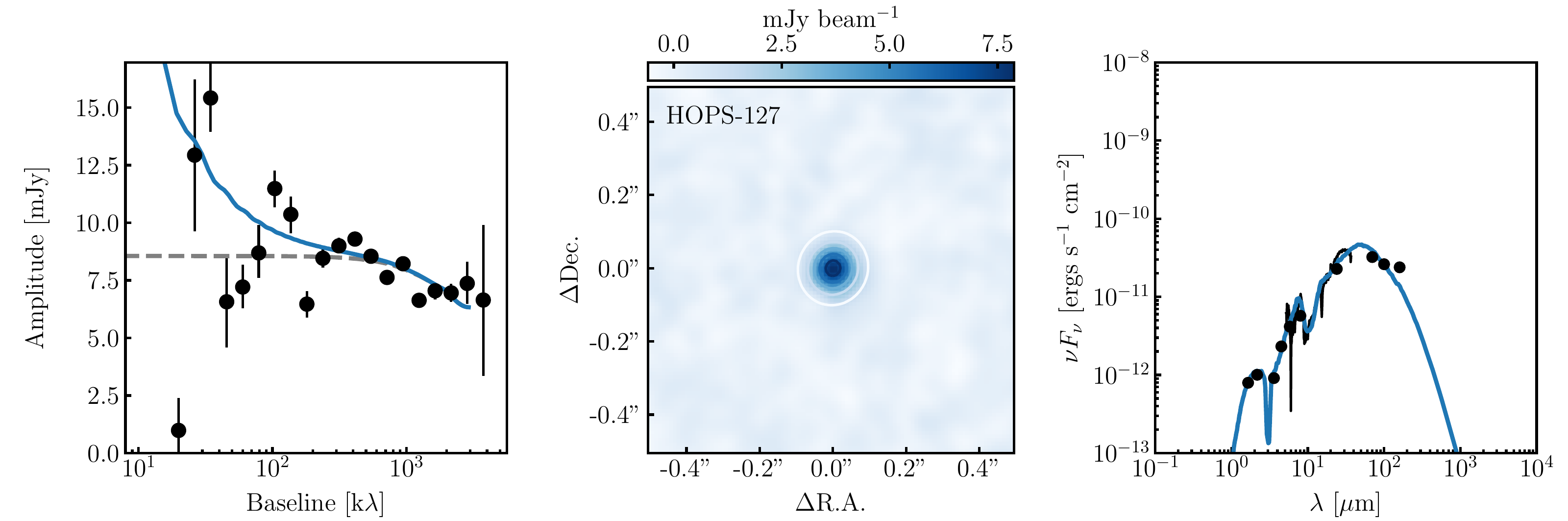}
\figsetgrpnote{A continuation of Figure \ref{fig:rt_fits} showing the best-fit model for HOPS-127. As in Figure \ref{fig:rt_fits}, black points ({\it left/right}) or color scale ({\it center}) show the data, while the blue lines ({\it left/right}) or contours ({\it center}) show the model, and the gray dashed line shows the disk contribution to the model.}
\figsetgrpend

\figsetgrpstart
\figsetgrpnum{1.29}
\figsetgrptitle{HOPS-131}
\figsetplot{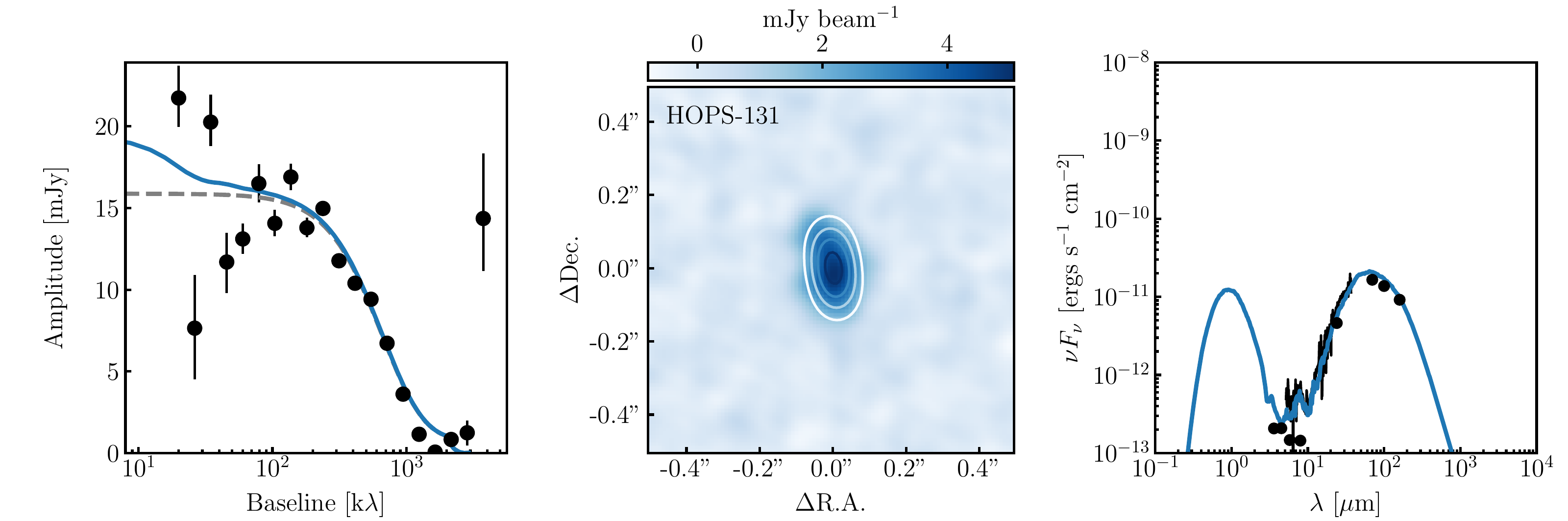}
\figsetgrpnote{A continuation of Figure \ref{fig:rt_fits} showing the best-fit model for HOPS-131. As in Figure \ref{fig:rt_fits}, black points ({\it left/right}) or color scale ({\it center}) show the data, while the blue lines ({\it left/right}) or contours ({\it center}) show the model, and the gray dashed line shows the disk contribution to the model.}
\figsetgrpend

\figsetgrpstart
\figsetgrpnum{1.30}
\figsetgrptitle{HOPS-133}
\figsetplot{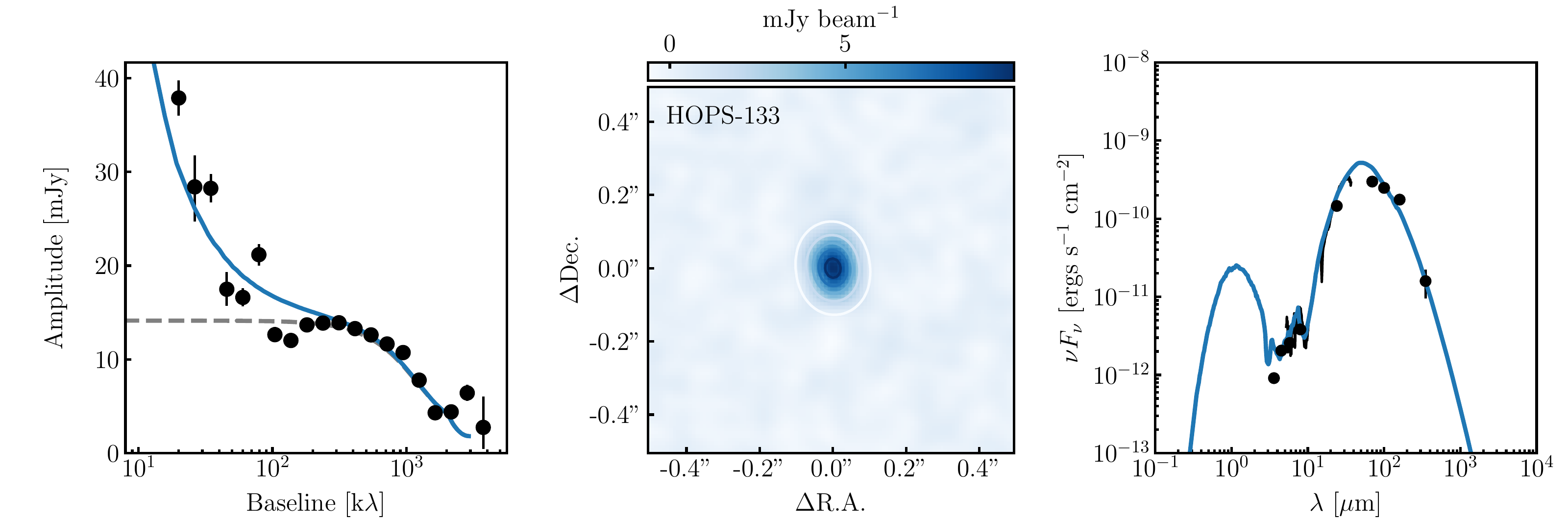}
\figsetgrpnote{A continuation of Figure \ref{fig:rt_fits} showing the best-fit model for HOPS-133. As in Figure \ref{fig:rt_fits}, black points ({\it left/right}) or color scale ({\it center}) show the data, while the blue lines ({\it left/right}) or contours ({\it center}) show the model, and the gray dashed line shows the disk contribution to the model.}
\figsetgrpend

\figsetgrpstart
\figsetgrpnum{1.31}
\figsetgrptitle{HOPS-134}
\figsetplot{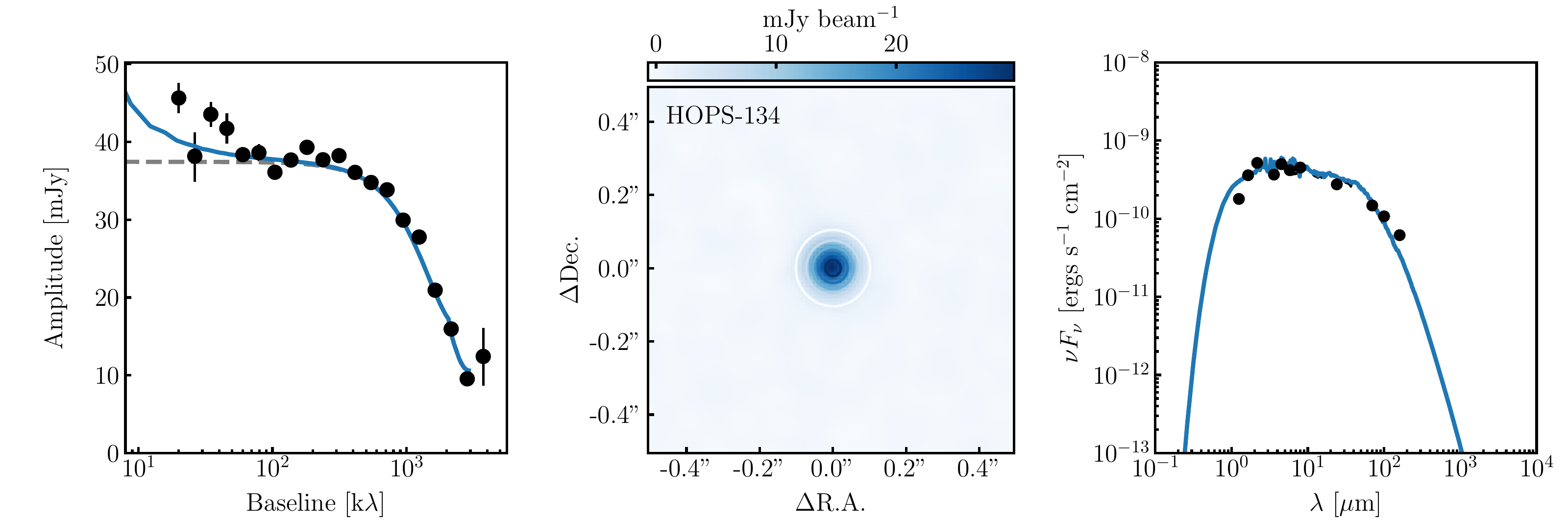}
\figsetgrpnote{A continuation of Figure \ref{fig:rt_fits} showing the best-fit model for HOPS-134. As in Figure \ref{fig:rt_fits}, black points ({\it left/right}) or color scale ({\it center}) show the data, while the blue lines ({\it left/right}) or contours ({\it center}) show the model, and the gray dashed line shows the disk contribution to the model.}
\figsetgrpend

\figsetgrpstart
\figsetgrpnum{1.32}
\figsetgrptitle{HOPS-136}
\figsetplot{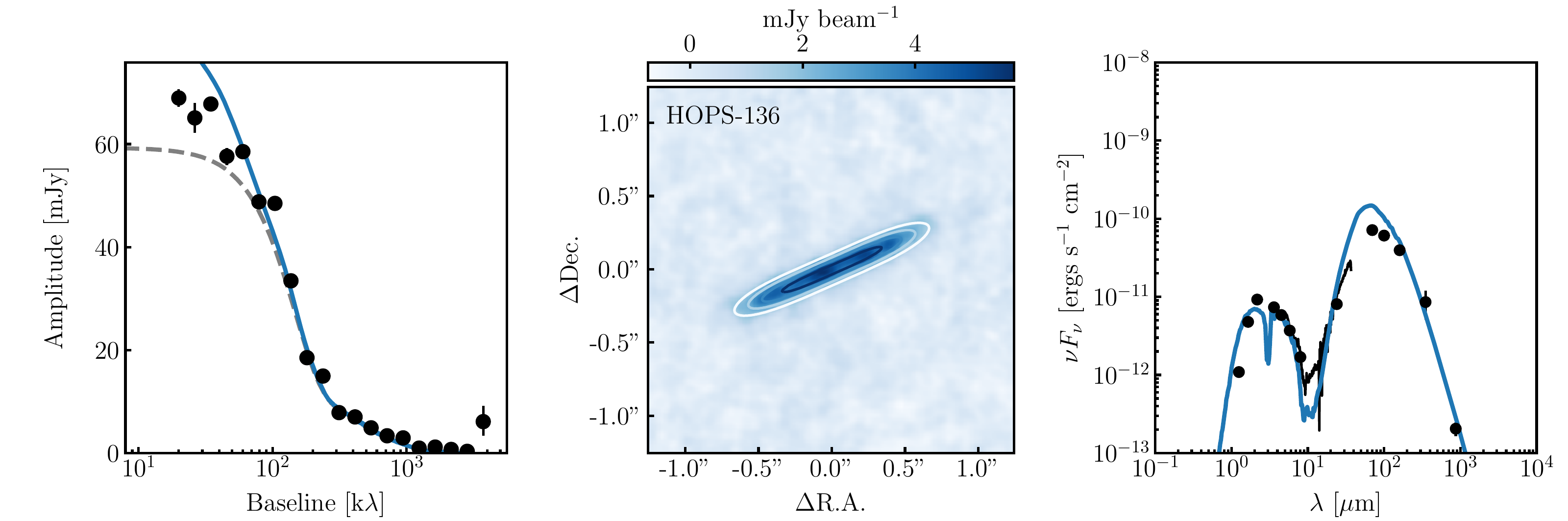}
\figsetgrpnote{A continuation of Figure \ref{fig:rt_fits} showing the best-fit model for HOPS-136. As in Figure \ref{fig:rt_fits}, black points ({\it left/right}) or color scale ({\it center}) show the data, while the blue lines ({\it left/right}) or contours ({\it center}) show the model, and the gray dashed line shows the disk contribution to the model.}
\figsetgrpend

\figsetgrpstart
\figsetgrpnum{1.33}
\figsetgrptitle{HOPS-139}
\figsetplot{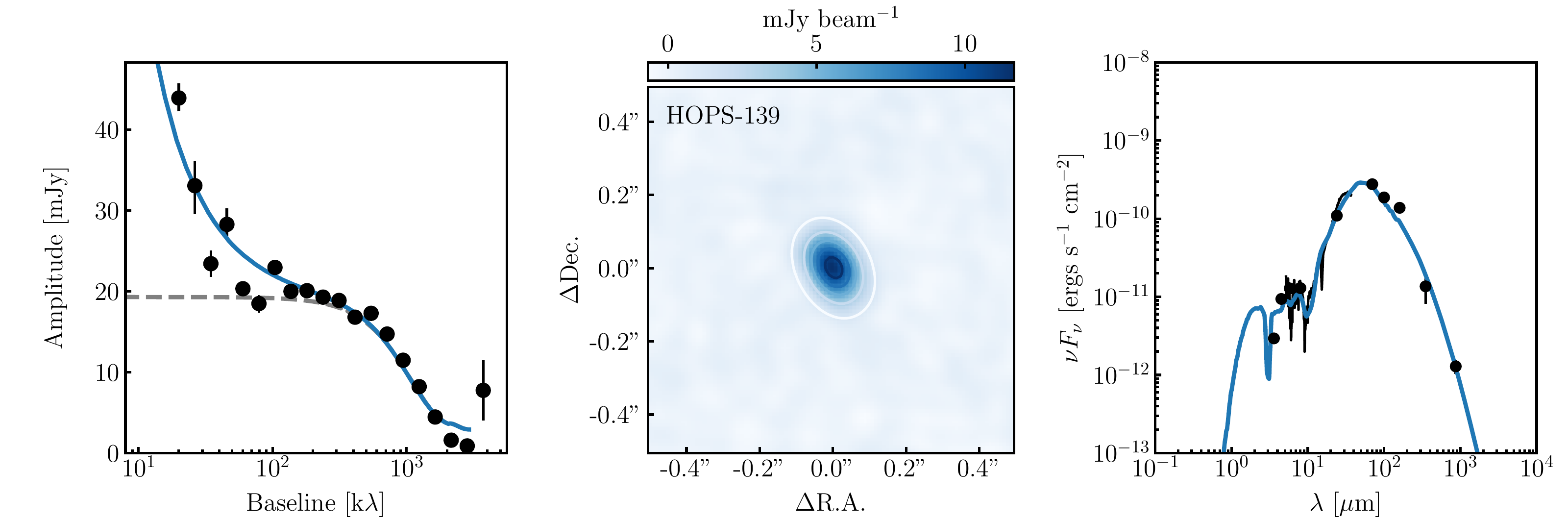}
\figsetgrpnote{A continuation of Figure \ref{fig:rt_fits} showing the best-fit model for HOPS-139. As in Figure \ref{fig:rt_fits}, black points ({\it left/right}) or color scale ({\it center}) show the data, while the blue lines ({\it left/right}) or contours ({\it center}) show the model, and the gray dashed line shows the disk contribution to the model.}
\figsetgrpend

\figsetgrpstart
\figsetgrpnum{1.34}
\figsetgrptitle{HOPS-143}
\figsetplot{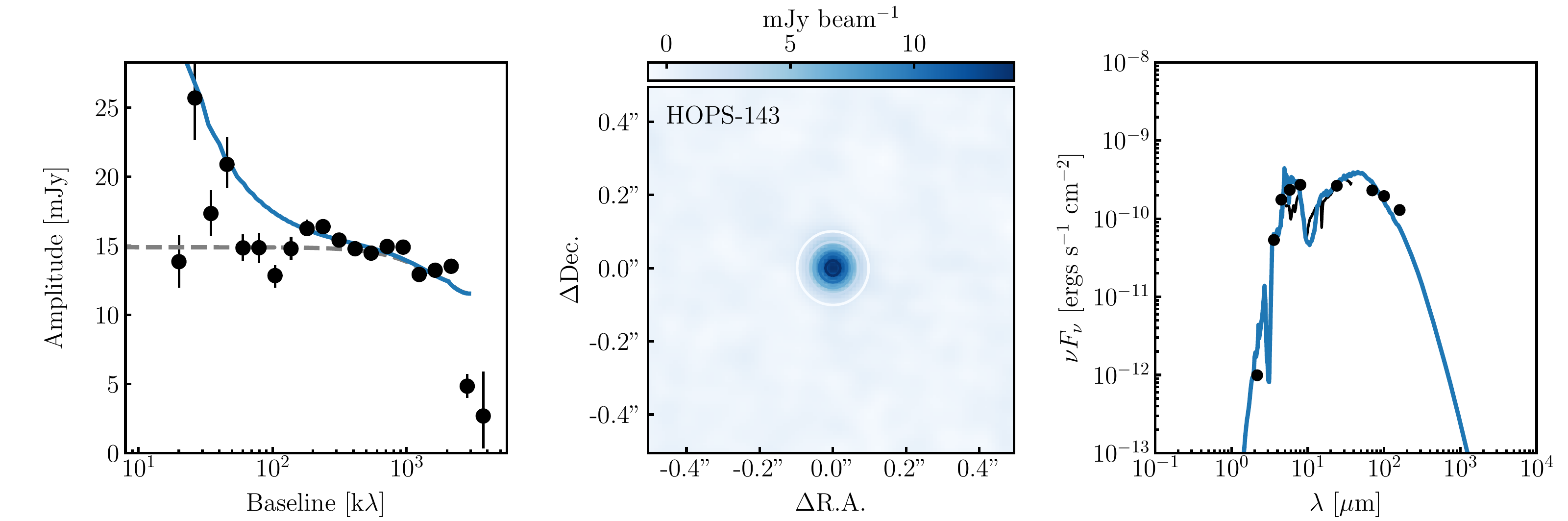}
\figsetgrpnote{A continuation of Figure \ref{fig:rt_fits} showing the best-fit model for HOPS-143. As in Figure \ref{fig:rt_fits}, black points ({\it left/right}) or color scale ({\it center}) show the data, while the blue lines ({\it left/right}) or contours ({\it center}) show the model, and the gray dashed line shows the disk contribution to the model.}
\figsetgrpend

\figsetgrpstart
\figsetgrpnum{1.35}
\figsetgrptitle{HOPS-144}
\figsetplot{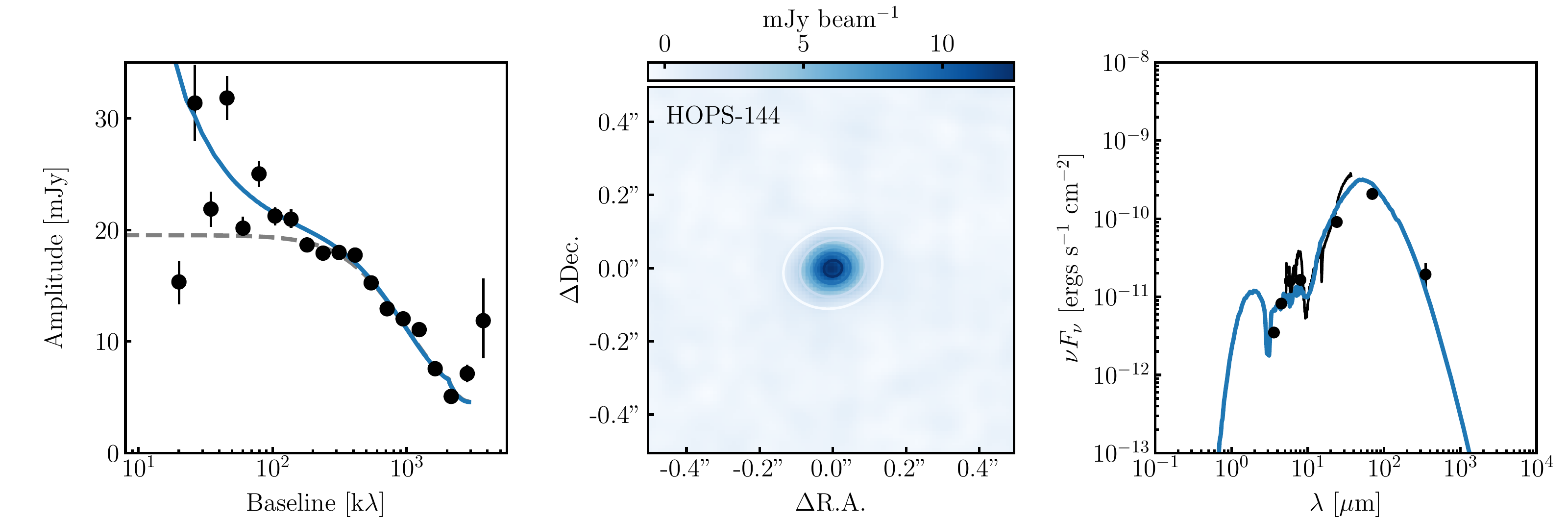}
\figsetgrpnote{A continuation of Figure \ref{fig:rt_fits} showing the best-fit model for HOPS-144. As in Figure \ref{fig:rt_fits}, black points ({\it left/right}) or color scale ({\it center}) show the data, while the blue lines ({\it left/right}) or contours ({\it center}) show the model, and the gray dashed line shows the disk contribution to the model.}
\figsetgrpend

\figsetgrpstart
\figsetgrpnum{1.36}
\figsetgrptitle{HOPS-152}
\figsetplot{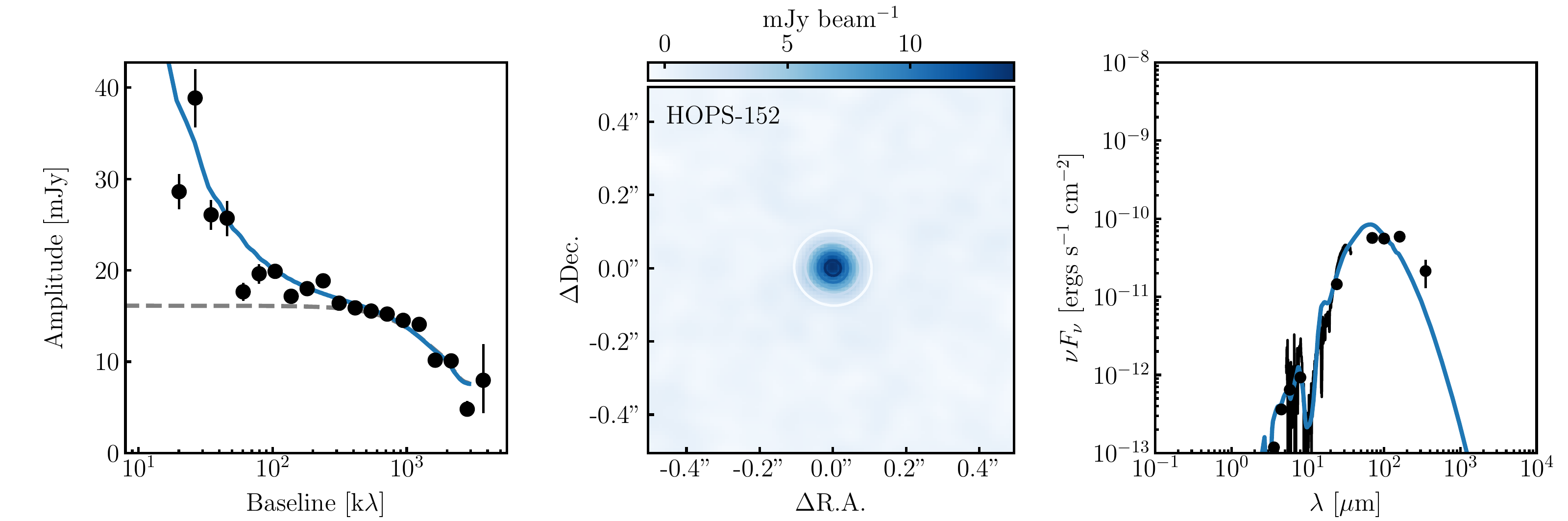}
\figsetgrpnote{A continuation of Figure \ref{fig:rt_fits} showing the best-fit model for HOPS-152. As in Figure \ref{fig:rt_fits}, black points ({\it left/right}) or color scale ({\it center}) show the data, while the blue lines ({\it left/right}) or contours ({\it center}) show the model, and the gray dashed line shows the disk contribution to the model.}
\figsetgrpend

\figsetgrpstart
\figsetgrpnum{1.37}
\figsetgrptitle{HOPS-154}
\figsetplot{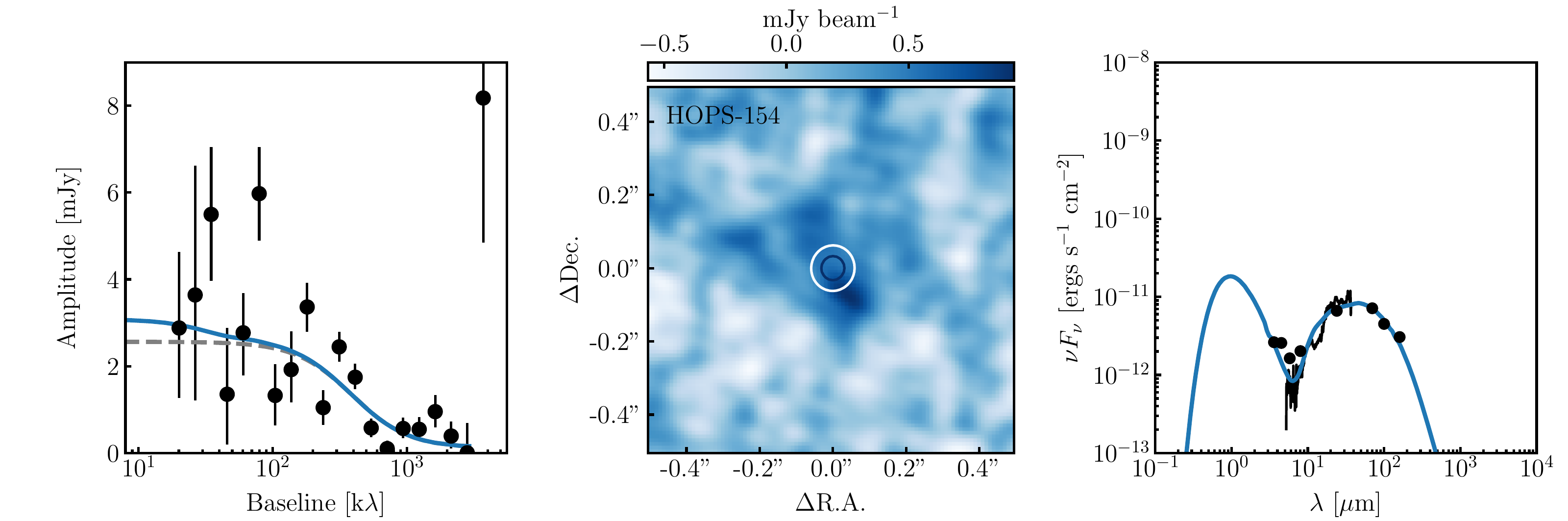}
\figsetgrpnote{A continuation of Figure \ref{fig:rt_fits} showing the best-fit model for HOPS-154. As in Figure \ref{fig:rt_fits}, black points ({\it left/right}) or color scale ({\it center}) show the data, while the blue lines ({\it left/right}) or contours ({\it center}) show the model, and the gray dashed line shows the disk contribution to the model.}
\figsetgrpend

\figsetgrpstart
\figsetgrpnum{1.38}
\figsetgrptitle{HOPS-160}
\figsetplot{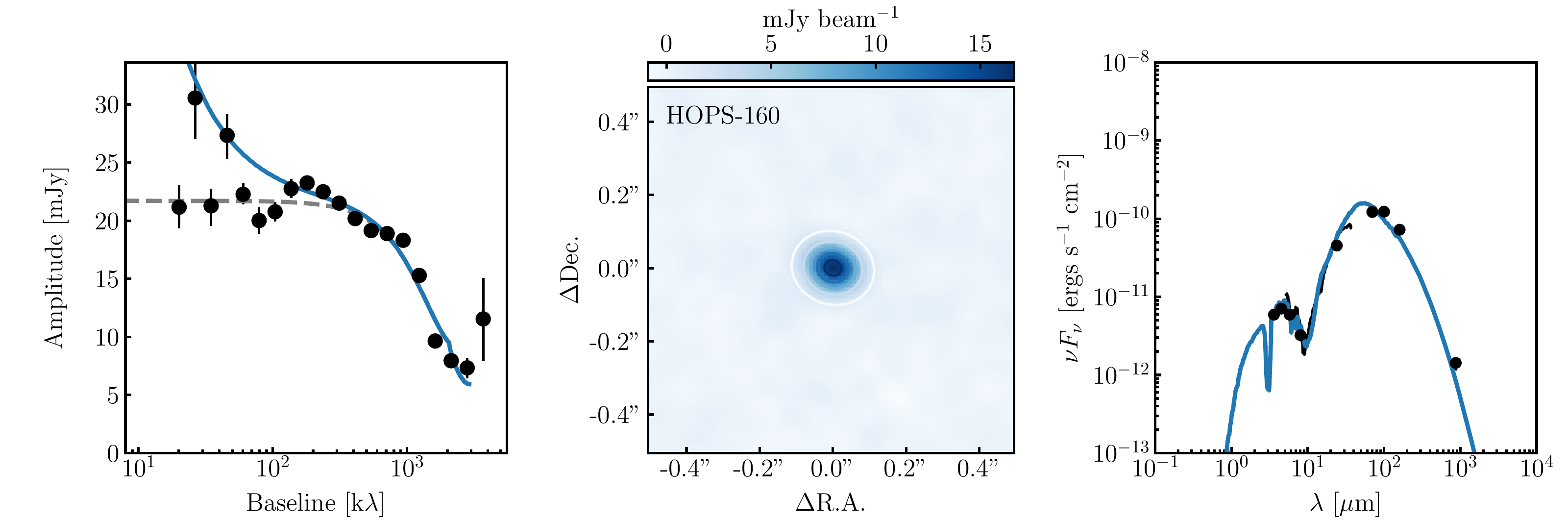}
\figsetgrpnote{A continuation of Figure \ref{fig:rt_fits} showing the best-fit model for HOPS-160. As in Figure \ref{fig:rt_fits}, black points ({\it left/right}) or color scale ({\it center}) show the data, while the blue lines ({\it left/right}) or contours ({\it center}) show the model, and the gray dashed line shows the disk contribution to the model.}
\figsetgrpend

\figsetgrpstart
\figsetgrpnum{1.39}
\figsetgrptitle{HOPS-165}
\figsetplot{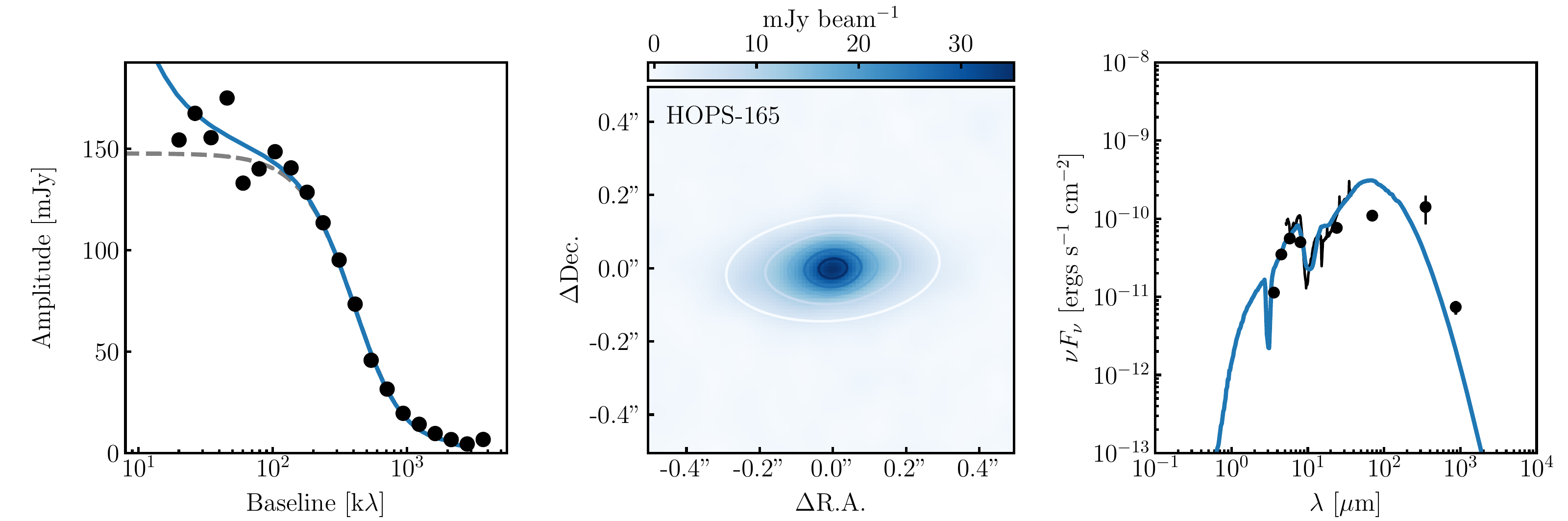}
\figsetgrpnote{A continuation of Figure \ref{fig:rt_fits} showing the best-fit model for HOPS-165. As in Figure \ref{fig:rt_fits}, black points ({\it left/right}) or color scale ({\it center}) show the data, while the blue lines ({\it left/right}) or contours ({\it center}) show the model, and the gray dashed line shows the disk contribution to the model.}
\figsetgrpend

\figsetgrpstart
\figsetgrpnum{1.40}
\figsetgrptitle{HOPS-166}
\figsetplot{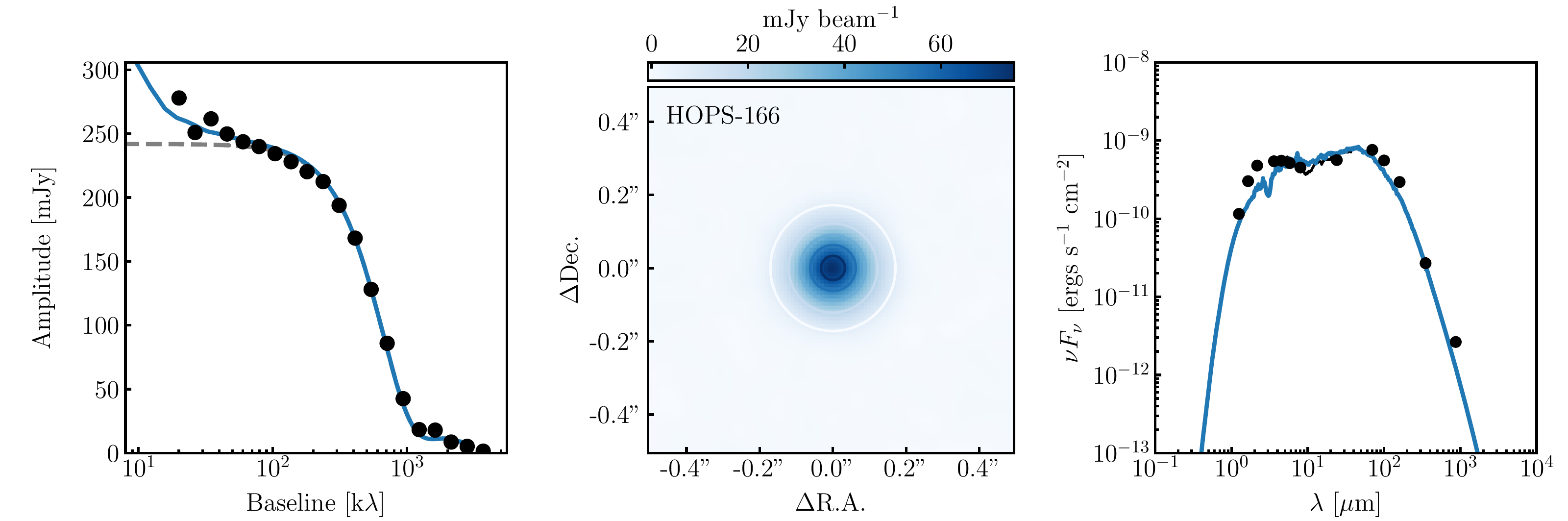}
\figsetgrpnote{A continuation of Figure \ref{fig:rt_fits} showing the best-fit model for HOPS-166. As in Figure \ref{fig:rt_fits}, black points ({\it left/right}) or color scale ({\it center}) show the data, while the blue lines ({\it left/right}) or contours ({\it center}) show the model, and the gray dashed line shows the disk contribution to the model.}
\figsetgrpend

\figsetgrpstart
\figsetgrpnum{1.41}
\figsetgrptitle{HOPS-171}
\figsetplot{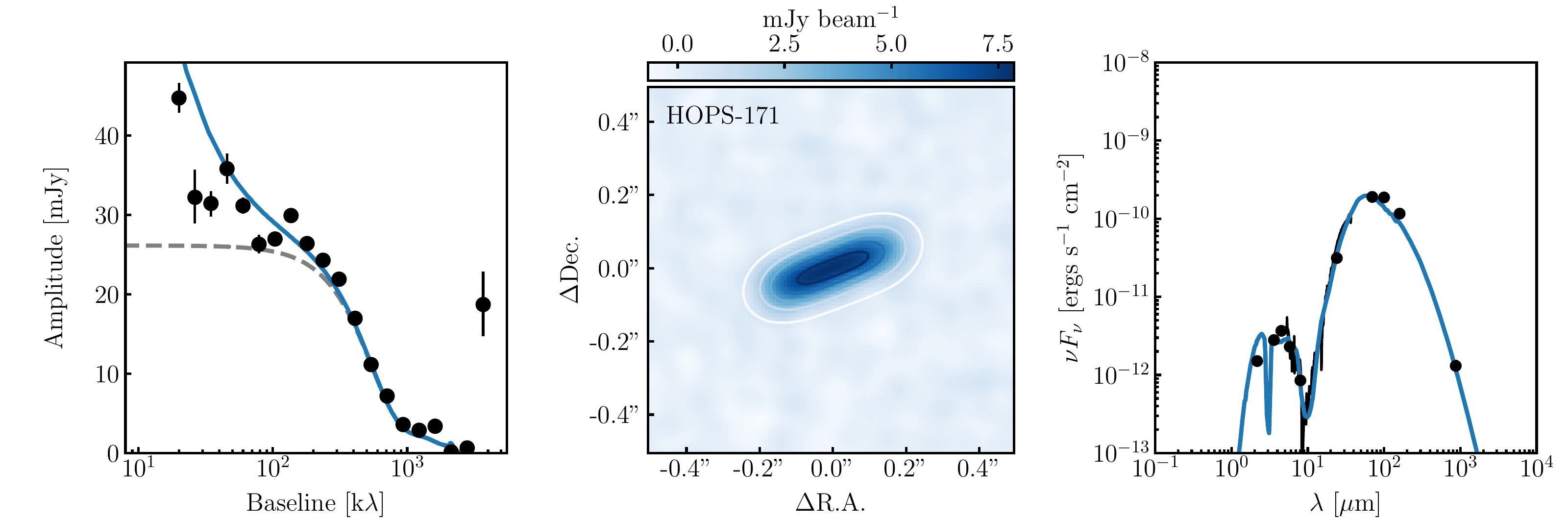}
\figsetgrpnote{A continuation of Figure \ref{fig:rt_fits} showing the best-fit model for HOPS-171. As in Figure \ref{fig:rt_fits}, black points ({\it left/right}) or color scale ({\it center}) show the data, while the blue lines ({\it left/right}) or contours ({\it center}) show the model, and the gray dashed line shows the disk contribution to the model.}
\figsetgrpend

\figsetgrpstart
\figsetgrpnum{1.42}
\figsetgrptitle{HOPS-177}
\figsetplot{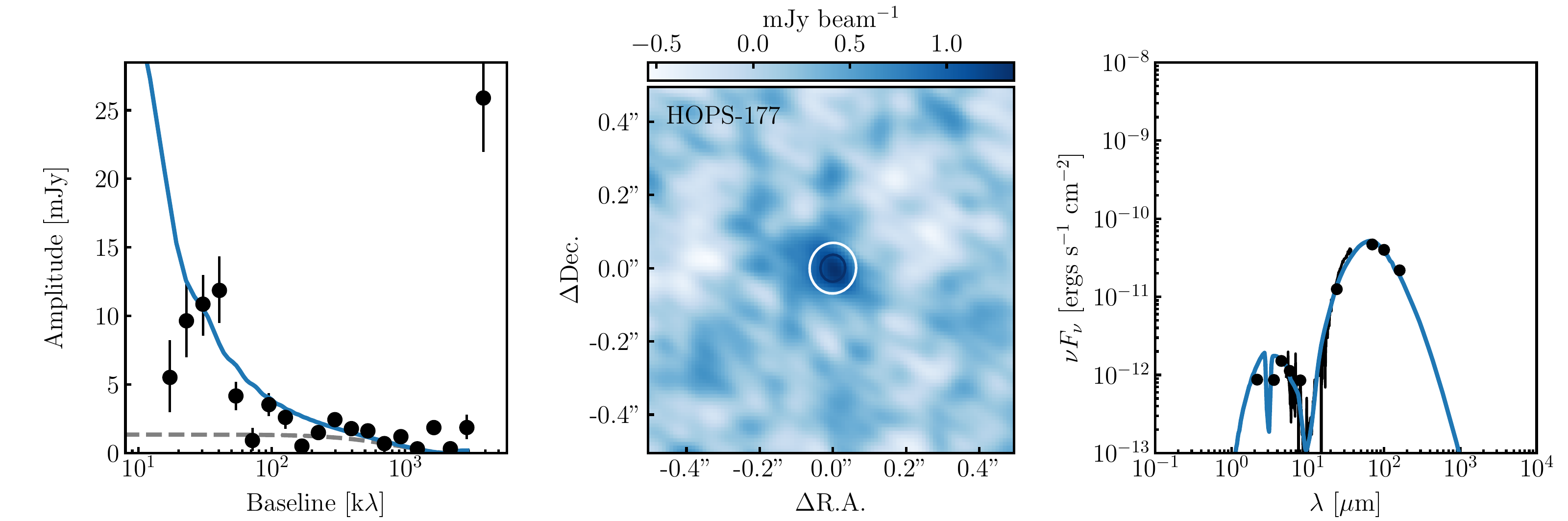}
\figsetgrpnote{A continuation of Figure \ref{fig:rt_fits} showing the best-fit model for HOPS-177. As in Figure \ref{fig:rt_fits}, black points ({\it left/right}) or color scale ({\it center}) show the data, while the blue lines ({\it left/right}) or contours ({\it center}) show the model, and the gray dashed line shows the disk contribution to the model.}
\figsetgrpend

\figsetgrpstart
\figsetgrpnum{1.43}
\figsetgrptitle{HOPS-178}
\figsetplot{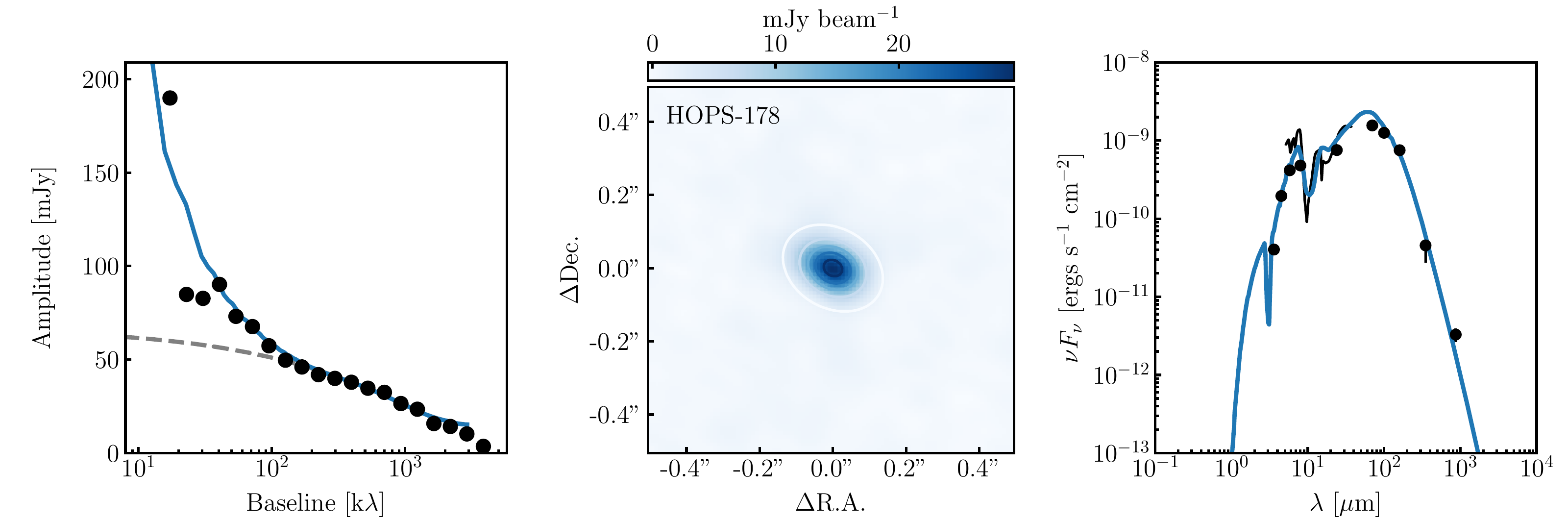}
\figsetgrpnote{A continuation of Figure \ref{fig:rt_fits} showing the best-fit model for HOPS-178. As in Figure \ref{fig:rt_fits}, black points ({\it left/right}) or color scale ({\it center}) show the data, while the blue lines ({\it left/right}) or contours ({\it center}) show the model, and the gray dashed line shows the disk contribution to the model.}
\figsetgrpend

\figsetgrpstart
\figsetgrpnum{1.44}
\figsetgrptitle{HOPS-179}
\figsetplot{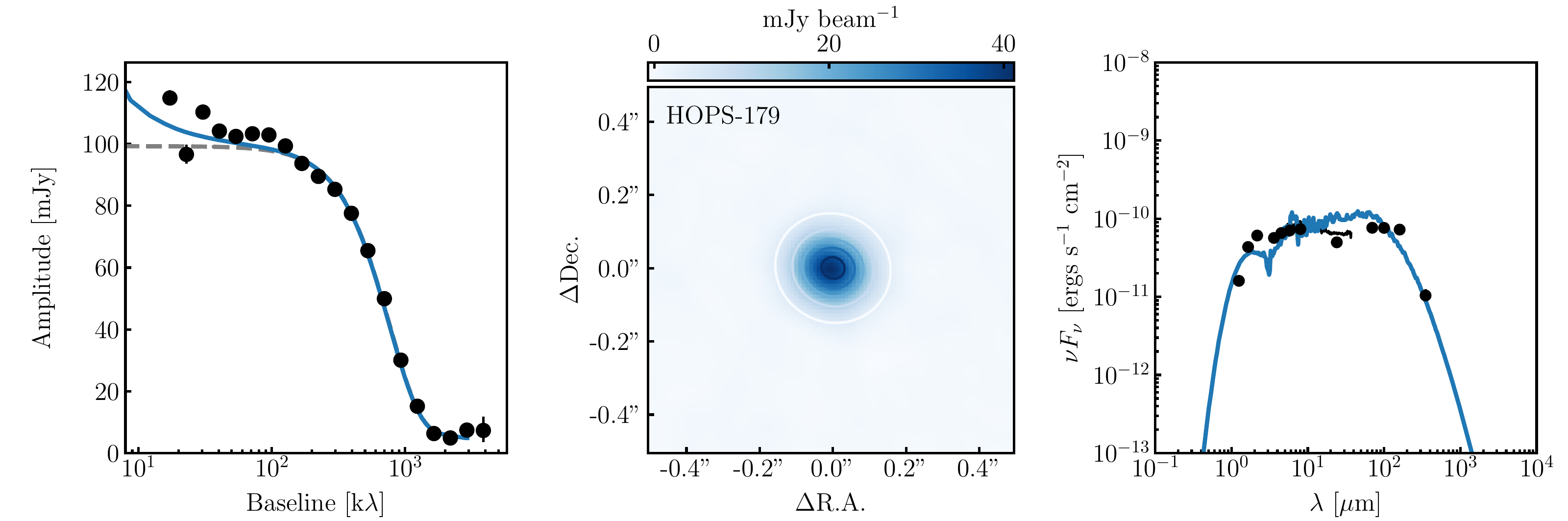}
\figsetgrpnote{A continuation of Figure \ref{fig:rt_fits} showing the best-fit model for HOPS-179. As in Figure \ref{fig:rt_fits}, black points ({\it left/right}) or color scale ({\it center}) show the data, while the blue lines ({\it left/right}) or contours ({\it center}) show the model, and the gray dashed line shows the disk contribution to the model.}
\figsetgrpend

\figsetgrpstart
\figsetgrpnum{1.45}
\figsetgrptitle{HOPS-188}
\figsetplot{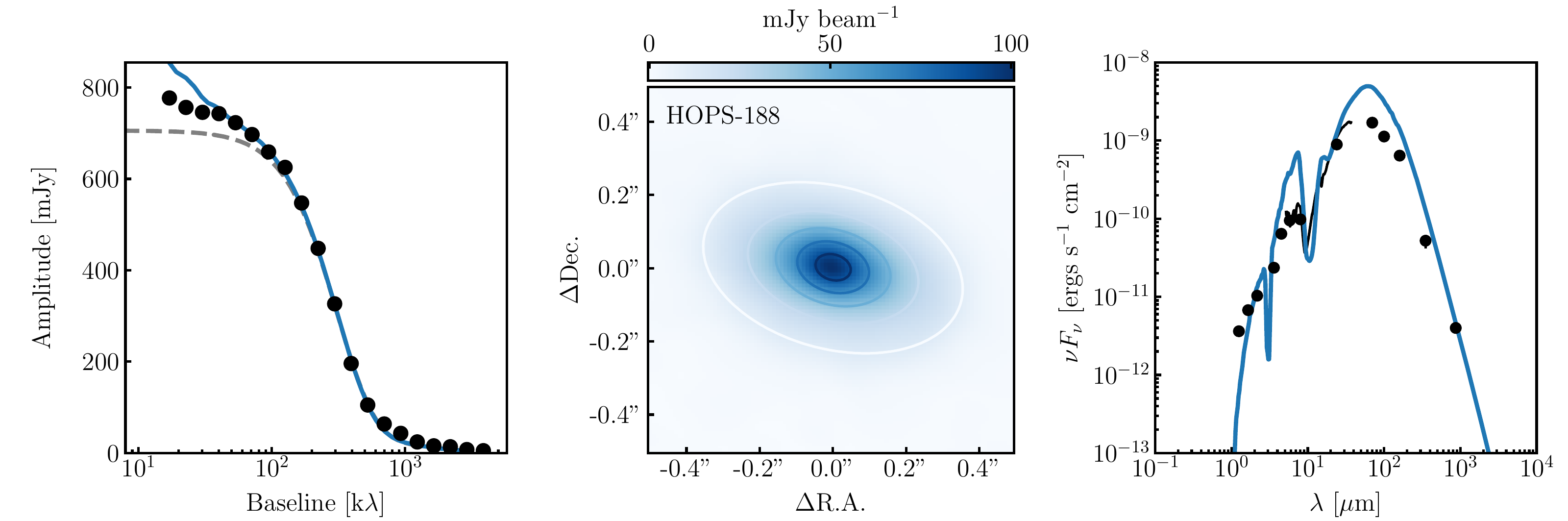}
\figsetgrpnote{A continuation of Figure \ref{fig:rt_fits} showing the best-fit model for HOPS-188. As in Figure \ref{fig:rt_fits}, black points ({\it left/right}) or color scale ({\it center}) show the data, while the blue lines ({\it left/right}) or contours ({\it center}) show the model, and the gray dashed line shows the disk contribution to the model.}
\figsetgrpend

\figsetgrpstart
\figsetgrpnum{1.46}
\figsetgrptitle{HOPS-191}
\figsetplot{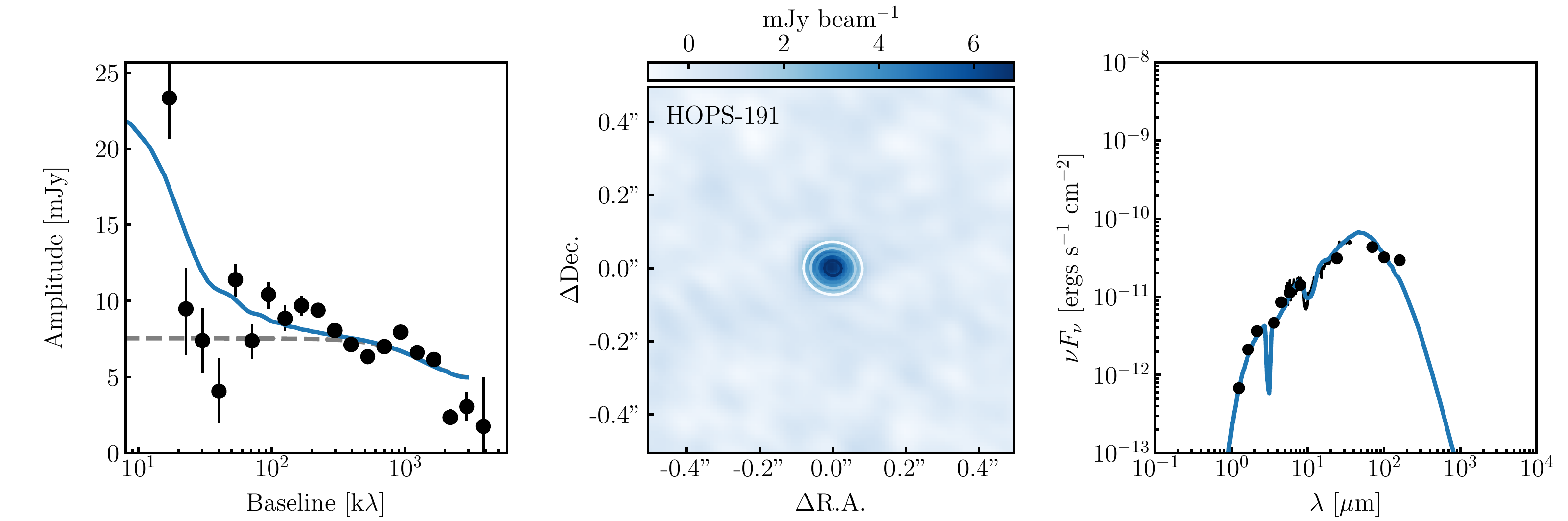}
\figsetgrpnote{A continuation of Figure \ref{fig:rt_fits} showing the best-fit model for HOPS-191. As in Figure \ref{fig:rt_fits}, black points ({\it left/right}) or color scale ({\it center}) show the data, while the blue lines ({\it left/right}) or contours ({\it center}) show the model, and the gray dashed line shows the disk contribution to the model.}
\figsetgrpend

\figsetgrpstart
\figsetgrpnum{1.47}
\figsetgrptitle{HOPS-194}
\figsetplot{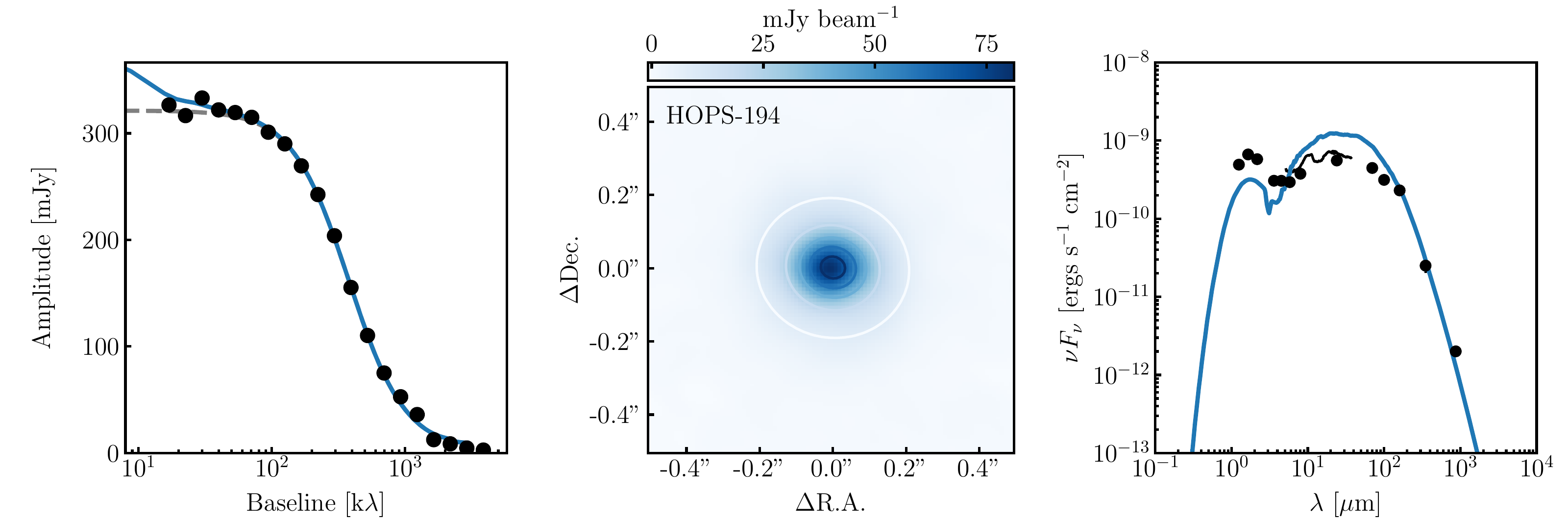}
\figsetgrpnote{A continuation of Figure \ref{fig:rt_fits} showing the best-fit model for HOPS-194. As in Figure \ref{fig:rt_fits}, black points ({\it left/right}) or color scale ({\it center}) show the data, while the blue lines ({\it left/right}) or contours ({\it center}) show the model, and the gray dashed line shows the disk contribution to the model.}
\figsetgrpend

\figsetgrpstart
\figsetgrpnum{1.48}
\figsetgrptitle{HOPS-204}
\figsetplot{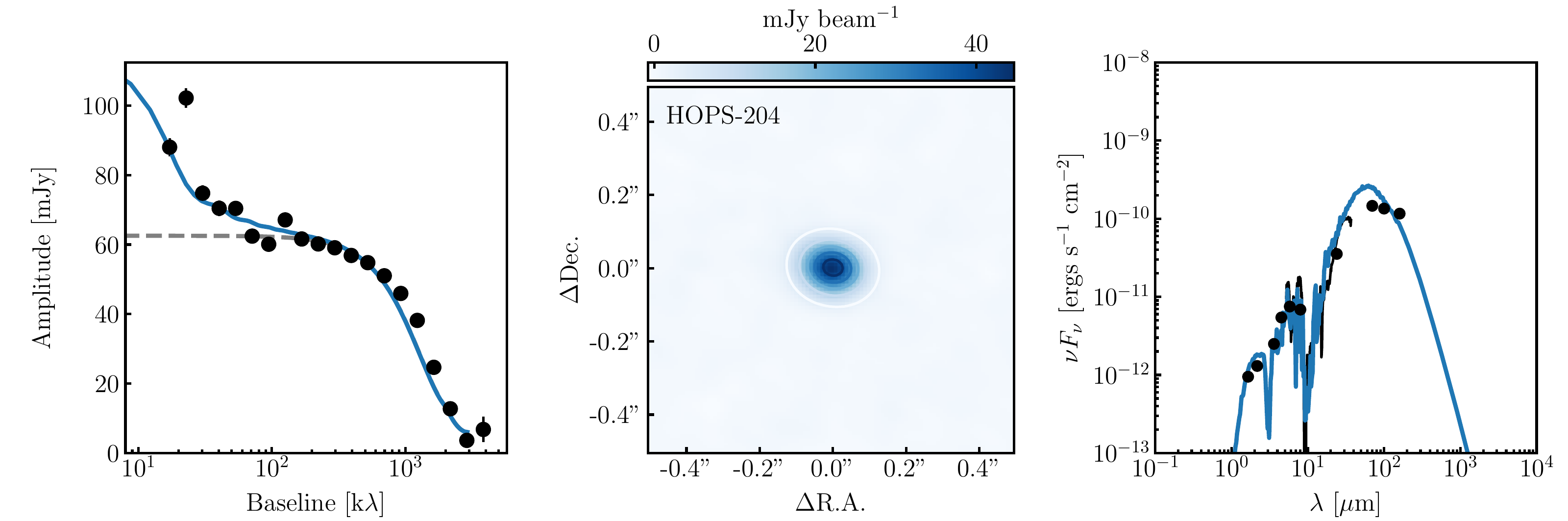}
\figsetgrpnote{A continuation of Figure \ref{fig:rt_fits} showing the best-fit model for HOPS-204. As in Figure \ref{fig:rt_fits}, black points ({\it left/right}) or color scale ({\it center}) show the data, while the blue lines ({\it left/right}) or contours ({\it center}) show the model, and the gray dashed line shows the disk contribution to the model.}
\figsetgrpend

\figsetgrpstart
\figsetgrpnum{1.49}
\figsetgrptitle{HOPS-206}
\figsetplot{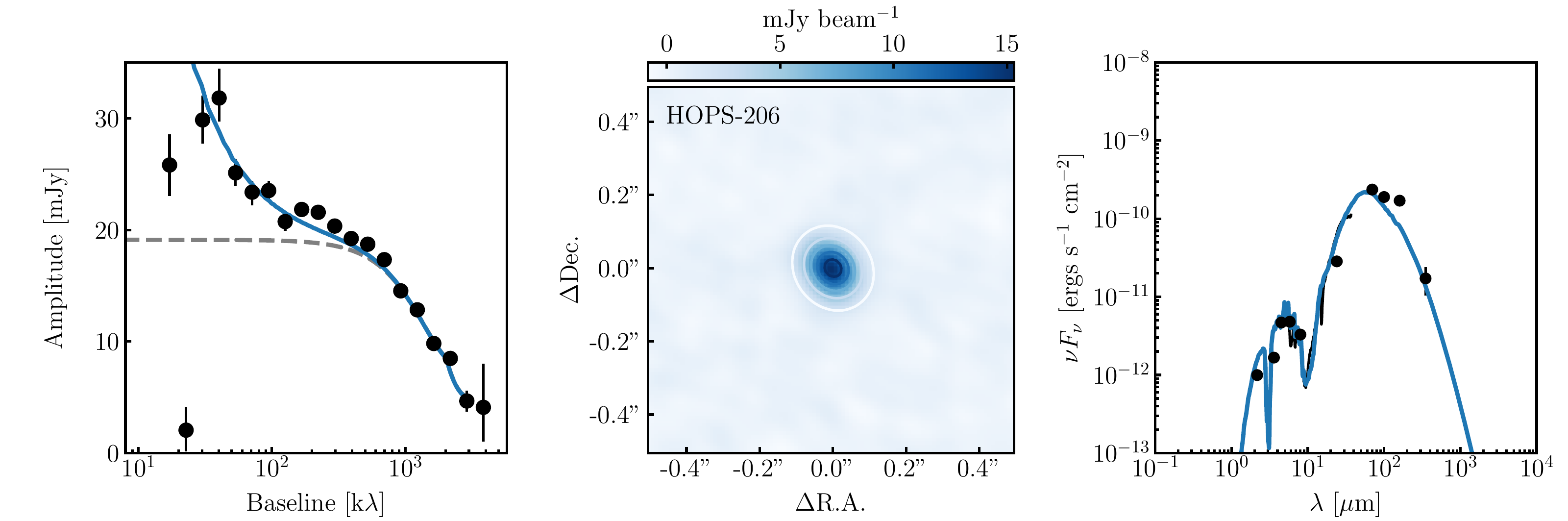}
\figsetgrpnote{A continuation of Figure \ref{fig:rt_fits} showing the best-fit model for HOPS-206. As in Figure \ref{fig:rt_fits}, black points ({\it left/right}) or color scale ({\it center}) show the data, while the blue lines ({\it left/right}) or contours ({\it center}) show the model, and the gray dashed line shows the disk contribution to the model.}
\figsetgrpend

\figsetgrpstart
\figsetgrpnum{1.50}
\figsetgrptitle{HOPS-207}
\figsetplot{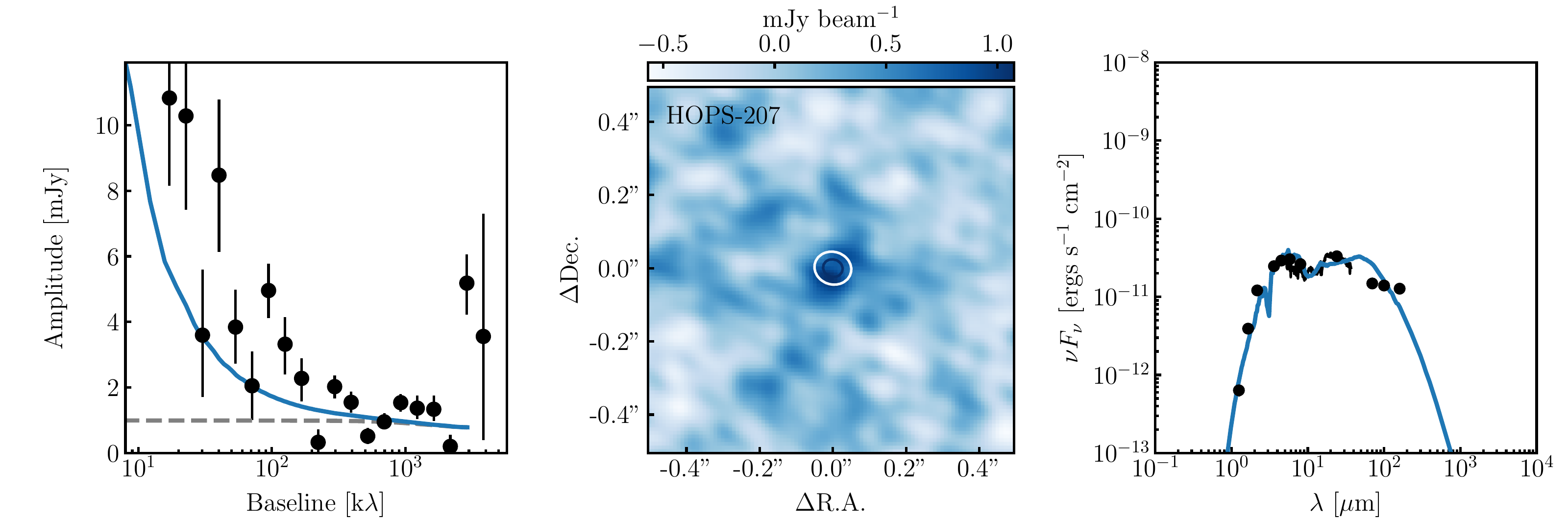}
\figsetgrpnote{A continuation of Figure \ref{fig:rt_fits} showing the best-fit model for HOPS-207. As in Figure \ref{fig:rt_fits}, black points ({\it left/right}) or color scale ({\it center}) show the data, while the blue lines ({\it left/right}) or contours ({\it center}) show the model, and the gray dashed line shows the disk contribution to the model.}
\figsetgrpend

\figsetgrpstart
\figsetgrpnum{1.51}
\figsetgrptitle{HOPS-211}
\figsetplot{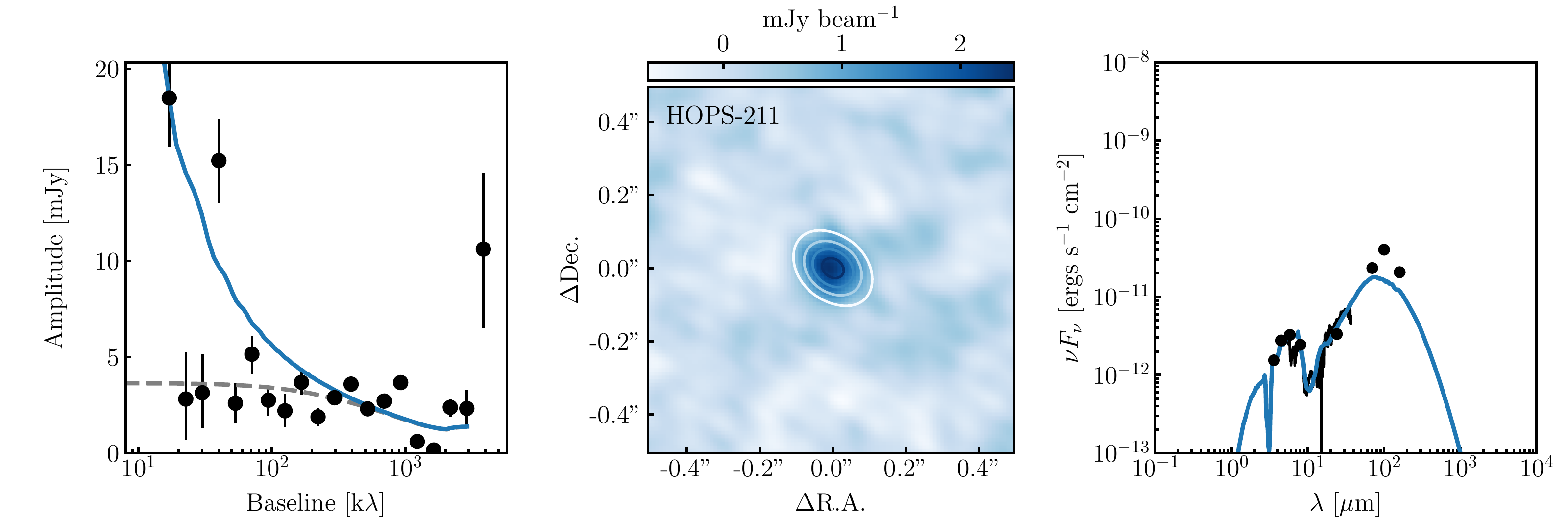}
\figsetgrpnote{A continuation of Figure \ref{fig:rt_fits} showing the best-fit model for HOPS-211. As in Figure \ref{fig:rt_fits}, black points ({\it left/right}) or color scale ({\it center}) show the data, while the blue lines ({\it left/right}) or contours ({\it center}) show the model, and the gray dashed line shows the disk contribution to the model.}
\figsetgrpend

\figsetgrpstart
\figsetgrpnum{1.52}
\figsetgrptitle{HOPS-215}
\figsetplot{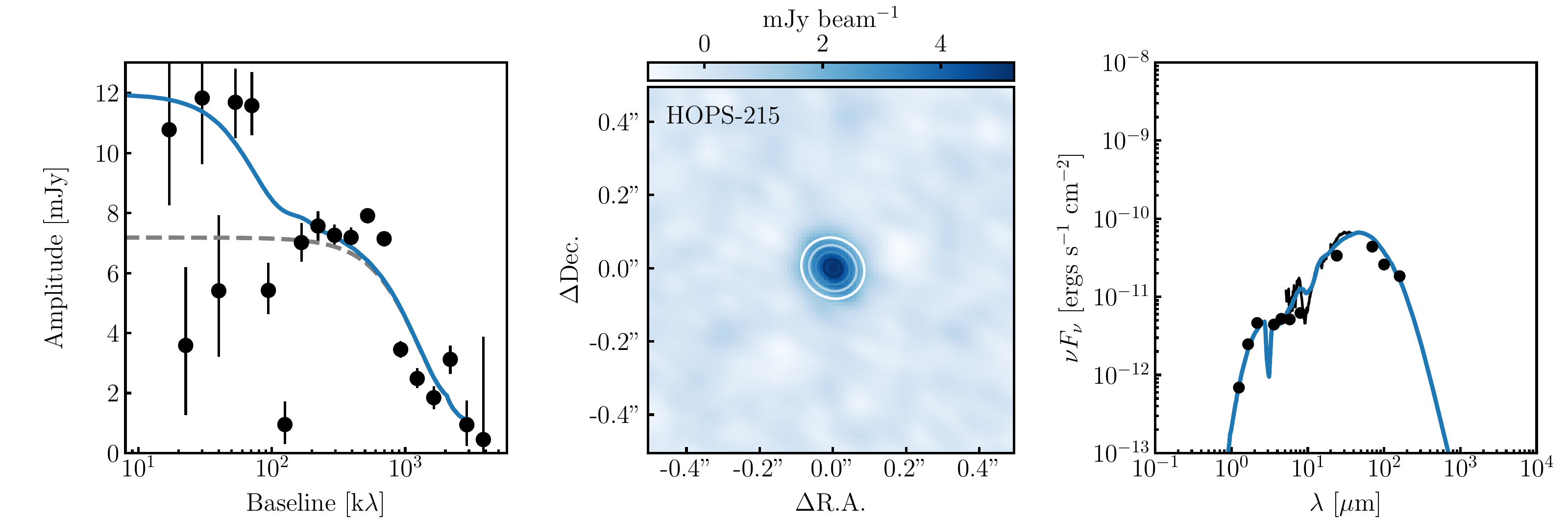}
\figsetgrpnote{A continuation of Figure \ref{fig:rt_fits} showing the best-fit model for HOPS-215. As in Figure \ref{fig:rt_fits}, black points ({\it left/right}) or color scale ({\it center}) show the data, while the blue lines ({\it left/right}) or contours ({\it center}) show the model, and the gray dashed line shows the disk contribution to the model.}
\figsetgrpend

\figsetgrpstart
\figsetgrpnum{1.53}
\figsetgrptitle{HOPS-216}
\figsetplot{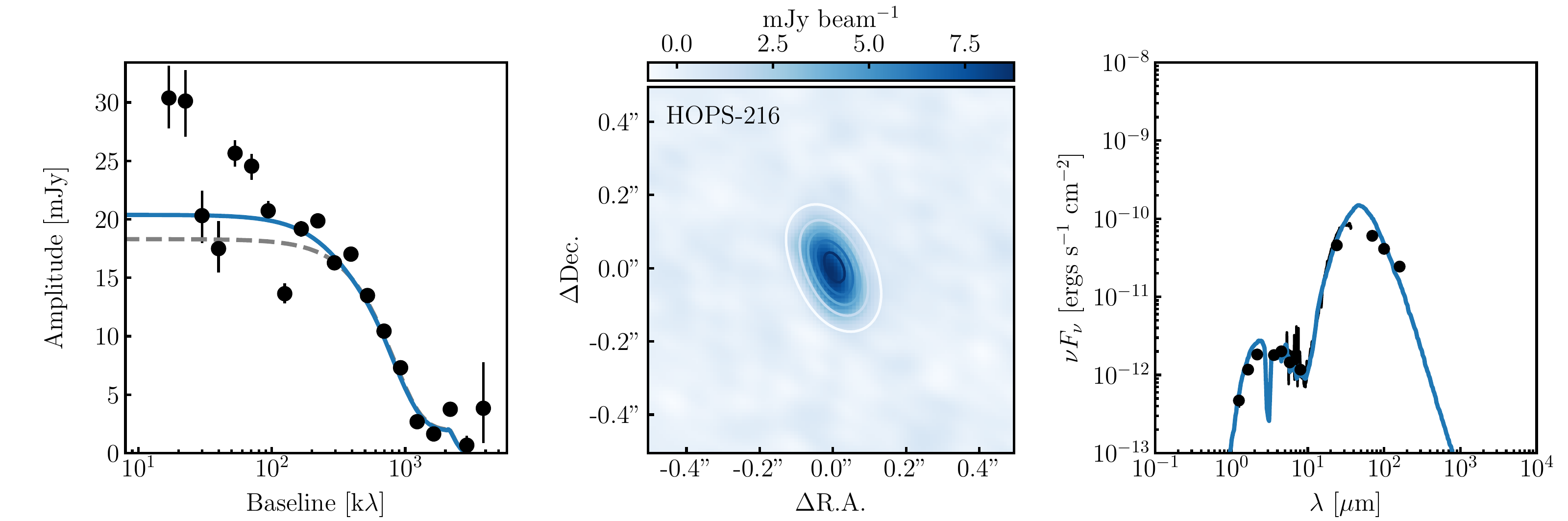}
\figsetgrpnote{A continuation of Figure \ref{fig:rt_fits} showing the best-fit model for HOPS-216. As in Figure \ref{fig:rt_fits}, black points ({\it left/right}) or color scale ({\it center}) show the data, while the blue lines ({\it left/right}) or contours ({\it center}) show the model, and the gray dashed line shows the disk contribution to the model.}
\figsetgrpend

\figsetgrpstart
\figsetgrpnum{1.54}
\figsetgrptitle{HOPS-220}
\figsetplot{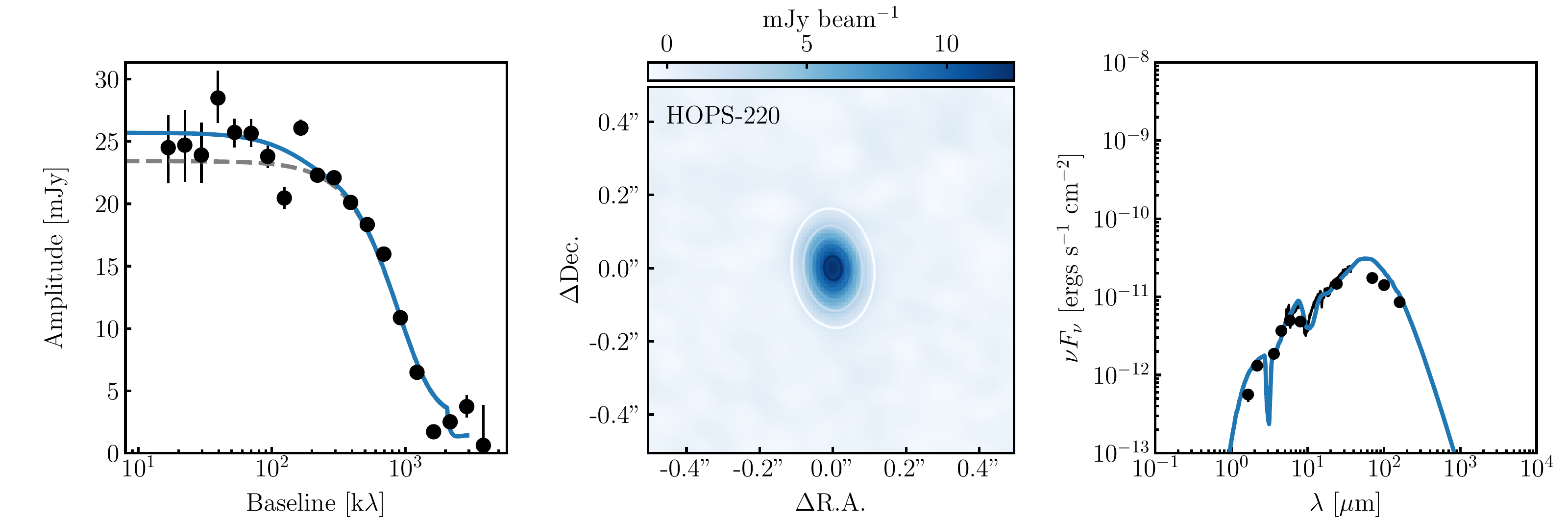}
\figsetgrpnote{A continuation of Figure \ref{fig:rt_fits} showing the best-fit model for HOPS-220. As in Figure \ref{fig:rt_fits}, black points ({\it left/right}) or color scale ({\it center}) show the data, while the blue lines ({\it left/right}) or contours ({\it center}) show the model, and the gray dashed line shows the disk contribution to the model.}
\figsetgrpend

\figsetgrpstart
\figsetgrpnum{1.55}
\figsetgrptitle{HOPS-223}
\figsetplot{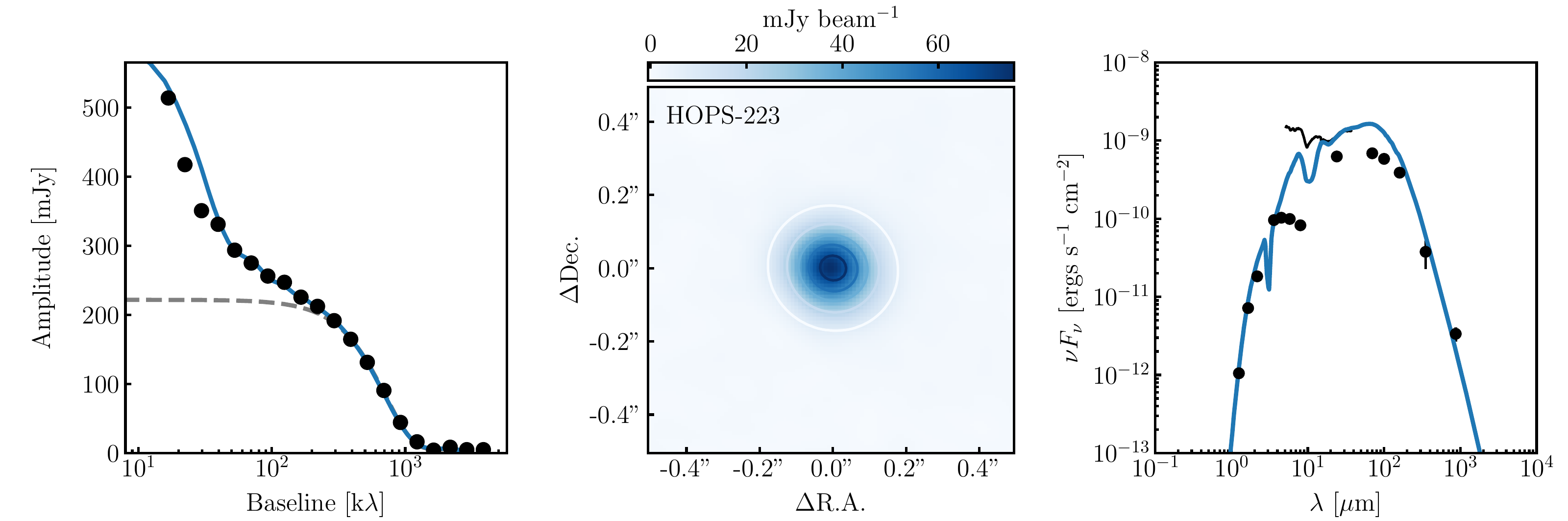}
\figsetgrpnote{A continuation of Figure \ref{fig:rt_fits} showing the best-fit model for HOPS-223. As in Figure \ref{fig:rt_fits}, black points ({\it left/right}) or color scale ({\it center}) show the data, while the blue lines ({\it left/right}) or contours ({\it center}) show the model, and the gray dashed line shows the disk contribution to the model.}
\figsetgrpend

\figsetgrpstart
\figsetgrpnum{1.56}
\figsetgrptitle{HOPS-224}
\figsetplot{HOPS-224_rt_model.pdf}
\figsetgrpnote{A continuation of Figure \ref{fig:rt_fits} showing the best-fit model for HOPS-224. As in Figure \ref{fig:rt_fits}, black points ({\it left/right}) or color scale ({\it center}) show the data, while the blue lines ({\it left/right}) or contours ({\it center}) show the model, and the gray dashed line shows the disk contribution to the model.}
\figsetgrpend

\figsetgrpstart
\figsetgrpnum{1.57}
\figsetgrptitle{HOPS-227}
\figsetplot{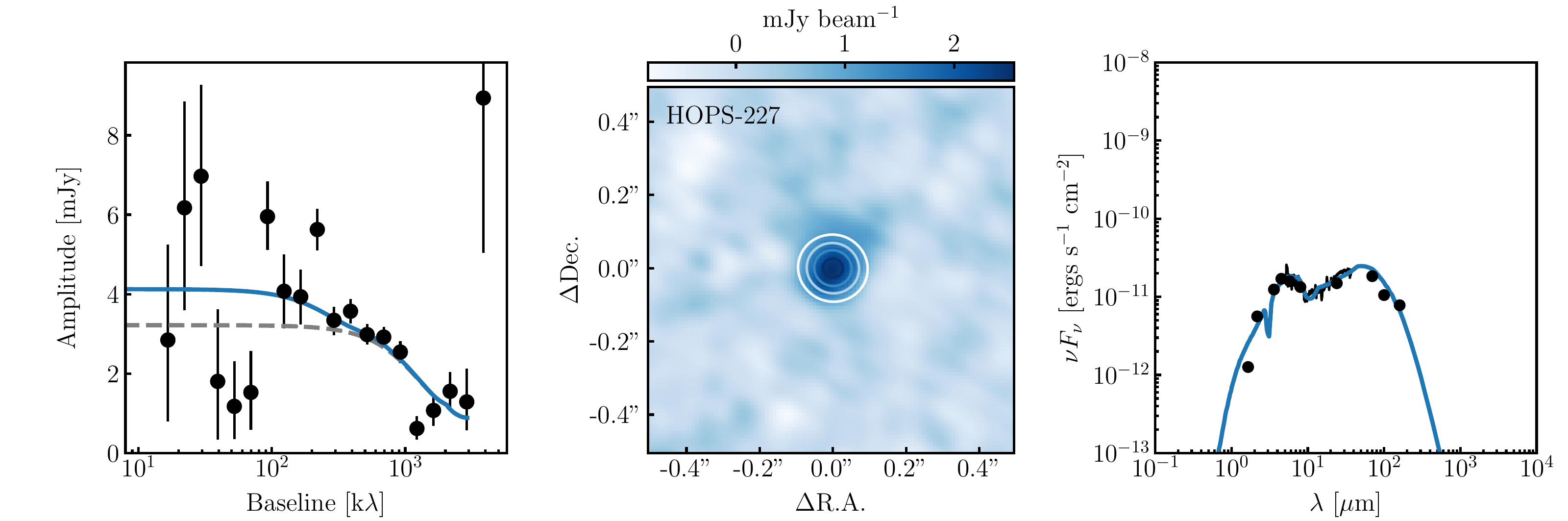}
\figsetgrpnote{A continuation of Figure \ref{fig:rt_fits} showing the best-fit model for HOPS-227. As in Figure \ref{fig:rt_fits}, black points ({\it left/right}) or color scale ({\it center}) show the data, while the blue lines ({\it left/right}) or contours ({\it center}) show the model, and the gray dashed line shows the disk contribution to the model.}
\figsetgrpend

\figsetgrpstart
\figsetgrpnum{1.58}
\figsetgrptitle{HOPS-232}
\figsetplot{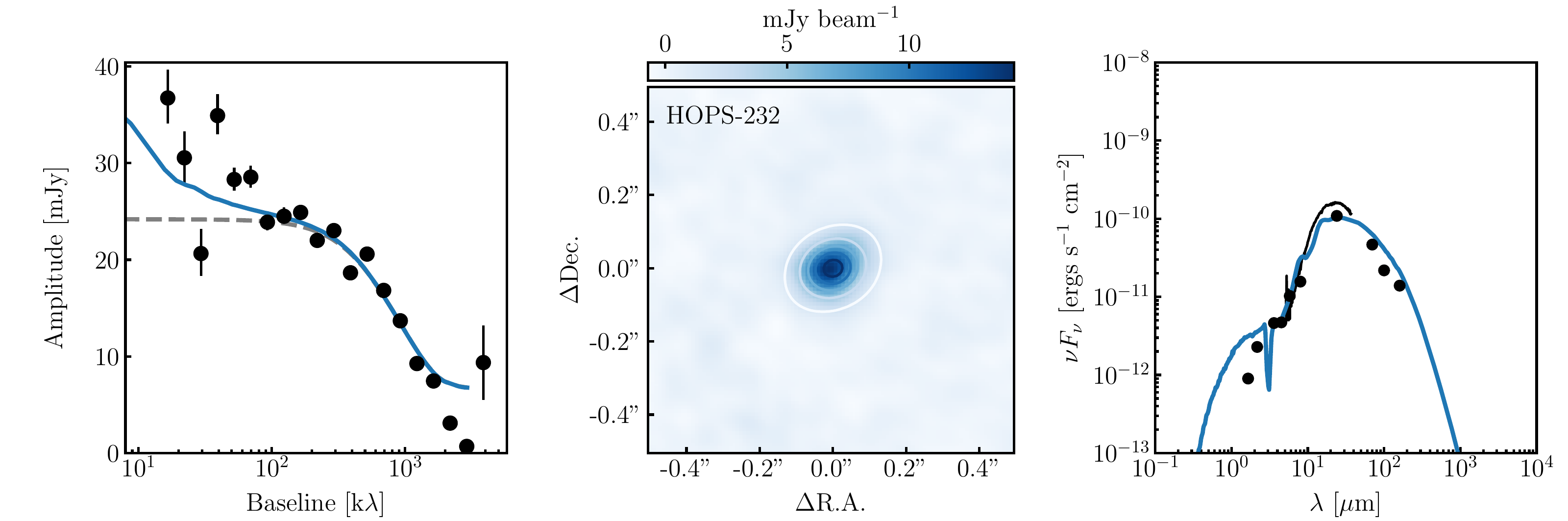}
\figsetgrpnote{A continuation of Figure \ref{fig:rt_fits} showing the best-fit model for HOPS-232. As in Figure \ref{fig:rt_fits}, black points ({\it left/right}) or color scale ({\it center}) show the data, while the blue lines ({\it left/right}) or contours ({\it center}) show the model, and the gray dashed line shows the disk contribution to the model.}
\figsetgrpend

\figsetgrpstart
\figsetgrpnum{1.59}
\figsetgrptitle{HOPS-234}
\figsetplot{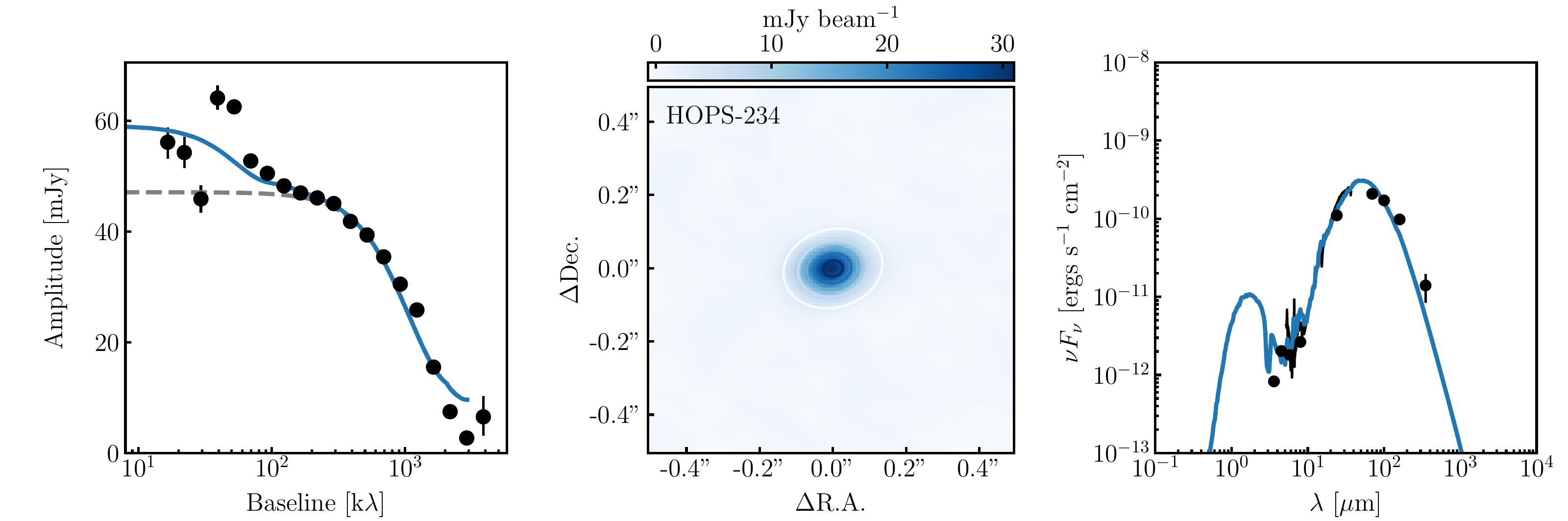}
\figsetgrpnote{A continuation of Figure \ref{fig:rt_fits} showing the best-fit model for HOPS-234. As in Figure \ref{fig:rt_fits}, black points ({\it left/right}) or color scale ({\it center}) show the data, while the blue lines ({\it left/right}) or contours ({\it center}) show the model, and the gray dashed line shows the disk contribution to the model.}
\figsetgrpend

\figsetgrpstart
\figsetgrpnum{1.60}
\figsetgrptitle{HOPS-235}
\figsetplot{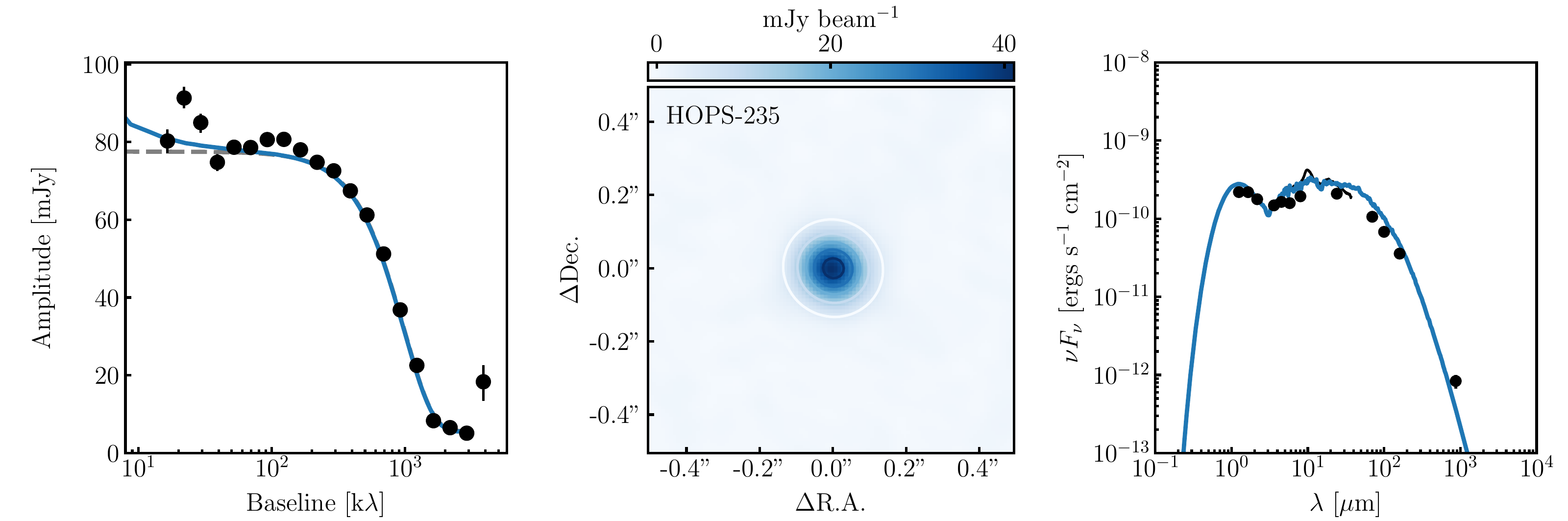}
\figsetgrpnote{A continuation of Figure \ref{fig:rt_fits} showing the best-fit model for HOPS-235. As in Figure \ref{fig:rt_fits}, black points ({\it left/right}) or color scale ({\it center}) show the data, while the blue lines ({\it left/right}) or contours ({\it center}) show the model, and the gray dashed line shows the disk contribution to the model.}
\figsetgrpend

\figsetgrpstart
\figsetgrpnum{1.61}
\figsetgrptitle{HOPS-238}
\figsetplot{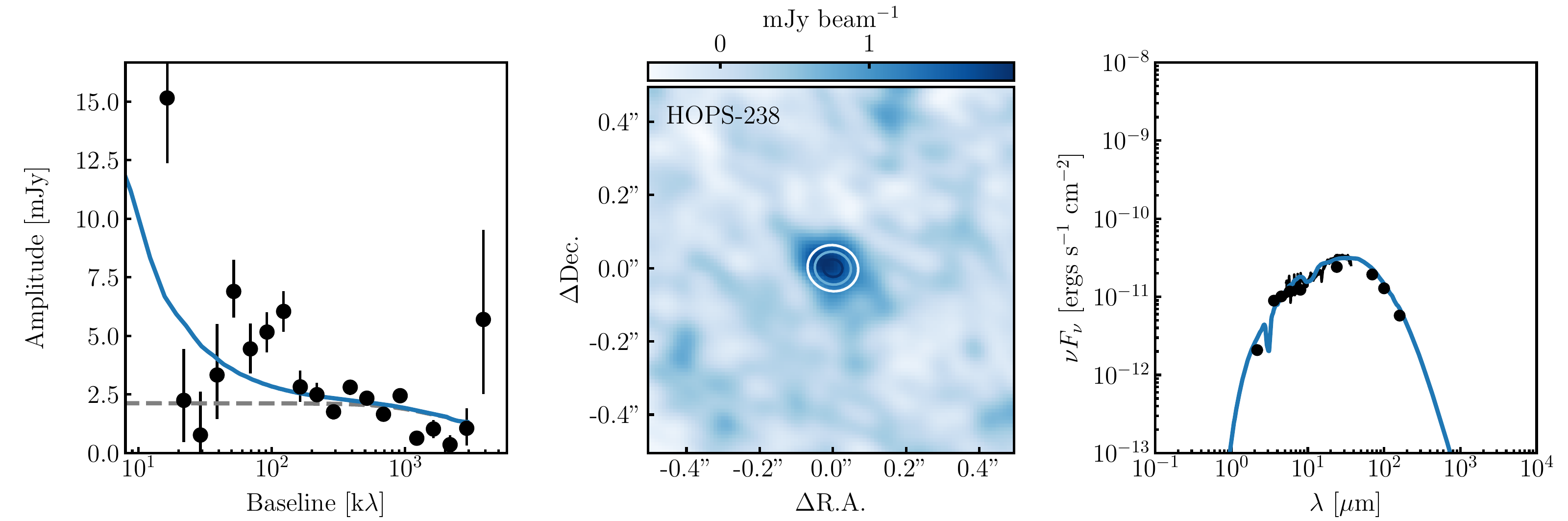}
\figsetgrpnote{A continuation of Figure \ref{fig:rt_fits} showing the best-fit model for HOPS-238. As in Figure \ref{fig:rt_fits}, black points ({\it left/right}) or color scale ({\it center}) show the data, while the blue lines ({\it left/right}) or contours ({\it center}) show the model, and the gray dashed line shows the disk contribution to the model.}
\figsetgrpend

\figsetgrpstart
\figsetgrpnum{1.62}
\figsetgrptitle{HOPS-240}
\figsetplot{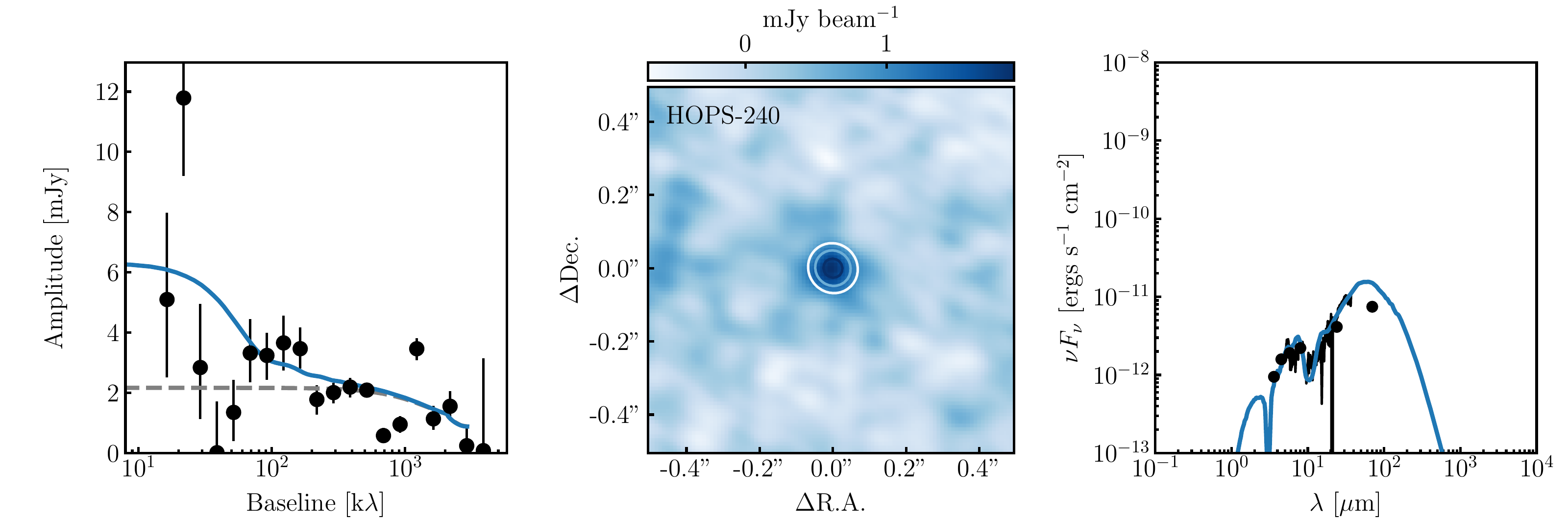}
\figsetgrpnote{A continuation of Figure \ref{fig:rt_fits} showing the best-fit model for HOPS-240. As in Figure \ref{fig:rt_fits}, black points ({\it left/right}) or color scale ({\it center}) show the data, while the blue lines ({\it left/right}) or contours ({\it center}) show the model, and the gray dashed line shows the disk contribution to the model.}
\figsetgrpend

\figsetgrpstart
\figsetgrpnum{1.63}
\figsetgrptitle{HOPS-243}
\figsetplot{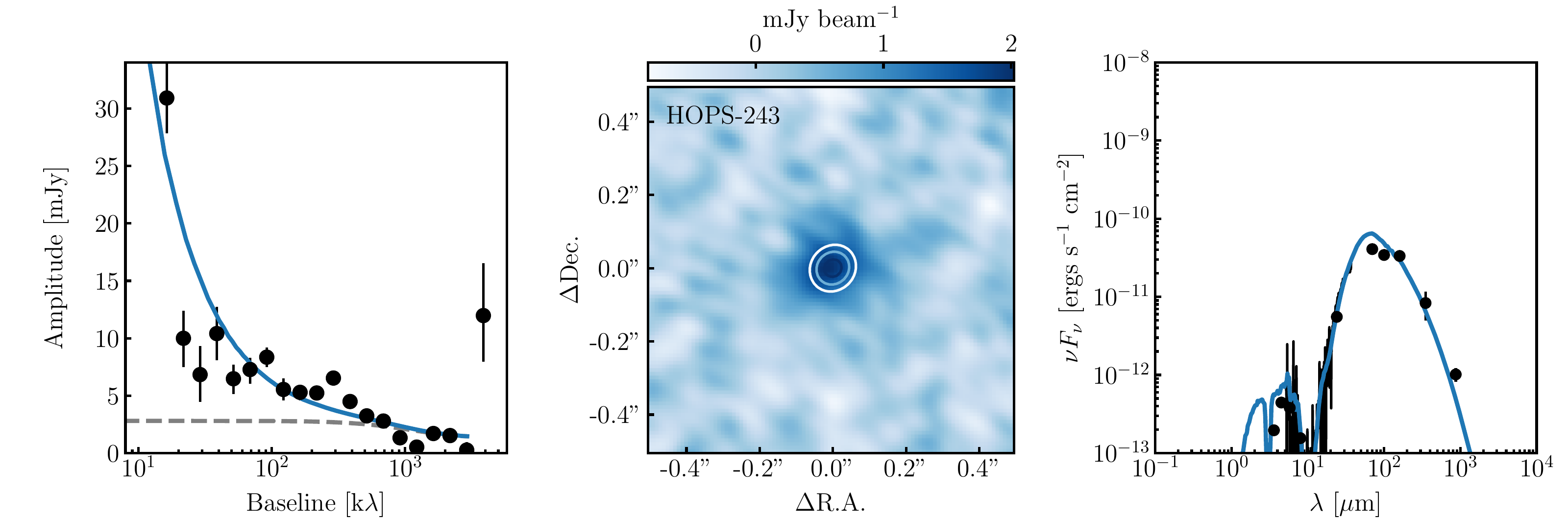}
\figsetgrpnote{A continuation of Figure \ref{fig:rt_fits} showing the best-fit model for HOPS-243. As in Figure \ref{fig:rt_fits}, black points ({\it left/right}) or color scale ({\it center}) show the data, while the blue lines ({\it left/right}) or contours ({\it center}) show the model, and the gray dashed line shows the disk contribution to the model.}
\figsetgrpend

\figsetgrpstart
\figsetgrpnum{1.64}
\figsetgrptitle{HOPS-249}
\figsetplot{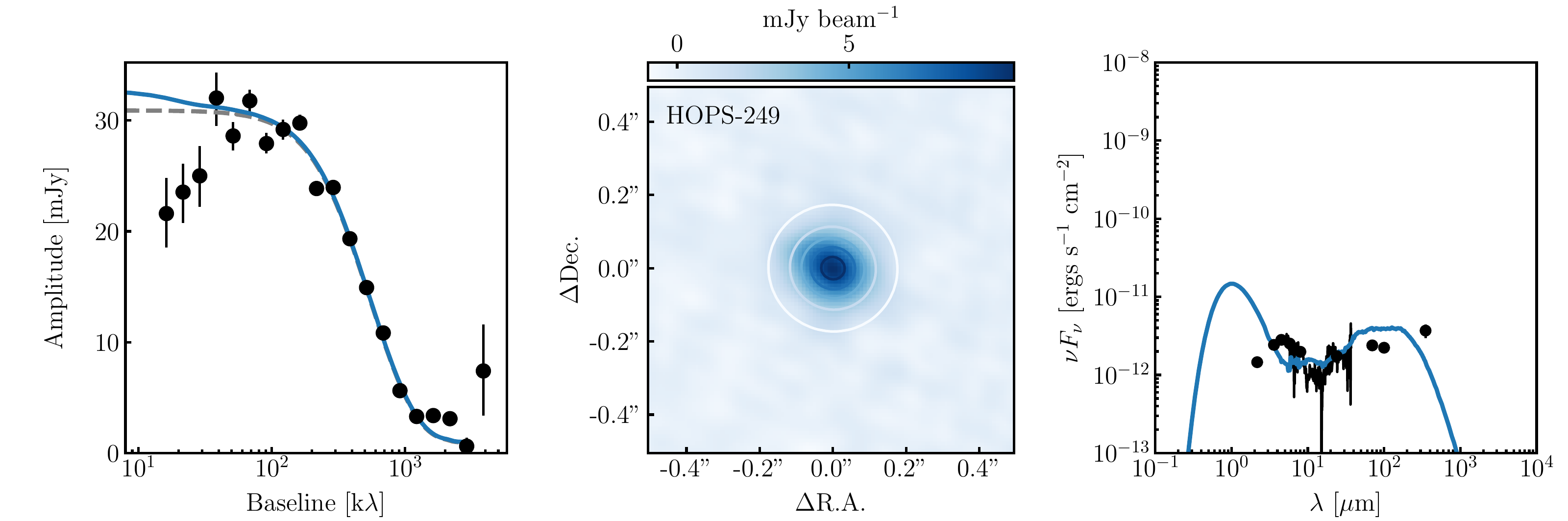}
\figsetgrpnote{A continuation of Figure \ref{fig:rt_fits} showing the best-fit model for HOPS-249. As in Figure \ref{fig:rt_fits}, black points ({\it left/right}) or color scale ({\it center}) show the data, while the blue lines ({\it left/right}) or contours ({\it center}) show the model, and the gray dashed line shows the disk contribution to the model.}
\figsetgrpend

\figsetgrpstart
\figsetgrpnum{1.65}
\figsetgrptitle{HOPS-250}
\figsetplot{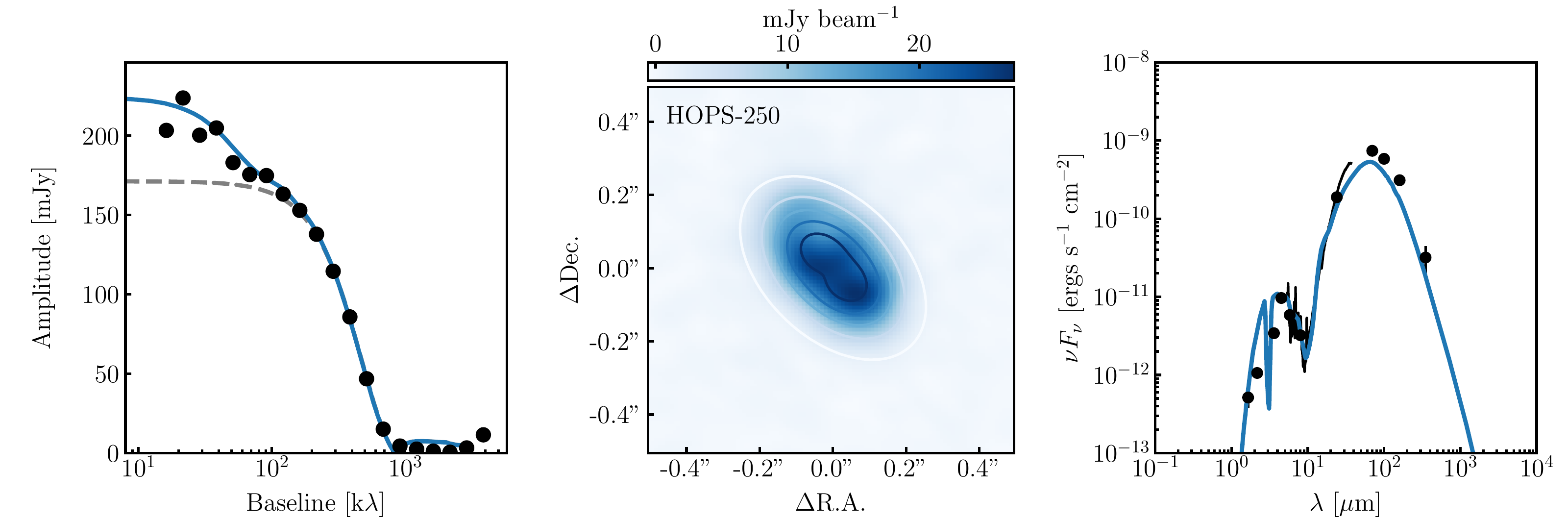}
\figsetgrpnote{A continuation of Figure \ref{fig:rt_fits} showing the best-fit model for HOPS-250. As in Figure \ref{fig:rt_fits}, black points ({\it left/right}) or color scale ({\it center}) show the data, while the blue lines ({\it left/right}) or contours ({\it center}) show the model, and the gray dashed line shows the disk contribution to the model.}
\figsetgrpend

\figsetgrpstart
\figsetgrpnum{1.66}
\figsetgrptitle{HOPS-252}
\figsetplot{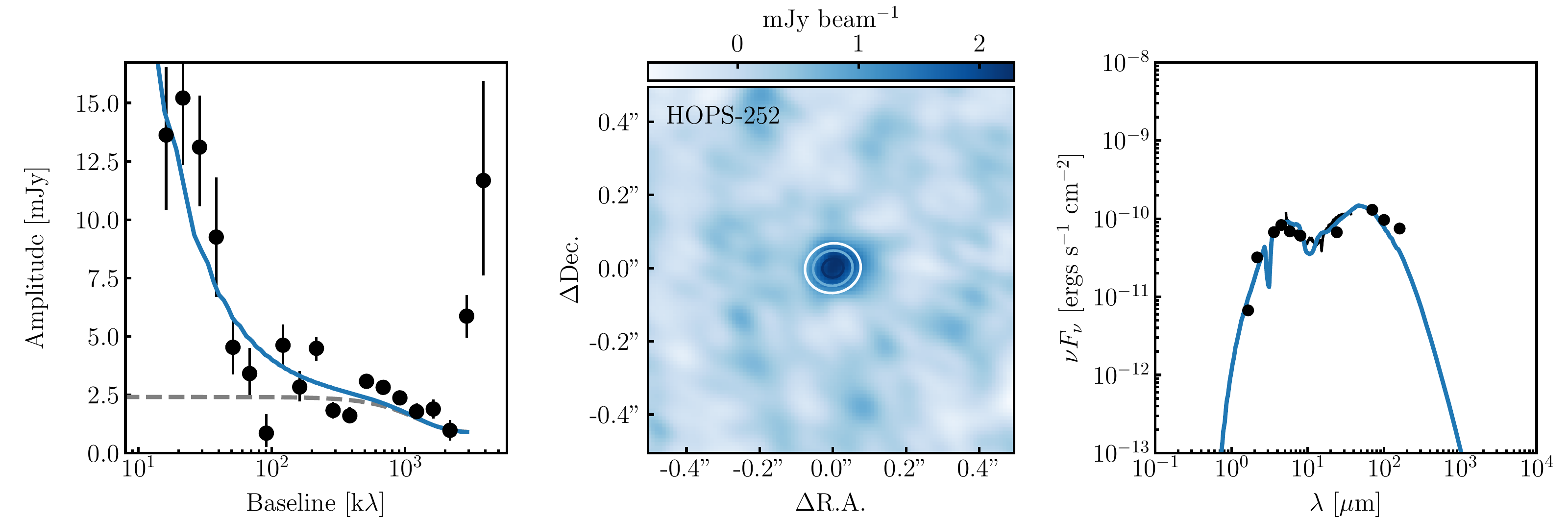}
\figsetgrpnote{A continuation of Figure \ref{fig:rt_fits} showing the best-fit model for HOPS-252. As in Figure \ref{fig:rt_fits}, black points ({\it left/right}) or color scale ({\it center}) show the data, while the blue lines ({\it left/right}) or contours ({\it center}) show the model, and the gray dashed line shows the disk contribution to the model.}
\figsetgrpend

\figsetgrpstart
\figsetgrpnum{1.67}
\figsetgrptitle{HOPS-253}
\figsetplot{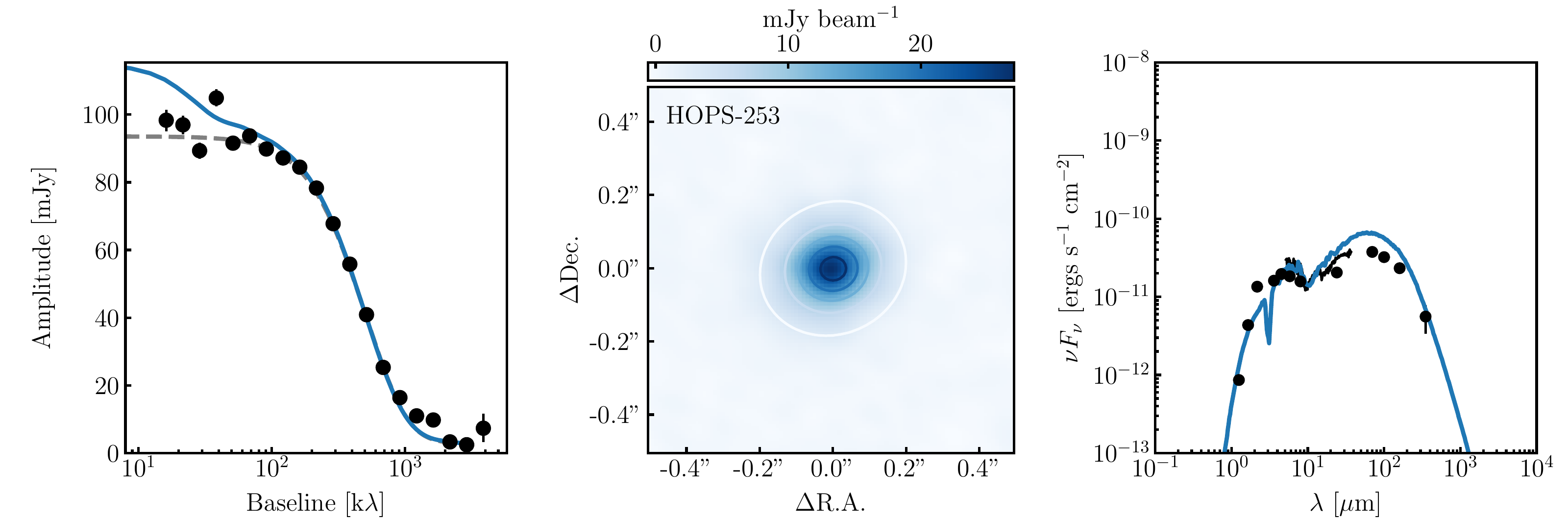}
\figsetgrpnote{A continuation of Figure \ref{fig:rt_fits} showing the best-fit model for HOPS-253. As in Figure \ref{fig:rt_fits}, black points ({\it left/right}) or color scale ({\it center}) show the data, while the blue lines ({\it left/right}) or contours ({\it center}) show the model, and the gray dashed line shows the disk contribution to the model.}
\figsetgrpend

\figsetgrpstart
\figsetgrpnum{1.68}
\figsetgrptitle{HOPS-258}
\figsetplot{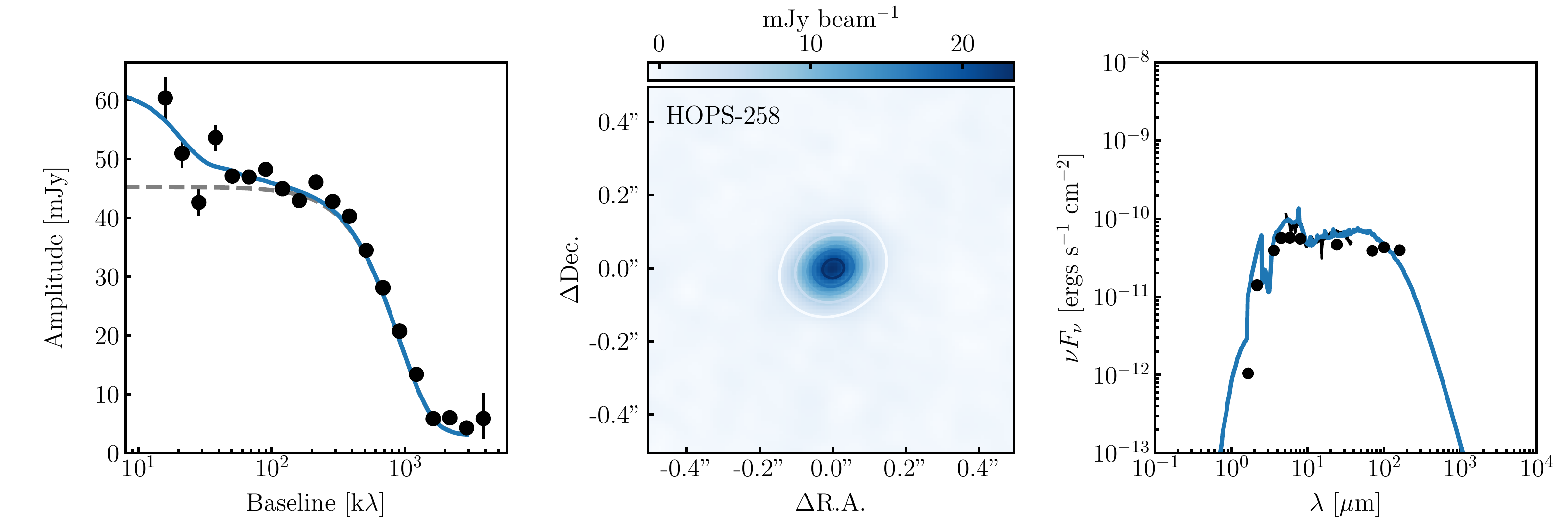}
\figsetgrpnote{A continuation of Figure \ref{fig:rt_fits} showing the best-fit model for HOPS-258. As in Figure \ref{fig:rt_fits}, black points ({\it left/right}) or color scale ({\it center}) show the data, while the blue lines ({\it left/right}) or contours ({\it center}) show the model, and the gray dashed line shows the disk contribution to the model.}
\figsetgrpend

\figsetgrpstart
\figsetgrpnum{1.69}
\figsetgrptitle{HOPS-260}
\figsetplot{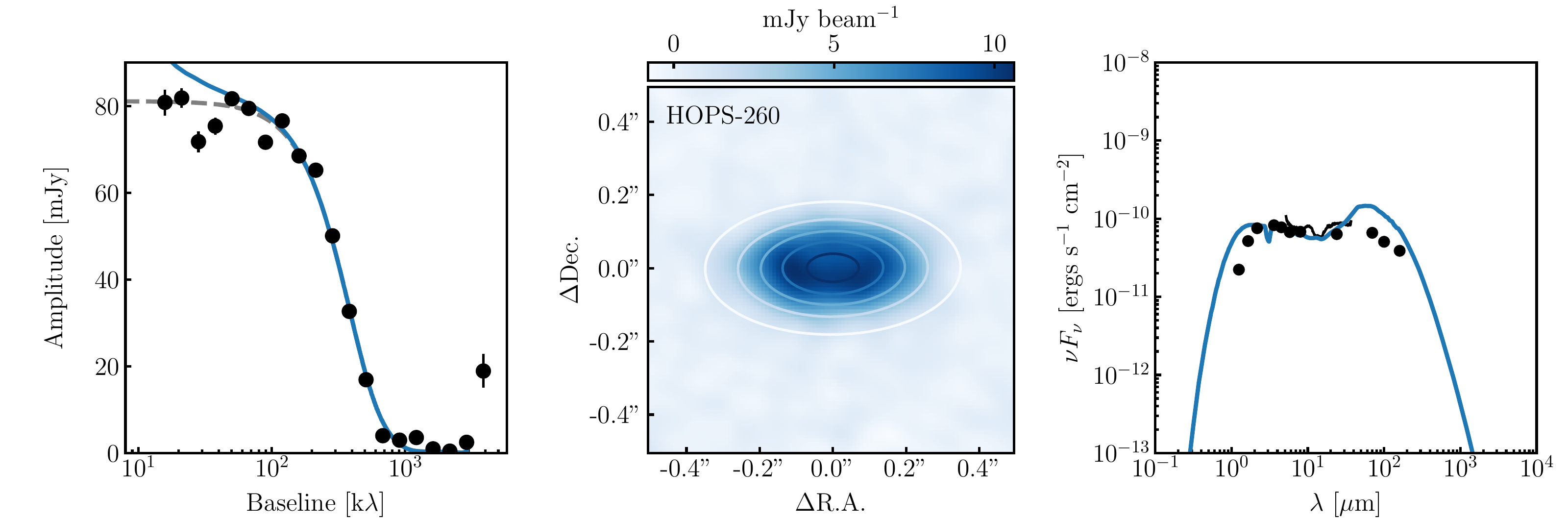}
\figsetgrpnote{A continuation of Figure \ref{fig:rt_fits} showing the best-fit model for HOPS-260. As in Figure \ref{fig:rt_fits}, black points ({\it left/right}) or color scale ({\it center}) show the data, while the blue lines ({\it left/right}) or contours ({\it center}) show the model, and the gray dashed line shows the disk contribution to the model.}
\figsetgrpend

\figsetgrpstart
\figsetgrpnum{1.70}
\figsetgrptitle{HOPS-265}
\figsetplot{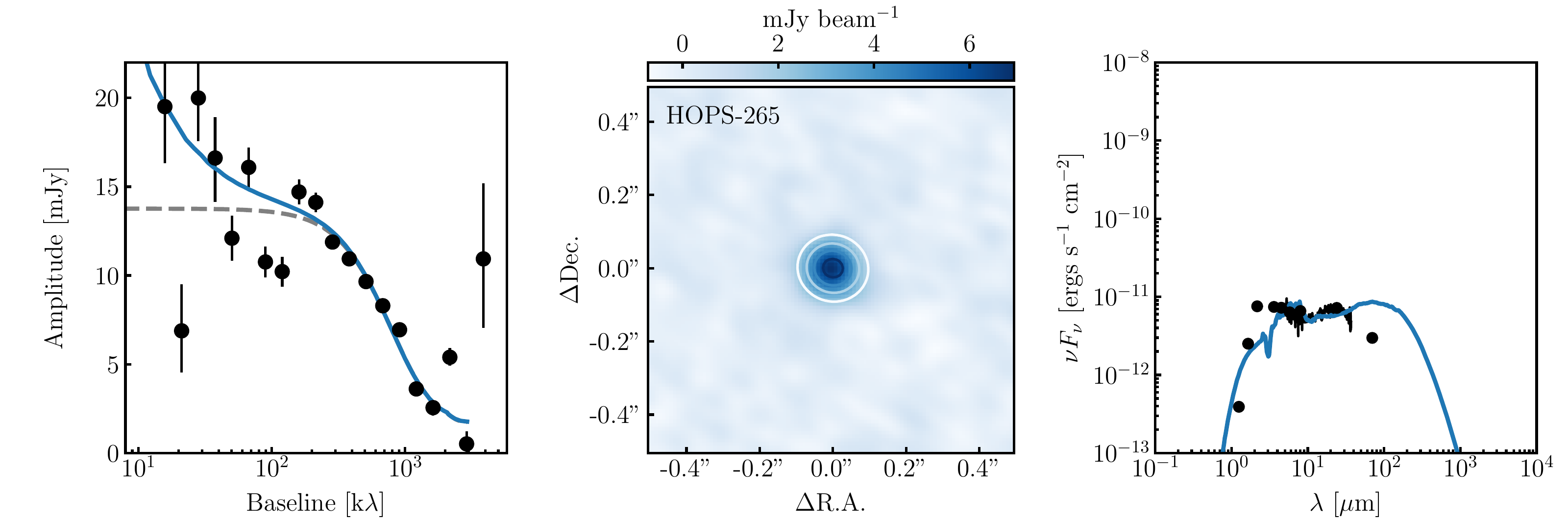}
\figsetgrpnote{A continuation of Figure \ref{fig:rt_fits} showing the best-fit model for HOPS-265. As in Figure \ref{fig:rt_fits}, black points ({\it left/right}) or color scale ({\it center}) show the data, while the blue lines ({\it left/right}) or contours ({\it center}) show the model, and the gray dashed line shows the disk contribution to the model.}
\figsetgrpend

\figsetgrpstart
\figsetgrpnum{1.71}
\figsetgrptitle{HOPS-268}
\figsetplot{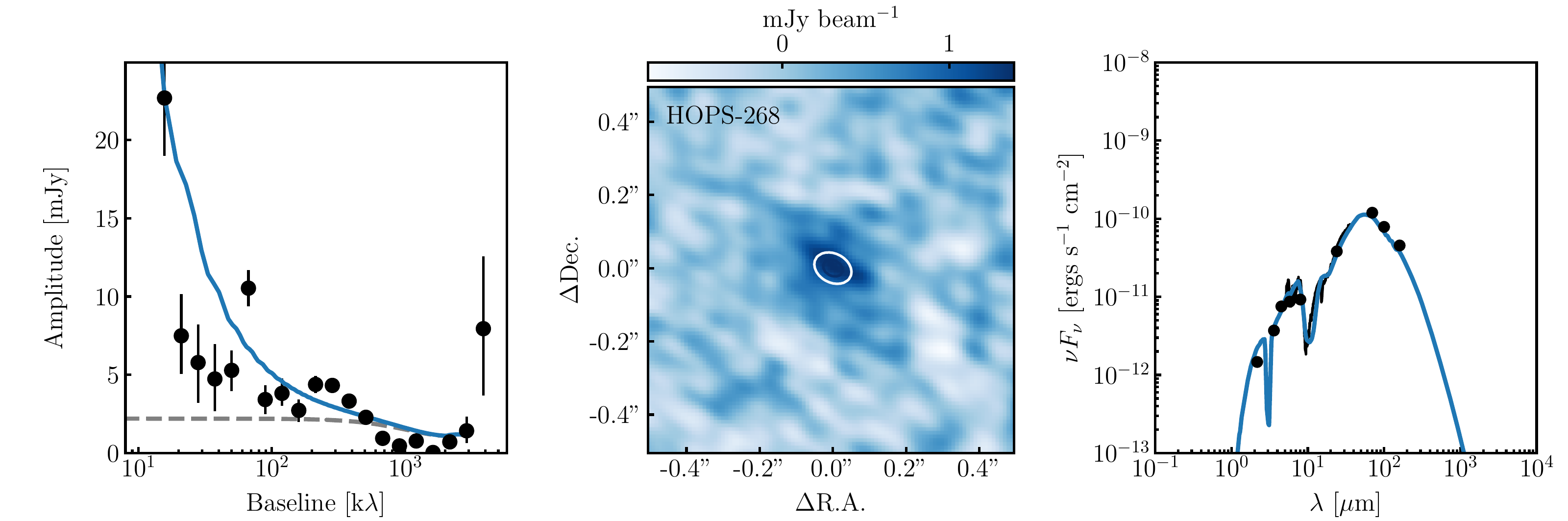}
\figsetgrpnote{A continuation of Figure \ref{fig:rt_fits} showing the best-fit model for HOPS-268. As in Figure \ref{fig:rt_fits}, black points ({\it left/right}) or color scale ({\it center}) show the data, while the blue lines ({\it left/right}) or contours ({\it center}) show the model, and the gray dashed line shows the disk contribution to the model.}
\figsetgrpend

\figsetgrpstart
\figsetgrpnum{1.72}
\figsetgrptitle{HOPS-273}
\figsetplot{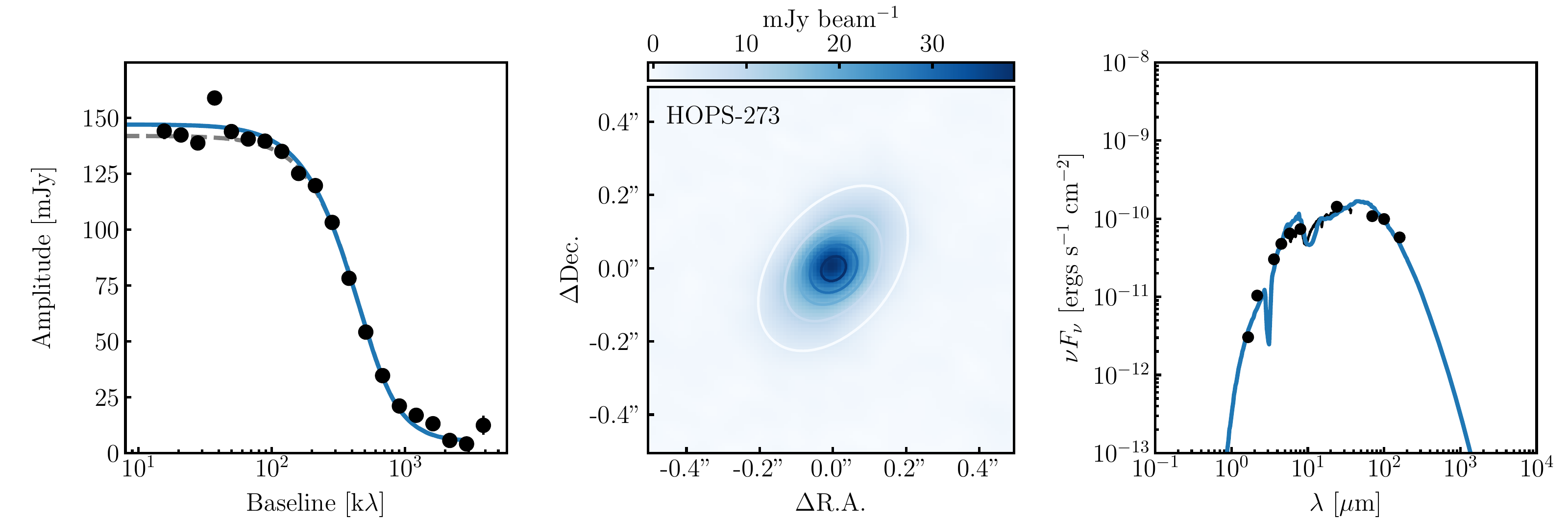}
\figsetgrpnote{A continuation of Figure \ref{fig:rt_fits} showing the best-fit model for HOPS-273. As in Figure \ref{fig:rt_fits}, black points ({\it left/right}) or color scale ({\it center}) show the data, while the blue lines ({\it left/right}) or contours ({\it center}) show the model, and the gray dashed line shows the disk contribution to the model.}
\figsetgrpend

\figsetgrpstart
\figsetgrpnum{1.73}
\figsetgrptitle{HOPS-287}
\figsetplot{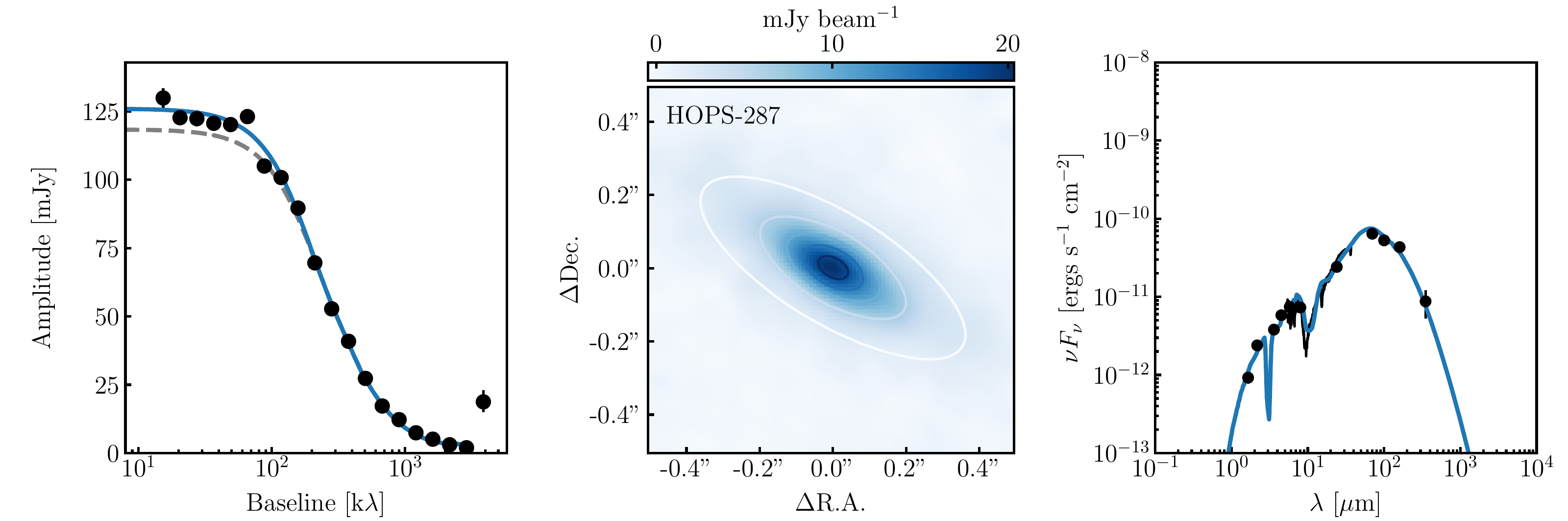}
\figsetgrpnote{A continuation of Figure \ref{fig:rt_fits} showing the best-fit model for HOPS-287. As in Figure \ref{fig:rt_fits}, black points ({\it left/right}) or color scale ({\it center}) show the data, while the blue lines ({\it left/right}) or contours ({\it center}) show the model, and the gray dashed line shows the disk contribution to the model.}
\figsetgrpend

\figsetgrpstart
\figsetgrpnum{1.74}
\figsetgrptitle{HOPS-294}
\figsetplot{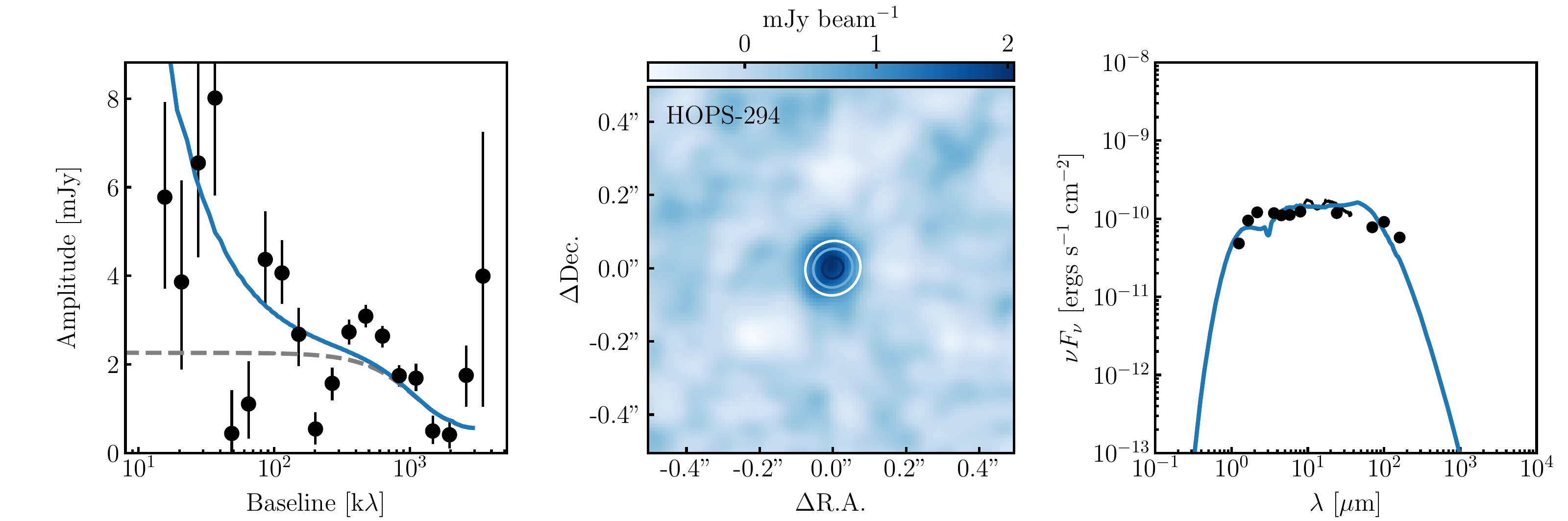}
\figsetgrpnote{A continuation of Figure \ref{fig:rt_fits} showing the best-fit model for HOPS-294. As in Figure \ref{fig:rt_fits}, black points ({\it left/right}) or color scale ({\it center}) show the data, while the blue lines ({\it left/right}) or contours ({\it center}) show the model, and the gray dashed line shows the disk contribution to the model.}
\figsetgrpend

\figsetgrpstart
\figsetgrpnum{1.75}
\figsetgrptitle{HOPS-297}
\figsetplot{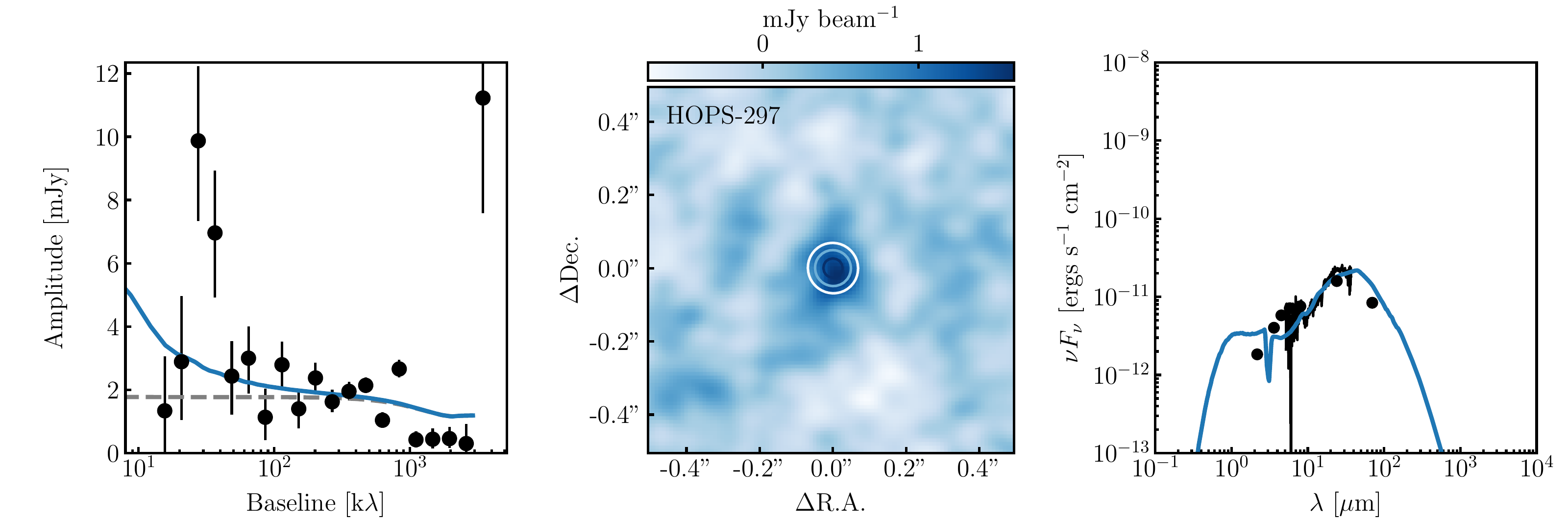}
\figsetgrpnote{A continuation of Figure \ref{fig:rt_fits} showing the best-fit model for HOPS-297. As in Figure \ref{fig:rt_fits}, black points ({\it left/right}) or color scale ({\it center}) show the data, while the blue lines ({\it left/right}) or contours ({\it center}) show the model, and the gray dashed line shows the disk contribution to the model.}
\figsetgrpend

\figsetgrpstart
\figsetgrpnum{1.76}
\figsetgrptitle{HOPS-305}
\figsetplot{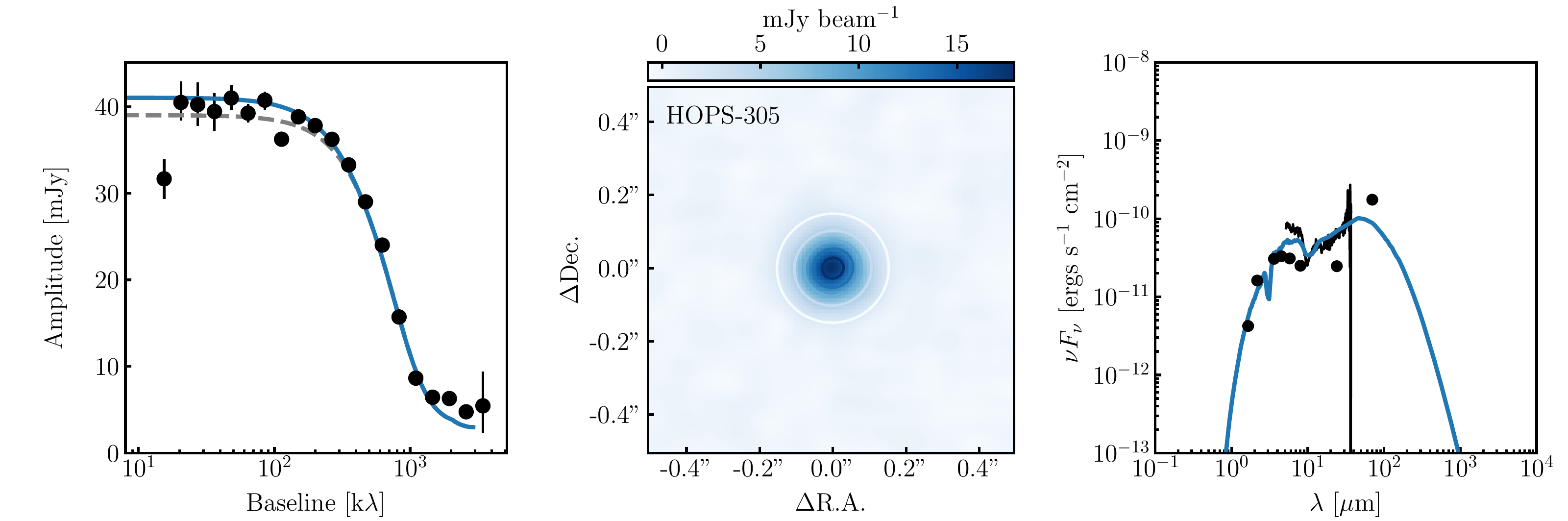}
\figsetgrpnote{A continuation of Figure \ref{fig:rt_fits} showing the best-fit model for HOPS-305. As in Figure \ref{fig:rt_fits}, black points ({\it left/right}) or color scale ({\it center}) show the data, while the blue lines ({\it left/right}) or contours ({\it center}) show the model, and the gray dashed line shows the disk contribution to the model.}
\figsetgrpend

\figsetgrpstart
\figsetgrpnum{1.77}
\figsetgrptitle{HOPS-310}
\figsetplot{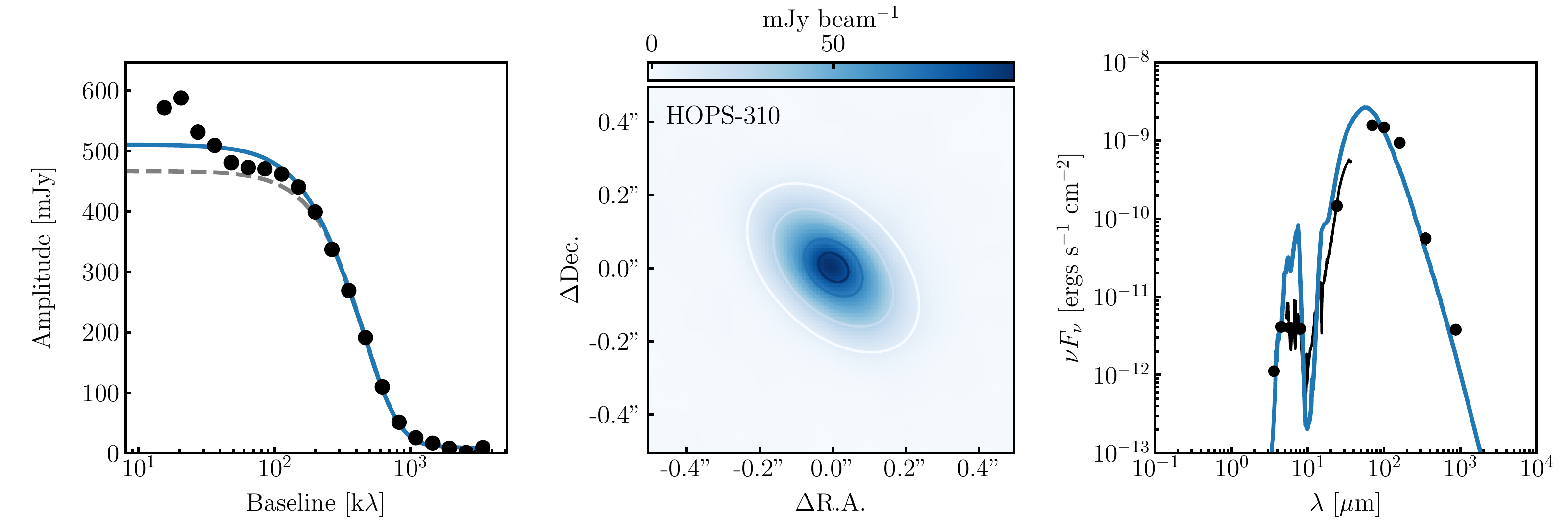}
\figsetgrpnote{A continuation of Figure \ref{fig:rt_fits} showing the best-fit model for HOPS-310. As in Figure \ref{fig:rt_fits}, black points ({\it left/right}) or color scale ({\it center}) show the data, while the blue lines ({\it left/right}) or contours ({\it center}) show the model, and the gray dashed line shows the disk contribution to the model.}
\figsetgrpend

\figsetgrpstart
\figsetgrpnum{1.78}
\figsetgrptitle{HOPS-315}
\figsetplot{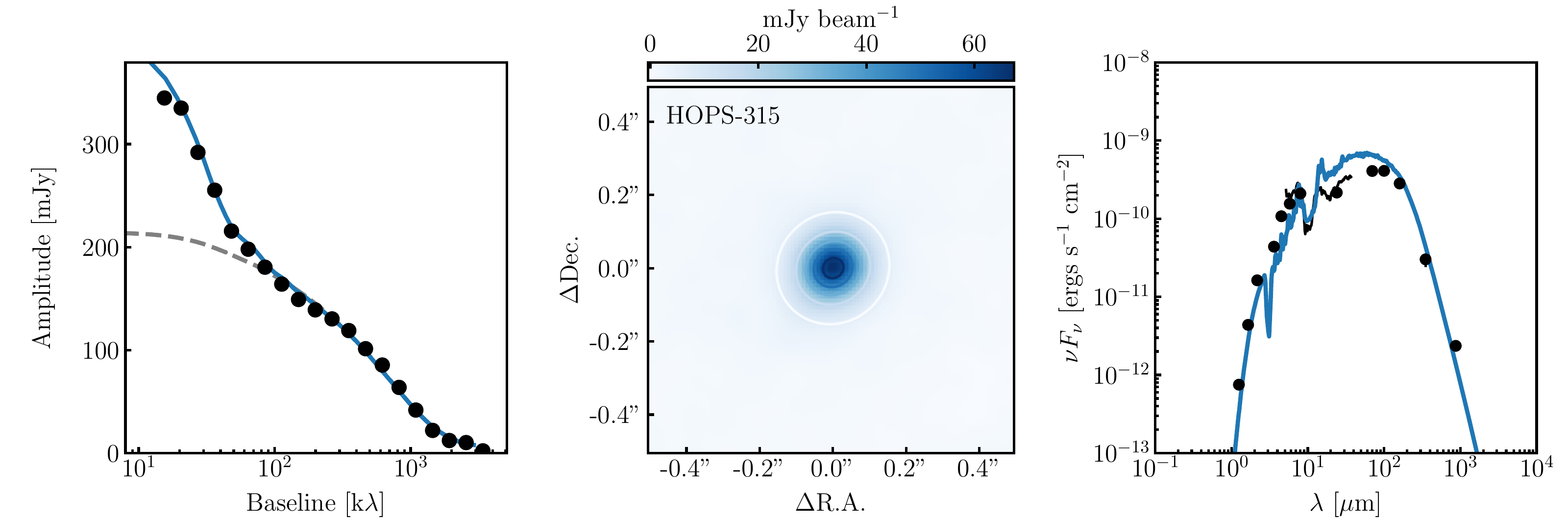}
\figsetgrpnote{A continuation of Figure \ref{fig:rt_fits} showing the best-fit model for HOPS-315. As in Figure \ref{fig:rt_fits}, black points ({\it left/right}) or color scale ({\it center}) show the data, while the blue lines ({\it left/right}) or contours ({\it center}) show the model, and the gray dashed line shows the disk contribution to the model.}
\figsetgrpend

\figsetgrpstart
\figsetgrpnum{1.79}
\figsetgrptitle{HOPS-322}
\figsetplot{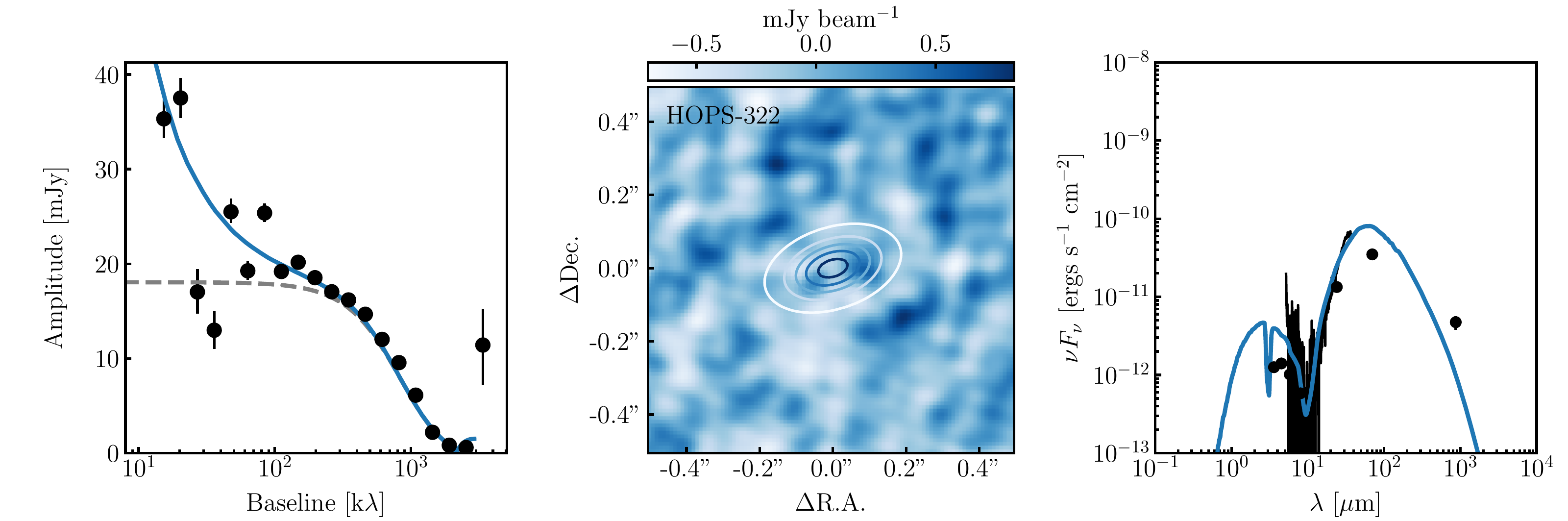}
\figsetgrpnote{A continuation of Figure \ref{fig:rt_fits} showing the best-fit model for HOPS-322. As in Figure \ref{fig:rt_fits}, black points ({\it left/right}) or color scale ({\it center}) show the data, while the blue lines ({\it left/right}) or contours ({\it center}) show the model, and the gray dashed line shows the disk contribution to the model.}
\figsetgrpend

\figsetgrpstart
\figsetgrpnum{1.80}
\figsetgrptitle{HOPS-325}
\figsetplot{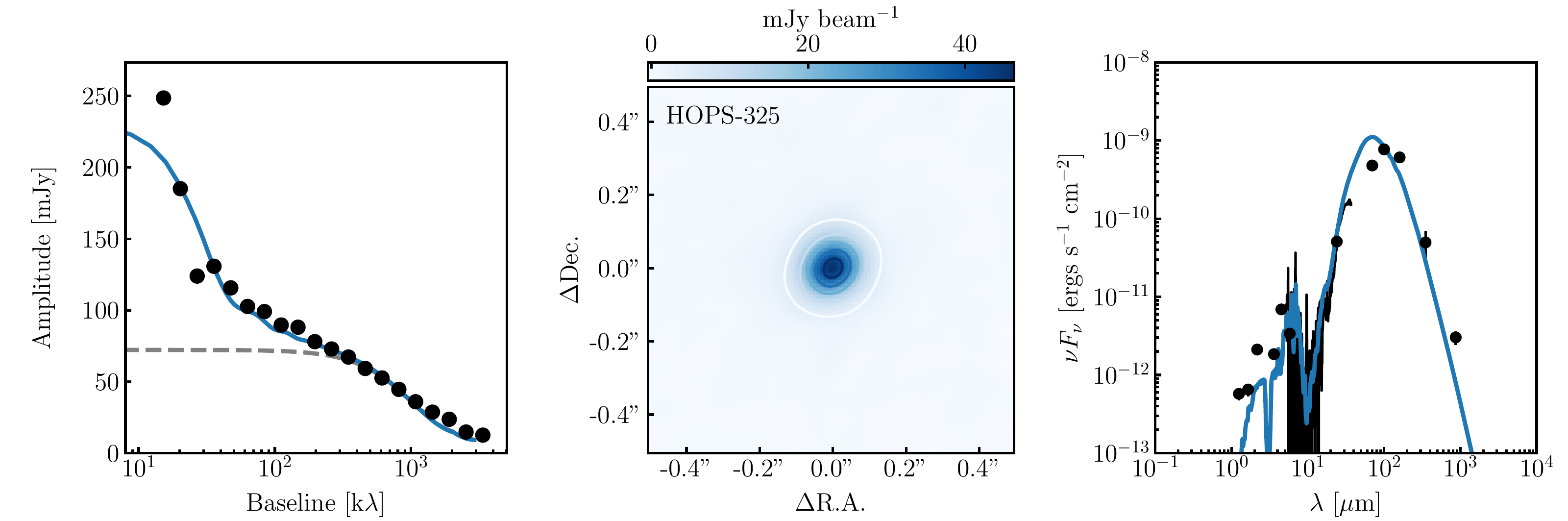}
\figsetgrpnote{A continuation of Figure \ref{fig:rt_fits} showing the best-fit model for HOPS-325. As in Figure \ref{fig:rt_fits}, black points ({\it left/right}) or color scale ({\it center}) show the data, while the blue lines ({\it left/right}) or contours ({\it center}) show the model, and the gray dashed line shows the disk contribution to the model.}
\figsetgrpend

\figsetgrpstart
\figsetgrpnum{1.81}
\figsetgrptitle{HOPS-343}
\figsetplot{HOPS-343_rt_model.pdf}
\figsetgrpnote{A continuation of Figure \ref{fig:rt_fits} showing the best-fit model for HOPS-343. As in Figure \ref{fig:rt_fits}, black points ({\it left/right}) or color scale ({\it center}) show the data, while the blue lines ({\it left/right}) or contours ({\it center}) show the model, and the gray dashed line shows the disk contribution to the model.}
\figsetgrpend

\figsetgrpstart
\figsetgrpnum{1.82}
\figsetgrptitle{HOPS-344}
\figsetplot{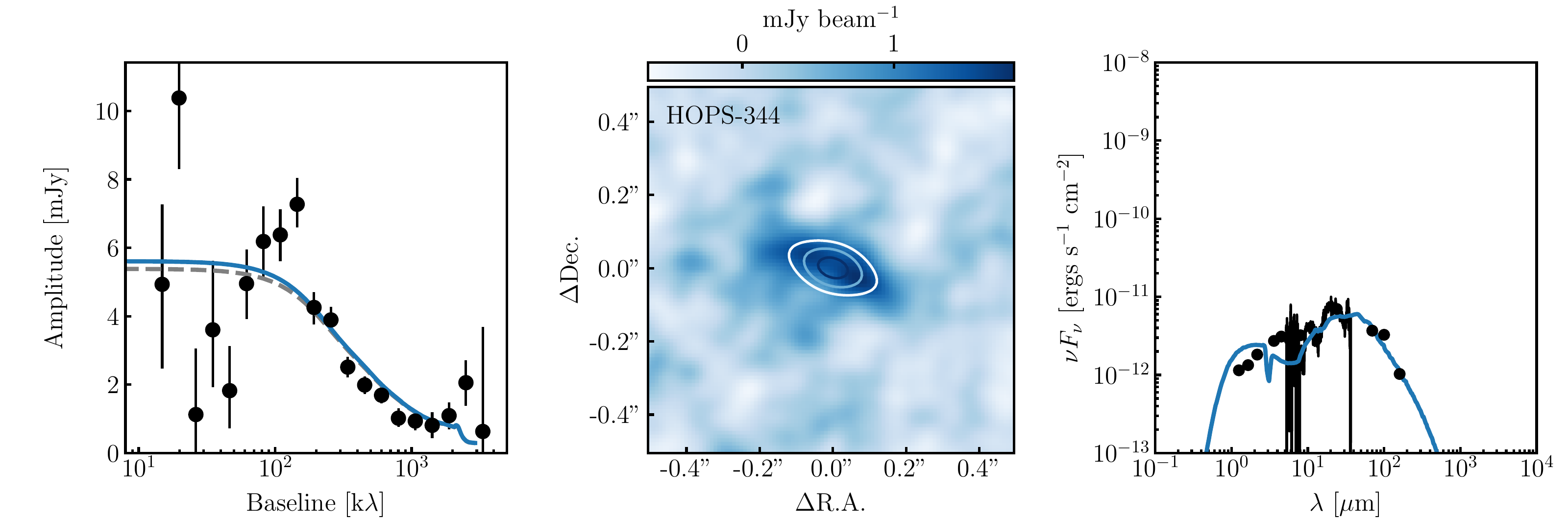}
\figsetgrpnote{A continuation of Figure \ref{fig:rt_fits} showing the best-fit model for HOPS-344. As in Figure \ref{fig:rt_fits}, black points ({\it left/right}) or color scale ({\it center}) show the data, while the blue lines ({\it left/right}) or contours ({\it center}) show the model, and the gray dashed line shows the disk contribution to the model.}
\figsetgrpend

\figsetgrpstart
\figsetgrpnum{1.83}
\figsetgrptitle{HOPS-345}
\figsetplot{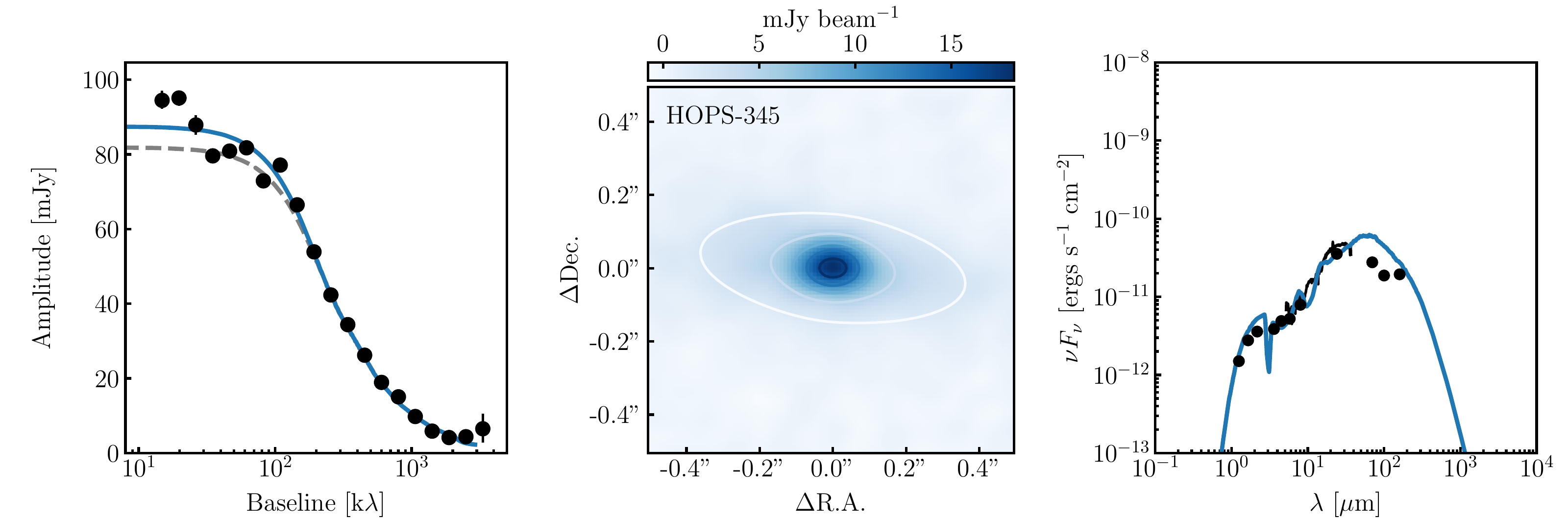}
\figsetgrpnote{A continuation of Figure \ref{fig:rt_fits} showing the best-fit model for HOPS-345. As in Figure \ref{fig:rt_fits}, black points ({\it left/right}) or color scale ({\it center}) show the data, while the blue lines ({\it left/right}) or contours ({\it center}) show the model, and the gray dashed line shows the disk contribution to the model.}
\figsetgrpend

\figsetgrpstart
\figsetgrpnum{1.84}
\figsetgrptitle{HOPS-347}
\figsetplot{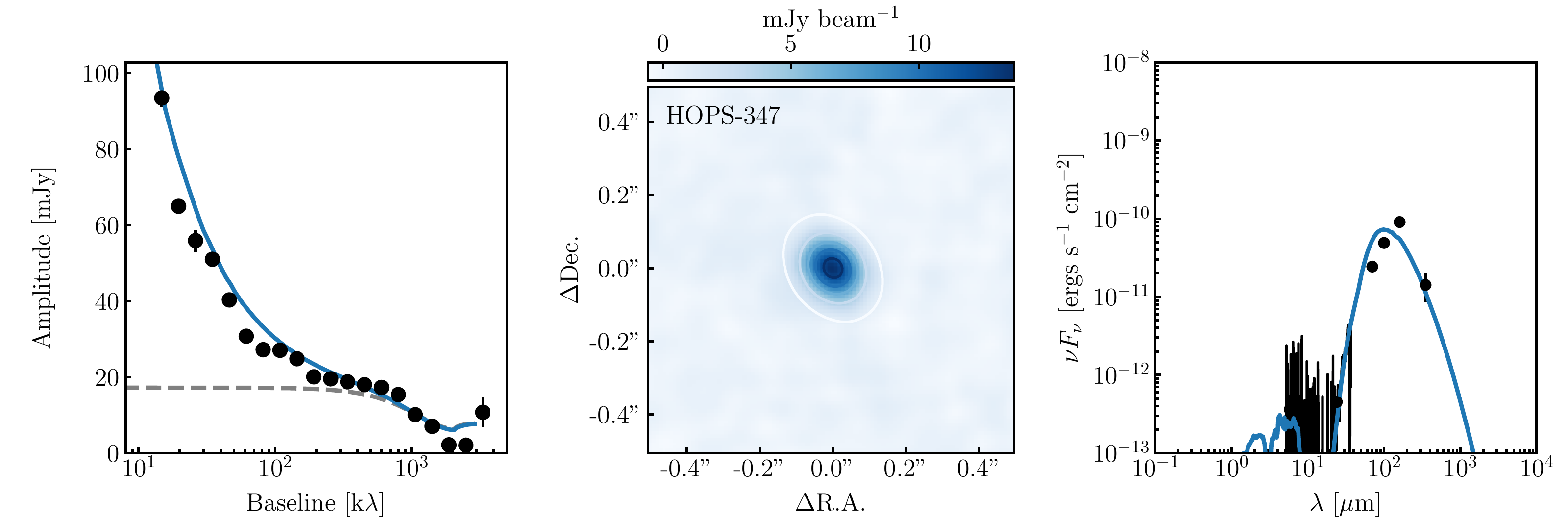}
\figsetgrpnote{A continuation of Figure \ref{fig:rt_fits} showing the best-fit model for HOPS-347. As in Figure \ref{fig:rt_fits}, black points ({\it left/right}) or color scale ({\it center}) show the data, while the blue lines ({\it left/right}) or contours ({\it center}) show the model, and the gray dashed line shows the disk contribution to the model.}
\figsetgrpend

\figsetgrpstart
\figsetgrpnum{1.85}
\figsetgrptitle{HOPS-354}
\figsetplot{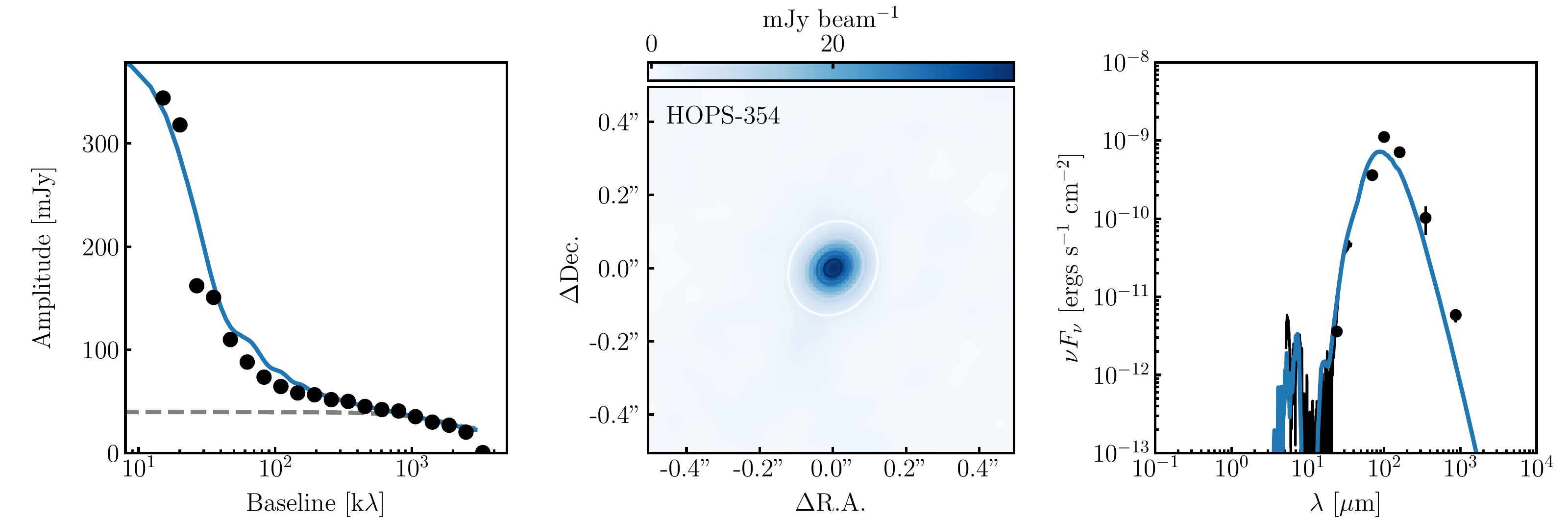}
\figsetgrpnote{A continuation of Figure \ref{fig:rt_fits} showing the best-fit model for HOPS-354. As in Figure \ref{fig:rt_fits}, black points ({\it left/right}) or color scale ({\it center}) show the data, while the blue lines ({\it left/right}) or contours ({\it center}) show the model, and the gray dashed line shows the disk contribution to the model.}
\figsetgrpend

\figsetgrpstart
\figsetgrpnum{1.86}
\figsetgrptitle{HOPS-355}
\figsetplot{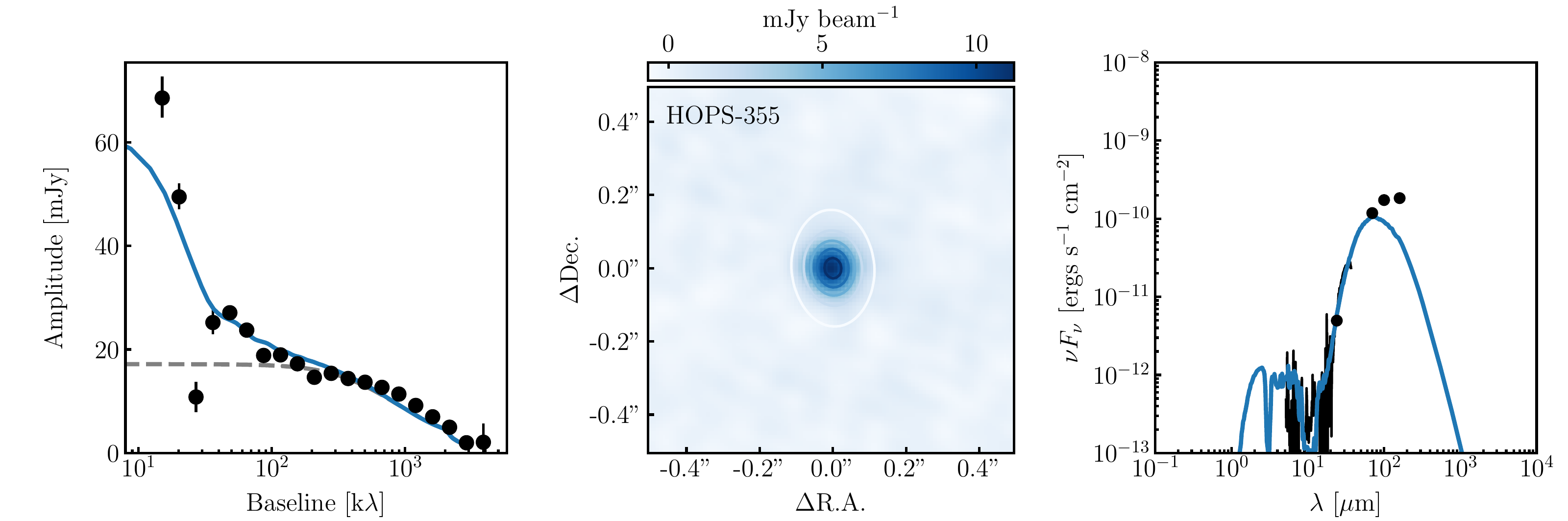}
\figsetgrpnote{A continuation of Figure \ref{fig:rt_fits} showing the best-fit model for HOPS-355. As in Figure \ref{fig:rt_fits}, black points ({\it left/right}) or color scale ({\it center}) show the data, while the blue lines ({\it left/right}) or contours ({\it center}) show the model, and the gray dashed line shows the disk contribution to the model.}
\figsetgrpend

\figsetgrpstart
\figsetgrpnum{1.87}
\figsetgrptitle{HOPS-365}
\figsetplot{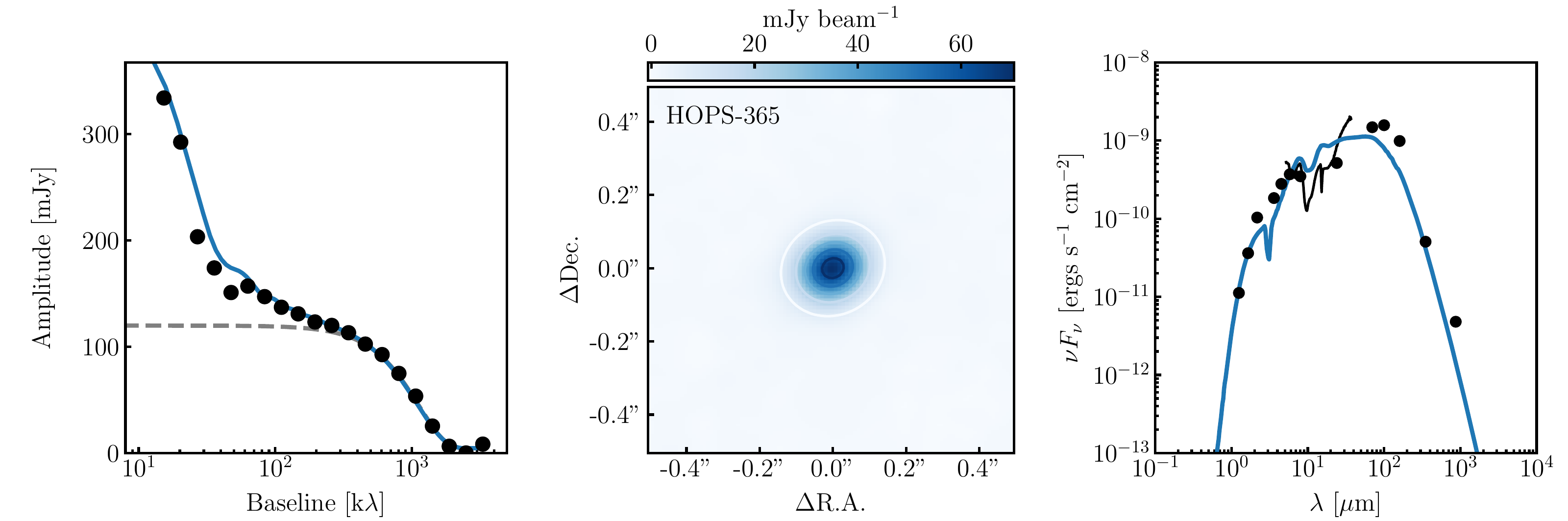}
\figsetgrpnote{A continuation of Figure \ref{fig:rt_fits} showing the best-fit model for HOPS-365. As in Figure \ref{fig:rt_fits}, black points ({\it left/right}) or color scale ({\it center}) show the data, while the blue lines ({\it left/right}) or contours ({\it center}) show the model, and the gray dashed line shows the disk contribution to the model.}
\figsetgrpend

\figsetgrpstart
\figsetgrpnum{1.88}
\figsetgrptitle{HOPS-367}
\figsetplot{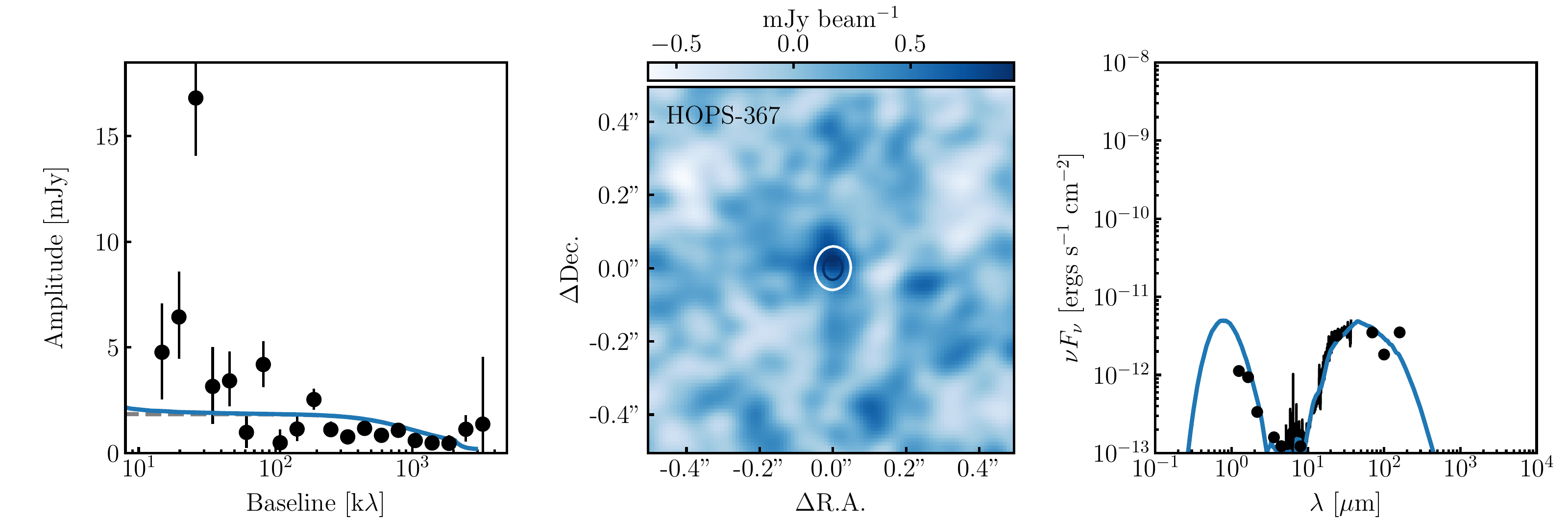}
\figsetgrpnote{A continuation of Figure \ref{fig:rt_fits} showing the best-fit model for HOPS-367. As in Figure \ref{fig:rt_fits}, black points ({\it left/right}) or color scale ({\it center}) show the data, while the blue lines ({\it left/right}) or contours ({\it center}) show the model, and the gray dashed line shows the disk contribution to the model.}
\figsetgrpend

\figsetgrpstart
\figsetgrpnum{1.89}
\figsetgrptitle{HOPS-372}
\figsetplot{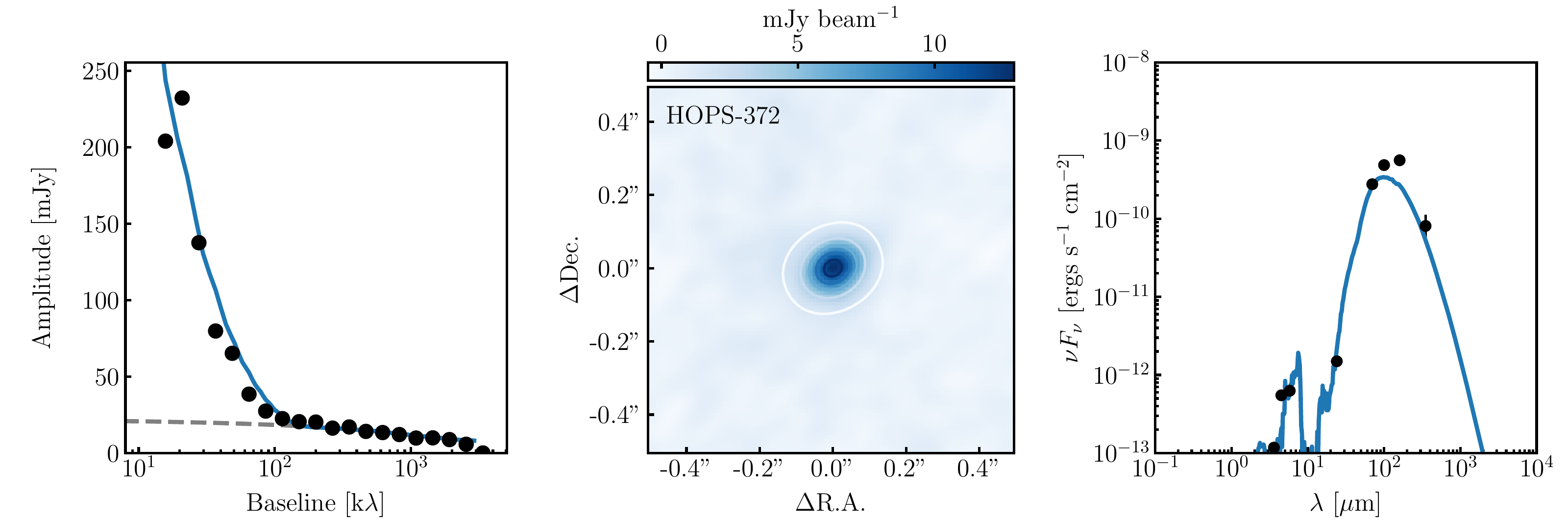}
\figsetgrpnote{A continuation of Figure \ref{fig:rt_fits} showing the best-fit model for HOPS-372. As in Figure \ref{fig:rt_fits}, black points ({\it left/right}) or color scale ({\it center}) show the data, while the blue lines ({\it left/right}) or contours ({\it center}) show the model, and the gray dashed line shows the disk contribution to the model.}
\figsetgrpend

\figsetgrpstart
\figsetgrpnum{1.90}
\figsetgrptitle{HOPS-376}
\figsetplot{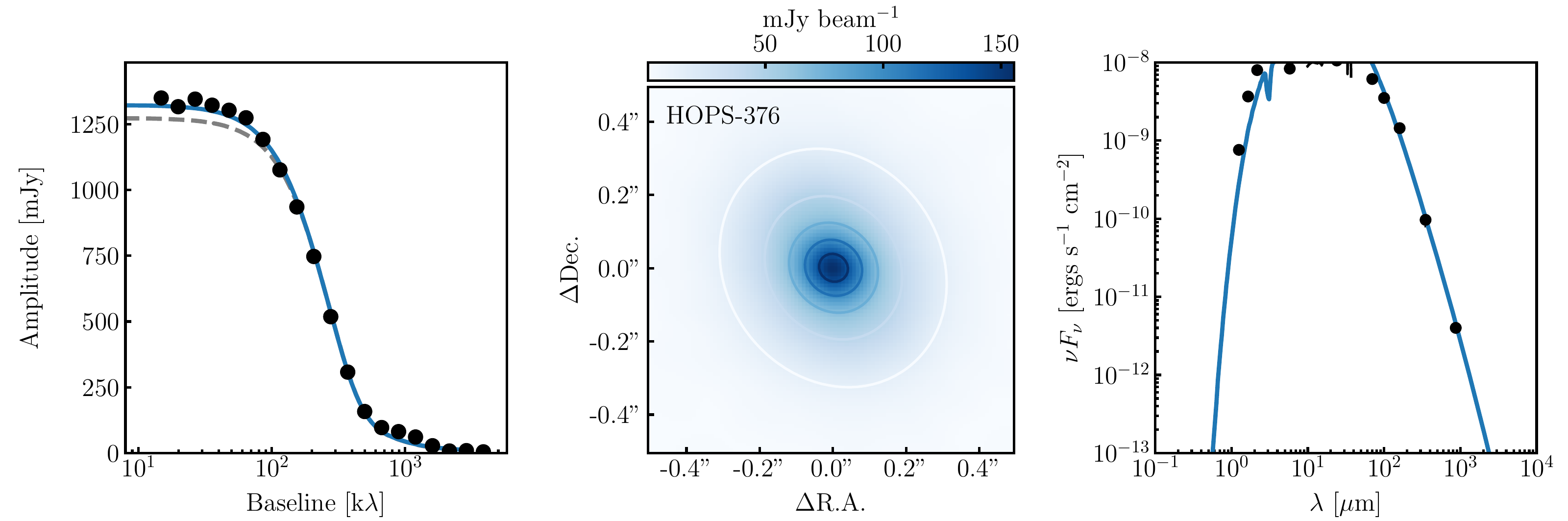}
\figsetgrpnote{A continuation of Figure \ref{fig:rt_fits} showing the best-fit model for HOPS-376. As in Figure \ref{fig:rt_fits}, black points ({\it left/right}) or color scale ({\it center}) show the data, while the blue lines ({\it left/right}) or contours ({\it center}) show the model, and the gray dashed line shows the disk contribution to the model.}
\figsetgrpend

\figsetgrpstart
\figsetgrpnum{1.91}
\figsetgrptitle{HOPS-377}
\figsetplot{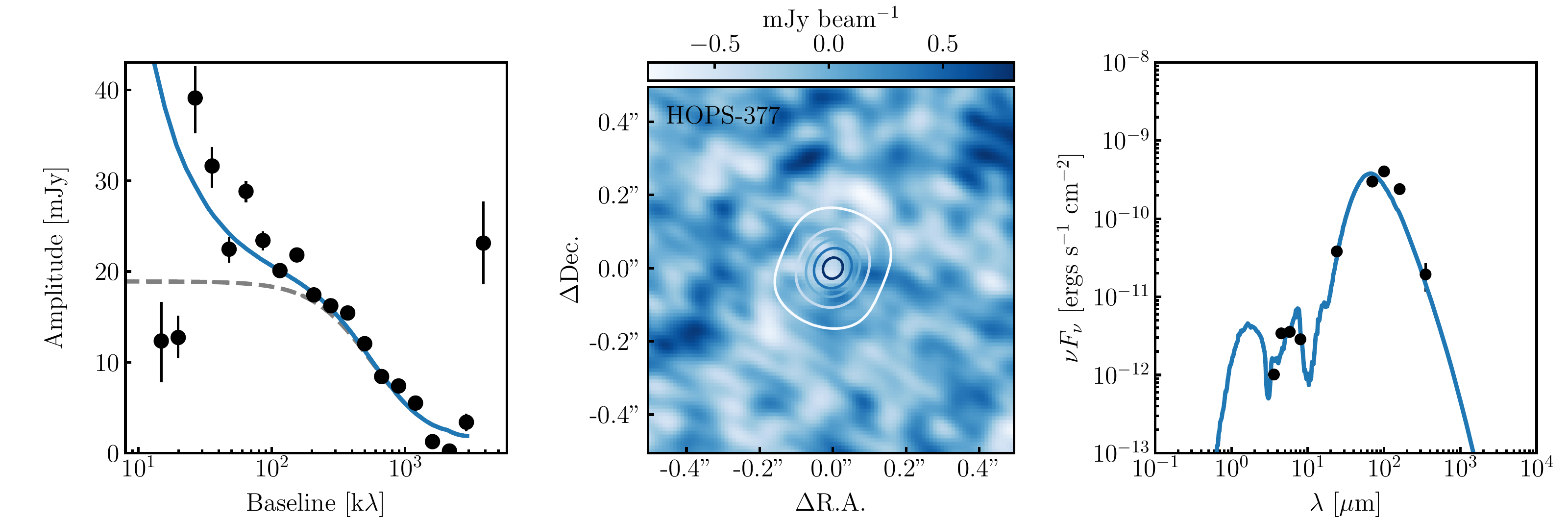}
\figsetgrpnote{A continuation of Figure \ref{fig:rt_fits} showing the best-fit model for HOPS-377. As in Figure \ref{fig:rt_fits}, black points ({\it left/right}) or color scale ({\it center}) show the data, while the blue lines ({\it left/right}) or contours ({\it center}) show the model, and the gray dashed line shows the disk contribution to the model.}
\figsetgrpend

\figsetgrpstart
\figsetgrpnum{1.92}
\figsetgrptitle{HOPS-385}
\figsetplot{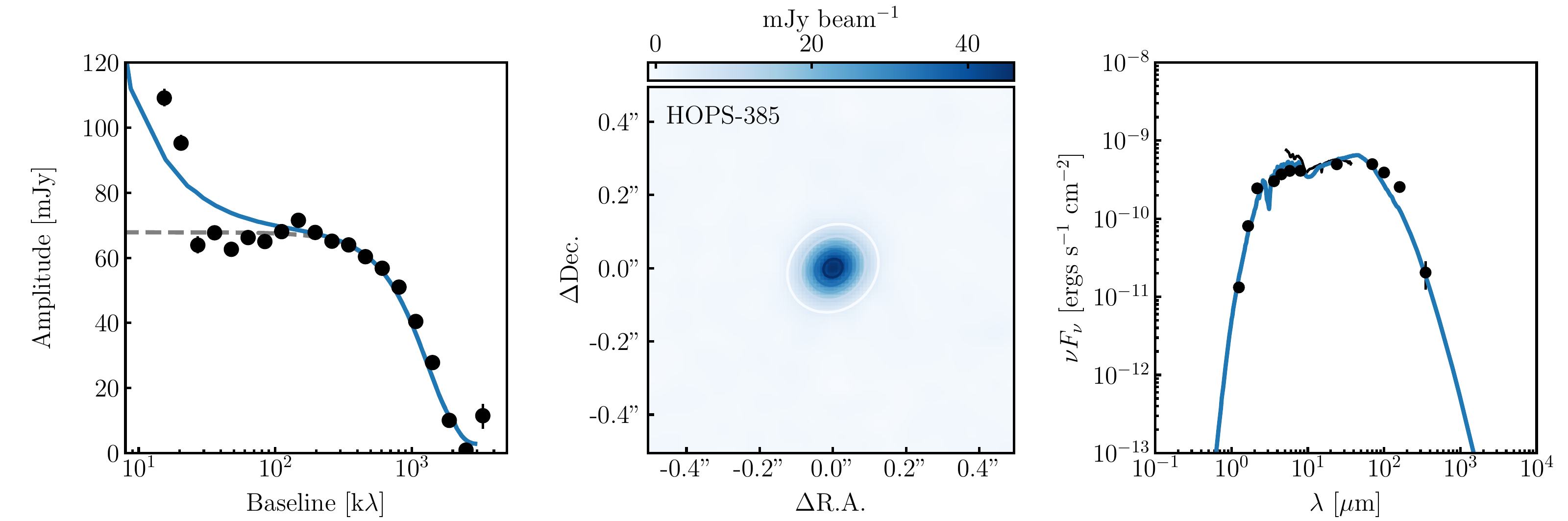}
\figsetgrpnote{A continuation of Figure \ref{fig:rt_fits} showing the best-fit model for HOPS-385. As in Figure \ref{fig:rt_fits}, black points ({\it left/right}) or color scale ({\it center}) show the data, while the blue lines ({\it left/right}) or contours ({\it center}) show the model, and the gray dashed line shows the disk contribution to the model.}
\figsetgrpend

\figsetgrpstart
\figsetgrpnum{1.93}
\figsetgrptitle{HOPS-388}
\figsetplot{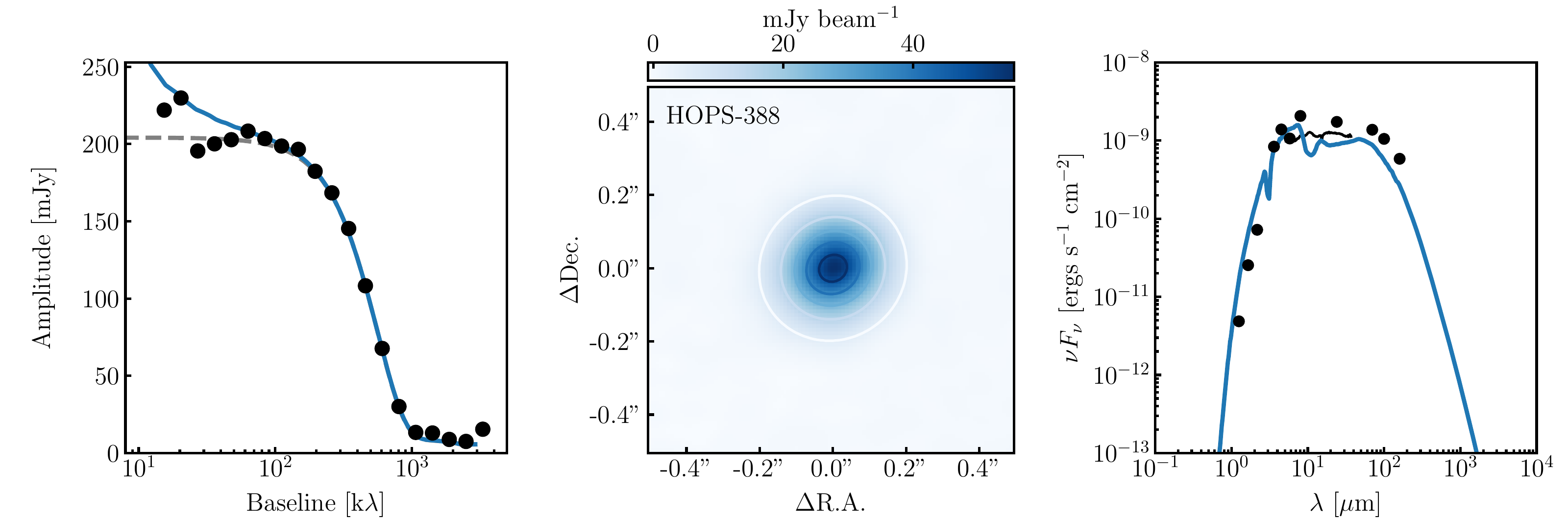}
\figsetgrpnote{A continuation of Figure \ref{fig:rt_fits} showing the best-fit model for HOPS-388. As in Figure \ref{fig:rt_fits}, black points ({\it left/right}) or color scale ({\it center}) show the data, while the blue lines ({\it left/right}) or contours ({\it center}) show the model, and the gray dashed line shows the disk contribution to the model.}
\figsetgrpend

\figsetgrpstart
\figsetgrpnum{1.94}
\figsetgrptitle{HOPS-393}
\figsetplot{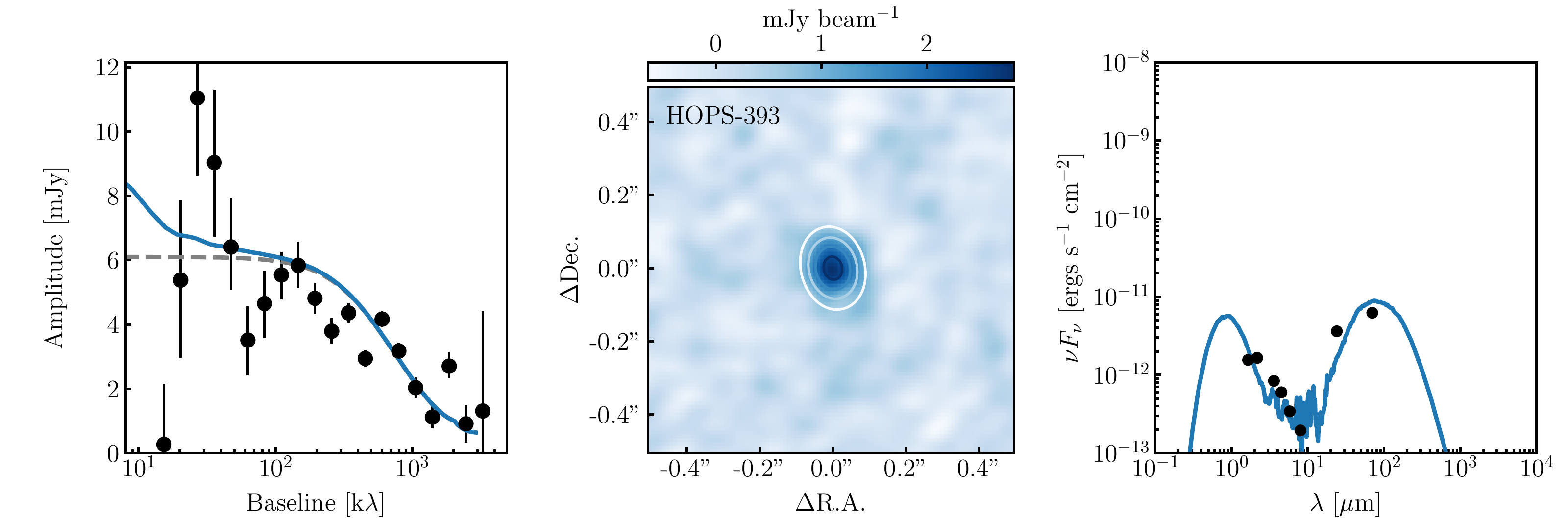}
\figsetgrpnote{A continuation of Figure \ref{fig:rt_fits} showing the best-fit model for HOPS-393. As in Figure \ref{fig:rt_fits}, black points ({\it left/right}) or color scale ({\it center}) show the data, while the blue lines ({\it left/right}) or contours ({\it center}) show the model, and the gray dashed line shows the disk contribution to the model.}
\figsetgrpend

\figsetgrpstart
\figsetgrpnum{1.95}
\figsetgrptitle{HOPS-399}
\figsetplot{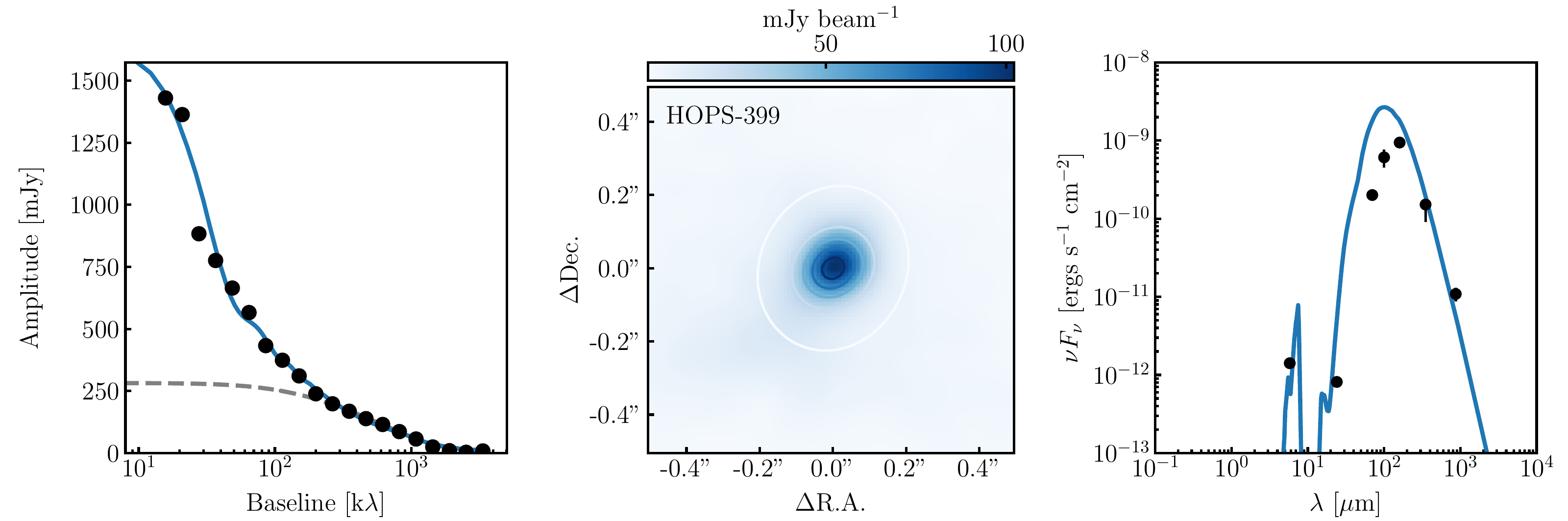}
\figsetgrpnote{A continuation of Figure \ref{fig:rt_fits} showing the best-fit model for HOPS-399. As in Figure \ref{fig:rt_fits}, black points ({\it left/right}) or color scale ({\it center}) show the data, while the blue lines ({\it left/right}) or contours ({\it center}) show the model, and the gray dashed line shows the disk contribution to the model.}
\figsetgrpend

\figsetgrpstart
\figsetgrpnum{1.96}
\figsetgrptitle{HOPS-401}
\figsetplot{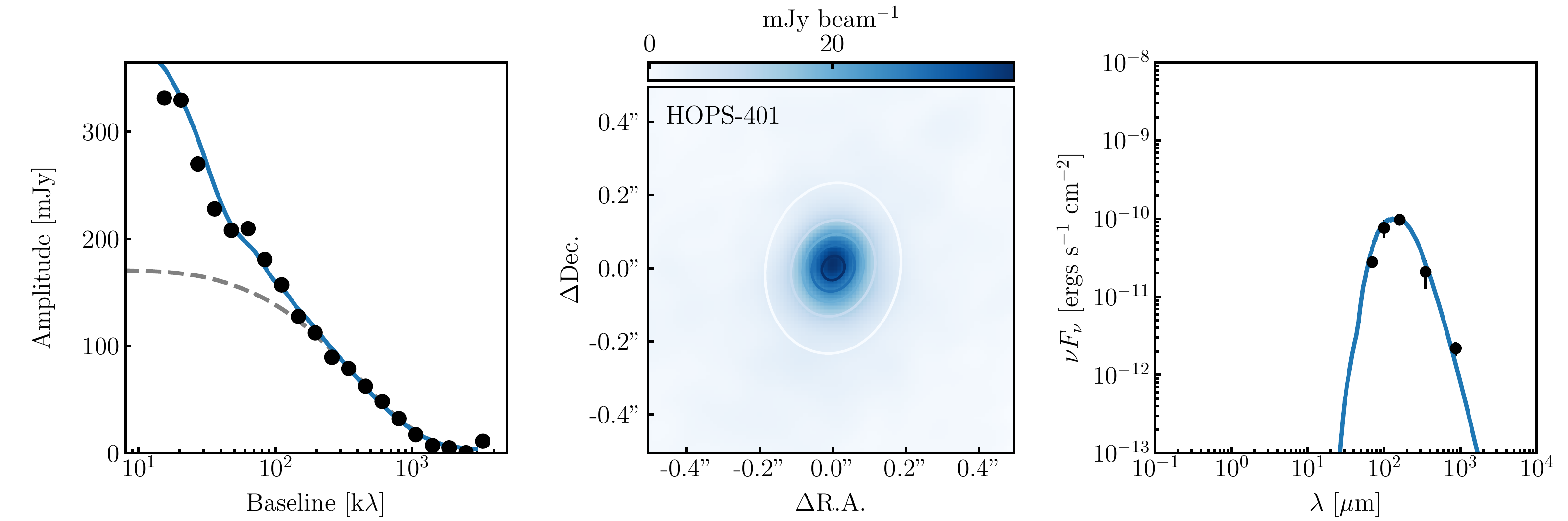}
\figsetgrpnote{A continuation of Figure \ref{fig:rt_fits} showing the best-fit model for HOPS-401. As in Figure \ref{fig:rt_fits}, black points ({\it left/right}) or color scale ({\it center}) show the data, while the blue lines ({\it left/right}) or contours ({\it center}) show the model, and the gray dashed line shows the disk contribution to the model.}
\figsetgrpend

\figsetgrpstart
\figsetgrpnum{1.97}
\figsetgrptitle{HOPS-403}
\figsetplot{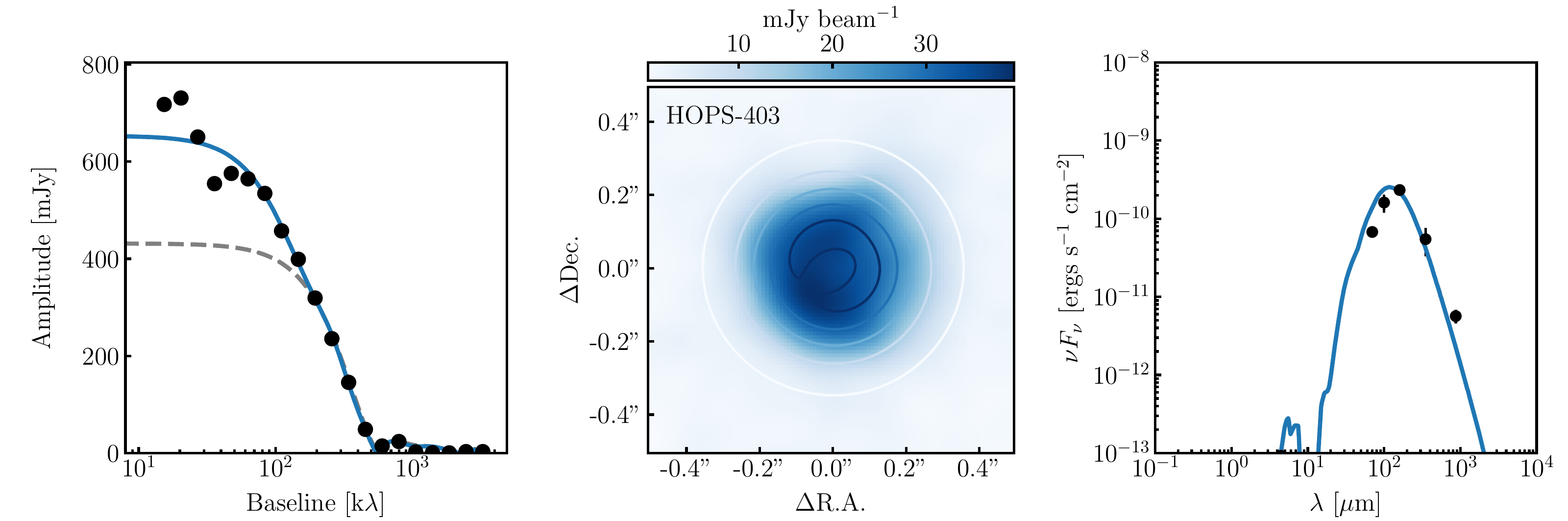}
\figsetgrpnote{A continuation of Figure \ref{fig:rt_fits} showing the best-fit model for HOPS-403. As in Figure \ref{fig:rt_fits}, black points ({\it left/right}) or color scale ({\it center}) show the data, while the blue lines ({\it left/right}) or contours ({\it center}) show the model, and the gray dashed line shows the disk contribution to the model.}
\figsetgrpend

\figsetend

%% file: best_fits_table.tex
\begin{deluxetable*}{l|cchchhhhchhchchhchchhcchhh}
\tablecaption{Best-fit Radiative Transfer Model Parameters}
\tablenum{1}
\tabletypesize{\scriptsize}
\label{table:rt_best_fits}
\tablehead{\colhead{Sources} & \colhead{$L_*$} & \colhead{$M_{disk,dust}$} & \nocolhead{$M_{disk}$} & \colhead{$R_{disk,dust}$} & \nocolhead{$R_{in,dust}$} & \nocolhead{$\gamma$} & \nocolhead{$h_0$} & \nocolhead{$\psi$} & \colhead{$F_{\nu,230GHz}$} & \nocolhead{$F_{\nu,345GHz}$} & \nocolhead{$Q_{min}$} & \colhead{$M_{env,dust}$} & \nocolhead{$M_{env}$} & \colhead{$R_{env,dust}$} & \nocolhead{$\xi$} & \nocolhead{$f_{cav}$} & \colhead{$a_{max}$} & \nocolhead{$p$} & \colhead{$i$} & \nocolhead{p.a.} & \nocolhead{$\kappa_{\nu,230GHz}$} & \colhead{$\kappa_{\nu,345GHz}$} & \colhead{$\beta$} & \nocolhead{$M_{tot}$} & \nocolhead{$M_{env} / M_{tot}$} & \nocolhead{Age}\\ \colhead{ } & \colhead{(L$_{\odot}$)} & \colhead{(M$_{\oplus}$)} & \nocolhead{($M_{\odot}$)} & \colhead{(au)} & \nocolhead{(au)} & \nocolhead{} & \nocolhead{(au)} & \nocolhead{} & \colhead{(mJy)} & \nocolhead{(mJy)} & \nocolhead{} & \colhead{(M$_{\oplus}$)} & \nocolhead{($M_{\odot}$)} & \colhead{(au)} & \nocolhead{} & \nocolhead{} & \colhead{($\mu$m)} & \nocolhead{} & \colhead{($^{\circ}$)} & \nocolhead{($^{\circ}$)} & \nocolhead{(cm$^2$ g$^{-1}$)} & \colhead{(cm$^2$ g$^{-1}$)} & \colhead{} & \nocolhead{(M$_{\odot}$)} & \nocolhead{} & \nocolhead{(yr)}}
\startdata
HOPS-2 & $0.43^{+1.32}_{-0.14}$ & $2.8^{+21.8}_{-1.5}$ & $0.00085^{+0.00655}_{-0.00045}$ & $11.3^{+8.4}_{-6.7}$ & $0.34^{+0.15}_{-0.23}$ & $0.86^{+0.70}_{-1.26}$ & $0.090^{+0.071}_{-0.036}$ & $0.98^{+0.20}_{-0.32}$ & $  2.876$ & $  8.534$ & $58^{+186}_{-39}$ & $0.30^{+23.78}_{-0.18}$ & $0.000089^{+0.007136}_{-0.000055}$ & $164^{+9867}_{-56}$ & $1.16^{+0.30}_{-0.62}$ & $0.62^{+0.35}_{-0.38}$ & $2.5^{+1060.7}_{-1.3}$ & $3.63^{+0.76}_{-1.08}$ & $54^{+11}_{-37}$ & $75^{+50}_{-34}$ & $2.000$ & $3.832$ & $1.604$ & $0.00077^{+0.01855}_{-0.00064}$ & $0.0023^{+0.1038}_{-0.0022}$ & $805962^{+141462}_{-431122}$ \\[2pt]
HOPS-3 & $2.30^{+0.81}_{-1.20}$ & $15.0^{+11.5}_{-4.7}$ & $0.0045^{+0.0035}_{-0.0014}$ & $98^{+22}_{-19}$ & $0.26^{+0.11}_{-0.15}$ & $0.70^{+0.36}_{-0.32}$ & $0.091^{+0.070}_{-0.059}$ & $0.62^{+0.13}_{-0.11}$ & $ 15.529$ & $ 40.825$ & $23^{+74}_{-16}$ & $0.132^{+0.240}_{-0.068}$ & $0.000039^{+0.000072}_{-0.000020}$ & $221^{+106}_{-21}$ & $0.87^{+0.17}_{-0.28}$ & $0.11^{+0.58}_{-0.10}$ & $397^{+692}_{-235}$ & $3.11^{+0.39}_{-0.48}$ & $77.8^{+1.0}_{-1.6}$ & $126.90^{+1.47}_{-0.98}$ & $6.553$ & $11.039$ & $1.286$ & $0.00070^{+0.01923}_{-0.00053}$ & $0.00067^{+0.00303}_{-0.00065}$ & $1035398^{+92974}_{-131495}$ \\[2pt]
HOPS-13 & $0.73^{+0.25}_{-0.19}$ & $0.54^{+0.30}_{-0.38}$ & $0.000161^{+0.000090}_{-0.000114}$ & $33^{+22}_{-21}$ & $0.1090^{+0.2084}_{-0.0074}$ & $0.08^{+1.27}_{-0.54}$ & $0.147^{+0.063}_{-0.078}$ & $0.97^{+0.22}_{-0.34}$ & $  0.605$ & $  2.410$ & $511^{+1658}_{-340}$ & $0.61^{+112.69}_{-0.35}$ & $0.00018^{+0.03382}_{-0.00011}$ & $167^{+10661}_{-64}$ & $0.568^{+0.894}_{-0.055}$ & $0.920^{+0.056}_{-0.733}$ & $4.1^{+1680.6}_{-2.9}$ & $4.03^{+0.43}_{-1.46}$ & $30^{+26}_{-21}$ & $102^{+64}_{-96}$ & $1.999$ & $3.829$ & $1.602$ & $0.00085^{+0.01875}_{-0.00072}$ & $0.0031^{+0.3929}_{-0.0030}$ & $778073^{+120966}_{-560963}$ \\[2pt]
HOPS-16 & $0.44^{+0.12}_{-0.10}$ & $15.4^{+11.7}_{-8.2}$ & $0.0046^{+0.0035}_{-0.0025}$ & $12.4^{+2.3}_{-3.3}$ & $0.157^{+0.229}_{-0.052}$ & $0.28^{+0.36}_{-0.74}$ & $0.158^{+0.051}_{-0.068}$ & $0.90^{+0.19}_{-0.37}$ & $  8.140$ & $ 21.047$ & $13.6^{+42.9}_{-9.2}$ & $0.35^{+45.56}_{-0.16}$ & $0.000105^{+0.013672}_{-0.000049}$ & $146^{+5402}_{-43}$ & $0.69^{+0.43}_{-0.18}$ & $0.36^{+0.52}_{-0.20}$ & $95^{+3054}_{-91}$ & $3.51^{+0.94}_{-0.72}$ & $26^{+12}_{-19}$ & $119^{+58}_{-110}$ & $2.105$ & $4.398$ & $1.818$ & $0.00077^{+0.01871}_{-0.00060}$ & $0.0021^{+0.1632}_{-0.0020}$ & $797207^{+94225}_{-479410}$ \\[2pt]
HOPS-18 & $1.93^{+0.33}_{-0.24}$ & $12.6^{+1.6}_{-5.5}$ & $0.00378^{+0.00047}_{-0.00165}$ & $83.1^{+5.4}_{-10.7}$ & $0.60^{+0.27}_{-0.34}$ & $0.19^{+0.42}_{-0.30}$ & $0.110^{+0.061}_{-0.040}$ & $0.91^{+0.18}_{-0.13}$ & $  9.097$ & $ 35.498$ & $36^{+114}_{-24}$ & $53^{+293}_{-32}$ & $0.0159^{+0.0880}_{-0.0097}$ & $2289^{+5836}_{-1118}$ & $1.16^{+0.30}_{-0.34}$ & $0.70^{+0.27}_{-0.47}$ & $144^{+670}_{-94}$ & $2.91^{+1.50}_{-0.34}$ & $77.5^{+1.5}_{-1.7}$ & $70.9^{+1.8}_{-1.1}$ & $2.205$ & $4.981$ & $2.010$ & $0.00088^{+0.01955}_{-0.00057}$ & $0.27^{+0.43}_{-0.26}$ & $384558^{+110505}_{-226783}$ \\[2pt]
HOPS-29 & $1.87^{+1.07}_{-0.28}$ & $5.4^{+4.0}_{-2.6}$ & $0.00161^{+0.00121}_{-0.00078}$ & $15.4^{+3.9}_{-2.2}$ & $0.66^{+0.27}_{-0.42}$ & $0.27^{+0.34}_{-0.42}$ & $0.038^{+0.110}_{-0.023}$ & $1.14^{+0.18}_{-0.59}$ & $  8.534$ & $ 27.207$ & $45^{+145}_{-30}$ & $165^{+632}_{-150}$ & $0.050^{+0.190}_{-0.045}$ & $2891^{+20971}_{-1025}$ & $0.973^{+0.373}_{-0.059}$ & $0.41^{+0.29}_{-0.27}$ & $81^{+86}_{-54}$ & $3.24^{+0.97}_{-0.62}$ & $50.4^{+12.7}_{-5.9}$ & $74.6^{+11.8}_{-9.9}$ & $2.055$ & $4.089$ & $1.696$ & $0.0013^{+0.0192}_{-0.0010}$ & $0.40^{+0.45}_{-0.39}$ & $415478^{+151863}_{-359692}$ \\[2pt]
HOPS-36 & $0.884^{+0.132}_{-0.090}$ & $18^{+67}_{-13}$ & $0.0055^{+0.0202}_{-0.0038}$ & $22.6^{+8.6}_{-9.5}$ & $0.147^{+0.144}_{-0.046}$ & $0.73^{+0.28}_{-0.88}$ & $0.061^{+0.040}_{-0.025}$ & $0.582^{+0.205}_{-0.077}$ & $ 16.079$ & $ 40.026$ & $15^{+51}_{-10}$ & $0.29^{+0.77}_{-0.12}$ & $0.000086^{+0.000232}_{-0.000035}$ & $114^{+246}_{-13}$ & $0.65^{+0.19}_{-0.14}$ & $0.36^{+0.52}_{-0.21}$ & $3989^{+45468}_{-3716}$ & $3.10^{+1.15}_{-0.55}$ & $55.4^{+2.7}_{-2.7}$ & $67.0^{+2.7}_{-3.1}$ & $4.228$ & $5.957$ & $0.846$ & $0.00071^{+0.01945}_{-0.00053}$ & $0.0012^{+0.0060}_{-0.0011}$ & $903011^{+57722}_{-143015}$ \\[2pt]
HOPS-41 & $1.27^{+0.40}_{-0.36}$ & $6.8^{+6.4}_{-2.2}$ & $0.00203^{+0.00194}_{-0.00065}$ & $8.1^{+1.4}_{-2.0}$ & $2.50^{+0.82}_{-0.94}$ & $-0.412^{+0.807}_{-0.074}$ & $0.163^{+0.091}_{-0.071}$ & $1.30^{+0.19}_{-0.28}$ & $  9.679$ & $ 25.644$ & $27^{+88}_{-18}$ & $95^{+908}_{-66}$ & $0.029^{+0.273}_{-0.020}$ & $3846^{+14624}_{-2027}$ & $1.465^{+0.028}_{-0.337}$ & $0.42^{+0.39}_{-0.38}$ & $1.26^{+68.65}_{-0.23}$ & $3.69^{+0.77}_{-1.02}$ & $27^{+12}_{-17}$ & $169.7^{+7.9}_{-161.2}$ & $1.996$ & $3.816$ & $1.597$ & $0.0015^{+0.0185}_{-0.0011}$ & $0.47^{+0.41}_{-0.45}$ & $313971^{+123902}_{-285760}$ \\[2pt]
HOPS-42 & $0.95^{+0.18}_{-0.23}$ & $26.6^{+5.3}_{-7.9}$ & $0.0080^{+0.0016}_{-0.0024}$ & $110^{+18}_{-22}$ & $0.41^{+1.23}_{-0.13}$ & $-0.06^{+0.25}_{-0.19}$ & $0.169^{+0.084}_{-0.112}$ & $0.62^{+0.20}_{-0.11}$ & $ 15.409$ & $ 45.070$ & $9.9^{+30.8}_{-6.7}$ & $0.80^{+911.78}_{-0.14}$ & $0.000239^{+0.273635}_{-0.000042}$ & $161^{+23152}_{-10}$ & $1.407^{+0.075}_{-0.728}$ & $0.910^{+0.086}_{-0.532}$ & $809^{+82310}_{-623}$ & $4.10^{+0.25}_{-0.42}$ & $77.53^{+0.97}_{-1.29}$ & $195.72^{+0.84}_{-1.10}$ & $2.814$ & $5.193$ & $1.511$ & $0.00082^{+0.01907}_{-0.00061}$ & $0.0033^{+0.7390}_{-0.0032}$ & $796840^{+24443}_{-776633}$ \\[2pt]
HOPS-43 & $2.17^{+0.61}_{-0.32}$ & $0.50^{+0.62}_{-0.26}$ & $0.000151^{+0.000187}_{-0.000078}$ & $88^{+816}_{-33}$ & $0.98^{+0.25}_{-0.38}$ & $1.61^{+0.29}_{-0.42}$ & $0.108^{+0.188}_{-0.052}$ & $0.578^{+0.374}_{-0.064}$ & $  0.683$ & $  2.581$ & $2420^{+7648}_{-1626}$ & $1545^{+591}_{-672}$ & $0.46^{+0.18}_{-0.20}$ & $28211^{+3222}_{-12816}$ & $1.19^{+0.30}_{-0.17}$ & $0.050^{+0.916}_{-0.040}$ & $49^{+10918}_{-48}$ & $4.24^{+0.23}_{-1.64}$ & $73.3^{+5.5}_{-8.8}$ & $173.3^{+5.9}_{-172.6}$ & $2.029$ & $3.965$ & $1.652$ & $0.0056^{+0.0185}_{-0.0022}$ & $0.84^{+0.14}_{-0.66}$ & $22457^{+45401}_{-11008}$ \\[2pt]
\enddata
\tablecomments{Table 1 is published in its entirety in the online content and as a machine-readable table. A portion is shown here for guidance regarding its form and content.}
\end{deluxetable*}

%% file: class_comps_table.tex
\begin{deluxetable*}{c|cccc|ccc}
\tablecaption{Comparison of Class 0/I/Flat Spectrum Properties}
\tablenum{2}
\tabletypesize{\normalsize}
\label{table:class_comps}
\tablehead{ \colhead{} & \multicolumn{4}{c}{Medians} & \multicolumn{3}{c}{Two Sample Tests\tablenotemark{a}} \\\colhead{Parameters} & \colhead{All} & \colhead{Class 0} & \colhead{Class I} & \colhead{Flat} & \colhead{0 vs. I} & \colhead{0 vs. Flat} & \colhead{I vs. Flat}\\ \colhead{ } & \colhead{(N = 97)} & \colhead{(N = 25)} & \colhead{(N = 44)} & \colhead{(N = 28)} & \colhead{ } & \colhead{ } & \colhead{ }}
\startdata
$L_*$ (L$_{\odot}$) & $1.90^{+0.27}_{-0.30}$ & $2.90^{+0.99}_{-0.46}$ & $1.84^{+0.11}_{-0.71}$ & $1.59^{+0.55}_{-0.66}$ & 6.4 & 0.0 & 0.0 \\[2pt]
$M_{disk,dust}$ (M$_{\oplus}$) & $  5.8^{+  4.6}_{-  2.7}$ & $  7.1^{+ 14.3}_{-  2.0}$ & $  4.9^{+  1.0}_{-  2.7}$ & $ 14.0^{+  1.3}_{-  7.0}$ & 69.0 & 0.0 & 0.0 \\[2pt]
$R_{disk,dust}$ (au) & $ 29.4^{+  4.1}_{-  3.2}$ & $ 35.6^{+ 17.1}_{- 10.0}$ & $ 26.9^{+  4.5}_{-  3.3}$ & $ 29.5^{+  6.1}_{-  4.4}$ & 0.0 & 0.0 & 0.0 \\[2pt]
$R_{in,dust}$ (au) & $  0.8^{+  0.2}_{-  0.1}$ & $  1.7^{+  0.7}_{-  0.3}$ & $  1.0^{+  0.5}_{-  0.3}$ & $  0.2^{+  0.1}_{-  0.0}$ & 18.5 & \bf{100.0} & \bf{100.0} \\[2pt]
$\gamma$ & $ 0.4^{+ 0.1}_{- 0.1}$ & $ 0.4^{+ 0.1}_{- 0.0}$ & $ 0.4^{+ 0.2}_{- 0.2}$ & $ 0.5^{+ 0.1}_{- 0.2}$ & 0.0 & 0.5 & 0.6 \\[2pt]
$h_0$ (au) & $0.11^{+0.01}_{-0.01}$ & $0.09^{+0.03}_{-0.02}$ & $0.10^{+0.02}_{-0.01}$ & $0.12^{+0.02}_{-0.01}$ & 2.5 & 7.5 & 1.9 \\[2pt]
$\psi$ & $0.93^{+0.05}_{-0.01}$ & $0.98^{+0.06}_{-0.05}$ & $0.99^{+0.06}_{-0.08}$ & $0.89^{+0.02}_{-0.06}$ & 1.4 & \bf{94.4} & \bf{93.5} \\[2pt]
$M_{env,dust}$ (M$_{\oplus}$) & $ 75.9^{+ 13.7}_{- 51.3}$ & $232.3^{+ 21.3}_{-102.5}$ & $ 24.5^{+ 33.6}_{- 13.8}$ & $ 13.7^{+ 25.7}_{-  8.0}$ & 15.5 & \bf{ 98.2} & 6.2 \\[2pt]
$R_{env,dust}$ (au) & $3429.4^{+ 602.3}_{-1116.5}$ & $2130.4^{+1492.6}_{- 294.3}$ & $3513.6^{+ 720.0}_{-1251.6}$ & $4031.6^{+1733.5}_{-2015.0}$ & 4.0 & 6.4 & 0.2 \\[2pt]
$\xi$ & $1.056^{+0.027}_{-0.031}$ & $1.077^{+0.037}_{-0.035}$ & $1.056^{+0.082}_{-0.041}$ & $1.020^{+0.051}_{-0.130}$ & 2.4 & 1.6 & 12.1 \\[2pt]
$f_{cav}$ & $0.48^{+0.04}_{-0.06}$ & $0.48^{+0.03}_{-0.11}$ & $0.51^{+0.06}_{-0.11}$ & $0.41^{+0.13}_{-0.01}$ & 2.9 & 3.7 & 1.3 \\[2pt]
$a_{max}$ ($\mu$m) & $  101^{+   34}_{-   17}$ & $   91^{+   45}_{-   12}$ & $   82^{+   50}_{-   24}$ & $  149^{+   60}_{-   71}$ & 0.3 & 4.4 & 2.4 \\[2pt]
$p$ & $3.55^{+0.07}_{-0.05}$ & $3.54^{+0.14}_{-0.10}$ & $3.53^{+0.14}_{-0.07}$ & $3.57^{+0.25}_{-0.15}$ & 0.8 & 1.0 & 2.0 \\[2pt]
$i$ ($^{\circ}$) & $57.2^{+ 3.2}_{- 2.0}$ & $57.2^{+ 7.8}_{- 2.9}$ & $69.4^{+ 2.6}_{- 4.3}$ & $29.9^{+ 3.1}_{- 4.9}$ & 0.1 & \bf{ 99.7} & \bf{100.0} \\[2pt]
p.a. ($^{\circ}$) & $ 96.6^{+  8.6}_{-  7.0}$ & $ 91.3^{+ 32.6}_{-  5.2}$ & $ 98.1^{+ 13.3}_{-  6.7}$ & $ 86.9^{+ 22.1}_{- 11.5}$ & 0.0 & 0.0 & 0.1 \\[2pt]
\hline
\multicolumn{8}{{c}}{{Derived Quantities}}\\[2pt]
\hline
$F_{\nu,345GHz}$ (mJy) & $26.353^{+11.239}_{-5.069}$ & $59.913^{+28.087}_{-30.849}$ & $19.194^{+4.236}_{-4.263}$ & $39.255^{+5.903}_{-8.303}$ & \bf{100.0} & 0.0 & 0.0 \\[2pt]
$\kappa_{\nu,345GHz}$ (cm$^2$ g$^{-1}$) & $ 4.2^{+ 0.2}_{- 0.1}$ & $ 4.0^{+ 0.3}_{- 0.0}$ & $ 4.1^{+ 0.8}_{- 0.1}$ & $ 4.4^{+ 0.3}_{- 0.3}$ & 0.0 & 0.0 & 0.0 \\[2pt]
$\beta$ & $1.6^{+0.0}_{-0.0}$ & $1.6^{+0.0}_{-0.0}$ & $1.6^{+0.0}_{-0.0}$ & $1.6^{+0.0}_{-0.0}$ & 0.0 & 0.0 & 0.0
\enddata
\tablenotetext{a}{Note: These columns report the percentage of 1000 randomly sampled realizations of the distribution of each parameter that have $p < 0.05$ from two sample tests comparing different protostar classes. Each realization is generated by randomly sampling a parameter value from the posterior for each source, for all sources in the sample. We bold parameter/Class combinations for which $>90\%$ of all realizations have $p < 0.05$, suggesting that the distributions are distinct.}
\end{deluxetable*}